\documentclass[11pt]{cernrepdbg}
\usepackage{graphicx}
\usepackage{here}
\usepackage{cite}
\usepackage{amsmath}
\usepackage{amssymb}
\usepackage{epsfig}
\usepackage{axodraw}
\usepackage{subfigure}

\begin{document}
\bibliographystyle{lesHouches}

  \title{\centering{THE QCD/SM WORKING GROUP:\\ \textbf{Summary Report}}}

  \author{\underline{Convenors}: M.~Dobbs$^1$, S.~Frixione$^2$, E.~Laenen$^3$, A.~De Roeck$^4$,
      K.~Tollefson$^5$
      \\
      \underline{Contributing authors}:
      J.~Andersen$^{6,7}$, C.~Bal\'azs$^8$, A.~Banfi$^3$, 
      S.~Berge$^9$,
      W.~Bernreuther$^{10}$, T.~Binoth$^{11}$,
      A.~Brandenburg$^{12}$, C.~Buttar$^{13}$, Q-H.~Cao$^5$, 
      G.~Corcella$^{4,14}$,
      A.~Cruz$^{15}$, I.~Dawson$^{13}$,
      V.~Del~Duca$^{16}$, A.~De~Roeck$^4$, V.~Drollinger$^{17,18}$,
L.~Dudko$^{19}$, T.~Eynck$^3$,
      R.~Field$^{15}$, S.~Frixione$^2$, M.~Grazzini$^4$,
J.P.~Guillet$^{20}$, G.~Heinrich$^{21}$,
      J.~Huston$^5$, N.~Kauer$^{10}$, N.~Kidonakis$^6$,
A.~Kulesza$^{22}$, E.~Laenen$^3$, K.~Lassila-Perini$^{23}$,
      L.~Magnea$^{16,24}$, F.~Mahmoudi$^{20}$, E.~Maina$^{16,24}$,
F.~Maltoni$^{25}$, M.~Nolten$^{26}$,
      A.~Moraes$^{13}$, S.~Moretti$^{26}$, S.~Mrenna$^{27}$,
 P.~Nadolsky$^9$,
      Z.~Nagy$^{28}$, F.~Olness$^9$, I.~Puljak$^{29}$, D.A.~Ross$^{26}$,
      A.~Sabio-Vera$^6$, G.P.~Salam$^{30}$, A.~Sherstnev$^{19}$, Z.G.~Si$^{31}$,
      T.~Sj\"{o}strand$^{32}$, P.~Skands$^{32}$, E.~Thom\'{e}$^{32}$, Z.~Tr\'ocs\'anyi$^{33}$,
      P.~Uwer$^4$, S.~Weinzierl$^{14}$, C.P.~Yuan$^5$,
      G.~Zanderighi$^{27}$\\\mbox{} }
\institute{\centering{\small
$^1$Lawrence Berkeley National Lab, Berkeley, CA 94720, USA. \\
$^2$INFN, Sezione di Genova, Via Dodecaneso 33, 16146 Genova, Italy\\
$^3$NIKHEF Theory Group, Kruislaan 409, 1098 SJ Amsterdam, The Netherlands\\
$^4$CERN, CH--1211 Geneva 23, Switzerland\\
$^5$Department of Physics and Astronomy, Michigan State University,
      East Lansing, MI 48824-1116, USA\\
$^6$Cavendish Laboratory, University of Cambridge, Madingley Road,
      Cambridge CB3 0HE, UK\\
$^7$DAMTP, Centre for Mathematical Science, Wilberforce Road, CB3 0WA,
 Cambridge, UK\\
$^8$HEP Division, Argonne National Laboratory, 9700 Cass Ave., Argonne IL
      60439,  USA\\
$^9$ Southern Methodist University, Department of Physics, Dallas, TX
      75275-0175, USA\\
$^{10}$Institut f\"ur Theoretische Physik, RWTH Aachen, 52056 Aachen,
      Germany\\
$^{11}$Institut f\"ur Theoretische Physik und Astrophysik, Universit\"at
W\"urzburg, D-97074 W\"urzburg, Germany\\
$^{12}$DESY-Theorie, 22603 Hamburg, Germany\\
$^{13}$ Department of Physics and Astronomy, University of Sheffield, UK\\
 $^{14}$Max Planck Institute f\"ur Physik, 80805 M\"unchen, Germany\\
$^{15}$ Department of Physics, University of Florida, Gainesville,
      Florida, 32611, USA\\
$^{16}$INFN, Sezione di Torino, Via P. Giuria 1, I-10125 Torino, Italy \\
$^{17}$New Mexico Center for Particle Physics, University of New Mexico, USA\\
$^{18}$ Dipartimento di Fisica "Galileo Galilei", Universit\a`a di Padova, Italy\\
$^{19}$Moscow State University, Moscow, Russia\\
$^{20}$LAPTH, F-79941 Annecy-le-Vieux, France\\
$^{21}$II Institut f\"ur Theoretische Physik,Universit\"at Hamburg, Luruper
Chaussee 149, D-22761 Hamburg, Germany\\
$^{22}$Institut fur Theoretische Teilchenphysik, Universit\"at Karlsruhe,
      Germany\\
$^{23}$Helsinki Institute of Physics, P.O. Box 64, Helsinki, Finland\\
$^{24}$Dipartimento di Fisica Teorica, Universit{\`a} di Torino, Via P. Giuria 1, I-10125 Torino, Italy\\
$^{25}$Centro Studi e Ricerche ``Enrico Fermi'', via Panisperna, 89/A - 00184 Rome, Italy\\
$^{26}$School of Physics and Astronomy, University of Southampton,
      Highfield, Southampton SO17 1BJ, UK\\
$^{27}$Fermi National Accelerator Laboratory, Batavia, IL 60510-500, USA\\
$^{28}$Institute of Theoretical Science, 5203 University of Oregon, Eugene,
      OR  97403-5203, USA\\
$^{29}$FESB, University of Split, Split, Croatia\\
$^{30}$LPTHE, Universities of Paris VI and VII and CNRS UMR 7589, 
  Paris, France\\
$^{31}$Department of Physics, Shandong University, Jinan, Shandong 250100,
      China\\
$^{32}$Department of Theoretical Physics, Lund University, S-223 62 Lund, Sweden\\
$^{33}$University of Debrecen and Institute of Nuclear Research of the Hungarian Academy of Sciences,
 H-4001 Debrecen, PO Box 51, Hungary\\
}}

 \maketitle

 \begin{center}
   \textit{Report of the Working Group on Quantum Chromodynamics and
     the Standard Model for the Workshop ``Physics at TeV Colliders'',
     Les Houches, France, 26 May - 6 June, 2003.  }
\end{center}
\newpage

\setcounter{tocdepth}{1}
\tableofcontents
\setcounter{footnote}{0}

\section[Foreword]{FOREWORD~\protect\footnote{M. Dobbs' work was supported in part by the Director, Office 
of Science, Office of Basic Energy Sciences, of the U.S. Department of 
Energy under  Contract No. DE-AC03-76SF00098.}}
Among the many physics processes at TeV hadron colliders, we look most
eagerly for those that display signs of the Higgs boson or of new
physics.  We do so however amid an abundance of processes that proceed
via Standard Model (SM) and in particular Quantum Chromodynamics (QCD)
interactions, and that are interesting in their own right.  Good
knowledge of these processes is required to help us distinguish the
new from the known. Their theoretical and experimental study teaches
us at the same time more about QCD/SM dynamics, and thereby
enables us to further improve such distinctions. This is important
because it is becoming increasingly clear that the success of finding
and exploring Higgs boson physics or other New Physics at the
Tevatron and LHC will depend significantly on precise understanding
of QCD/SM effects for many observables.

To improve predictions and deepen the study of QCD/SM signals and backgrounds
was therefore the ambition for our QCD/SM working
group at this Les Houches workshop. Members of the working group made 
significant progress towards this on a number of fronts.
A variety of tools were further developed,
from methods to perform higher order perturbative
calculations or various types of resummation, to improvements
in the modelling of underlying events and parton showers. 
Furthermore, various precise studies of important specific processes
were conducted.

A signficant part of the activities in Les Houches revolved around
Monte Carlo simulation of collision events. A number of contributions
in this report reflect the progress made in this area. At present
a large number of Monte Carlo programs exist, each written 
with a different purpose and employing different techniques. 
Discussions in Les Houches revealed the need for an accessible primer on
Monte Carlo programs, featuring a listing of various codes, each
with a short description,  but also providing a low-level explanation
of the underlying methods. This primer has now been compiled 
and a synopsis of it is included here as the first contribution to
this report (see below for where to obtain
the full document). 

This report reflects the hard and creative work by the
many contributors which took place in the working group.
After the MC guide description, the 
next contributions report on progress in describing multiple
interactions, important for the LHC, and underlying events. An
announcement of a Monte Carlo database, under construction, is
followed by a number of contributions improving parton shower
descriptions. Subsequently, a large number of contributions address
resummations in various forms, after which follow studies of QCD
effects in pion pair, top quark pair and photon pair plus jet production. 
After a study of electroweak corrections to hadronic precision
observables, the report ends by presenting recent progress in
methods to compute finite order corrections at one-loop with many
legs, and at two-loop.

\label{introduction}
%

\section[Les Houches Guidebook to Monte Carlo Generators for Hadron 
       Collider Physics]{LES HOUCHES GUIDEBOOK TO MONTE CARLO GENERATORS FOR HADRON 
       COLLIDER PHYSICS}

{ \em
\underline{Editors:} M.~Dobbs, S.~Frixione, E.~Laenen, K.~Tollefson
\\ \underline{Contributing Authors:}
H.~Baer,
E.~Boos,
B.~Cox,
M.~Dobbs,
R.~Engel,
S.~Frixione,
W.~Giele,
J.~Huston,
S.~Ilyin,
B.~Kersevan,
F.~Krauss,
Y.~Kurihara,
E.~Laenen,
L.~L\"onnblad,
F.~Maltoni,
M.~Mangano,
S.~Odaka,
P.~Richardson,
A.~Ryd,
T.~Sj\"ostrand,
P.~Skands,
Z.~Was,
B.R.~Webber,
D.~Zeppenfeld
}
\vspace{0.2cm}
\begin{abstract}
Recently the collider physics community has seen significant advances
in the formalisms and implementations of event generators. 
This review is a primer of the methods commonly used
for the simulation of high energy physics events at particle
colliders. We provide brief descriptions, references, and links to the
specific computer codes which implement the methods.
The aim is to provide an overview of the available tools, allowing the
reader to ascertain which tool is best for a particular application,
but also making clear the limitations of each tool.

Due to its long length and stand-alone nature, the Monte Carlo
Guidebook entry in the Les Houches proceedings has been published as a
separate document (\texttt{hep-ph/0403045}). The table of contents follows.
\end{abstract}

\begin{minipage}[c]{0.9\textwidth}
\contentsline {section}{\numberline {1.}Introduction}{}
\contentsline {section}{\numberline {2.}The Simulation of Hard Processes}{}
\contentsline {section}{\numberline {3.}Tree Level Matrix Element Generators}{}
\contentsline {subsection}{\numberline {3.1}Matrix Element Generators for Specific Processes}{}
\contentsline {subsection}{\numberline {3.2}Matrix Element Generators for Arbitrary Processes}{}
\contentsline {section}{\numberline {4.}Higher Order Corrections -- Perturbative QCD Computations}{}
\contentsline {section}{\numberline {5.}Parton Distribution Functions}{}
\contentsline {section}{\numberline {6.}Higher Order Corrections -- Showering and Hadronization Event Generators}{}
\contentsline {subsection}{\numberline {6.1}General Purpose Showering and Hadronization Event Generators}{}
\contentsline {subsection}{\numberline {6.2}Specialised Initial and Final State Radiation Programs}{}
\contentsline {subsection}{\numberline {6.3}Programs for Diffractive Collisions}{}
\contentsline {subsection}{\numberline {6.4}Specialised Decay Programs}{}
\contentsline {section}{\numberline {7.}Resummation}{}
\contentsline {section}{\numberline {8.}Combining Matrix Elements with Showering}{}
\contentsline {subsection}{\numberline {8.1}Programs using NLO Matrix Elements with Showering}{}
\contentsline {section}{\numberline {9.}Conclusions}{}
\contentsline {section}{\numberline {10.}Acknowledgments}{}
\end{minipage}

\section[Multiple Interactions and Beam 
Remnants]{MULTIPLE INTERACTIONS AND BEAM 
REMNANTS~\protect\footnote{Contributed by: {T.~Sj\"ostrand and P.~Skands }}}




\subsection{Introduction} 

Hadrons are composite systems of quarks and gluons. A direct 
consequence is the possibility to have hadron--hadron collisions 
in which several distinct pairs of partons collide with each other, i.e. 
multiple interactions, a.k.a.\ multiple scatterings. 
At first glance, the divergence of the perturbative $t$-channel 
one-gluon-exchange graphs in the $p_{\perp} \to 0$ limit implies an
infinity of interactions per event. However, the perturbative framework 
does not take into account screening from the fact that a hadron is in 
an overall colour singlet state. Therefore an effective cutoff 
$p_{\perp\mathrm{min}}$ of the order of one to a few GeV is introduced, 
representing an inverse colour correlation distance inside the hadron.
For realistic $p_{\perp\mathrm{min}}$ values most inelastic events in 
high-energy hadronic collisions should then contain several 
perturbatively calculable interactions, in addition to whatever 
nonperturbative phenomena may be present.   

Although most of this activity is not hard enough to play a significant 
role in the description of high--$p_{\perp}$ jet physics, it can be
responsible for a large fraction of the total multiplicity (and large
\emph{fluctuations} in it) for semi-hard (mini-)jets in the event, for 
the details of jet profiles and for the jet pedestal effect, leading to 
random as well as systematic shifts in the jet energy scale. Thus, a
good understanding of multiple interactions would seem prerequisite 
to carrying out precision studies involving jets and/or the underlying 
event in hadronic collisions. 

In an earlier study \cite{Sjostrand:1987su}, it was argued that \emph{all} 
the underlying event activity is triggered by the multiple interactions 
mechanism. However, while the origin of underlying events is thus 
assumed to be perturbative, many nonperturbative aspects still need to be
considered and understood:\\
\textit{(i)} What is the detailed mechanism and functional form of the 
dampening of the perturbative cross section at small $p_{\perp}$?
(Certainly a smooth dampening is more realistic than a sharp 
$p_{\perp\mathrm{min}}$ cutoff.)\\ 
\textit{(ii)} Which energy dependence would this mechanism have?\\
\textit{(iii)} How is the internal structure of the proton reflected in an
impact-parameter-dependent multiple interactions rate, as manifested
e.g.\ in jet pedestal effects?\\
\textit{(iv)} How can the set of colliding partons from a hadron be
described in terms of correlated multiparton distribution functions
of flavours and longitudinal momenta?\\
\textit{(v)} How does a set of initial partons at some low perturbative 
cutoff scale, `initiators', evolve into such a set of colliding partons?
(Two colliding partons could well have a common initiator.)
Is standard DGLAP evolution sufficient, or must BFKL/CCFM effects be 
taken into account?\\ 
\textit{(vi)} How would the set of initiators correlate with the flavour 
content of, and the longitudinal momentum sharing inside, the left-behind 
beam remnant?\\
\textit{(vii)} How are the initiator and remnant partons correlated by 
confinement effects (`primordial $k_{\perp}$')?\\
\textit{(viii)} How are all produced partons, both the 
interacting and the beam-remnant ones, 
correlated in colour? Is the large number-of-colours limit 
relevant, wherein partons can be hooked up into strings (with quarks as 
endpoints and gluons as intermediate kinks) representing a linear 
confinement force \cite{Andersson:1983ia}?\\
\textit{(ix)} How is the original baryon number of an incoming proton 
reflected in the colour topology?\\ 
\textit{(x)} To what extent would a framework with independently 
fragmenting  string systems, as defined from the colour topology, be 
modified by the space--time overlap of several strings?

Needless to say, we should not expect to find a perfect solution to any
of these issues, but only successively improved approximations. The 
framework in \cite{Sjostrand:1987su} is very primitive in a number of 
respects. Nevertheless, it has turned out to be quite successful. Thus 
the \textsc{Pythia} Tune A of R.D.~Field \cite{TuneAField} is capable 
of describing a host of jet and minimum-bias event data at the Tevatron. 
The model appears inadequate to fully describe correlations and 
fluctuations, however, and we would expect a poor performance for several 
topics not yet studied experimentally.

In particular, only very simple beam remnant structures could 
technically be dealt with in \cite{Sjostrand:1987su}. One recent 
development was the extension of the standard Lund string framework
\cite{Andersson:1983ia} to include a junction fragmentation description 
\cite{Sjostrand:2002ip} that allows the hadronization of nontrivial 
colour topologies containing non-zero baryon number. In the context of 
multiple interactions, this improvement means that almost arbitrarily 
complicated baryon beam remnants may now be dealt with, hence many of 
the restrictions present in the old model are no longer necessary.
 
Here, we report on the development of a new model for the flavour-, 
colour-, and momentum-correlated partonic structure involved in a 
hadron--hadron collision, i.e.\ partly addressing several of the points 
above. We first present the main work on flavour and momentum space 
correlations, and thereafter separately the very thorny issue of colour 
correlations, before concluding. A more complete description of the model, 
also including references to experimental data and other theoretical 
ideas, and with comments on all the issues, may be found in 
\cite{Sjostrand:2004pf}. A toy model study of the first two points 
is found in \cite{Dischler:2000pk}. The \textsc{Pythia} manual
\cite{Sjostrand:2003wg} contains some complementary information.
                   
\subsection{Correlated Parton Densities}

Consider a hadron undergoing multiple interactions in a collision. 
Such an object should be described by multi-parton densities, 
giving the joint probability of simultaneously finding $n$ partons with 
flavours $f_1,\ldots,f_n$, carrying momentum fractions $x_1,\ldots,x_n$ 
inside the hadron, when probed by interactions at scales 
$Q_1^2,\ldots,Q_n^2$. However, we are nowhere near having sufficient 
experimental information to pin down such distributions. Therefore, and
wishing to make maximal use of the information that we \emph{do} have, namely
the standard one-parton-inclusive parton densities, we propose the following
strategy. 

As described in \cite{Sjostrand:1987su}, the interactions may be generated in
an ordered sequence of falling $p_{\perp}$. For the hardest interaction, all
smaller $p_{\perp}$ scales may be effectively integrated out of the (unknown) 
fully correlated  distributions, leaving an object described by the standard
one-parton distributions, by definition. For the second and subsequent
interactions, again all lower--$p_{\perp}$ scales can be integrated out, but 
the correlations with the first cannot, and so on. 
Thus, we introduce modified parton densities, that correlate the $i$'th
interaction and its shower evolution to what happened in the $i-1$ previous
ones. 

The first and most trivial observation is that each interaction $i$
removes a momentum fraction $x_i$ from the hadron remnant. Already in
\cite{Sjostrand:1987su} this momentum loss was taken into account by assuming
a simple scaling ansatz for the parton distributions, $f(x) \to f(x/X)/X$,
where $X = 1 - \sum_{i=1}^n x_i$ is the momentum remaining in the beam hadron
after the $n$ first interactions. Effectively, the PDF's are simply `squeezed' 
into the range $x\in[0,X]$. 

Next, for a given baryon, the valence distribution of flavour $f$ after $n$ 
interactions, $q_{f\mathrm{v} n}(x,Q^2)$, should integrate to the number
$N_{f\mathrm{v} n}$ of valence quarks of flavour $f$ remaining in the hadron 
remnant. This rule may be enforced by scaling the original distribution down, 
by the ratio of remaining to original valence quarks 
$N_{f\mathrm{v} n}/N_{f\mathrm{v} 0}$, in addition to the $x$ scaling 
mentioned above.

Also, when a sea quark is knocked out of a hadron, it must leave behind a
corresponding antisea parton in the beam remnant. We call this a companion 
quark. In the perturbative approximation the sea quark 
$\mathrm{q}_{\mathrm{s}}$ and its companion $\mathrm{q}_{\mathrm{c}}$ 
come from a gluon branching
$\mathrm{g} \to \mathrm{q}_{\mathrm{s}} + \mathrm{q}_{\mathrm{c}}$ 
(it is implicit that if $\mathrm{q}_{\mathrm{s}}$ is a quark, 
$\mathrm{q}_{\mathrm{c}}$ is its antiquark). Starting from this 
perturbative ansatz, and neglecting other interactions and 
any subsequent perturbative evolution of the $\mathrm{q}_{\mathrm{c}}$, 
we obtain the $q_{\mathrm{c}}$ distribution from
the probability that a sea quark $\mathrm{q}_{\mathrm{s}}$, carrying a
momentum fraction $x_{\mathrm{s}}$, is produced by the branching of a
gluon with momentum fraction $y$, so that the 
companion has a momentum fraction $x=y-x_{\mathrm{s}}$, 
\begin{equation}
q_{\mathrm{c}}(x;x_{\mathrm{s}}) \propto \int_0^1 g(y) \, 
P_{\mathrm{g}\to\mathrm{q}_{\mathrm{s}}\mathrm{q}_{\mathrm{c}}}(z) \, 
\delta(x_{\mathrm{s}}-zy)~\mathrm{d} z =
\frac{g(x_{\mathrm{s}}+x)}{x_{\mathrm{s}}+x} \, 
P_{\mathrm{g}\to\mathrm{q}_{\mathrm{s}}\mathrm{q}_{\mathrm{c}}}
\left(\frac{x_{\mathrm{s}}}{x_{\mathrm{s}}+x}\right), 
\end{equation}
with $P_{\mathrm{g}\to\mathrm{q}_{\mathrm{s}}\mathrm{q}_{\mathrm{c}}}$ 
the usual DGLAP gluon splitting kernel. A simple ansatz 
$g(x) \propto (1-x)^n/x$ is here used for the gluon. Normalizations
are fixed so that a sea quark has exactly one companion.
Qualitatively, $xq_{\mathrm{c}}(x;x_s)$ is peaked around 
$x \approx x_{\mathrm{s}}$, by virtue of the symmetric 
$P_{\mathrm{g}\to\mathrm{q}_{\mathrm{s}}\mathrm{q}_{\mathrm{c}}}$ 
splitting kernel.

Without any further change, the reduction of the valence
distributions and the introduction of companion distributions, in the 
manner described above, would result in a violation of the total 
momentum sum rule, that the $x$-weighted parton densities should 
integrate to $X$: by removing a valence quark from the parton 
distributions we also remove a total amount of momentum corresponding 
to $\langle x_{f\mathrm{v}} \rangle$, the average momentum fraction 
carried by a valence quark of flavour $f$,
\begin{equation}
\langle x_{f\mathrm{v} n} \rangle \equiv 
\frac{\int_0^X xq_{f\mathrm{v} n}(x,Q^2)~\mathrm{d} x}%
{\int_0^X q_{f\mathrm{v} n}(x,Q^2)~\mathrm{d} x} = X \, 
\langle x_{f\mathrm{v} 0} \rangle ~,
\end{equation}
and by adding a companion distribution we add an analogously defined
momentum fraction. 

To ensure that the momentum sum rule is still respected, we assume that 
the sea+gluon normalizations fluctuate up when a valence distribution is 
reduced and down when a companion distribution is added, by a 
multiplicative factor
\begin{equation}
a = \frac{1-\sum_fN_{f\mathrm{v} n}\langle x_{f\mathrm{v} 0} \rangle
-\sum_{f,j} \langle x_{f\mathrm{c}_j 0} \rangle}{1- \sum_fN_{f\mathrm{v} 0}
\langle x_{f\mathrm{v} 0} \rangle} ~. 
\end{equation}
The requirement of a physical $x$ range is of course still maintained by 
`squeezing' all distributions into the interval $x\in[0,X]$. The full parton 
distributions after $n$ interactions thus take the forms
\begin{eqnarray}
q_{f n}\left(x, Q^2\right) & = &  \frac{1}{X} \left[ 
\frac{N_{f\mathrm{v} n}}{N_{f\mathrm{v} 0}}
{q_{f\mathrm{v} 0}\left(\frac{x}{X}, Q^2\right)} + 
a\, q_{f \mathrm{s} 0} \left(\frac{x}{X}, Q^2\right) + 
\sum_{j} q_{f\mathrm{c}_j} \left(\frac{x}{X};x_{s_j}\right) \right]  ~,
\\
\displaystyle {g_n(x)} &=&\frac{a}{X} g_0\left(\frac{x}{X}, Q^2\right) ~,
\end{eqnarray}
where $q_{f\mathrm{v} 0}$ ($q_{f \mathrm{s}0}$) denotes the original 
valence (sea) distribution of flavour $f$, and the index $j$ on the 
companion distributions $q_{f\mathrm{c}_j}$ counts different companion 
quarks of the same flavour $f$. 

After the perturbative interactions have each taken their fraction of
longitudinal momentum, the remaining momentum is to be shared between 
the beam remnant partons. Here, valence quarks receive an $x$ picked at 
random according to a small-$Q^2$ valence-like parton density, while sea 
quarks must be companions of one of the initiator quarks, and hence should 
have an $x$ picked according to the $q_{\mathrm{c}}(x ; x_{\mathrm{s}})$ 
distribution introduced above. In the rare case that no valence quarks 
remain and no sea quarks need be added for flavour conservation, the beam 
remnant is represented by a gluon, carrying all of the beam remnant 
longitudinal momentum. 

Further aspects of the model include the possible formation of composite
objects in the beam remnants (e.g.\ diquarks) and the addition
of non-zero primordial $k_{\perp}$ values to the parton shower
initiators. Especially the latter introduces some complications, to
obtain consistent kinematics. Details on these aspects 
are presented in \cite{Sjostrand:2004pf}. 

\subsection{Colour Correlations}

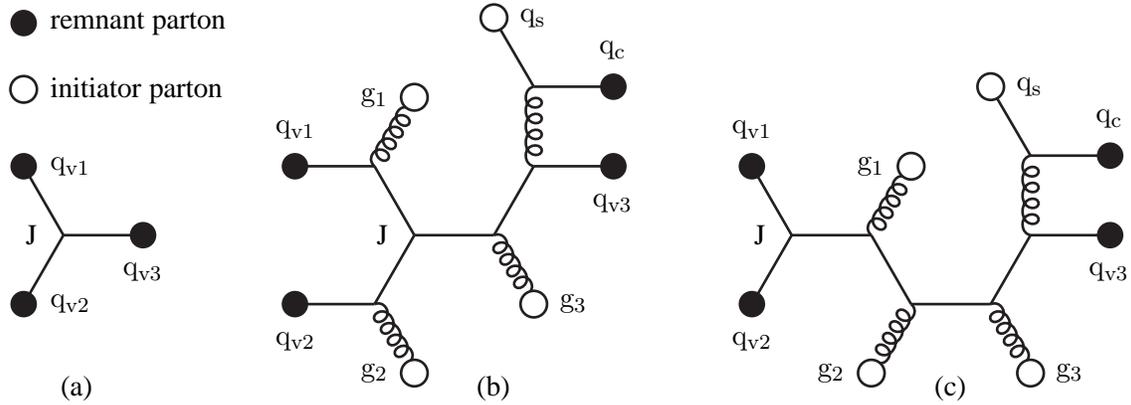
\begin{figure}[t]
\begin{center}
\begin{picture}(120,150)(-30,-60)
\SetWidth{1}
\Text(-10,0)[r]{J}
\Vertex(-15,80){5}\Text(-5,80)[l]{remnant parton}
\BCirc(-15,55){5}\Text(-5,55)[l]{initiator parton}
\Line(0,0)(30,0)\Vertex(30,0){5}
\Text(30,-14)[]{$\mathrm{q}_{\mathrm{v}3}$}
\Line(0,0)(-15,26)\Vertex(-15,26){5}
\Text(-5,26)[l]{$\mathrm{q}_{\mathrm{v}1}$}
\Line(0,0)(-15,-26)\Vertex(-15,-26){5}
\Text(-5,-26)[l]{$\mathrm{q}_{\mathrm{v}2}$}
\Text(5,-58)[]{(a)}
\end{picture}
\begin{picture}(150,150)(-40,-60)
\SetWidth{1}
\Text(-10,0)[r]{J}
\Line(0,0)(-15,26)
\Line(-15,26)(-45,26)\Vertex(-45,26){5}
\Text(-45,40)[]{$\mathrm{q}_{\mathrm{v}1}$}
\Gluon(-15,26)(-2,48){3}{4}\BCirc(0,52){5}
\Text(-10,52)[r]{$\mathrm{g}_1$}
\Line(0,0)(-15,-26)
\Line(-15,-26)(-45,-26)\Vertex(-45,-26){5}
\Text(-45,-40)[]{$\mathrm{q}_{\mathrm{v}2}$}
\Gluon(-15,-26)(-2,-48){3}{4}\BCirc(0,-52){5}
\Text(-10,-52)[r]{$\mathrm{g}_2$}
\Line(0,0)(30,0)
\Gluon(30,0)(43,-22){3}{4}\BCirc(45,-26){5}
\Text(55,-26)[l]{$\mathrm{g}_3$}
\Line(30,0)(45,26)
\Line(45,26)(75,26)\Vertex(75,26){5}
\Text(75,12)[]{$\mathrm{q}_{\mathrm{v}3}$}
\Gluon(45,26)(45,56){3}{4}
\Line(45,56)(75,56)\Vertex(75,56){5}
\Text(75,70)[]{$\mathrm{q}_{\mathrm{c}}$}
\Line(45,56)(31,80)\BCirc(30,82){5}
\Text(40,82)[l]{$\mathrm{q}_{\mathrm{s}}$}
\Text(30,-58)[]{(b)}
\end{picture}
\begin{picture}(160,150)(-30,-60)
\SetWidth{1}
\Text(-10,0)[r]{J}
\Line(0,0)(-15,26)\Vertex(-15,26){5}
\Text(-15,40)[]{$\mathrm{q}_{\mathrm{v}1}$}
\Line(0,0)(-15,-26)\Vertex(-15,-26){5}
\Text(-15,-40)[]{$\mathrm{q}_{\mathrm{v}2}$}
\Line(0,0)(30,0)
\Gluon(30,0)(43,22){3}{4}\BCirc(45,26){5}
\Text(35,26)[r]{$\mathrm{g}_1$}
\Line(30,0)(45,-26)
\Gluon(45,-26)(32,-48){3}{4}\BCirc(30,-52){5}
\Text(20,-52)[r]{$\mathrm{g}_2$}
\Line(45,-26)(75,-26)
\Gluon(75,-26)(88,-48){3}{4}\BCirc(90,-52){5}
\Text(100,-52)[l]{$\mathrm{g}_3$}
\Line(75,-26)(90,0)
\Line(90,0)(120,0)\Vertex(120,0){5}
\Text(120,-14)[]{$\mathrm{q}_{\mathrm{v}3}$}
\Gluon(90,0)(90,30){3}{4}
\Line(90,30)(120,30)\Vertex(120,30){5}
\Text(120,44)[]{$\mathrm{q}_{\mathrm{c}}$}
\Line(90,30)(76,54)\BCirc(75,56){5}
\Text(85,56)[l]{$\mathrm{q}_{\mathrm{s}}$}
\Text(60,-58)[]{(c)}
\end{picture}
\end{center}
\caption{(a) The initial state of a baryon, with the valence quarks
colour-connected via a central string junction J. 
(b) Example of a topology with initiators connected at random.
(c) Alternative with the junction in the remnant.}
\label{fig:initialstate}
\end{figure}

The initial state of a baryon may be represented by three valence quarks,
connected antisymmetrically in colour via a central junction, which acts 
as a switchyard for the colour flow and carries the net baryon number, 
Fig.~\ref{fig:initialstate}a. 

The colour-space evolution of this state into the initiator and remnant 
partons actually found in a given event is not predicted by perturbation 
theory, but is crucial in determining how the system hadronizes; in the 
Lund string model \cite{Andersson:1983ia}, two colour-connected final 
state partons together define a string piece, which hadronizes by 
successive non-perturbative breakups along the string. Thus, the colour 
flow of an event determines the topology of the hadronizing strings, 
and consequently where and how~many hadrons will be produced. 
The question can essentially be reduced to one of choosing a fictitious
sequence of gluon emissions off the initial valence topology, since sea 
quarks together with their companion partners are associated with parent 
gluons, by construction.

The simplest solution is to assume that gluons are attached to the initial
quark lines in a random order, see Fig.~\ref{fig:initialstate}b. 
If so, the junction would rarely be colour-connected directly to two 
valence quarks in the beam remnant, and the initial-state baryon number 
would be able to migrate to large $p_{\perp}$ and small $x_F$ values.
While such a mechanism should be present, there are reasons to believe
that a purely random attachment exaggerates the migration effects.  
Hence a free parameter is introduced to suppress gluon attachments onto 
colour lines that lie entirely within the remnant, so that topologies
such as Fig.~\ref{fig:initialstate}c become more likely. 

This still does not determine the order in which gluons are attached to 
the colour line between a valence quark and the junction. We consider a 
few different possibilities: 1) random, 2) gluons are ordered according 
to the rapidity of the hard scattering subsystem they are associated with, 
and 3) gluons are ordered so as to give rise to the smallest possible 
total string lengths in the final state. The two latter possibilities 
correspond to a tendency of nature to minimize the total potential 
energy of the system, i.e.\ the string length. Empirically such a 
tendency among the strings formed by multiple interactions is supported 
e.g.\ by the observed rapid increase of 
$\langle p_{\perp} \rangle$ with $n_{\mathrm{charged}}$. It appears, 
however, that a string minimization in the initial state is not enough,
and that also the colours inside the initial-state cascades and hard 
interactions may be nontrivially correlated. These studies are still 
ongoing, and represent the major open issues in the new model. 

\subsection{Conclusion}

A new model for the underlying event in hadron--hadron collisions 
\cite{Sjostrand:2004pf} has been introduced. This model extends the multiple 
interactions mechanism proposed in \cite{Sjostrand:1987su} with the
possibility of non-trivial flavour and momentum correlations, with
initial- and final-state showers for all interactions, and with several 
options for colour correlations between initiator and remnant partons. 
Many of these improvements rely on the development of junction 
fragmentation in \cite{Sjostrand:2002ip}. 

This is not the end of the line. Rather we see that many issues remain
to understand better, such as colour correlations between partons
in interactions and beam remnants, whereas others have not yet been 
studied seriously, such as the extent to which two interacting partons
stem from the same initiator. Theoretical advances alone cannot solve
all problems; guidance will have to come from experimental information. 
The increased interest in such studies bodes well for the future.



\label{torbjorn_multint3}
\section[Describing Minimum Bias and the Underlying Event at the
  LHC in PYTHIA and PHOJET]{DESCRIBING MINIMUM BIAS AND THE UNDERLYING EVENT AT THE
  LHC IN PYTHIA AND PHOJET~\protect\footnote{Contributed by: {A.~Moraes, C.~Buttar, and I.~Dawson}}}
\label{amoraes-LH2003}










\subsection{Introduction}

Our ability to describe parton scatterings through QCD depends on the
amount of transverse momenta with respect to the collision axis (p$_{t}$)
involved in a given scattering \cite{Collins:1982pi}.
QCD has been fairly successful in describing quark, anti-quark and
gluon scatterings involving large amounts of transverse momenta
(p$_{t} >> \Lambda_{\text{QCD}}$), also known as ``hard''
interactions.  
On the other hand, QCD simply cannot be applied to interactions with
small transverse momenta (or ``soft'' interactions) because the strong
coupling constant, $\alpha_{s}\left( \text{Q}^{2}\right) $, becomes
too large for perturbation theory to be applied and QCD models suffer
from divergent cross sections as $\text{p}_{t}\rightarrow 0$
\cite{Collins:1982pi}.  
Most high-energy hadron collisions are dominated by soft partonic
interactions.

A full picture of high-energy hadron collisions will typically combine
perturbative QCD to explain parton interactions where it is
applicable (high-p$_{t}$ scatterings), with an alternative
phenomenological approach to describe soft processes.
Examples of these are the Dual Parton Model (DPM) \cite{Capella:1994yb}
and modified versions of QCD in which the divergencies presented by
the running coupling constant are phenomenologically corrected to
reproduce experimental observations \cite{Sjostrand:1987su}. 

In this article we investigate two Monte Carlo (MC) event
generators, PYTHIA6.214 \cite{Sjostrand:2000wi,Sjostrand:2001yu} and PHOJET1.12
\cite{Engel:1995vs,PHOJET112}, focusing on 
their models for soft interactions in hadron-hadron collisions.
Aiming to check the consistency of these models, we compare their
predictions to wide range of data for minimum bias and the underlying
event. A tuning for PYTHIA6.214 is presented and examples of its
predictions are compared to those generated with PHOJET1.12 .
Predictions for levels of particle production and event
activity at the LHC for interactions dominated by soft processes such
as minimum bias interactions and the underlying event associated to
jet production are also discussed.

\subsection{PYTHIA Model for Hadron Collisions}

A comprehensive description of PYTHIA can be found at
\cite{Sjostrand:2001yu} and references therein.
The evolution of a hadronic event generated by PYTHIA is based on
parton-parton scatterings \cite{Sjostrand:2000wi,Sjostrand:2001yu}. 
In this model the total rate of parton interactions,
$N_{parton-parton}$, as a function of the transverse momentum scale
p$_{t}$, is assumed to be given by perturbative QCD.  At reasonably
large p$_{t}$ values (p$_{t} \gtrsim$ 2 GeV) parton scatterings can be
correctly described by the standard perturbative QCD, but to extend
the parton-parton scattering framework to the low-p$_{t}$ region a
regularisation to correct the divergence in the cross-section is
introduced. 

In order to deal with low-p$_{t}$ interactions, PYHTIA introduces a
cut-off parameter p$_{t_{min}}$ given by 
\begin{equation}
\text{p}_{t_{min}}(s)=(1.9~ \text{GeV}) \left( \frac{s}{1 ~ \text{TeV}^{2}}
\right)^{0.08} \label{eq:01}
\end{equation}
which can be interpreted as the inverse of some colour screening
length in the hadron \cite{Dischler:2000pk}.
There are two strategies, or scenarios, to
implement the cut-off parameter defined by equation \ref{eq:01}.

In the first one, labelled ``simple'' scenario, an
effective cut-off is established at  $\text{p}_{t_{min}}$, which means
that $d\sigma/dp_{t}^{2} = 0$ for $\text{p}_{t} <
\text{p}_{t_{min}}$. 
This model assumes that different pairwise interactions take place
essentially independent of each other, and that therefore the number
of interactions in an event is given by a Poissonian
distribution \cite{Sjostrand:1987su}. 
In the second approach, called the `complex' scenario, the probability
associated with each interacting parton depends on the assumed matter
distribution inside the colliding hadrons. In the `complex' scenario
an impact parameter dependent approach is therefore introduced
\cite{Sjostrand:1987su}. 

The parameters defining p$_{t_{min}}$ are PARP(81), PARP(82), PARP(89)
and PARP(90). The factor 1.9~GeV is defined in the simple scenario by
PARP(81) and by PARP(82) in the complex scenario. The energy scale 1
~TeV is defined by PARP(89) and is included in equation (1) to be a
convenient tuning parameter rather than a parameter with physical
meaning. PARP(90) gives the power with which p$_{t_{min}}$ varies with
the centre of mass energy, $\sqrt{s}$. The default option is set as
PARP(90)=0.16 \cite{Sjostrand:2000wi,Sjostrand:2001yu}. 

\subsection{PHOJET}

The physics model used in the MC event generator PHOJET
combines the ideas of the DPM \cite{Capella:1994yb} with perturbative QCD
\cite{Collins:1982pi} to give an almost complete picture
of high-energy hadron collisions \cite{Engel:1995vs,PHOJET112,Engel:97}.

PHOJET is formulated as a two-component model containing contributions
from both soft and hard interactions. The DPM is used describe the
dominant soft processes and perturbative QCD is applied to generate
hard interactions \cite{PHOJET112}. 

The model employed by PHOJET is based on the calculation of scattering
amplitudes, taking into account the unitarization principle.
Comparisons between the calculated results for
cross-sections and the available data are used to determine the
unknown model parameters (couplings, Pomeron intercepts and slope
parameters), which are needed to generate multiparticle
final states produced in inelastic interactions
\cite{Engel:1995vs,PHOJET112}.     

The soft, $\sigma_{soft}$, and hard, $\sigma_{hard}$, cross sections
are inclusive cross sections and the average multiplicities of soft
and hard scatterings in an inelastic event are 
\begin{eqnarray}
\langle n_{s} \rangle = \frac{\sigma_{s}}{\sigma_{inel}},~~~~~~~~
\langle n_{h} \rangle = \frac{\sigma_{h}}{\sigma_{inel}}, \label{eq:02}
\end{eqnarray}
respectively.
The hard scatterings are mostly independent of each
other, being related only by the sharing of energy and momentum of the
incoming protons.
These multiplicities increase with the colliding centre-of-mass
energy. For pp collisions at $\sqrt{s} = 14$ TeV a considerable part
of interactions is expected to have more than one hard or
soft scattering.   

\subsection{Minimum Bias Interactions \label{sec:minbias}}

Throughout this article, we will
associate minimum bias events with non-single diffractive inelastic
(NSD) interactions, following the experimental definition used in
\cite{Breakstone:1984ns,Alner:1987wb,Abe:1990td,Alexopoulos:1998bi,Matinyan:1998ja}. 
In the language of the MC event generators 
used in this work, this means that subprocesses 94 and 95 are switched
on in PYTHIA6.214 (MSUB(94)=1 and MSUB(95)=1), and processes IPRON(1,1),
IPRON(4,1) and IPRON(7,1) in PHOJET1.12. 
For both generators, we also adapt the MC distributions to the data by
setting $\pi^{0}, K_{s}$ and $\Lambda^{0}$ as stable particles. 

\subsubsection{KNO Distribution}

The KNO distribution \cite{Koba:1972ng} has been widely used as an
important tool for studying multiple particle production in inelastic
and NSD events.
The observed KNO scaling violation for p$\overline{\text{p}}$ collisions
at energies higher than those achieved at ISR
\cite{Breakstone:1984ns,Alner:1987wb} has been explained by the rising
number of multiple parton scatterings as $s \rightarrow \infty$
\cite{Alexopoulos:1998bi,Matinyan:1998ja}. KNO distributions are therefore
good tools to exploit how well hadronic models can describe the event
properties associated to multiple parton scattering. 

By default PYTHIA is set to use multiple parton
interactions. Nevertheless, one still has to define how the divergency
for scatterings with p$_{t} < \text{p}_{t_{min}}$ will be treated by
the event generator. 
PYTHIA allows two different phenomenological approaches: simple
(MSTP(82)=1) and complex scenarios (MSTP(82)=2, 3 or 4). Selecting the
complex scenario one has also the choice of selecting different matter
distributions for the colliding hadrons: uniform (MSTP(82)=2), single
Gaussian (MSTP(82)=3) and double Gaussian (MSTP(82)=4) matter
distributions.  
\begin{figure}[htbp]
\begin{center}
\begin{tabular}{cc}
\scalebox{0.335}[0.3]{\includegraphics*[0.0cm,0.0cm][20cm,22cm]{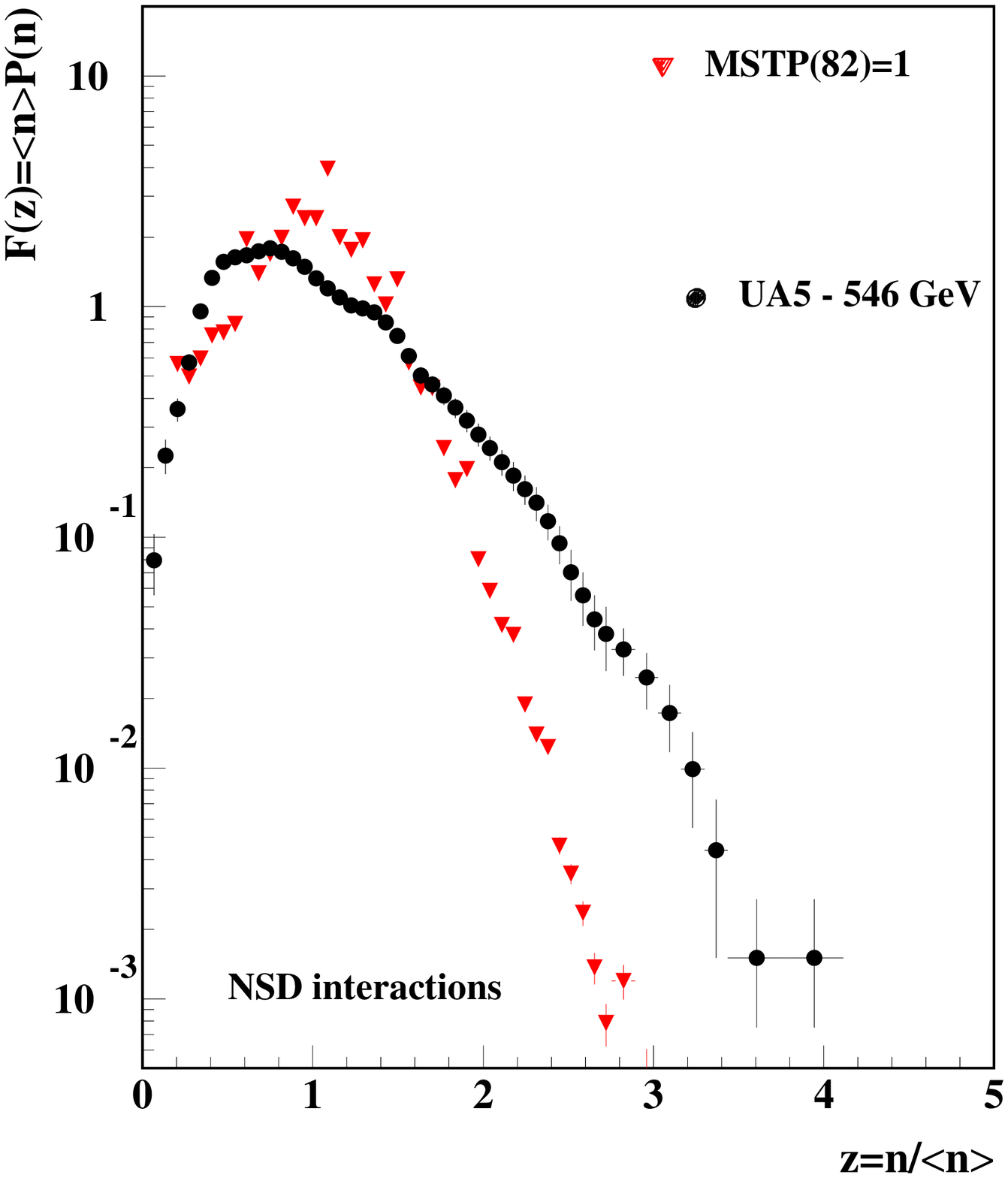}} 
&
\scalebox{0.335}[0.3]{\includegraphics*[0.0cm,0.0cm][20cm,22cm]{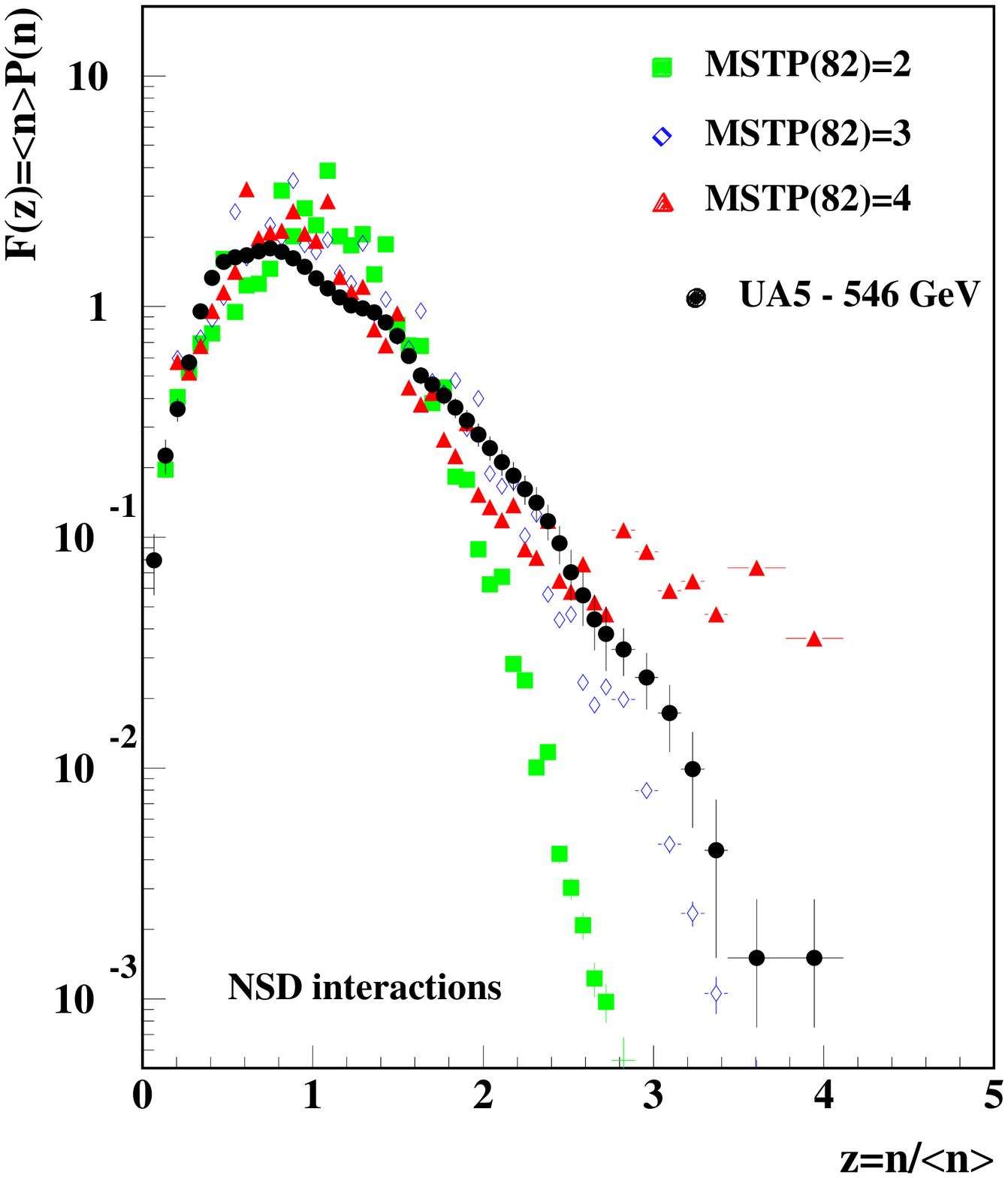}} 
\\
(a) & (b)

\end{tabular}
\caption{KNO distributions for NSD p$\overline{\text{p}}$
collisions at $\sqrt{\text{s}}$ = 546 GeV: (a) simple scenario and (b)
complex scenario distributions compared to data.}
\label{fig:kno-scenarios}
\end{center}
\end{figure}

Figure \ref{fig:kno-scenarios} shows KNO distributions for NSD
p$\overline{\text{p}}$ collisions at $\sqrt{\text{s}}$ = 546 GeV. 
We compare distributions generated by PYTHIA's simple and
complex scenarios to UA5 data \cite{Alner:1987wb}. Apart from the mentioned
changes in the setting MSTP(82), all other parameters are set to use
PYTHIA's default options, as described in \cite{Sjostrand:2001yu}.
Figure \ref{fig:kno-scenarios}(a) shows that using the simple
scenario (MSTP(82)=1), which is the default PYTHIA6.214 option
\cite{Sjostrand:2001yu}, the generated distributions fail to reproduce the data,
especially in the region of high z (z$>$1.5). This is the region of
events with particle multiplicities several times greater than the
average multiplicity. Distributions generated using the complex
scenario vary with the hadronic matter distribution selected for each
case, as can be seen in figure \ref{fig:kno-scenarios}(b) .

The comparisons of KNO distributions shown in figure
\ref{fig:kno-scenarios}(b) indicate that the matter
distribution used to describe the colliding hadrons does affect the
probability of particle production in minimum bias events.
Although in the comparisons shown in figure \ref{fig:kno-scenarios}
the best agreement to the data was obtained by selecting the
complex scenario with the single Gaussian matter distribution option, 
we shall adopt the complex scenario with a double Gaussian matter
distribution (MSTP(82)=4) as our preferred choice.

This is done because by choosing the double Gaussian option, the user
is able to control some of the properties of this matter
distribution. 
Hadrons described by this distribution have a small core region of
radius a$_{2}$ containing a fraction $\beta$ of the total hadronic
matter. This core is embedded in a larger volume of radius a$_{1}$
containing the remaining fraction of matter, i.e., (1 - $\beta$) of
the total hadronic matter. The parameter PARP(83) controls the portion
$\beta$ of the total hadronic matter assigned to the core of the
hadron. The ratio a$_{2}$/a$_{1}$ is given by the parameter
PARP(84). By default, PYTHIA sets PARP(83)=0.5 and PARP(84)=0.2
describing any given hadron as a body with half of its matter
concentrated within a core which is limited by a radius a$_{2}$ = 20\%
of the hadron radius a$_{1}$ \cite{Sjostrand:2001yu}. 
\begin{figure}[htbp]
\begin{center}
\begin{tabular}{cc}
\scalebox{0.335}[0.3]{\includegraphics*[0.0cm,0.0cm][20cm,22cm]{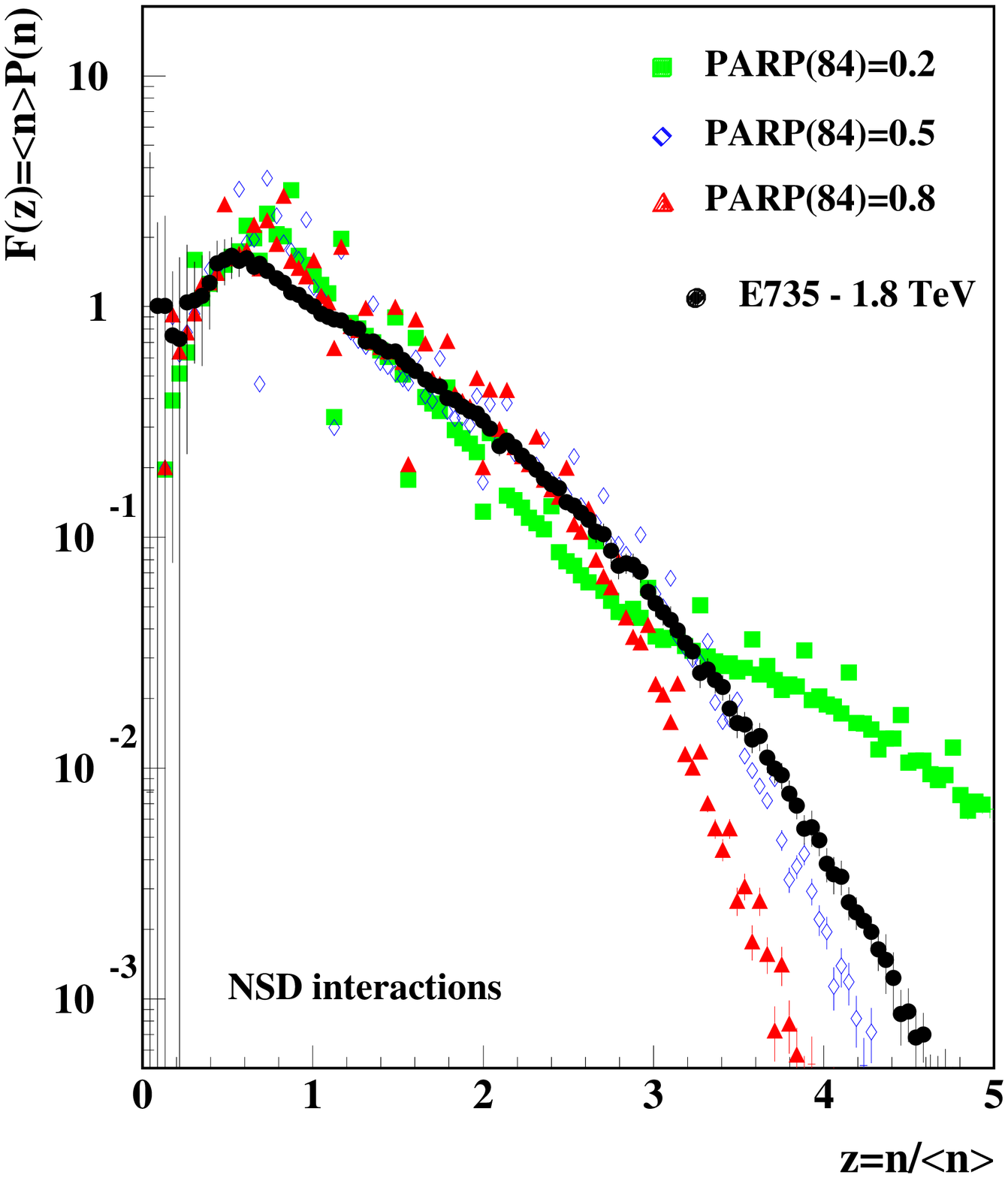}} 
&
\scalebox{0.335}[0.3]{\includegraphics*[0.0cm,0.0cm][20cm,22cm]{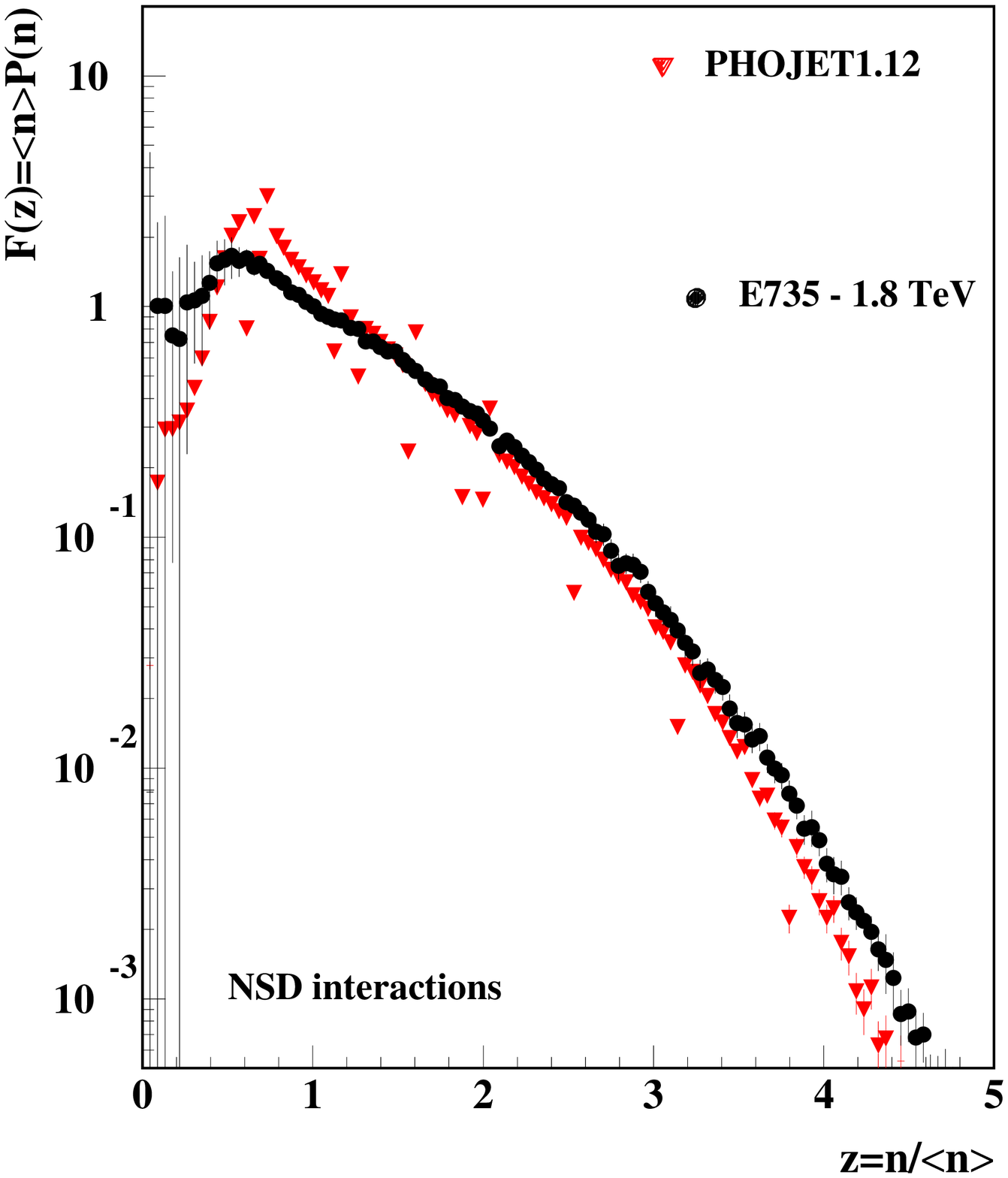}}
\\
(a) & (b)
\end{tabular}
\caption{KNO distributions for NSD p$\overline{\text{p}}$
collisions at $\sqrt{\text{s}}$ = 1.8 TeV: (a) double
Gaussian model with different core-sizes, and (b) PHOJET1.12 compared
to the data \cite{Alexopoulos:1998bi,Matinyan:1998ja}.} 
\label{fig:kno-core}
\end{center}
\end{figure}

As shown in figure \ref{fig:kno-core}(a) considerable changes in the
high-z tale of the KNO distributions are observed as the core radius
varies from 20\% to 50\% and 80\% of the radius of the colliding
hadrons. As the core is made harder and denser
(smaller core-radius) the overlap between two colliding cores makes
high-p$_{t}$ partonic scatterings more likely, yielding higher
multiplicity events more often. When two relatively softer cores
(larger radius) overlap in a collision, the generated activity will be
smaller and softer, hence producing high-multiplicity events less
frequently. 

Figure \ref{fig:kno-core}(b) shows a comparison between PHOJET1.12 and
the KNO distribution measured by E735
\cite{Alexopoulos:1998bi,Matinyan:1998ja} for NSD p$\overline{\text{p}}$
collisions at $\sqrt{\text{s}}$ = 1.8 TeV. Describing hadron
collisions using the multiple Pomeron exchange mechanism proposed by
the DPM \cite{Capella:1994yb,PHOJET112} and the QCD picture for high-p$_{t}$
interactions, PHOJET1.12 is in good agreement to the data.

\subsubsection{Pseudorapidity Distribution}

The rate of parton-parton scattering in a hadronic collision is
strongly correlated to the observed particle multiplicity and the
pseudorapidity distribution of produced particles. This
happens because multiple parton interactions convert part of the
collision energy that would otherwise be carried by the fast moving
system of beam-remnants in the forward regions, into low-p$_{t}$
particles which populate the central region.

In PYTHIA, one of the main parameters used to regulate the rate of
parton-parton interactions is p$_{t_{min}}$ given by equation
\ref{eq:01}. Low values of p$_{t_{min}}$ imply in high rates of
parton-parton scatterings and hence in high levels of particle
multiplicity. Increasing p$_{t_{min}}$ the opposite is expected.
\begin{figure}[htbp]
\begin{center}
\begin{tabular}{cc}
\scalebox{0.35}[0.325]{\includegraphics*[0.0cm,0.0cm][20cm,22cm]{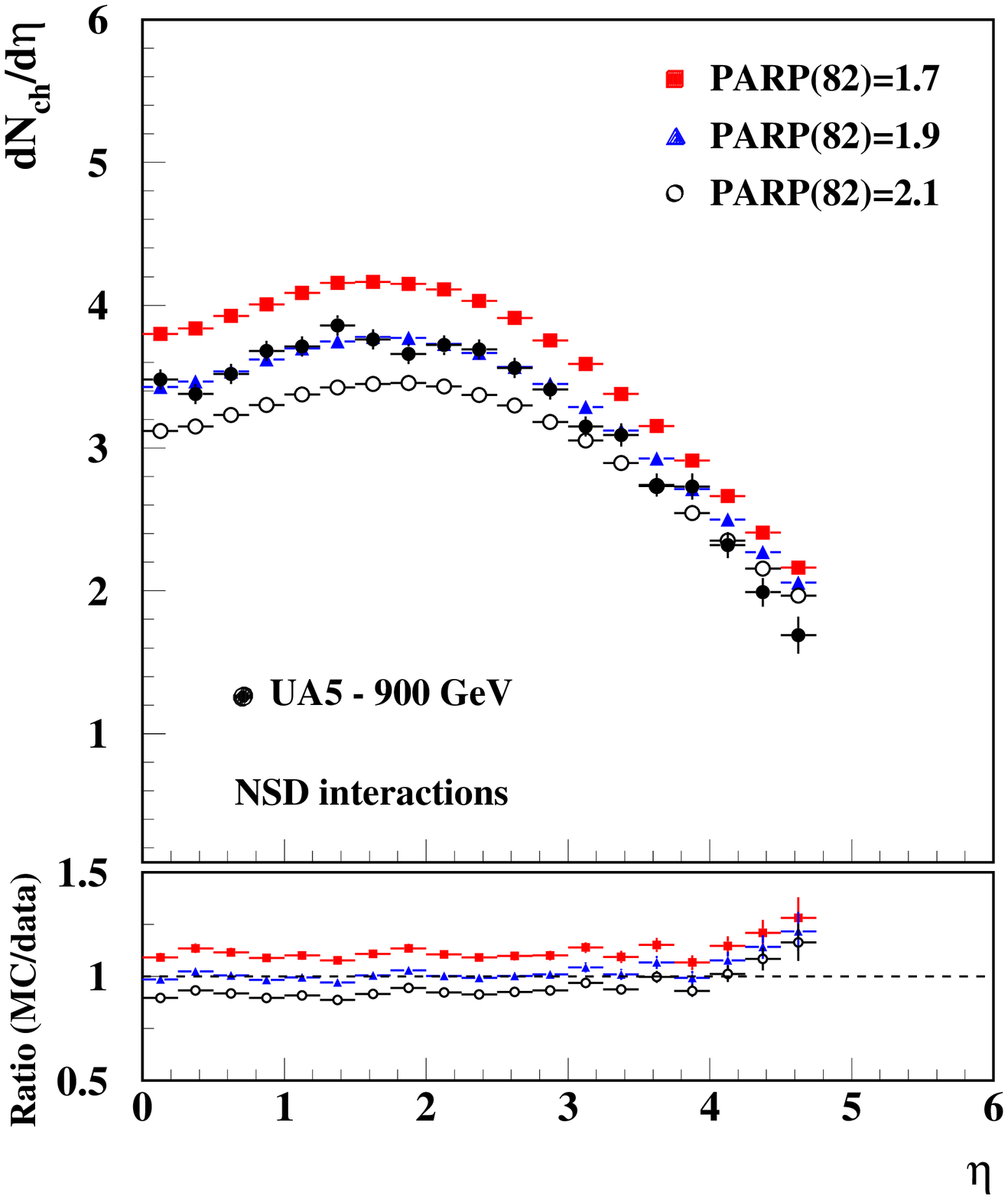}} 
&
\scalebox{0.35}[0.325]{\includegraphics*[2.5cm,0.0cm][20cm,22cm]{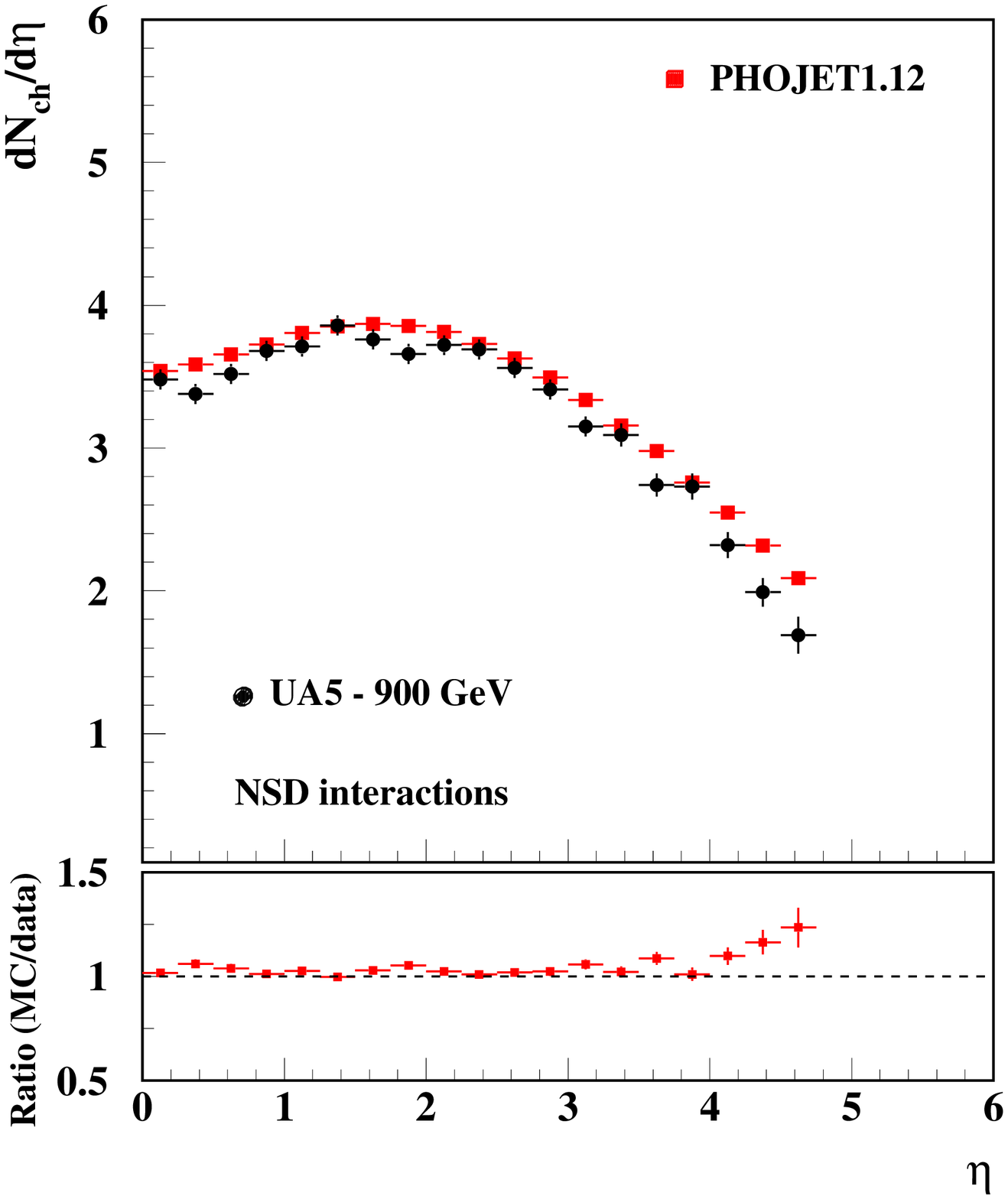}} 
\\
(a) & (b)
\end{tabular}
\caption{Charged particle density distributions,dN$_{ch}$/d$\eta$,
for NSD p$\overline{\text{p}}$ at $\sqrt{\text{s}}$ = 900 GeV
collisions comparing the data \cite{Ansorge:1989kn} to (a) PYTHIA6.214 with
various p$_{t_{min}}$ and (b) PHOJET1.12.} 
\label{fig:dndeta-parp82}
\end{center}
\end{figure}

As can be seen in figure \ref{fig:dndeta-parp82}(a),
increasing PARP(82) from 1.7 to 1.9 and 2.1, which effectively
increases the p$_{t_{min}}$ used by PYTHIA6.214,
the charged particle density, dN$_{ch}$/d$\eta$, decreases. Notice
that relatively small changes in PARP(82) ($\sim 10\%$) can cause
significant variations in the plateau of dN$_{ch}$/d$\eta$.  

In PHOJET, multiple Pomeron exchanges predicted by the
DPM control the plateau of dN$_{ch}$/d$\eta$. Similarly to PYTHIA,
this model also depends on a p$_{t}^{\text{cut-off}}$ which is used to
connect the soft and hard components of a hadronic
interaction. PHOJET1.12 has its default options tuned for
p$_{t}^{\text{cut-off}}=2.5$ GeV. Figure \ref{fig:dndeta-parp82}(b)
shows dN$_{ch}$/d$\eta$ generated by PHOJET1.12 with its default cuts,
compared to UA5 data \cite{Ansorge:1989kn}. There is a good agreement between
PHOJET1.12 predictions and the data. 

\subsection{The Underlying Event \label{sec:ue}}

In a hadronic event containing jets, the underlying event (UE)
consists of all event activity except the two outgoing hard scattered
jets \cite{Affolder:2001xt}. As for minimum bias events, soft interactions
and the mechanism of multiple parton interaction play
an important role in the structure of the underlying event and ought
to be carefully considered by any model attempting to describe the
underlying event. 

The conditions applied to particle selection and to the event region
to be investigated are described in Ref. \cite{Affolder:2001xt}.
The region transverse to the leading jet is used to study the UE and
is defined by $ 60^{\circ} < \left| \Delta \phi \right| <
120^{\circ}$, where the angular difference in the azimuthal angle
$\phi$ is given by $\Delta \phi = \phi_{\text{particle}} -
\phi_{\text{ljet}}$.  

\begin{figure}[htbp]
\begin{center}
\begin{tabular}{cc}
\scalebox{0.335}[0.3]{\includegraphics*[0.0cm,0.0cm][22cm,22cm]{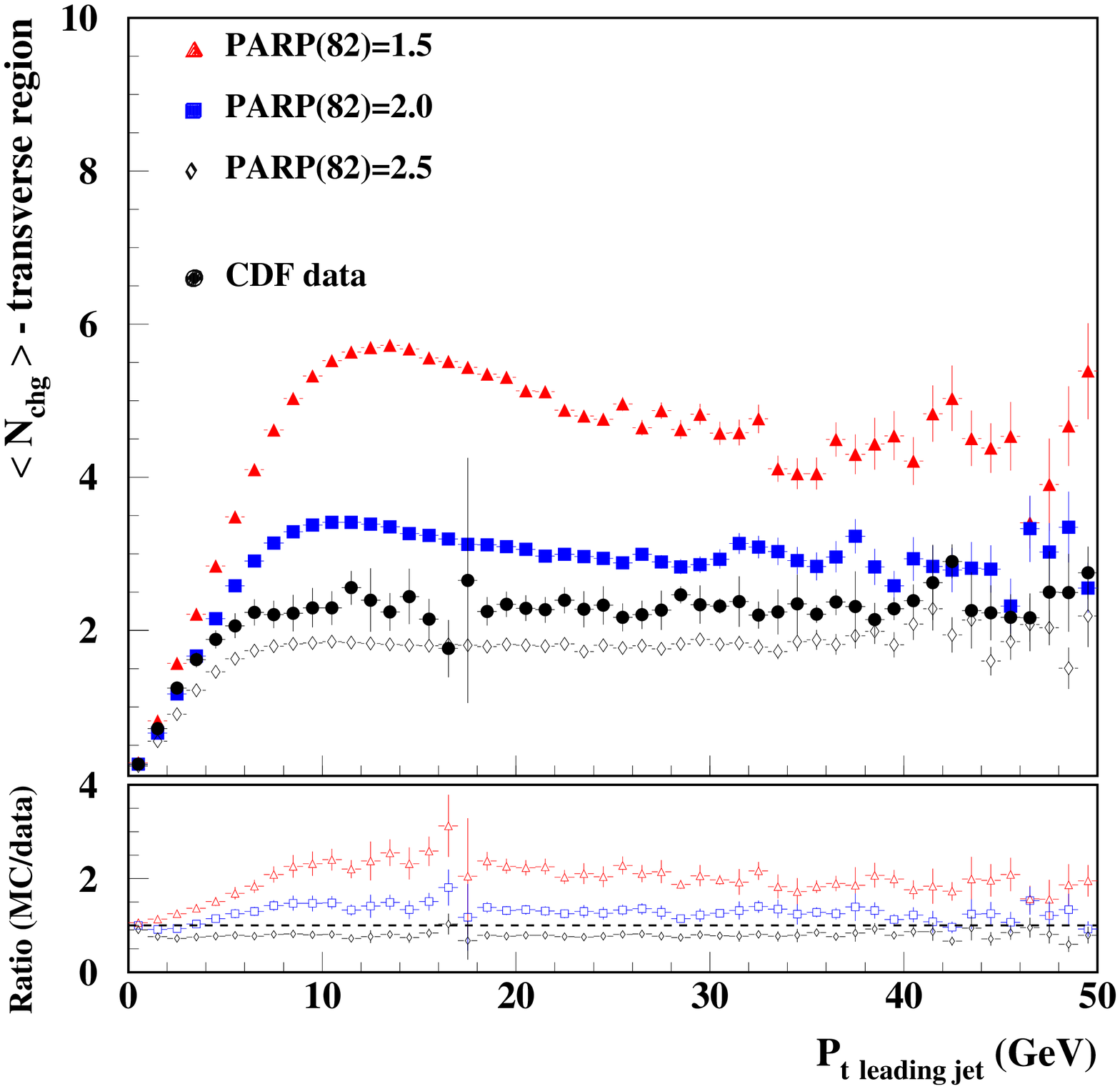}} 
&
\scalebox{0.335}[0.3]{\includegraphics*[0.0cm,0.0cm][22cm,22cm]{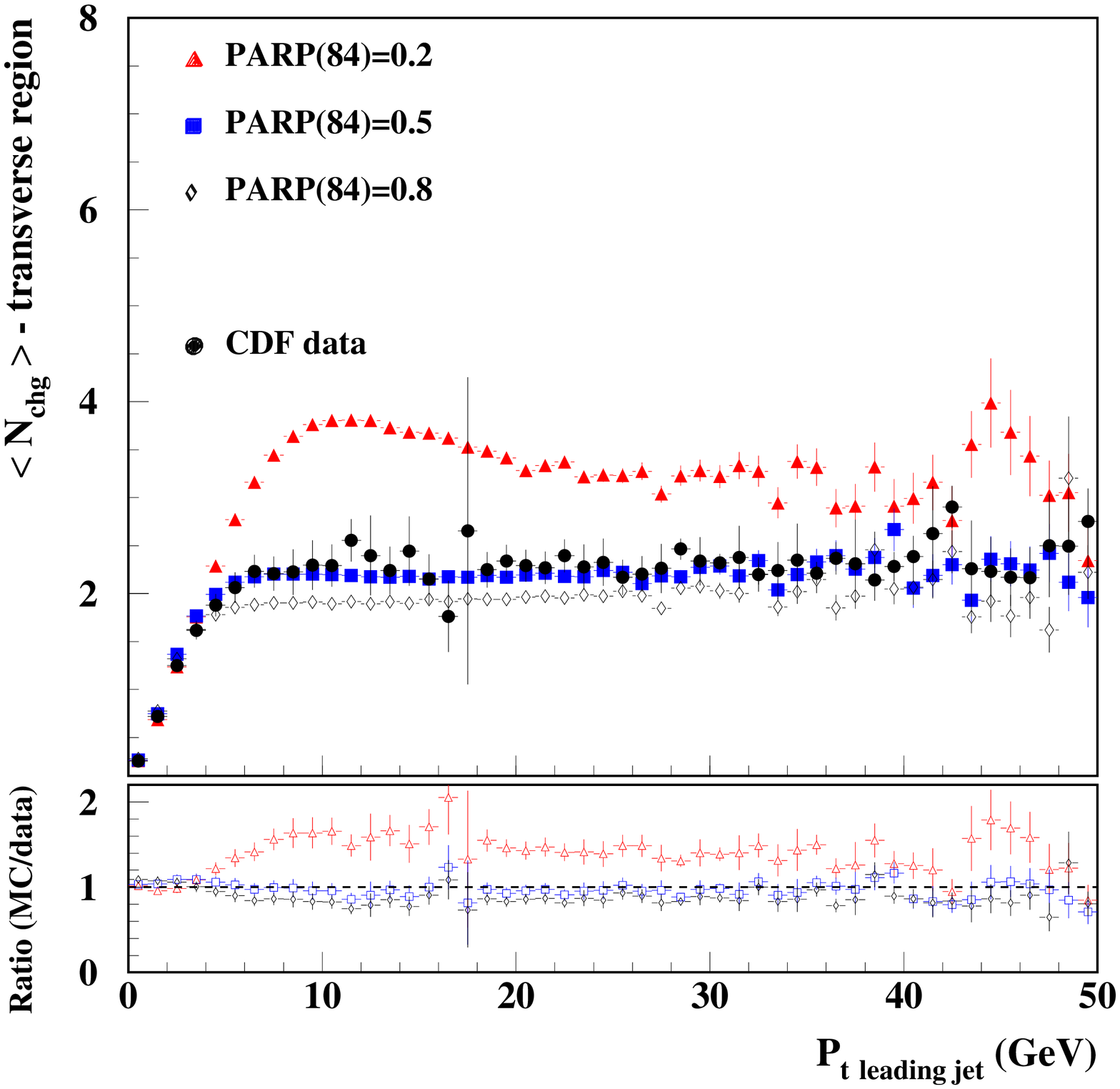}} 
\\
(a) & (b)

\end{tabular}
\caption{Average charged particles multiplicity in the transverse
  region showing PYTHIA6.214 - MSTP(82)=4 with (a) different values of
  PARP(82) (i.e. p$_{t_{min}}$) and (b) different values of PARP(84)
  (core-size). }  
\label{fig:trans-pythia}
\end{center}
\end{figure}

Figure \ref{fig:trans-pythia}(a) shows PYTHIA6.214 - MSTP(82)=4
distributions generated with different values of PARP(82), i.e. different
p$_{t_{min}}$, compared to the data for the average charged particle
multiplicity in the transverse region. Increasing p$_{t_{min}}$, which
corresponds to a decrease on the rate of semi-hard parton scatterings,
$<N_{\text{chg}}>$ decreases. This effect is similar to the one
observed in figure \ref{fig:dndeta-parp82}(a) for minimum bias charged
particle density distributions dN$_{ch}$/d$\eta$. 

As shown in figure \ref{fig:trans-pythia}(b), depending on the core
size variation (PARP(84)) the plateau level of $<N_{chg}>$ can suffer
severe changes. For example, changing PARP(84) from 0.2 to 0.5 reduces
the plateau of $<N_{chg}>$ by nearly a factor of two, while a further
increase in PARP(84) from 0.5 to 0.8 only reduces the plateau by $\sim
15$\%. 

Jets are likely to be produced when there is a core overlap in
the hadronic collision. Smaller and dense cores imply that events with
a core overlap have also a large overlap of less dense matter regions
which surround the core, and when overlapped generate high rates of
soft interactions causing the higher plateaus observed in the
$<N_{chg}>$ distributions shown in figure
\ref{fig:trans-pythia}(b). Larger cores also imply in smaller soft
surrounding regions in the colliding hadrons, hence producing lower
multiplicity (and $\text{p}_{t}$) distributions in the UE. 

\begin{figure}[htbp]
\begin{center}
\begin{tabular}{cc}
\scalebox{0.335}[0.3]{\includegraphics*[0.0cm,0.0cm][22cm,22cm]{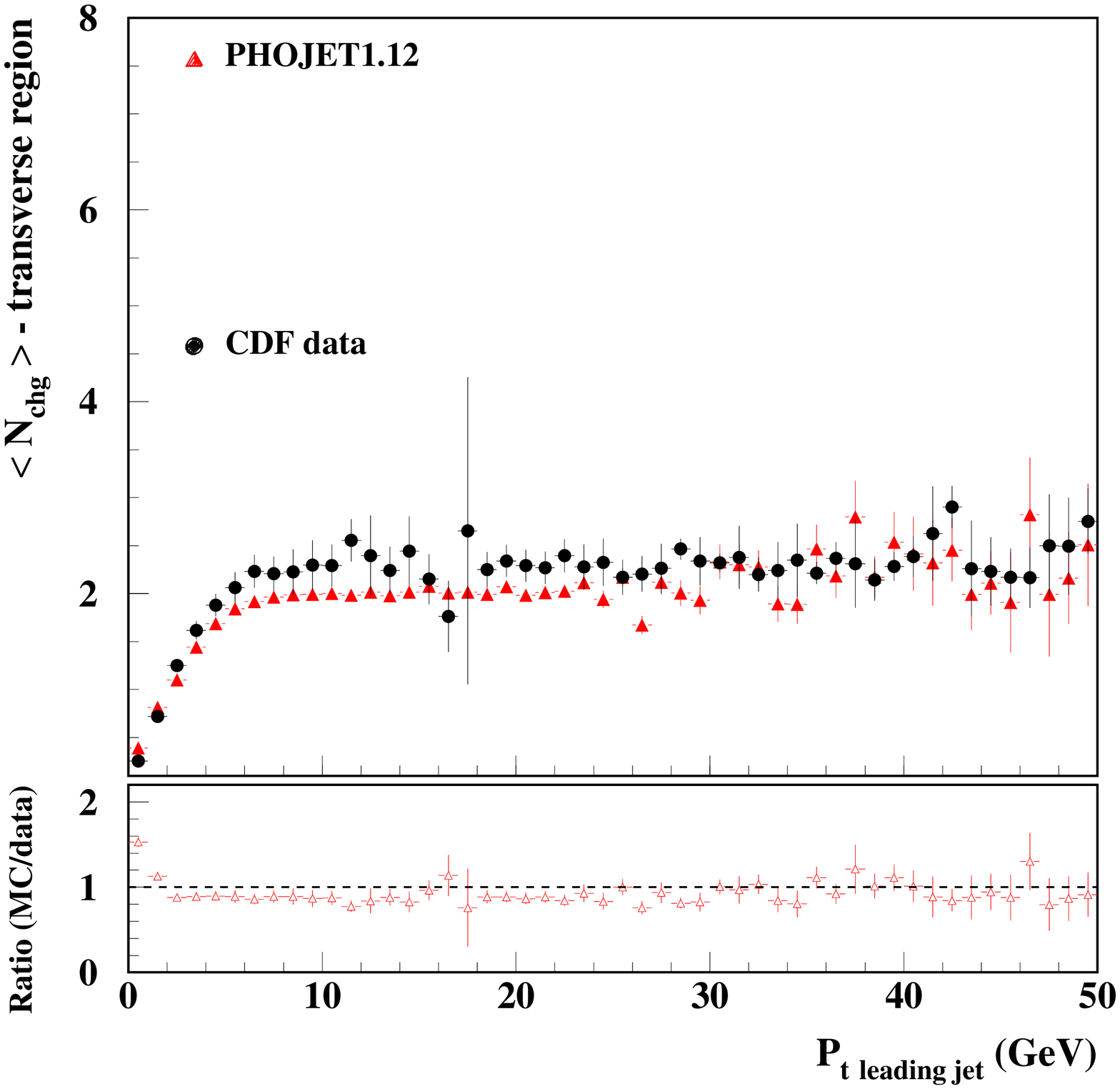}} 
&
\scalebox{0.335}[0.3]{\includegraphics*[0.0cm,0.0cm][22cm,22cm]{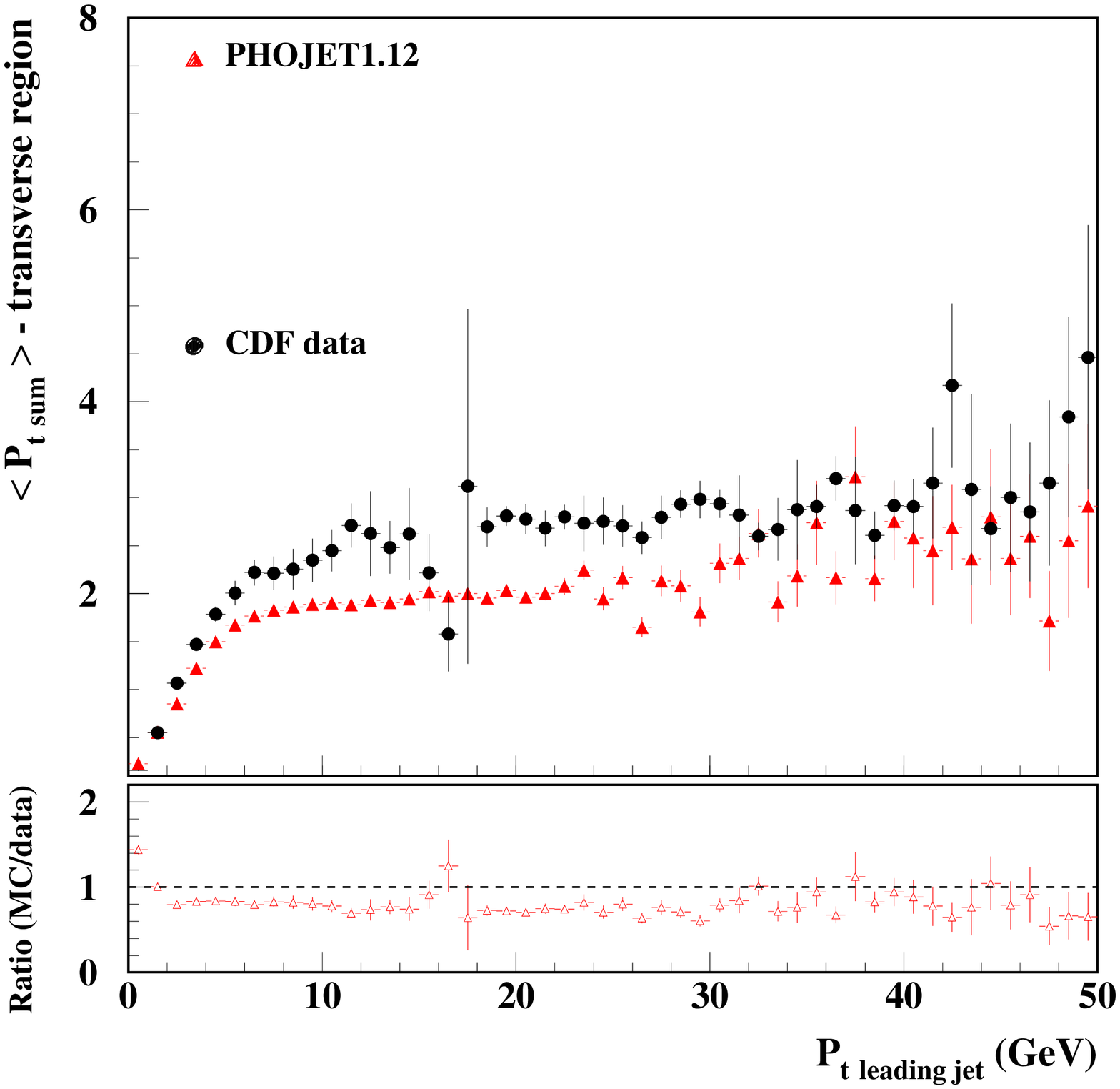}} 
\\
(a) & (b)

\end{tabular}
\caption{PHOJET1.12 predictions compared to CDF data for: (a) average
  multiplicity in the transverse region and (b) average
  $\text{p}_{t_{\text{sum}}}$ in the transverse region.}

\label{fig:ue-phojet}
\end{center}
\end{figure}

Figure \ref{fig:ue-phojet} shows PHOJET1.12 predictions compared to
data for: (a) average multiplicity in the transverse region and (b)
average $\text{p}_{t_{\text{sum}}}$ in the transverse region.
PHOJET1.12 reproduces reasonably well the data for the UE
multiplicity distribution, as displayed in figure \ref{fig:ue-phojet}
(a). 
However, it underestimates the average $\text{p}_{t_{\text{sum}}}$
distribution (figure \ref{fig:ue-phojet}(b) ). The measured
$<\text{p}_{t_{\text{sum}}}>$ distribution is underestimated by
PHOJET1.12 by $\sim 20$\%. 

\subsubsection{UE vs. Minimum Bias}

The CDF measurement shows that the underlying event multiplicity forms
a plateau for events with P$_{t_{\text{ljet}}} \gtrsim 5$ GeV at
$<N_{\text{chg}}> \sim 2.3$. Supposing that the transverse region in
events with P$_{t_{\text{ljet}}} \gtrsim 5$ GeV is uniform in
azimuthal angle $\phi$ and in pseudorapidity $\eta$, this multiplicity
corresponds to 3.45 particle per unit pseudorapidity. Further
corrections to detector effects and low-p$_{t}$ extrapolation
\cite{Affolder:2001xt} implies that there are roughly 10 charged particles
per pseudorapidity unit with p$_{t} > 0$ GeV in the underlying
event. 

In p$\overline{\text{p}}$ collision at 1.8 TeV, the minimum bias
density, which has also been
measured by CDF, gives dN$_{ch}$/d$\eta \sim 4$ for $|\eta| < 1$
\cite{Abe:1990td}, while the equivalent density for the underlying event is
at least a factor of two larger.
This comparison, though not highly accurate due to the uncertainties
in estimating the particle density for the underlying event 
(i.e. extrapolation to low-p$_{t}$ and several assumptions made on the
particle distribution in $\phi$ and $\eta$), clearly shows that the
underlying event in hard scattering processes (P$_{t_{\text{ljet}}}
\gtrsim 5$ GeV) has much more activity
than an average minimum bias event.

\subsection{PYTHIA6.214 - Tuned VS. PHOJET1.12 \label{sec:tuning}}

Combining the effects of variations in p$_{t_{\text{min}}}$ and in the
core-size we obtained a set of PYTHIA6.214 parameters which considerably
improves PYTHIA's description of minimum bias and underlying event
distributions. Our tuned parameters for PYTHIA6.214 are displayed in
table \ref{tab:PYTHIA-tuning}.    
\begin{table}[htbp]
\begin{center}
\begin{tabular}{|c|c|}
\hline
\multicolumn{2}{|c|}{\textbf{PYTHIA6.214 - tuned}} \\
\hline

\scriptsize ISUB: 11,12,13,28,53,68~ & \scriptsize QCD $2 \rightarrow
 2$ partonic scattering \\ 
\scriptsize 94,95,96 & \scriptsize + non-diffractive + double
 diffractive \\ 

\scriptsize MSTP(51)=7 & \scriptsize CTEQ5L - selected p.d.f. \\

\scriptsize MSTP(81)=1 & \scriptsize multiple interactions \\

\scriptsize MSTP(82)=4 & \scriptsize complex scenario \\
 & \scriptsize + double Gaussian matter distribution\\

\scriptsize PARP(82)=1.8 & \scriptsize p$_{t_{\text{min}}}$ parameter \\

\scriptsize PARP(84)=0.5 & \scriptsize core radius: 50\% of the \\
 & \scriptsize hadronic radius \\

\hline
\end{tabular}
\caption{ PYTHIA6.214 tuned parameters for minimum bias and the underlying
  event. }
\label{tab:PYTHIA-tuning}
\end{center}
\end{table}                    

\begin{figure}[htbp]
\begin{center}
\begin{tabular}{cc}
\scalebox{0.35}[0.3]{\includegraphics*[0.0cm,0.0cm][22cm,22cm]{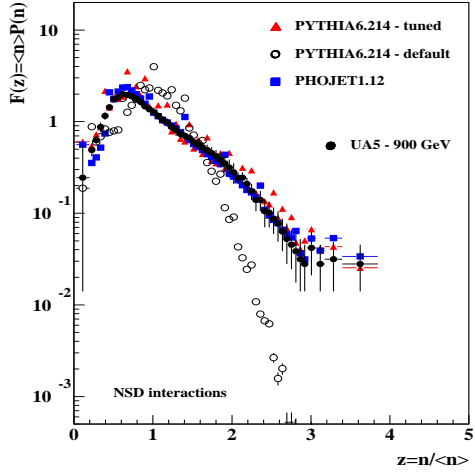}} 
&
\scalebox{0.35}[0.3]{\includegraphics*[0.0cm,0.0cm][22cm,22cm]{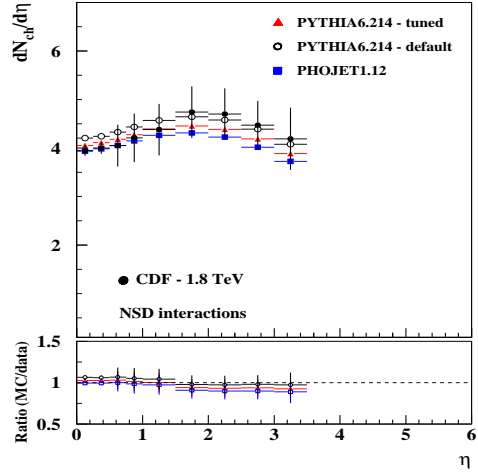}} 
\\
(a) & (b)
\\
\scalebox{0.335}[0.3]{\includegraphics*[0.0cm,0.0cm][22cm,22cm]{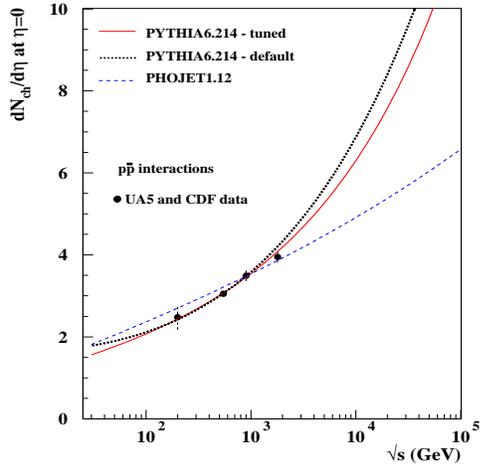}}
&
\scalebox{0.335}[0.3]{\includegraphics*[4.0cm,0.0cm][22cm,22cm]{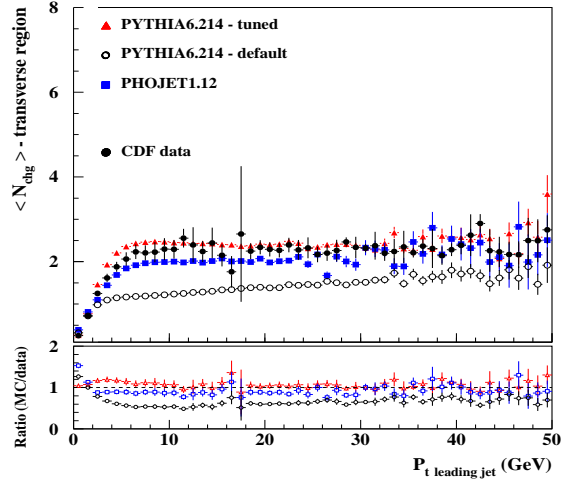}} 
\\
(c) & (d)

\end{tabular}
\caption{(a) KNO distributions for NSD p$\overline{\text{p}}$
collisions at $\sqrt{\text{s}}$ = 900 GeV; (b) dN$_{ch}$/d$\eta$
for NSD p$\overline{\text{p}}$ at $\sqrt{\text{s}}$ = 1.8 TeV; (c)
dN$_{ch}$/d$\eta$ at $\eta =0$ for a wide range of $\sqrt{\text{s}}$;
and (d) $<N_{chg}>$ in the transverse region.}
\label{fig:tuned}
\end{center}
\end{figure}

Figure \ref{fig:tuned} shows predictions generated by PYTHIA6.214 -
tuned and default, and PHOJET1.12 compared to some minimum bias and
underlying event distributions. 
The description of both minimum bias and underlying event
distributions is improved by using PYTHIA6.214 - tuned
compared to the predictions generated by PYTHIA's default
settings. Notice that PYTHIA6.214 - tuned and PHOJET1.12
can generate very different predictions when extrapolated to higher
energies, as shown in fig. \ref{fig:tuned}(c).  

\subsection{LHC Predictions \label{sec:lhc}}

\begin{figure}[htbp]
\begin{center}
\begin{tabular}{cc}
\scalebox{0.325}[0.3]{\includegraphics*[0.0cm,0.0cm][22cm,22cm]{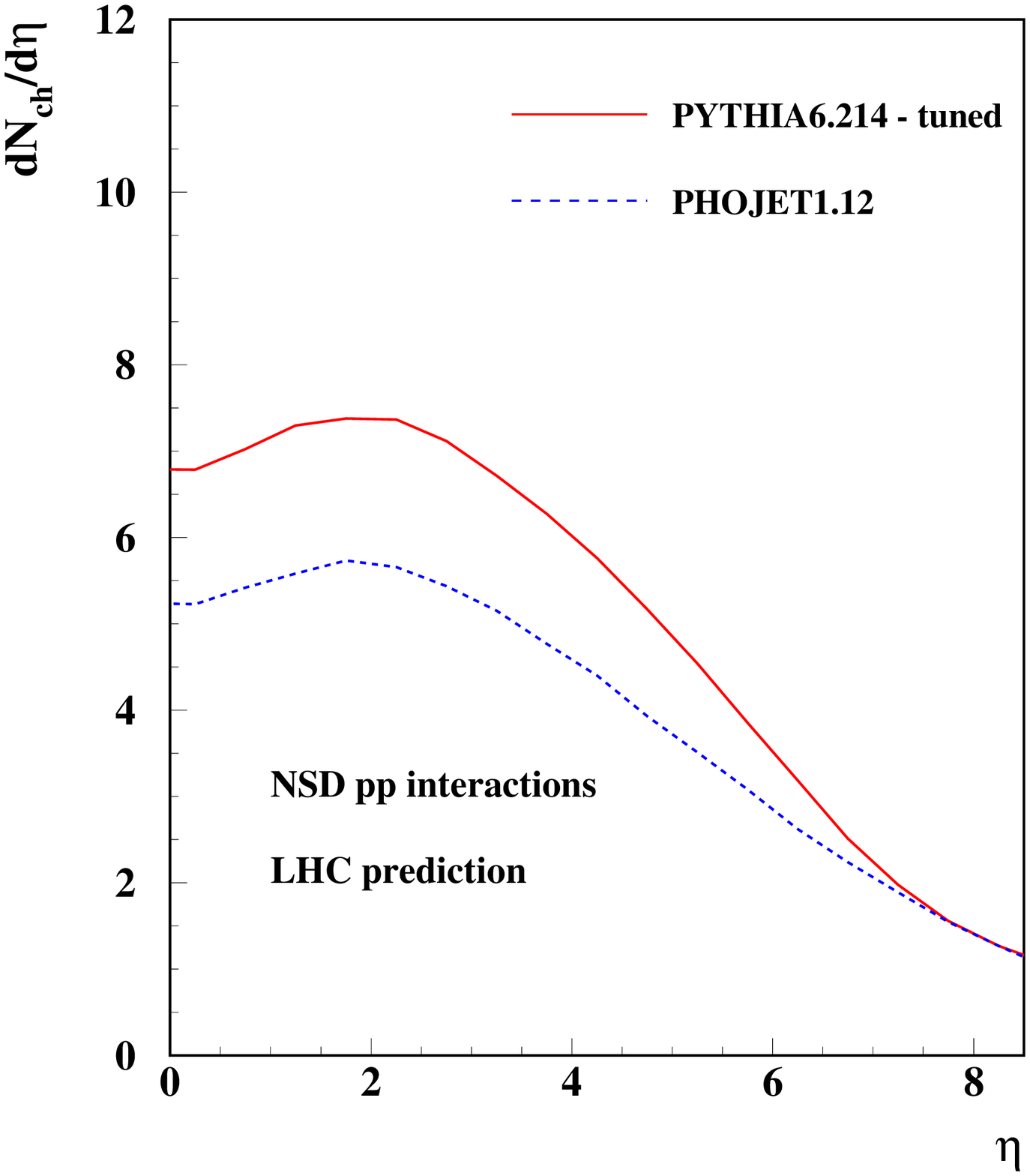}} 
&
\scalebox{0.375}[0.335]{\includegraphics*[1.0cm,0.0cm][22cm,22cm]{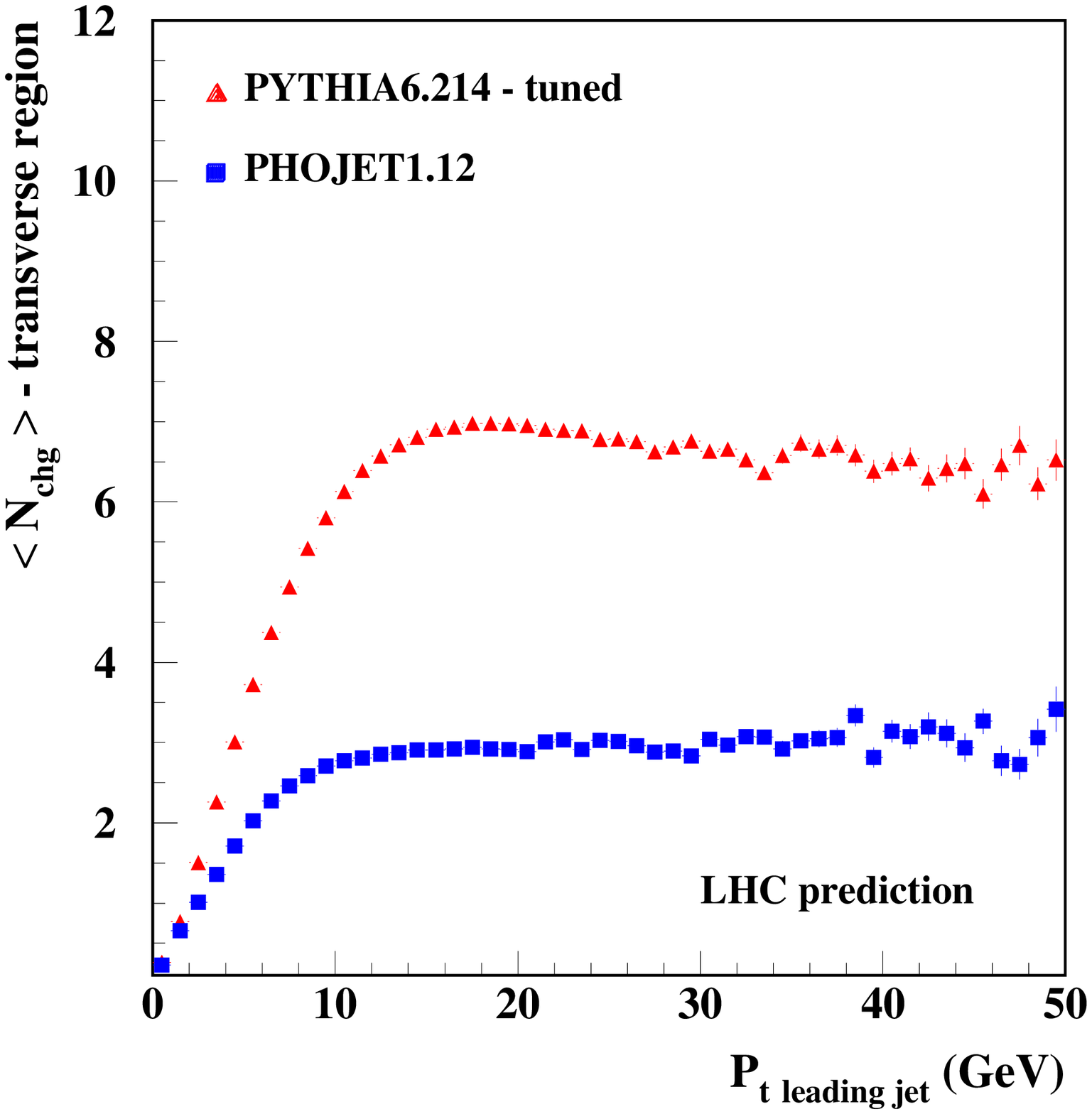}} 
\\
(a) & (b)

\end{tabular}
\caption{(a) Charged particle density distributions, dN$_{ch}$/d$\eta$,
  for NSD pp collisions at $\sqrt{\text{s}}$ = 14 TeV; (b) Average
  multiplicity in the underlying event for jet events in pp collisions
  at $\sqrt{\text{s}}$ = 14 TeV. }
\label{fig:lhc}
\end{center}
\end{figure}

Figure \ref{fig:lhc}(a) shows dN$_{ch}$/d$\eta$
distributions for minimum bias pp collisions at
$\sqrt{\text{s}}$ = 14 TeV generated by PHOJET1.12 and PYTHIA6.214 -
tuned. The charged particle density generated by PHOJET1.12 and
PYTHIA6.214 - tuned at $\eta = 0$ is 5.13 and 6.82, respectively.
In the central region ($|\eta| < 2.5$) dN$_{ch}$/d$\eta$ is $\sim
5.5$ and $\sim 7$, respectively for PHOJET1.12 and PYTHIA6.214 -
tuned. Contrasting to the agreement shown
for p$\overline{\text{p}}$ collisions at $\sqrt{\text{s}}$ = 1.8 TeV in
figure \ref{fig:tuned}(b), at the LHC PYTHIA6.214 - tuned
generates $\sim 27 \%$ more charged particle density in the central
region than PHOJET1.12. 

Compared to the charged particle density dN$_{ch}$/d$\eta$ measured by
CDF at 1.8 TeV (figure \ref{fig:tuned}(b) ), PYTHIA6.214 -
tuned indicates a plateau rise of $\sim 70\%$ at the LHC in the central
region while PHOJET1.12 suggests a smaller rise of $\sim 35\%$. 

Figure \ref{fig:lhc}(b) displays PYTHIA6.214 - tuned and PHOJET1.12
predictions for the average particle multiplicity in the UE for pp
collisions at the LHC (charged particles with $p_{t} >0.5~$GeV and
$\left| \eta \right| <1$). 
The distributions generated by the two models are fundamentally
different. Excepting the events with P$_{t_{\text{ljet}}} \lesssim
3$ GeV, PYTHIA6.214 - tuned generates greater activity than
PHOJET1.12 in the UE.
The average multiplicity in the UE for P$_{t_{\text{ljet}}} > 10$
GeV reaches a plateau at $\sim 6.5$ charged particles according to
PYTHIA6.214 - tuned and $\sim 3.0$ according to PHOJET1.12. 
Compared to the UE distributions measured by CDF at 1.8
TeV (figure \ref{fig:tuned}(d) ), PYTHIA6.214 - tuned indicates
a plateau rise of $\sim 200\%$ at the LHC while PHOJET1.12 suggests a
much smaller rise of $\sim 40\%$. 

At the LHC, the minimum bias predictions generated by PYTHIA6.214 -
tuned and PHOJET1.12 for the central plateau of dN$_{ch}$/d$\eta$,
indicate a rise of $\sim 70\%$ and $\sim
35\%$, respectively. These are smaller than the predicted increase
for the UE suggested by both models. As discussed
previously, at the Tevatron, for events with
P$_{t_{\text{ljet}}} > 10$ GeV the particle density in the underlying
event is at least a factor of two larger than the equivalent minimum
bias prediction. Using similar assumptions as those adopted in the
analysis for the CDF data, LHC events with
P$_{t_{\text{ljet}}} > 10$ GeV are predicted to have a charged
particle density dN$_{ch}$/d$\eta$ of $\sim 29$ charged particles per
pseudorapidity unit according to PYTHIA6.214 - tuned and $\sim 13$
according to PHOJET1.12. In other words, for P$_{t_{\text{ljet}}} >
10$ GeV the UE at the LHC is predicted to have a
particle density $\sim 4$ times larger than its equivalent minimum
bias prediction according to PYTHIA6.214 - tuned, and $\sim 2$ times
larger according to PHOJET1.12. 

Therefore PYTHIA6.214 - tuned predicts not only that the UE particle
density will increase at the LHC, but it will also
increase its activity compared to the equivalent minimum bias
distribution. On the other hand, PHOJET1.12 estimates that the
increase in charged particle density in the UE at the
LHC will follow the same rate to the minimum bias density measured at
the Tevatron. In both cases however, the underlying event density is
greater than its equivalent minimum bias counterpart. 


\section[Using Correlations in the Transverse Region to Study the
  Underlying Event in Run 2 at the Tevatron]{USING CORRELATIONS IN THE TRANSVERSE REGION TO STUDY THE
  UNDERLYING EVENT IN RUN 2 AT THE TEVATRON~\protect\footnote{Contributed by: {A.~Cruz and R.~Field}}}
\label{LH2003_field}


%
\newcommand*{\eg}{\textit{e.g.},\ }
\newcommand*{\ie}{\textit{i.e.},\ }
\newcommand*{\etal}{\textit{et al.\ }}
\newcommand*{\gev}{\,\textrm{GeV}}
\newcommand*{\gevc}{\,\textrm{GeV/c}}
\newcommand*{\mevc}{\,\textrm{MeV/c}}
\newcommand*{\tev}{\,\textrm{TeV}}
\newcommand*{\pt}{$p_T$}
\newcommand*{\bigpt}{$P_T$}
\newcommand*{\UE}{``underlying event"}
\newcommand*{\BBR}{``beam-beam remnants"}
\newcommand*{\MB}{``min-bias"}
\newcommand*{\TR}{``transverse"}
\newcommand*{\LJ}{``leading jet"}
\newcommand*{\BB}{``back-to-back"}
\newcommand*{\Tone}{``transverse $1$"}
\newcommand*{\Ttwo}{``transverse $2$"}
\newcommand*{\Jone}{jet\# $1$}
\newcommand*{\Jtwo}{jet\# $2$}
\newcommand*{\binA}{$30<E_T({\rm jet}\#1)<70\gev$}
\newcommand*{\binB}{$130<E_T({\rm jet}\#1)<250\gev$}
\newcommand*{\pthard}{$p_T({\rm hard})$}
\newcommand*{\ptcut}{$p_T\!>\!0.5\,{\rm GeV/c}$}
\newcommand*{\etacut}{$|\eta|\!<\!1$}
\newcommand*{\aveN}{$\langle\!N_{\rm chg}\!\rangle$}
\newcommand*{\avePT}{$\langle\!p_T\!\rangle$}
\newcommand*{\etaphi}{$\eta$-$\phi$}
\newcommand*{\delphi}{$\Delta\phi$}
\newcommand*{\absdelphi}{$|\Delta\phi|$}
\newcommand*{\delphicut}{$|\Delta\phi|>150^\circ$}

\subsection{Introduction}

\begin{figure}[htbp]
\begin{center}
\includegraphics[scale=0.6]{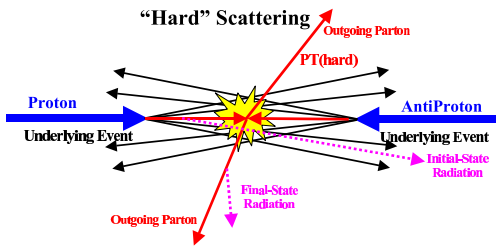}
\caption{Illustration of the way QCD Monte-Carlo models simulate a proton-antiproton 
collision in which a ``hard" $2$-to-$2$ parton scattering with transverse momentum, \pthard, 
has occurred.  The resulting event contains particles that originate from the two 
outgoing partons (plus initial and final-state radiation) and particles that come 
from the breakup of the proton and antiproton (\ie \BBR).  The \UE\ is everything 
except the two outgoing hard scattered ``jets" and consists of the \BBR\ plus initial 
and final-state radiation. The ``hard scattering" component consists of the outgoing 
two jets plus initial and final-state radiation.}\label{LH2003_fig1}
\end{center}
\end{figure}

Fig.~\ref{LH2003_fig1} illustrates the way QCD Monte-Carlo models simulate a proton-antiproton collision 
in which a ``hard" $2$-to-$2$ parton scattering with transverse momentum, \pthard, has 
occurred.  The resulting event contains particles that originate from the two outgoing 
partons (plus initial and final-state radiation) and particles that come from the breakup 
of the proton and antiproton (\ie \BBR).  The \UE\ is 
everything except the two outgoing hard scattered ``jets" and receives contributions from 
the \BBR\ plus initial and final-state radiation. The ``hard scattering" 
component consists of the outgoing two jets plus initial and final-state radiation.

\begin{figure}[htbp]
\begin{center}
\includegraphics[scale=0.6]{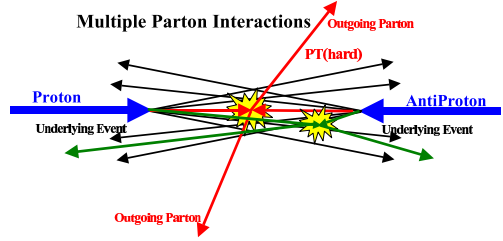}
\caption{
Illustration of the way PYTHIA models the \UE\ in proton-antiproton 
collisions by including multiple parton interactions. In addition to the hard $2$-to-$2$ 
parton-parton scattering with transverse momentum, \pthard, there is a 
second ``semi-hard" $2$-to-$2$ parton-parton scattering that contributes particles to 
the \UE.}\label{LH2003_fig2}
\end{center}
\end{figure}

The \BBR\ are what is left over after a parton is knocked out of each of 
the initial two beam hadrons.  It is the reason hadron-hadron collisions are more ``messy" 
than electron-positron annihilations and no one really knows how it should be modeled.  
For the QCD Monte-Carlo models the \BBR\ are an important component of 
the \UE.  Also, it is possible that multiple parton scattering contributes 
to the \UE.  Fig.~\ref{LH2003_fig2} shows the way PYTHIA \cite{Sjostrand:1985xi,Bengtsson:1986gz,Sjostrand:1987su} models the \UE\ 
in proton-antiproton collisions by including multiple parton interactions. In addition to 
the hard $2$-to-$2$ parton-parton scattering and the \BBR, sometimes there is 
a second ``semi-hard" $2$-to-$2$ parton-parton scattering that contributes particles to the 
\UE.  

Of course, from a certain point of view there is no such thing as an \UE\ 
in a proton-antiproton collision.  There is only an ``event" and one cannot say where a 
given particle in the event originated.  On the other hand, hard scattering collider 
``jet" events have a distinct topology.  On the average, the outgoing hadrons ``remember" 
the underlying the $2$-to-$2$ hard scattering subprocess.  An average hard scattering event 
consists of a collection (or burst) of hadrons traveling roughly in the direction of the 
initial beam particles and two collections of hadrons (\ie ``jets") with large transverse 
momentum.  The two large transverse momentum ``jets" are roughly \BB\ in azimuthal 
angle.  One can use the topological structure of hadron-hadron collisions to study the 
\UE\ \cite{Affolder:2001xt,hustontalk,tano,pyA1,pyA2,cernfield}.  Here we study the \UE\ in the Run $2$ using the 
direction of the leading calorimeter jet (JetClu, $R = 0.7$) to isolate regions of \etaphi\ 
space that are sensitive to the \UE.  

\begin{figure}[htbp]
\begin{center}
\includegraphics[scale=0.6]{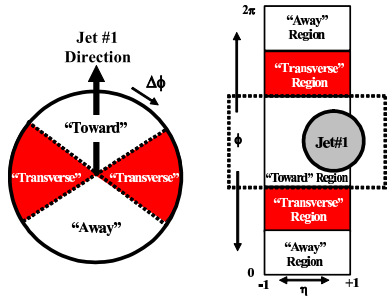}
\caption{
Illustration of correlations in azimuthal angle $\phi$ relative to the direction of the 
leading jet (JetClu, $R = 0.7$) in the event, jet\#$1$.  The angle 
$\Delta\phi=\phi-\phi_{\rm jet1}$ is the relative azimuthal angle between charged 
particles and the direction of \Jone.  The ``toward" region is defined by 
$|\Delta\phi| < 60^\circ$ and \etacut, while the ``away" region 
is $|\Delta\phi| > 120^\circ$ and \etacut.   The \TR\ region is defined 
by  $60^\circ<|\Delta\phi|< 120^\circ$ and \etacut.  Each of the three 
regions ``toward", \TR, and ``away" have an area in \etaphi space 
of $\Delta\eta\Delta\phi=4\pi/3$.  We examine charged particles in the 
range \ptcut\ and \etacut,  but allow the leading 
jet to be in the region $|\eta({\rm jet}\# 1)|<2$.
}\label{LH2003_fig3}
\end{center}
\end{figure}

As illustrated in Fig.~\ref{LH2003_fig3}, the direction of the leading jet, \Jone, is used to define 
correlations in the azimuthal angle, $\phi$.  The angle 
$\Delta\phi=\phi-\phi_{\rm jet1}$ is the relative 
azimuthal angle between a charged particle and the direction of \Jone.  The ``toward" 
region is defined by $|\Delta\phi| < 60^\circ$ and \etacut, while the ``away" region 
is $|\Delta\phi| > 120^\circ$ 
and \etacut.   The \TR\ region is defined by  $60^\circ<|\Delta\phi|< 120^\circ$ 
and \etacut.  The three regions ``toward", ``transverse", and ``away" are shown in Fig.~\ref{LH2003_fig3}.  
Each region has an area in \etaphi space of $\Delta\eta\Delta\phi=4\pi/3$. The \TR\ region is 
perpendicular to the plane of the hard $2$-to-$2$ scattering and is therefore very 
sensitive to the \UE. We restrict ourselves to charged particles in the range 
\ptcut\ and \etacut,  but allow the leading jet that is used to define the 
\TR\ region to have $|\eta({\rm jet}\# 1)|<2$.  

\begin{figure}[htbp]
\begin{center}
\includegraphics[scale=0.6]{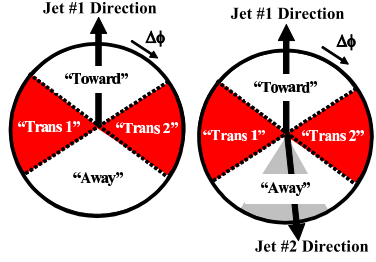}
\caption{
Illustration of correlations in azimuthal angle $\phi$ relative to the direction of the 
leading jet (highest ET jet) in the event, jet\#$1$. The angle 
$\Delta\phi=\phi-\phi_{\rm jet1}$ is the 
relative azimuthal angle between charged particles and the direction of jet\#$1$. The 
two \TR\ regions $60^\circ<\Delta\phi<120^\circ$ and $60^\circ<-\Delta\phi<120^\circ$ are 
referred to as \Tone\ and \Ttwo.  Each of the two \TR\ regions have an area in \etaphi\ space 
of $\Delta\eta\Delta\phi=4\pi/6$.  The overall \TR\ region defined in Fig.~\ref{LH2003_fig3} 
corresponds to combining the \Tone\ and \Ttwo\ regions.  Events in which 
there are no restrictions placed on the on the second highest $E_T$ jet, jet\#$2$, are referred 
to as \LJ\ events ({\it left}).  Events with at least two jets with $|\eta({\rm jet})|<2$, where 
the leading two jets are nearly \BB\ (\delphicut) with $E_T(jet\#2)/E_T(jet\#1) > 0.8$ are referred to as 
\BB\ events. ({\it right}).
}\label{LH2003_fig4}
\end{center}
\end{figure}

In this analysis we look in more detail at the two \TR\ regions defined in Fig.~\ref{LH2003_fig4}.  
The overall \TR\ region in Fig.~\ref{LH2003_fig3} corresponds to combining the \Tone\ 
and \Ttwo\ regions.   Comparing these two \TR\ regions on an 
event-by-event basis provides a closer look at the \UE.  Here we refer 
to events in which there are no restrictions placed on the second highest $E_T$ jet, jet\#$2$, 
as \LJ\ events.  Our previous analysis of the \UE we only 
considered \LJ\ events \cite{Affolder:2001xt,pyA1,pyA2,cernfield}.  In this analysis we define a second class of events.  
Events with at least two jets with $|\eta({\rm jet})|<2$, where the leading two 
jets are nearly \BB\ (\delphicut) with $E_T(jet\#2)/E_T(jet\#1) > 0.8$ are referred to as 
\BB\ events.  ``Back-to-back" events are a subset of the \LJ\ events.  
The idea here is to suppress hard initial and final-state radiation thus increasing the 
sensitivity of the \TR\ region to the \BBR\ and the multiple parton 
scattering component of the \UE.  

As in our published Run 1 analysis \cite{Affolder:2001xt} we consider charged particles only in 
the region 
\ptcut\ and \etacut\ where the COT efficiency is high and compare uncorrected 
data with PYTHIA Tune A \cite{pyA1,pyA2} and HERWIG \cite{Marchesini:1988cf,Knowles:1988vs,Catani:1991rr} after detector corrections 
(\ie CDFSIM).  
Systematic errors are calculated in the same way as in our Run 1 analysis.  We 
generate every plot twice, once with our chosen track selection cuts and again with 
the very tight track cuts.  The change in each point in every plot due to this tighter 
cut is used as a measure of the systematic error and is added in quadrature with the 
statistical error to form the overall error.  

\subsection{\boldmath Transverse Average $P_T$ vs $N_{CHG}$}

\subsubsection{Definition}

We study the average transverse momentum of charged particles in the \TR\ 
region as a function of the number of charged particles in the \TR\ region 
for \ptcut\ and \etacut.  The average transverse momentum, \avePT, is formed, 
on an event-by-event basis, and then plotted as a function of the charged multiplicity.  
The idea here is to look for correlations between multiplicity and \avePT.  If, for 
example, there is a mixture of ``hard" and ``soft" events then one expects that \avePT\ 
will increase with multiplicity because demanding a large multiplicity will 
preferentially select the ``hard" process that also has a larger \avePT.  On the other 
hand, it may be possible to get a high multiplicity in a ``soft" collision so the rate 
that \avePT\ rises with multiplicity is a rough measure of the ``hard" and ``soft" mixture.  
The steeper the slope the larger the ``hard" component.  There is a very nice published 
CDF Run 1 analysis that looks at this in \MB\ collisions \cite{Acosta:2001rm}, but it has never 
previously been studied in the \TR\ region of a ``hard" scattering process.

\begin{figure}[htbp]
\begin{center}
\includegraphics[scale=0.6]{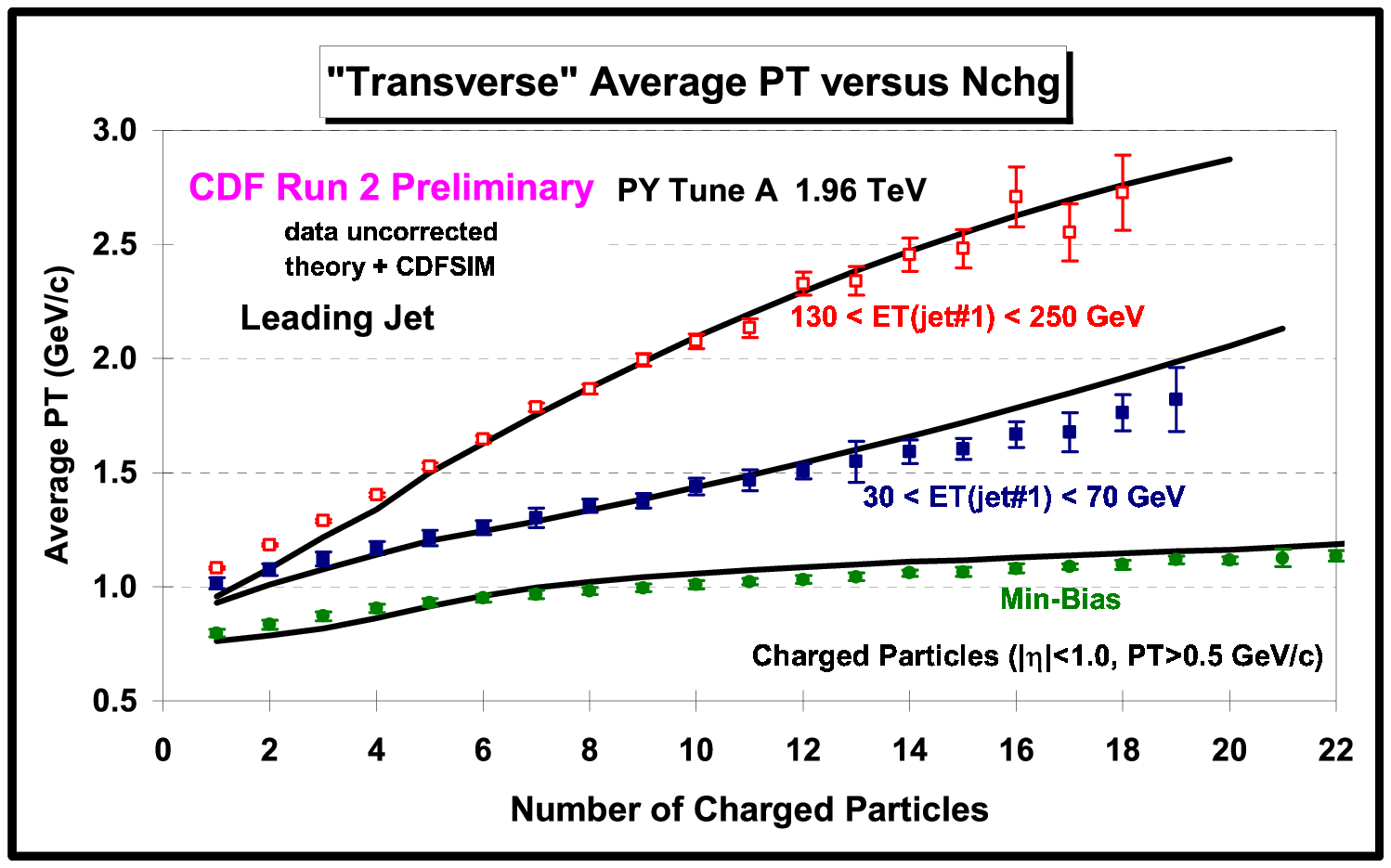}
\caption{
Run 2 data on the average transverse momentum as a function number of particles for 
charged particles with \ptcut\ and \etacut\ in the \TR\ region 
for \LJ\ events defined in Fig.~\ref{LH2003_fig4} with \binA\ and \binB.  Also shown are the 
data on the average transverse momentum as a function of the number particles for 
charged particles with \ptcut\ and \etacut\  for \MB\ collisions at $1.96\tev$.  
The theory curves correspond to PYTHIA Tune A at 1.96 TeV (after CDFSIM).
}
\end{center}
\label{LH2003_fig5}
\end{figure}

\begin{figure}[htbp]
\begin{center}
\includegraphics[scale=0.6]{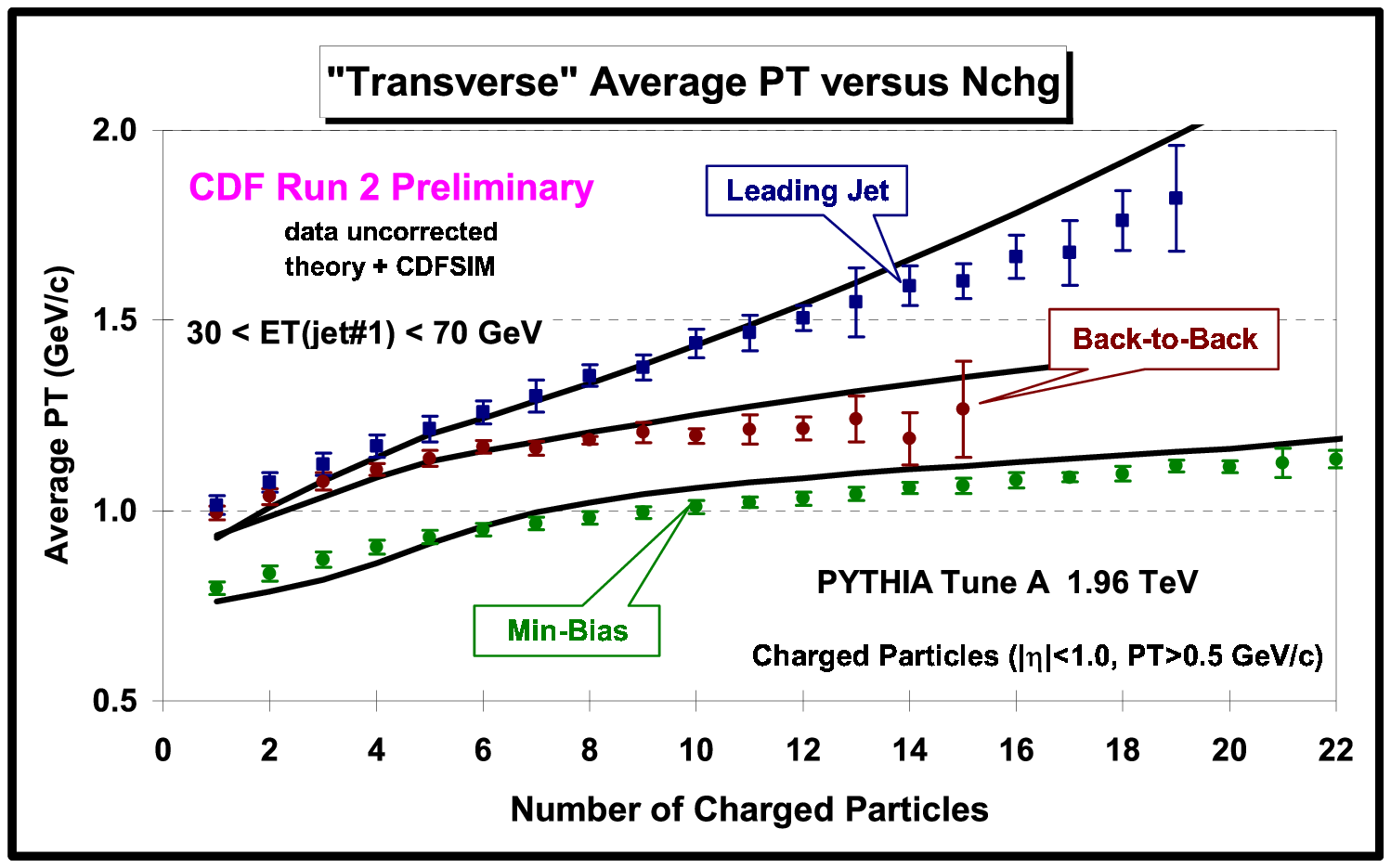}
\caption{
Run 2 data on the average transverse momentum as a function of the number of particles 
for particles for charged particles with \ptcut\ and \etacut\ in the 
\TR\ region for \LJ\ events and for \BB\ events defined 
in Fig.~\ref{LH2003_fig4} with \binA.  Also shown are the 
data on the average transverse momentum as a function of the number particles for 
charged particles with \ptcut\ and \etacut\ for \MB\ collisions at $1.96\tev$.  
The theory curves correspond to PYTHIA Tune A at 1.96 TeV (after CDFSIM).
}\label{LH2003_fig6}
\end{center}
\end{figure}

\subsubsection{Overall Transverse Region}

Fig.~\ref{LH2003_fig5} shows uncorrected Run 2 data on the \avePT\ of charged particles versus the 
number of charged particles in "min-bias" collisions and in the \TR\ region 
for  \LJ\ events with  \binA\ and \binB\ 
compared with PYTHIA Tune A (after CDFSIM).  The data suggest that there is more ``hard" 
scattering in the \TR\ region (\ie initial and final-state radiation) than there 
is in an average \MB\ collision.

Fig.~\ref{LH2003_fig6} shows the \avePT\ of charged particles versus the number of charged particles in 
\MB\ collisions and in the \TR\ region for  \LJ\ and \BB\ 
events with \binA\ compared with PYTHIA Tune A (after CDFSIM).   The 
\TR\ region for the \BB\ events looks more like an average \MB\ 
collision, which is exactly what one expects since the \BB\ requirement suppress 
hard initial and final-state radiation.

\begin{figure}[htbp]
\begin{center}
\includegraphics[scale=0.6]{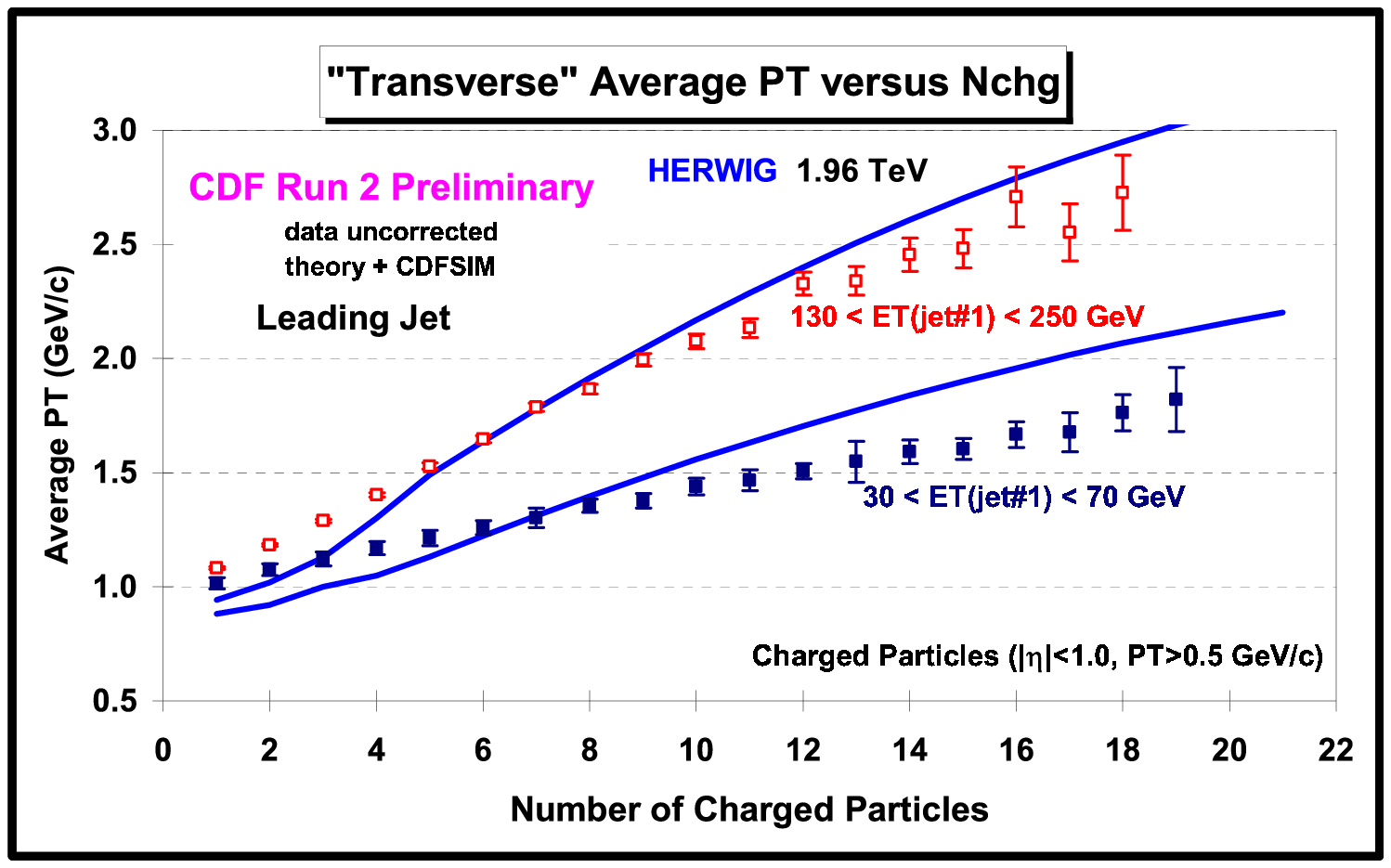}
\caption{
Run 2 data on the average transverse momentum as a function of the number of particles 
for charged particles with \ptcut\ and \etacut\  in the \TR\ region 
for \LJ\ events defined in Fig.~\ref{LH2003_fig4} with \binA\ and \binB\ compared to 
HERWIG at $1.96\tev$ (after CDFSIM).
}\label{LH2003_fig7}
\end{center}
\end{figure}

\begin{figure}[htbp]
\begin{center}
\includegraphics[scale=0.6]{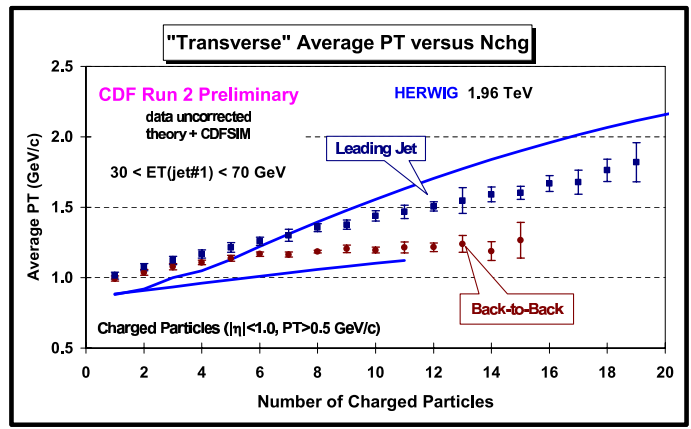}
\caption{
Run 2 data on the average transverse momentum as a function of the number of particles for 
charged particles with \ptcut\ and \etacut\ in the \TR\ region for 
\LJ\ and \BB\ events defined in Fig.~\ref{LH2003_fig4} with \binA\ compared to HERWIG at $1.96\tev$ (after CDFSIM).
}\label{LH2003_fig8}
\end{center}
\end{figure}

Fig.~\ref{LH2003_fig7} compares HERWIG (after CDFSIM) with the data on the \avePT\ of charged particles 
in the \TR\ region versus the number of charged particles in the \TR\ 
region for  \LJ\ events with  \binA\ and \binB.  

Fig.~\ref{LH2003_fig8} compares HERWIG (after CDFSIM) with the data on the \avePT\ of charged particles 
in the \TR\ region versus the number of charged particles in the \TR\ 
region for  \LJ\ and \BB\ events with \binA.  
HERWIG (without multiple parton interactions) does not describe the data as well as 
PYTHIA Tune A (with multiple parton interactions).

\begin{figure}[htbp]
\begin{center}
\includegraphics[scale=0.6]{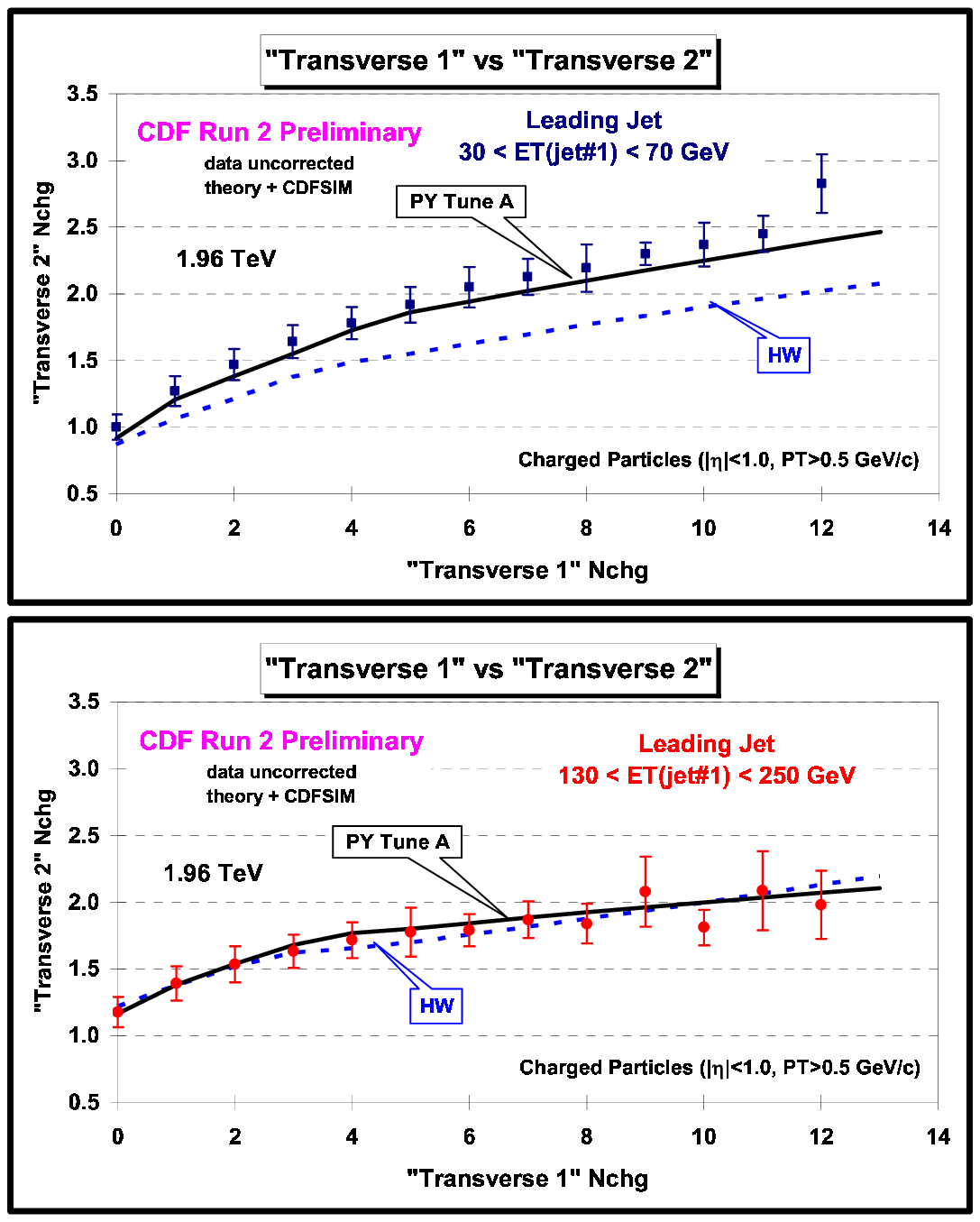}
\caption{
Run 2 data on the average number of particles in the \Ttwo\ region defined in 
Fig.~\ref{LH2003_fig4} as a function of the number of particles in the \Tone\ region for charged 
particles with \ptcut\ and \etacut\ for \LJ\ events with \binA\ ({\it top}) and  
\binB\ ({\it bottom}). The theory curves correspond to PYTHIA Tune A 
and HERWIG at $1.96\tev$ after CDFSIM.
}\label{LH2003_fig9}
\end{center}
\end{figure}

\begin{figure}[htbp]
\begin{center}
\includegraphics[scale=0.6]{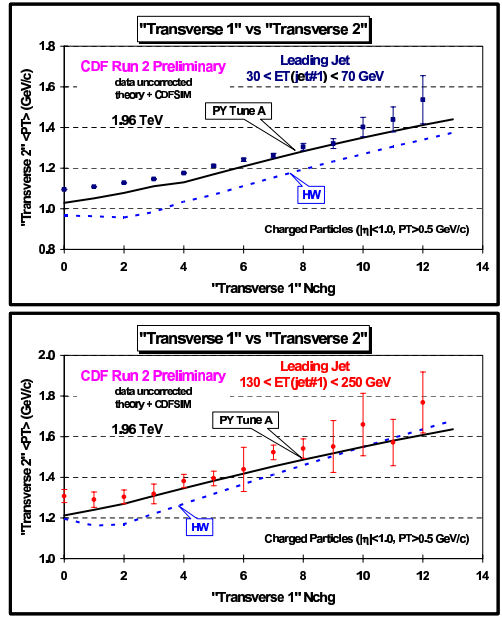}
\caption{
Run 2 data on the average transverse momentum of particles in the \Ttwo\ region 
defined in Fig.~\ref{LH2003_fig4} as a function of the number of particles in the \Tone\ region for 
charged particles with \ptcut\ and \etacut\ for \LJ\ events with \binA\ ({\it top}) and  
\binB\ ({\it bottom}). The theory curves correspond to PYTHIA Tune A 
and HERWIG at $1.96\tev$ after CDFSIM.
}\label{LH2003_fig10}
\end{center}
\end{figure}

\begin{figure}[htbp]
\begin{center}
\includegraphics[scale=0.6]{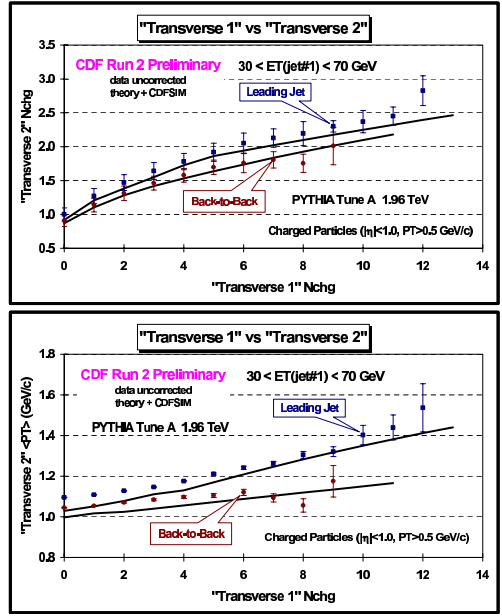}
\caption{
Run 2 data on the average number of particles ({\it top}) and the average transverse 
momentum of particles ({\it bottom}) in the \Ttwo\ region defined in Fig.~\ref{LH2003_fig4} as a 
function of the number of particles in the \Tone\ region for charged particles 
with \ptcut\ and \etacut\ for \LJ\ events and \BB\ events with \binA.   The theory curves 
correspond to PYTHIA Tune A at $1.96\tev$ (after CDFSIM).
}\label{LH2003_fig11}
\end{center}
\end{figure}

\subsubsection{Transverse 1 versus transverse 2}

Fig.~\ref{LH2003_fig9} shows the number of charged particles in the \Ttwo\ region versus 
the number of charged particles in the \Tone\ region for  \LJ\ 
events with  \binA\ and \binB\ compared with PYTHIA Tune A  and HERWIG after CDFSIM.

Fig.~\ref{LH2003_fig10} shows the \avePT\ of charged particles in the \Ttwo\ region versus 
the number of charged particles in the \Tone\ region for  \LJ\ 
events with \binA\ and \binB\ compared with PYTHIA Tune A and HERWIG after CDFSIM.  

Fig.~\ref{LH2003_fig11} shows the number and \avePT\ of charged particles in the \Ttwo\ 
region versus the number of charged particles in the \Tone\ region for  
\LJ\ and \BB\ events with \binA\ compared with PYTHIA Tune A (after CDFSIM).  

\begin{figure}[htbp]
\begin{center}
\includegraphics[scale=0.6]{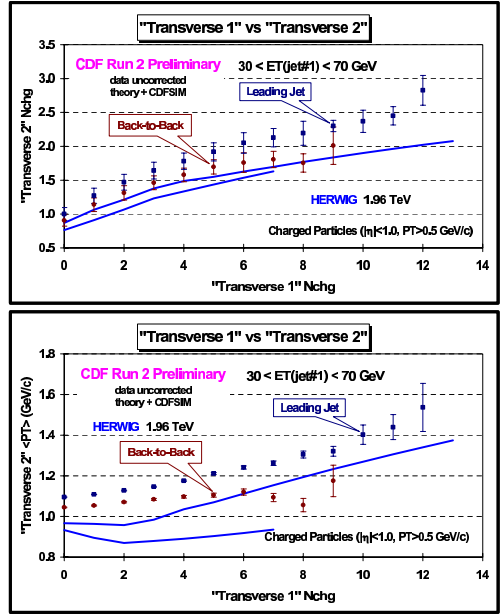}
\caption{
Run 2 data on the average number of particles ({\it top}) and the average transverse 
momentum of particles ({\it bottom}) in the \Ttwo\ region defined in Fig.~\ref{LH2003_fig4} as a 
function of the number of particles in the \Tone\ region for charged particles 
with \ptcut\ and \etacut\ for \LJ\ events and \BB\ events with \binA.   The theory curves 
correspond to HERWIG at $1.96\tev$ (after CDFSIM).
}\label{LH2003_fig12}
\end{center}
\end{figure}

Fig.~\ref{LH2003_fig12} shows the number and \avePT\ of charged particles in the \Ttwo\ region 
versus the number of charged particles in the \Tone\ region for  \LJ\ 
and \BB\ events with \binA\ compared with HERWIG (after CDFSIM).  HERWIG (without multiple
parton interactions) does not do nearly as well describing the \Ttwo\ versus
\Tone\ correlations seen in the data as does PYTHIA Tune A (with multiple parton interactions). 

\subsection{Summary}

This analysis takes a closer look at the \UE\ in hard scattering 
proton-antiproton collisions at $1.96\tev$. We look only at the charged particle 
component of the \UE\ and restrict the charged particles to be in 
the range \ptcut\ and \etacut.  We use the direction of the leading calorimeter 
jet in each event to define two \TR\ regions of \etaphi\ space that are very sensitive 
to the \UE.  Comparing these two \TR\ regions on an event-by-event 
basis provides more details about the \UE.   In addition, by selecting 
events with at least two jets that are nearly \BB\ (\delphicut) we are able to 
look closer at the \BBR\ and multiple parton interaction components of 
the \UE.  PYTHIA Tune A (with multiple parton interactions) does a good 
job in describing the \UE\ (\ie \TR\ regions) for both \LJ\ 
and \BB\ events.  HERWIG (without multiple parton interactions) does not have 
enough activity in the \UE, which was also observed in our published Run 1 
analysis \cite{Affolder:2001xt}.

The data presented here show interesting correlations between the two \TR\ 
regions defined in Fig.~\ref{LH2003_fig4}. The charged multiplicity and the \avePT\ in the \Ttwo\ 
region increases with the charged multiplicity in the \Tone\ region.  This is a 
new type of correlation.   It might simply be due to a high multiplicity in \Tone\ 
biasing in favor of a harder $2$-to-$2$ scattering (\ie higher \pthard) which would 
result in a higher multiplicity and larger \avePT\ in \Ttwo.  However, we have seen 
in previous studies \cite{Affolder:2001xt,hustontalk,tano,pyA1,pyA2,cernfield} that the average charged particle density in the \TR\ 
region does not change much as one increases $E_T({\rm jet}\#1)$.   It is possible 
that the \Tone\ versus \Ttwo\ correlations arise from multiple parton interactions.
A large multiplicity in the \Tone\ region would indicate a small impact 
parameter collision has occurred with several multiple parton scatterings which would 
then cause an increased multiplicity and \avePT\ in the \Ttwo\ region.   The fact 
that PYTHIA Tune A (with multiple parton interactions) agrees with the data better than 
HERWIG (without multiple parton interactions) is very interesting.  However, much more 
work is necessary to actually pinpoint the source of the \Tone\ 
versus \Ttwo\ correlations.

\section[Simulation of the QCD Background for $t\bar{t}$ Analyses at the Tevatron
with a $l^\pm +$ Jets Final State]{SIMULATION OF THE QCD BACKGROUND
  FOR $t\bar{t}$ ANALYSES AT THE TEVATRON 
WITH A $l^\pm +$ JETS FINAL STATE~\protect\footnote{Contributed by:
  {V.~Drollinger}}}
\label{volker_droll}

\renewcommand\labelitemi{\boldmath $\diamond$}
\renewcommand\labelitemii{\boldmath $\bullet$}
\renewcommand\labelitemiii{\boldmath $-$}

\newcommand{\bc}       {\begin{center}}
\newcommand{\ec}       {\end{center}}
\newcommand{\met}      {\mbox{$\not\!\!E_T$}}
\renewcommand{\arraystretch}{1.0}




%

\subsection{Introduction}

The top quark mass and the cross section of top quark pair production
are important measurements at the Tevatron. Compared to Run I, the
precision of these measurements is expected to be much higher in Run II
Ref.~\cite{ref_cdf2tdr}. This requires an accurate understanding of
all the important backgrounds. In the semileptonic top decay channel
with one lepton and four jets plus missing energy 
in the final state, the QCD background
is one of the main backgrounds. The cross section of this background
is large compared to the cross section of $t\bar{t}$ production, but
can be reduced strongly with appropriate selection criteria. In the
past, this background was modeled using $l^\pm + jets$ events with
non-isolated leptons coming from experimental data. 
It is however important that in $t\bar{t}$ analyses with $l^\pm
+ jets$ final states the lepton is isolated in order to
reduce backrounds with non-isolated leptons. It is not possible to get a clean
sample of QCD events with isolated leptons directly from the data
since these would be mixed with $t\bar{t}$ and $W+ jets$ events.
Therefore, in the following, we instead model the QCD background
via Monte Carlo with $b\bar{b}q\bar{q}$ and $b\bar{b}gg$  production
and subsequent leptonic $B$-decay.

\subsection{Simulation Procedure}

\vspace*{3mm}
\begin{itemize}

\item Hard Process: the hard $2 \rightarrow 4$ process is generated
  with ALPGEN Ref.~\cite{Mangano:2002ea}. The final state consists of a
  $b\bar{b}$ pair together with two other partons (light quarks or
  gluons). In the following, these events are referred to as ``bb2j''
  events. They are generated with $\sqrt{s_{p\bar{p}}} = 1960\ GeV$,
  $CTEQ5L$, and $m_b = 4.75\ GeV$. In order to produce the events in
  kinematically interesting regions, we apply the following phase
  space cuts for all four partons, including the $b$ and $\bar{b}$: 
  $p_T(j) > 10\ GeV$, $|\eta(j)| < 3$
  and $\Delta R(j,j) > 0.4$. In addition, we require at least one
  $b$-quark with a transverse momentum bigger than 30~$GeV$ in order
  to provide phase space for the leptonic $B$-decay. The bb2j cross
  section including these cuts is about 30 $nb$. 

\item Fragmentation: during the fragmentation with PYTHIA
  Ref.~\cite{Sjostrand:2000wi} a high $p_T$ electron or muon is
  produced in a leptonic $B$-decay. In order to increase the number of
  events with energetic leptons, the fragmentation is repeated exactly
  forty times. For this number of repetitions, we obtain about one
  event with a high $p_T$ lepton. A constant number of repetitions
  should not bias the $p_T$ spectrum of the lepton. Events preselected
  with a lepton with $p_T$ of at least 15 $GeV$ are written to a
  HEPEVT file. The high $p_T$ lepton requirement reduces the cross
  section to 14.5 $pb$ (288864 events, the first number in
  Table~\ref{table_parms} below), presumably with a large theoretical
  uncertainty\footnote{The cross section of bb2j is calculated at
    leading order. Additional background contributions from similar
    processes like cc2j are not included.}. 

\item CDF detector simulation: The standard CDF software package reads
  the HEPEVT file, and performs a detailed simulation of the CDF
  detector response. After the simulation, the events are
  reconstructed with the same algorithms as used to reconstruct the
  Run II data. Finally, the top working group`s standard ntuple is
  written which we examine in the following. 

\end{itemize}

\subsection{Comparison with Data}

\begin{figure}[htbp]
\centering
\includegraphics[width=\textwidth]{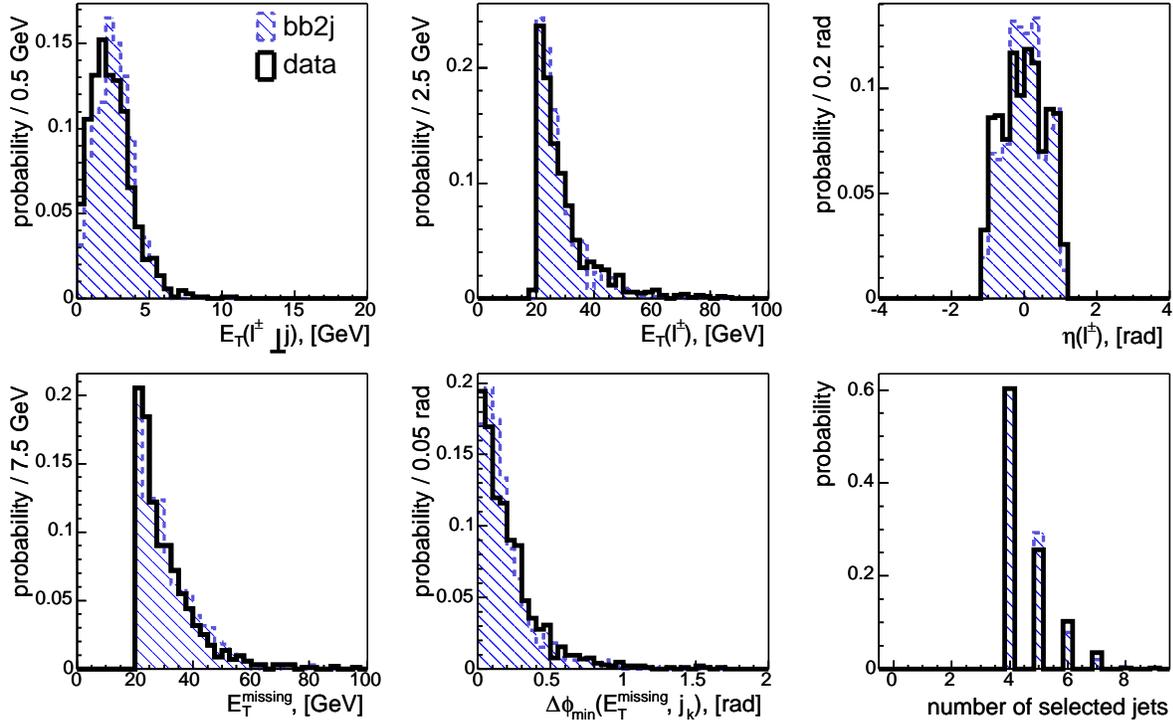}
\caption{Comparison of bb2j events with 126 $pb^{-1}$ of CDF Run II
  data: normalized distributions of $E_T$ of the lepton in respect to
  the closest jet axis, $E_T$ and $\eta$ of the lepton, missing $E_T$,
  $\Delta \phi$ between missing $E_T$ and closest jet, and
  multiplicities for all selected jets.} 
\label{fig_lept}
\end{figure}

\begin{figure}[htbp]
\centering
\includegraphics[width=\textwidth]{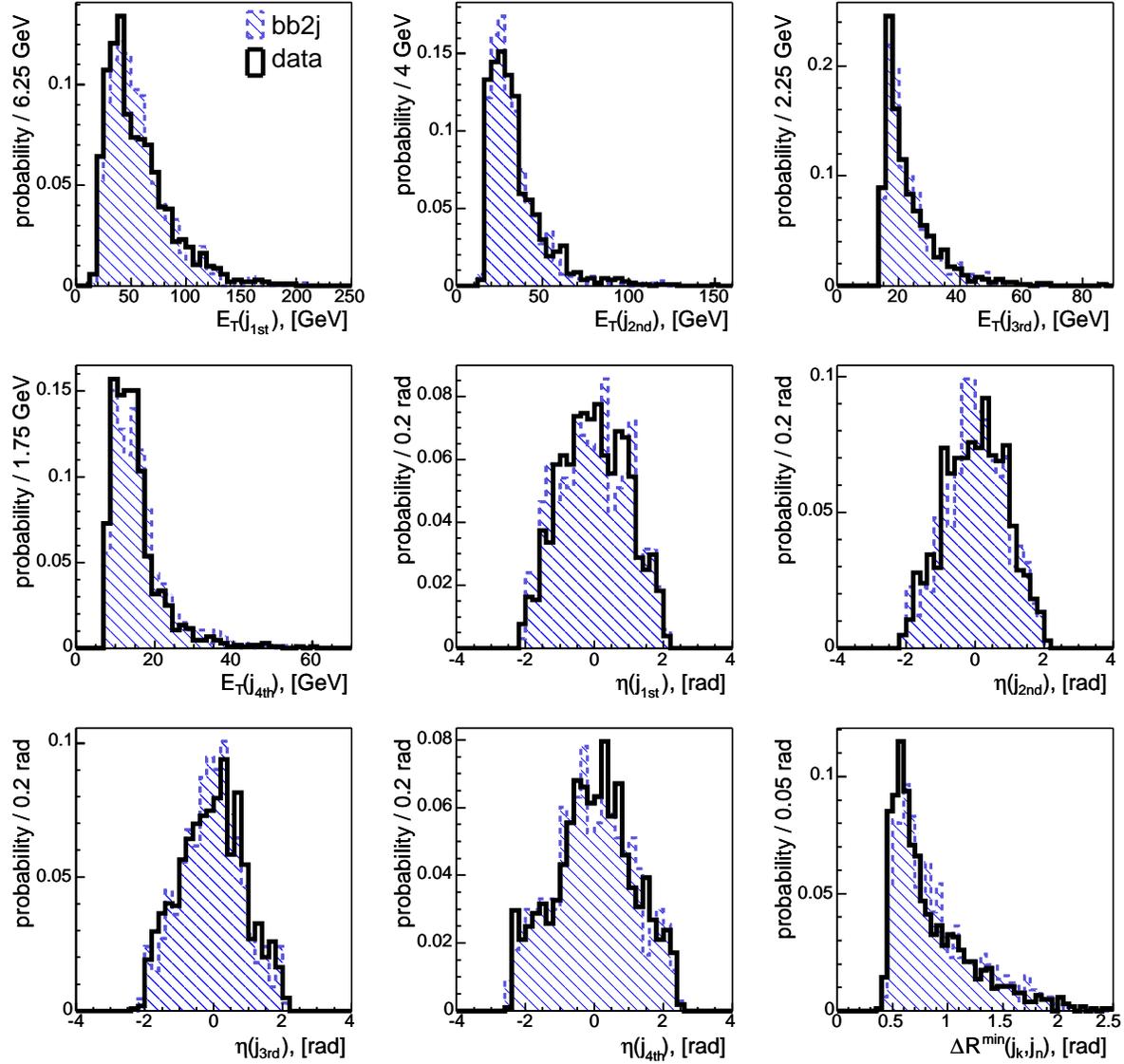}
\caption{Comparison of bb2j events with 126 $pb^{-1}$ of CDF Run II
  data: normalized $E_T$ and $\eta$ distributions of the four leading
  jets, and the $\Delta R$ distribution of the pair of closest jets to
  each other. All jet energies shown, are corrected.} 
\label{fig_jets}
\end{figure}

\vspace*{3mm}
\begin{itemize}

\item Preselection: we select events with exactly one
  lepton\footnote{We use the CDF baseline selection criteria for high
    quality lepton identification without the isolation cut. Events
    with cosmic muons or more than one lepton are rejected as well.}
  with $E_T > $ 20 $GeV$ and with missing transverse energy of at
  least 20 $GeV$. Furthermore, we ask for at least four jets with
  $|\eta(j)| < 2.4$ and $E_T(j) > 8\ GeV$. Three jets per event are
  required to be high quality jets with $|\eta(j)| < 2.0$ and $E_T(j)
  > 15\ GeV$. In addition, we remove events with more than two
  $b$-quarks in case of simulated events. This mitigates a potential
  bias\footnote{The repetition of the fragmentation would cause an
    artificial enhancement of additional $b\bar{b}$ pairs coming from
    gluon splitting with subsequent leptonic $B$-decay.} from the
  repetition of the fragmentation. In case of the data, we use good
  runs only. 

\item QCD region: in the plane of the lepton isolation
  variable $iso$ \footnote{The lepton isolation variable $iso$ is defined as the
    fractional calorimeter isolation $E_T$ in a cone of $R = $ 0.4
    around the lepton.} and the distance of the lepton to the closest
  jet $\Delta R_{min}(l^\pm,j)$ we select a region, where we expect
  almost purely QCD events in the data. In this region the lepton is
  non-isolated and is well inside a jet: $iso > 0.15$ and $\Delta
  R_{min}(l^\pm,j) < 0.15\ rad$. Isolated leptons coming from real
  $W^\pm \rightarrow l^\pm \nu$ decays, are found typically at low
  $iso$ and at any value of $\Delta R_{min}(l^\pm,j)$ which means,
  those leptons are usually isolated and also separated from jets. 

\item Comparison: in order to evaluate the simulated bb2j event
  simulation, we compare the events in the QCD region with events from
  the data which have undergone the same preselection. The data sets
  are called btop0g, btop1g, btop0j, and btop1j with the number of
  events listed in Table~\ref{table_parms}. All events correspond to
  an integrated luminosity of 126 $pb^{-1}$. In case of the simulated
  bb2j events ''all events'' is the number of events after the full
  simulation and reconstruction. We compare the most relevant
  kinematic variables in Figs.~\ref{fig_jets} and \ref{fig_lept}. 

\end{itemize}

\renewcommand{\arraystretch}{1.5}
\begin{table}[h]
\bc
\begin{tabular}{r|c|c}
  event type                &   bb2j &   data \\ \hline
  all events                & 288864 & 955870 \\
  preselection              &    846 &   1669 \\
  QCD region                &    666 &   1043 \\
\end{tabular}
\ec
\caption{Number of events after each major selection step.}
\label{table_parms}
\end{table}

\noindent
In order to validate the simulation of bb2j events, we have studied
the shapes of kinematic distributions. Whereas the jets
(Fig.~\ref{fig_jets}) are more related to the hard QCD process, the
lepton and the neutrino, seen as missing $E_T$, (Fig.~\ref{fig_lept})
depend rather on the $B$-decay and fragmentation,
respectively. However, the kinematic properties of the jets, the
lepton, and the neutrino are correlated. There are no major
differences between simulated bb2j events and the CDF data. 

\subsection{Summary}

Top physics at the Tevatron has entered the Run II phase with high
luminosity and upgraded detectors, and therefore the measurements
obtained in the top sector will reach much higher precision. This
requires a good understanding of all relevant background
processes. The use of new Monte Carlo tools enable us to simulate the
QCD background.  
 
We have described the bb2j event generation procedure. After the hard
process with four partons in the final state is generated with ALPGEN,
the high $p_T$ lepton is obtained from a leptonic $B$-decay during the
fragmentation with PYTHIA. 

We define a region, where we expect to have a clean QCD sample and
compare the fully simulated events with the CDF Run II data. All of
the kinematic variables studied, compare well. From this comparison we
conclude, that bb2j events can be used to model the QCD background
with a $l^\pm + jets$ final state.

\section*{Acknowledgments}
We would like to thank Michelangelo Mangano and Torbj\"orn Sj\"ostrand
for very useful comments and suggestions concerning the generation of
bb2j events. Many thanks to Michael Gold and Dmitri Smirnov for
helping to prepare this article.



\section[Monte-Carlo Database]{MONTE-CARLO 
DATABASE~\protect\footnote{Contributed by: {L.~Dudko and A.~Sherstnev}}}
\label{lev_dudko_mcdb}
  




\newcommand{\bi}{\begin{itemize}}
\newcommand{\ei}{\end{itemize}}
  

\subsection{Problem description}
One of the most general problems for the experimental high energy physics community
is Monte-Carlo (MC) simulation of physics processes.
There are numerous publicly available MC generators. However, 
the correct MC simulation of complicated processes requires in general
rather sophisticated expertise on the part of their users. Often, a physics group
in an experimental collaboration requests experts and/or authors of MC generators
to create MC samples for a particular process. Furthermore, it is common that
the same physics process is investigated by various physics groups
in need of the the same MC event samples. The main motiviation behind the Monte-Carlo 
Database (MCDB) project
is to make MC event samples, as prepared by experts, available for various
physics groups. In this contribution we  present a version of 
the MCDB that is already operative in the CMS collaboration,
and discuss future plans.

There are a number of useful aspects to a central MC Database that motivate
its establishment.

\begin{enumerate}
\item
Correct and reliable MC event generation of most processes of interest 
requires considerable expertise. Moreover,
most MC generators, in particular those calculating higher order
perturbative corrections, require significant amounts of computer resources. 
By means of the MCDB, samples prepared by experts can be distributed easily 
and used as many times as needed.

\item
Public availability of common event files helps speed up
the validation procedure of the events.

\item
A central and public location where well-documented
MC events and MC generators can be found would be very useful.
It would also allow rapid communication between authors of MC events and their users.

\item
The same MC samples for SM processes can be used for multiple purposes, e.g.
to study backgrounds to various new physics processes.

\item
Files containing detector and beam-related backgrounds can also be kept 
in a common location.

\end{enumerate}

\subsection{History}

The first MC Database (version 0) -- named PEVLIB~\cite{pevlib} --
was established at CERN, on the AFS file-system. This database provided 
CompHEP~\cite{Pukhov:1999gg} parton level events
for CMS users, but lacked a special interface. Rather, it was built as a set
of directories where event
samples were stored. The sample documentation consisted of ASCII files 
(README) located in the same directories as the event files.

The next version (version 0.5) of the MC Database was established at Fermilab.
This database was split in two independent parts:
\bi
\item[-] MC event files, stored via the FNAL tape system ENSTORE~\cite{enstore}
\item[-] the events' documentation, publicly available via the web~\cite{fnal_mcdb}
\ei

\noindent The latest 
version of the MCDB (version 1.0), described in this contribution, 
CMS MCDB~\cite{cms_mcdb},
was created at first for use by the CMS collaboration. This database includes web interfaces 
both for event files (enabling download and upload) and documentation. Its main goal is 
to store events, only at the parton level, generated by MC experts, for use by the LHC community.
Note that all files from PEVLIB have now been moved to the CMS MCDB.

\subsection{General concepts and practical realization of CMS MCDB.}
The MCDB must provide persistent storage of event samples, with  
public and convenient interfaces for users -- mainly consisting of LHC experimentalists
--  and authors of MC generators, or other experts.
The main features present in the CMS MCDB are:
\bi
\item the MCDB is based on web technologies.

\item the MCDB stores parton level events with a standard interface to the next 
      level of simulation (based on Les Houches Accord I).

\item the MCDB stores detailed documentation for each set of event samples.

\item the MCDB make rapid communication between users and
experts possible via its web pages.

\item the MCDB is divided in two zones:
\vspace{0.2cm}
\bi
\item[$\vartriangleright$]
a public area, for all users interested in using MC events. Users 
can find all necessary information about available event samples  and their generators,
and download the corresponding event files.

\item[$\vartriangleright$]
a restricted area for authors. In this area, authors of MC generators or 
 MC experts can change the content of MCDB dynamically 
-- e.g. upload events for new processes, create and edit documents, reply to user's 
comments and questions and upload new event files and generators.
\ei
\vspace{0.2cm}
\item the MCDB requires users to reference the authors of the event
sample in case the events are used in a physics analysis.
\ei
 
The CMS MCDB at CERN is realized as a dedicated web site~\cite{cms_mcdb}.
The structure of the site corresponds to the stored physics processes.
For example, all event files relevant for studies of top quark
production are collected in the category ``TOP''. This category contains 
files with events involving top quark production via QCD and the electroweak interaction. 
When a user clicks on a reference in
a certain category, he/she will see all so-called ''articles'' related to the selected
physics process.

The main units of CMS MCDB are HTML documents called articles that
describe the event samples. The articles are created by authors;
these are authorized users 
who can upload new event samples to the CMS MCDB. There are only a 
few restrictions on the articles
in the CMS MCDB. An author can create a new article on the basis of a template
-- this is a web form, where the author fills out various fields: author name, abstract,
category name, article body, etc. The body and the abstract of the article
may include HTML tags for more flexible visualization. 

It is easy for users to understand which process is present in the CMS MCDB,
via a click on a reference to  the article that describes the corresponding event samples.
Users can download these event files directly from the article web page,
together with all other files which authors have uploaded to the article.

The web intereface of the present version of the MCDB has
the following features, in brief:

\noindent {\bf A.} Authorized authors can
\bi 
\item upload files (with events or a MC generator code). There are two different methods
 to upload file (through  a web browser and/or directly from the AFS file system).
\item document each set of files in a new article.
\item reply to comments from users (MCDB allows one to organize a special
      forum for each article).
\item change any of the parameters and content of MCDB, according to permissions 
 (including the web design).
\ei
\noindent {\bf B.} Users can
\bi
\item[$\circ$] read documentation (MCDB articles) for files uploaded by authors.
\item[$\circ$] download files.
\item[$\circ$] search the CMS MCDB web site (enabled for articles only, not for
 the event samples themselves).
\item[$\circ$] ask questions about a particular document.
\item[$\circ$] send new articles to moderators.
\ei

Event files in the CMS MCDB are stored on the AFS filesystem~\cite{afs_link}. If a 
user has access to AFS, he/she may download files from AFS directly. 
To become an authorized author one should send a request (by e-mail) to 
the administrators \footnote{A.~De~Roeck (\texttt{Albert.de.Roeck@cern.ch}),
L.~Dudko (\texttt{Lev.Doudko@cern.ch}), S.~Slabospitsky (\texttt{Sergey.Slabospitsky@cern.ch})
and A.~Sherstnev (\texttt{Alexandre.Cherstnev@cern.ch})} of the CMS MCDB.

\subsection{Future plans}
The CMS MCDB is designed to store parton level events. This implies that the size
of event files should not be too large (typically smaller then 100Mb). 
The expected number of physics processes in the CMS MCDB is several hundred.
Note that CMS MCDB is not a SQL database, so that its search engine cannot 
serve complex queries, only keyword
phonetic searches.
These aspects of the CMS MCDB are not a limitation at 
present. However, we expect that
in a few years users will request 
a more powerful MCDB where these restrictions will be removed.
The next version of the MCDB now under development for use,
within the CERN LCG framework, by all CERN collaborations. It will
be described in the near future~\cite{lcg_note}.

\subsection*{Acknowledgements}
L.~Dudko would like to thank INTAS for their
financial support (YSF-2002-239 and RF President grant MK-1954.2003.02) 
of the MCDB project.


\clearpage

\section[Resummation and Shower Studies]{RESUMMATION AND SHOWER
  STUDIES~\protect\footnote{Contributed by: {J. Huston, I. Puljak, T. Sj\"ostrand, E. Thom\'e}}} 
\label{torbjorn_resumandshower7}



\subsection{Introduction}

The transverse momentum of a colour-singlet massive particle produced 
in a hadronic collision provides important information on perturbative 
and nonperturbative effects. A process like 
$\mathrm{q} \overline{\mathrm{q}} \to \mathrm{Z}^0$ corresponds to 
$p_{\perp\mathrm{Z}} = 0$, while higher-order processes 
provide $p_{\perp}$ kicks as the $\mathrm{Z}^0$ recoils against quarks
and gluons. At large $p_{\perp\mathrm{Z}}$ values the bulk of the
$p_{\perp}$ comes from one hard emission, and perturbation theory is a 
reasonable approach. In the small-$p_{\perp\mathrm{Z}}$ region, on the 
other hand, many emissions can contribute with $p_{\perp}$ kicks of 
comparable size, and so the order-by-order approach is rather poorly 
convergent. Furthermore, in this region nonperturbative effects may 
start to become non-negligible relative to the perturbative ones.

The traditional solution has been to apply either an analytical 
resummation approach or a numerical parton-shower one. These methods 
to some extent are complementary. The norm today is for showers to be
based on an improved leading-log picture, while resummation is carried 
out to next-to-leading logs. 
However, resummation gives no information on the partonic system 
recoiling against the $\mathrm{Z}^0$, while showers do, and therefore 
can be integrated into full-fledged event generators, allowing accurate 
experimental studies. In both approaches the high-$p_{\perp}$ tail is 
constrained by fixed-order perturbation theory, so the interesting and 
nontrivial region is the low-to-medium-$p_{\perp}$ one. Both also require 
nonperturbative input to handle the low-$p_{\perp}$ region, e.g.\ in 
the form of a primordial $k_{\perp}$ carried by the initiator of a 
shower. 

One of the disconcerting aspects of the game is that a large primordial 
$k_{\perp}$ seems to be required and that the required value of this
primordial $k_{\perp}$ can be dependent on the kinematics of the process 
being considered. Confinement of partons inside the 
proton implies a $\langle k_{\perp} \rangle \approx 0.3$~GeV, while 
fits to $\mathrm{Z}^0$ data  at the Tevatron favour $\approx 2$~GeV 
\cite{Balazs:2000sz} (actually as a root-mean-square value, assuming 
a Gaussian distribution). Also resummation approaches tend to require 
a non-negligible nonperturbative contribution,
but that contribution can be determined from fixed-target data and then
automatically evolved to the kinematical region of interest. 
In this note we present updated comparisons and study 
possible shower modifications that might alleviate the problem. We will 
use the two cases of $\mathrm{q} \overline{\mathrm{q}} \to \mathrm{Z}^0$ 
and $\mathrm{g} \mathrm{g} \to \mathrm{H}^0$ (in the infinite-top-mass
limit) to illustrate differences in quark and gluon evolution, and
the Tevatron and the LHC to quantify an energy dependence.

\subsection{Comparison Status}

A detailed comparison of analytic resummation and parton showers 
was presented in \cite{Balazs:2000sz}. For many physical quantities, 
the predictions from parton shower Monte Carlo programs should be nearly
as precise as those from analytical theoretical calculations. In 
particular, both analytic and parton shower Monte Carlos should accurately 
describe the effects of the emission of multiple soft gluons from the 
incoming partons. 

Parton showers resum primarily the leading logs, which are universal, i.e.
process-independent, depending only on the initial state. An analytic 
resummmation calculation, in principle, can resum all logs, but in practice
the number of towers of logarithms included in the analytic Sudakov 
exponent depends on the level to which a fixed-order calculation was 
performed for a given process. Generally, if a NNLO calculation is 
available, then the $B^{(2)}$ coefficient (using the CSS formalism
\cite{Collins:1981uk}) can be extracted and incorporated. If we try 
to interpret parton showering in the same language then we say that 
the Monte Carlo Sudakov exponent always contains a term analogous to 
$A^{(1)}$ and $B^{(1)}$ and that an approximation to $A^{(2)}$ is also 
present in some kinematic regions. 

In Ref.~ \cite{Balazs:2000sz}, predictions were made for both 
$\mathrm{Z}^0$ and Higgs production at the Tevatron and the  LHC, 
using both resummation and parton shower Monte Carlo programs. In 
general, the shapes for the $p_{\perp}$ distributions agreed well, 
although the \textsc{Pythia} showering algorithm typically caused 
the Higgs $p_{\perp}$ distribution to peak at somewhat lower values 
of transverse momentum. 

\subsection{Shower Algorithm Constraints}

While customarily classified as leading log, shower algorithms
tend to contain elements that go beyond the conventional leading-log
definition. Specifically, some emissions allowed by leading log
are forbidden in the shower description. Taking the \textsc{Pythia}
\cite{Sjostrand:2000wi,Sjostrand:2003wg} initial-state shower algorithm 
\cite{Sjostrand:1985xi,Bengtsson:1986gz,Miu:1998ju} as an example,
the following aspects may be noted (see \cite{ptgeninprep} for
further details):\\
\textit{(i)} Emissions are required to be angularly ordered, such that 
opening angles increase on the way in to the hard scattering subprocess.
That is, non-angularly-ordered emissions are vetoed.\\
\textit{(ii)} The $z$ and $Q^2$ of a branching $a \to b c$ are required 
to fulfill the condition $\hat{u} = Q^2 - \hat{s}(1-z) < 0$. Here 
$\hat{s} = (p_a + p_d)^2 = (p_b + p_d)^2/z$, for $d$ the incoming parton 
on the other side of the event. In the case that $b$ and $d$ form a 
$\mathrm{Z}^0$, say, and $c$ is the recoiling parton, $\hat{u}$ 
coincides with the standard Mandelstam variable for the 
$a + d \to (\mathrm{Z}^0=b+d) + c$ process. In general, it may be 
viewed as a kinematics consistency constraint.\\
\textit{(iii)} The evolution rate is proportional to
$\alpha_{\mathrm{s}}((1-z) Q^2) \approx 
\alpha_{\mathrm{s}}(p_{\perp}^2)$ rather than 
$\alpha_{\mathrm{s}}(Q^2)$. Since $p_{\perp}^2 < Q^2$ this 
implies by itself a larger $\alpha_{\mathrm{s}}$ and thus an increased rate of 
evolution. However, one function of the $Q_0 \approx 1$~GeV 
nonperturbative cutoff parameter is to avoid the 
divergent-$\alpha_{\mathrm{s}}$ region, so now one must require
$(1-z) Q^2 > Q_0^2$ rather than $Q^2 > Q_0^2$. The net result again
is a reduced emission rate.\\
\textit{(iv)} One of the partons of a branching may develop a timelike
parton shower. The more off-shell this parton, the less the $p_{\perp}$
of the branching. The evolution rate in $x$ is unaffected, however.\\
\textit{(v)} There are some further corrections, that in practice appear
to have negligible influence: the non-generation of very soft gluons to 
avoid the divergence of the splitting kernel, the possibility of photon
emission off quarks, and extra kinematical constraints when heavy quarks 
are produced.\\
\textit{(vi)} The emission rate is smoothly merged with the first-order 
matrix elements at large $p_{\perp}$. This is somewhat separate from
the other issues studied, and the resulting change only appreciably 
affects a small fraction of the total cross section, so it will not be 
considered further here.

The main consequence of the first three points is a lower rate of $x$
evolution. That is, starting from a set of parton densities 
$f_i(x, Q_0^2)$ at some low $Q_0^2$ scale, and a matching $\Lambda$, 
tuned such that standard DGLAP evolution provides a reasonable fit to 
data at $Q^2 > Q_0^2$, the constraints above lead to $x$ distributions 
less evolved and thus harder than data. If we e.g.\ take the CTEQ5L tune 
\cite{Lai:1999wy} with $\Lambda^{(4)} = 0.192$~GeV, the $\Lambda^{(4)}$ 
would need to be raised to about 0.23~GeV in the shower to give the same 
fit to data as CTEQ5L when the angular-ordering cut in \textit{(i)} is 
imposed. Unfortunately effects from points \textit{(ii)} and  
\textit{(iii)} turn out to be process-dependent, presumably reflecting 
kinematical differences between $\mathrm{q}\to\mathrm{q}\mathrm{g}$ and
$\mathrm{g} \to \mathrm{g}\mathrm{g}$. There is also some energy 
dependence. The net result of the first three points suggests that 
\textsc{Pythia} should be run with a $\Lambda^{(4)}$ of about 0.3~GeV
(0.5~GeV) for $\mathrm{Z}^0$ ($\mathrm{H}^0$) production in order to 
compensate for the restrictions on allowed branchings.

One would expect the increased perturbative evolution to allow the 
primordial $k_{\perp}$ to be reduced. Unfortunately, while the total
radiated transverse energy, $\sum_i |\mathbf{p}_{\perp i}|$, comes up 
by about 10\% at the Tevatron, this partly cancels in the vector sum, 
$\mathbf{p}_{\perp\mathrm{Z}} = - \sum_i \mathbf{p}_{\perp i}$. For a
2~GeV primordial $k_{\perp}$ the shift of the peak position of the 
$p_{\perp\mathrm{Z}}$ spectrum is negligible. Results are more visible 
for $p_{\perp\mathrm{H}}$ at the LHC.

Note that a primordial $k_{\perp}$ assigned to the initial parton at
the low $Q_0^2$ scale is shared between the partons at each shower 
branching, in proportion to the longitudinal momentum fractions a 
daughter takes. Only a fraction $x_{\mathrm{hard}}/x_{\mathrm{initial}}$
of the initial $k_{\perp}$ thus survives to the hard-scattering parton.
Since the typical $x$ evolution range is much larger at the LHC than at
the Tevatron, a tuning of the primordial $k_{\perp}$ is hardly an option
for $\mathrm{H}^0$ at the LHC, while it is relevant for $\mathrm{Z}^0$ at 
the Tevatron. Therefore an increased $\Lambda$ value is an interesting 
option. 
 
We now turn to the point \textit{(iv)} above. By coherence 
arguments, the main chain of spacelike branchings sets the maximum
virtuality for the emitted timelike partons, i.e.\ the timelike 
branchings occur at longer timescales than the related spacelike ones. 
In a dipole-motivated language, one could therefore imagine that the 
recoil, when a parton acquires a timelike mass, is not taken by a 
spacelike parton but by other final-state colour-connected partons. A 
colour-singlet particle, like the $\mathrm{Z}^0$ or $\mathrm{H}^0$, 
would then be unaffected by the timelike showers. 
    
The consequences for $p_{\perp\mathrm{Z}}$ and $p_{\perp\mathrm{H}}$
of such a point of view can be studied by switching off timelike showers 
in \textsc{Pythia}, but there is then no possibility to fully simulate 
the recoiling event. A new set of shower routines is in preparation 
\cite{newshowerhere}, however, based on $p_{\perp}$-ordered emissions 
in a hybrid between conventional showers and the dipole approach. It is 
well suited for allowing final-state radiation at later times, leaving 
$p_{\perp\mathrm{Z}}$ and $p_{\perp\mathrm{H}}$ unaffected at that 
stage. Actually, without final-state radiation, the two approaches give 
surprisingly similar results overall. Both are lower in the peak region 
than the algorithm with final-state radiation, in better agreement with 
CDF data \cite{Affolder:1999jh}. The new one is slightly lower, i.e.\ 
better relative to data, at small $p_{\perp\mathrm{Z}}$ values. 

\begin{figure}[htbp]
\begin{center}
\vspace*{-5mm}
\includegraphics[width=12cm]{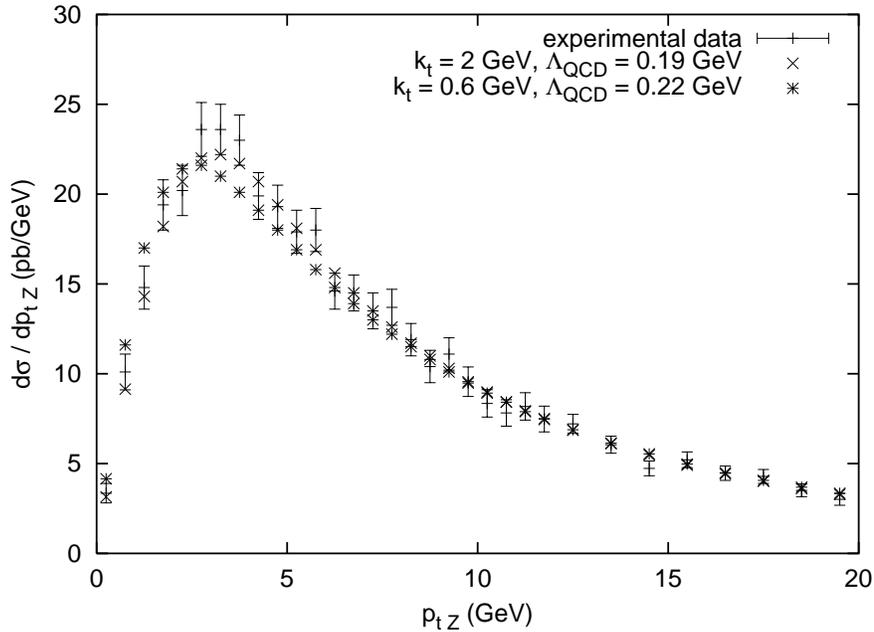}
\vspace*{-5mm}
\end{center}
\caption{Comparison of the CDF $p_{\perp\mathrm{Z}}$ spectrum with the 
new shower algorithm for two parameter sets.}
\label{resumandshower:figa}
\end{figure}

A combined study \cite{ptgeninprep}, leaving both the primordial 
$k_{\perp}$ and the $\Lambda$ value free, still gives some preference 
to $\langle k_{\perp} \rangle = 2$~GeV and the standard 
$\Lambda^{(4)} = 0.192$~GeV, but differences relative 
to an alternative with $\langle k_{\perp} \rangle = 0.6$~GeV 
and $\Lambda^{(4)} = 0.22$~GeV are not particularly large, 
Fig.~\ref{resumandshower:figa}. 

\subsection{Further Comparisons}

\begin{figure}[htbp]
\begin{center}
\vspace*{-5mm}
\includegraphics[width=12cm]{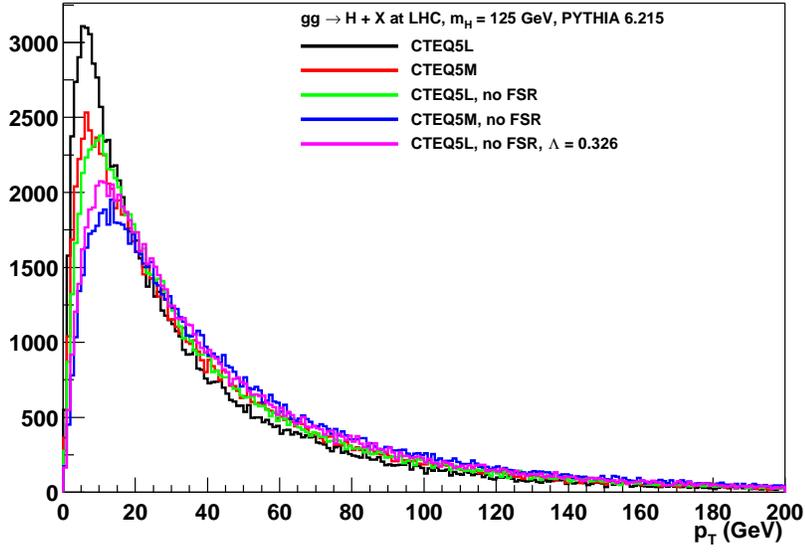}
\vspace*{-5mm}
\end{center}
\caption{Comparison of the \textsc{Pythia} $p_{\perp}$ distributions for 
Higgs production at the LHC using LO and NLO pdf's, with and without (no FSR) timelike 
parton showering.}  
\label{resumandshower:figb}
\end{figure}

Returning to Higgs production at the LHC, in Fig.~\ref{resumandshower:figb} 
are shown a number of predictions for the current standard \textsc{Pythia} 
shower routines. Using CTEQ5M rather than CTEQ5L results in more  gluon 
radiation and a broader $p_{\perp}$ distribution due to the 
large value of $\Lambda$. Likewise turning off timelike showers for gluons
radiated from the initial state also results in the peak of the $p_{\perp}$ 
distribution moving outwards. 

\begin{figure}[htbp]
\begin{center}
\vspace*{-3mm}
\includegraphics[width=12cm]{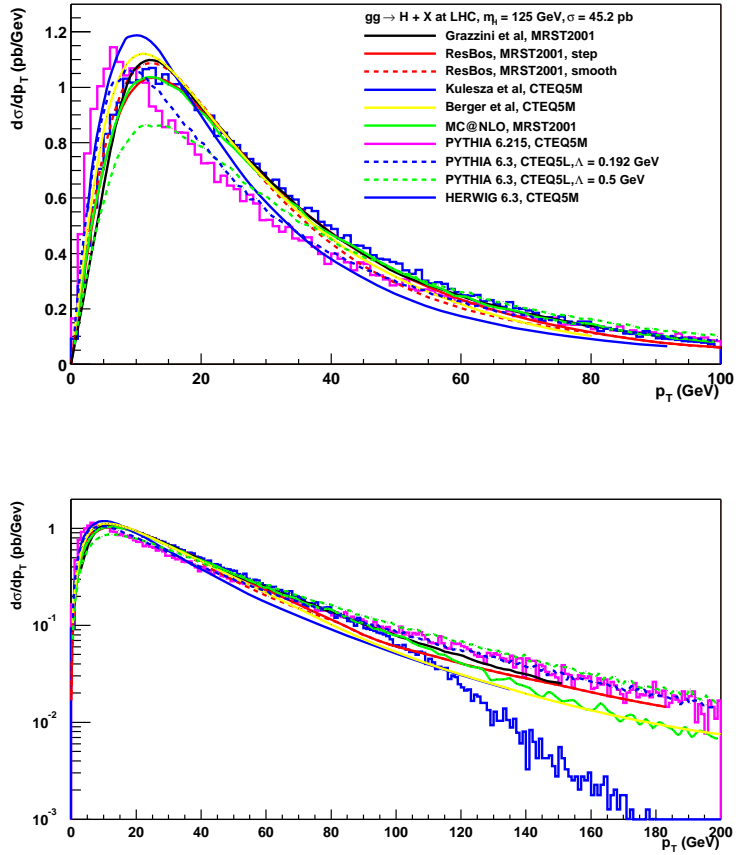}
\vspace*{-6mm}
\end{center}
\caption{Comparison of various $p_{\perp}$ distributions for Higgs 
production at the LHC. The curves denoted Grazzini \cite{Bozzi:2003jy}, 
ResBos \cite{Balazs:2000wv}, Kulesza \cite{Kulesza:2003wn} and Berger 
\cite{Berger:2002ut} are resummation descriptions, while 
MC\@NLO \cite{Frixione:2002ik,Frixione:2003ei},
\textsc{Herwig} \cite{Corcella:2000bw} and \textsc{Pythia} are 
generators, \textsc{Pythia} 6.3 refers to the new algorithm 
outlined above. }  
\label{resumandshower:figc}
\end{figure}

We can now compare the results with resummation descriptions and other
generators, Fig.~\ref{resumandshower:figc} \cite{higgshere}. As we see, 
the new \textsc{Pythia} routines agree  better with resummation 
descriptions than in the past \cite{Balazs:2000sz}, attesting to the 
importance of various minor technical details of the Monte Carlo approach. 
One must note, however, that some spread remains, and that it is not 
currently possible to give an unambiguous prediction.

\subsection{Conclusions}

We have studied $p_{\perp\mathrm{Z}}$ and $p_{\perp\mathrm{H}}$ spectra,
as a way of exploring perturbative and nonperturbative effects in hadronic
physics. Specifically, we have pointed out a number of ambiguities that
can exist in a shower approach, e.g.\ that the shower goes beyond the
simpleminded leading-log evolution and kinematics, while still making use 
of leading-log parton densities. Attempts to correct for mismatches in
general tend to increase the perturbative $p_{\perp\mathrm{Z}}$.
The need for an unseemly large primordial $k_{\perp}$ in the shower 
approach is thus reduced, but not eliminated. There is still room for,
possibly even a need of, perturbative evolution beyond standard DGLAP at 
small virtuality scales.  



\section[New Showers with Transverse-Momentum-Ordering]{NEW SHOWERS
  WITH TRANSVERSE-MOMENTUM-ORDERING~\protect\footnote{Contributed by: {T. Sj\"ostrand}}} 
\label{torbjorn_shower3}



\subsection{Introduction}

The initial- \cite{Sjostrand:1985xi,Bengtsson:1986gz,Miu:1998ju} 
and final-state \cite{Bengtsson:1987et,Norrbin:2000uu} showers 
in the \textsc{Pythia} event generator 
\cite{Sjostrand:2000wi,Sjostrand:2003wg} are based on 
virtuality-ordering, i.e.\ uses spacelike $Q^2$ and timelike $M^2$, 
respectively, as evolution variables. Other algorithms in common 
use are the angular-ordered ones in \textsc{Herwig} 
\cite{Marchesini:1984bm,Corcella:2000bw} and the 
$p_{\perp}$-ordered dipole-based ones in \textsc{Ariadne/Ldc}
\cite{Lonnblad:1992tz,Kharraziha:1998dn}. All three have been 
comparably successful, in terms of ability to predict or describe 
data, and therefore have offered useful cross-checks. Some 
shortcomings of the virtuality-ordering approach, with respect to
coherence conditions, have been compensated (especially relative to
\textsc{Herwig}) by a better coverage of phase space and more 
efficient possibilities to merge smoothly with first-order matrix 
elements.

Recently, the possibility to combine matrix elements of several 
orders consistently with showers has been raised 
\cite{Catani:2001cc,Lonnblad:2001iq}, e.g.\ $\mathrm{W} + n$ jets,
$n= 0, 1, 2, 3, \ldots$. In such cases, a $p_{\perp}$-ordering 
presumably offers the best chance to provide a sensible definition 
of hardness. It may also tie in better e.g.\ with the 
$p_{\perp}$-ordered approach to multiple interactions 
\cite{multinthere}. This note therefore is a study of
how the existing \textsc{Pythia} algorithms can be reformulated 
in $p_{\perp}$-ordered terms, while retaining their strong points. 

The main trick that will be employed is to pick formal definitions of 
$p_{\perp}$, that simply and unambiguously can be translated into 
the older virtuality variables, e.g.\ for standard matrix-element 
merging. These definitions are based on lightcone kinematics, wherein 
a timelike branching into two massless daughters corresponds to 
$p_{\perp}^2 = z(1-z)M^2$ and the branching of a massless mother into 
a spacelike and a massless daughter to $p_{\perp}^2 = (1-z)Q^2$.
The actual $p_{\perp}$ of a branching will be different, and e.g.\
depend on the subsequent shower history, but should normally not 
deviate by much.

\subsection{Timelike Showers}

The new timelike algorithm is a hybrid between the traditional 
parton-shower and dipole-emission approaches, in the sense that 
the branching process is associated with the evolution of a single 
parton, like in a shower, but recoil effects occur inside dipoles. 
That is, a dipole partner is assigned for each branching, and energy 
and momentum is `borrowed' from this partner to give mass to the 
parton about to branch, while preserving the invariant mass of the 
dipole. (Thus four-momentum is not preserved locally for each
parton branching $a \to b c$. It was in the old algorithm, where
the kinematics of a branching was not constructed before the
off- or on-shell daughter masses had been found.)
Often the two partners are colour-connected, i.e.\ the 
colour of one matches the anticolour of the other, as defined by
the preceding showering history, but this need not be the case. 
In particular, intermediate resonances normally have masses that
should be preserved by the shower, e.g., in 
$\mathrm{t}\to\mathrm{b}\mathrm{W}^+$ the $\mathrm{W}^+$ 
takes the recoil when the $\mathrm{b}$ radiates a gluon.

The evolution variable is approximately the $p_{\perp}^2$ of a branching,
where $p_{\perp}$ is the transverse momentum for each of the two daughters 
with respect to the direction of the mother, in the rest frame of the 
dipole. (The recoiling dipole partner does not obtain any $p_{\perp}$ 
kick in this frame; only its longitudinal momentum is affected.) For 
the simple case of massless radiating partons and small virtualities 
relative to the kinematically possible ones, and in the limit that 
recoil effects from further emissions can be neglected, it agrees with
the $d_{ij}$ $p_{\perp}$-clustering distance defined in the 
\texttt{PYCLUS} algorithm \cite{Moretti:1998qx}.

All emissions are ordered in a single sequence $p_{\perp\mathrm{max}} > 
p_{\perp 1} > p_{\perp 2} > \ldots > p_{\perp\mathrm{min}}$. That is, 
each initial parton is evolved from the input $p_{\perp\mathrm{max}}$ 
scale downwards, and a hypothetical  branching $p_{\perp}$ is thereby 
found for it. The one with the largest $p_{\perp}$ is chosen to undergo 
the first actual branching. Thereafter, all partons now existing are 
evolved downwards from $p_{\perp 1}$, and a $p_{\perp 2}$ is chosen, 
and so on, until $p_{\perp\mathrm{min}}$ is reached. (Technically, the 
$p_{\perp}$ values for partons not directly or indirectly affected by 
a branching need not be reselected.) The evolution of a gluon is  
split in evolution on two separate sides, with half the branching
kernel each, but with different kinematical constraints since the
two dipoles have different masses. The evolution of a quark is also
split, into one $p_{\perp}$ scale for gluon emission and one for photon 
one, in general corresponding to different dipoles. 

With the choices above, the evolution factorizes. That is, a set of 
successive calls, where the $p_{\perp\mathrm{min}}$ of one call becomes 
the $p_{\perp\mathrm{max}}$ of the next, gives the same result (on the 
average) as one single call for the full $p_{\perp}$ range. This is the 
key element to allow Sudakovs to be conveniently obtained from trial 
showers \cite{Lonnblad:2001iq}, and to veto emissions above some 
$p_{\perp}$ scale, as required to combine different $n$-parton 
configurations efficiently.

The formal $p_{\perp}$ definition is
$p_{\perp\mathrm{evol}}^2 = z(1-z)(M^2 - m_0^2)$,
where $p_{\perp\mathrm{evol}}$ is the evolution variable, $z$ gives the 
energy sharing in the branching, as selected from the branching 
kernels, $M$ is the off-shell mass of the branching parton and 
$m_0$ its on-shell value. This $p_{\perp\mathrm{evol}}$ is also used as
$\alpha_{\mathrm{s}}$ scale.

When a $p_{\perp\mathrm{evol}}$ has been selected, this is translated 
to a $M^2 = m_0^2 + p_{\perp\mathrm{evol}}^2/(z(1-z))$. Note that the 
Jacobian factor is trivial:
$\mathrm{d}M^2/(M^2 - m_0^2) \; \mathrm{d}z = \mathrm{d}%
p_{\perp\mathrm{evol}}^2/p_{\perp\mathrm{evol}}^2 \; \mathrm{d}z$.  From 
there on, the three-body kinematics of a 
branching is  constructed as in the old routine. This includes the 
detailed interpretation of $z$ and the related handling of nonzero 
on-shell masses for branching and recoiling partons, which leads to the 
physical $p_{\perp}$ not agreeing with the $p_{\perp\mathrm{evol}}$ 
defined here. In this sense, $p_{\perp\mathrm{evol}}$ becomes a formal 
variable, while $M$ really is a well-defined mass of a parton.

Also the corrections to $b\to b\mathrm{g}$ branchings ($b$ being a 
generic coloured particle) by merging with first-order 
$a\to bc\mathrm{g}$ matrix elements closely follows the existing 
machinery \cite{Norrbin:2000uu}, once the $p_{\perp\mathrm{evol}}$ has 
been converted to a mass of the branching parton. In general, the other 
parton $c$ used to define the matrix element need not be the same as 
the recoiling partner. To illustrate, consider a 
$\mathrm{Z}^0 \to \mathrm{q}\overline{\mathrm{q}}$ decay. Say the 
$\mathrm{q}$ branches first, $\mathrm{q} \to \mathrm{q}\mathrm{g}_1$. 
Obviously the $\overline{\mathrm{q}}$ then takes the recoil, and the new 
$\mathrm{q}$, $\mathrm{g}_1$ and $\overline{\mathrm{q}}$ momenta are used 
to match to the $\mathrm{Z}^0 \to \mathrm{q}\overline{\mathrm{q}}\mathrm{g}$ 
matrix element. The next time $\mathrm{q}$ branches, 
$\mathrm{q} \to \mathrm{q}\mathrm{g}_2$, the recoil is taken by 
the colour-connected $\mathrm{g}_1$ gluon, but the matrix element 
corrections are based on the newly created $\mathrm{q}$ and $\mathrm{g}_2$ 
momenta together with the $\overline{\mathrm{q}}$ (not the 
$\mathrm{g}_1$!) momentum. That way one may expect to achieve the most 
realistic description of mass effects in the collinear and soft regions.  

The shower inherits some further elements from the old algorithm, such as 
azimuthal anisotropies in gluon branchings from polarization effects.

The relevant parameters will have to be retuned, since the shower is 
quite different from the old mass-ordered one. In particular, it appears 
that the five-flavour $\Lambda_{\mathrm{QCD}}$ value has to be reduced 
relative to the current default, roughly by a factor of two (from 0.29 
to 0.14~GeV). 

\subsection{Spacelike Showers}

Initial-state showers are constructed by backwards evolution
\cite{Sjostrand:1985xi}, starting at the hard interaction and 
successively reconstructing preceding branchings. To simplify 
the merging with first-order matrix elements, $z$ is defined by 
the ratio of $\hat{s}$ before and after an emission. For a massless 
parton branching into one spacelike with virtuality $Q^2$ and 
one with mass $m$, this gives
$p_{\perp}^2 = Q^2 - z (\hat{s} + Q^2)(Q^2 + m^2)/\hat{s}$, or
$p_{\perp}^2 = (1-z) Q^2 - z Q^4/\hat{s}$ for $m=0$.
Here $\hat{s}$ is the squared invariant mass after the emission, 
i.e.\ excluding the emitted on-mass-shell parton. 

The last term, $z Q^4/\hat{s}$, while normally expected to be small, 
gives a nontrivial relationship between $p_{\perp}^2$ and $Q^2$, 
e.g.\ with two possible $Q^2$  solutions for a given $p_{\perp}^2$.
To avoid the resulting technical problems, the evolution variable 
is picked to be $p_{\perp\mathrm{evol}}^2 = (1-z) Q^2$. Also here 
$p_{\perp\mathrm{evol}}$ sets the scale for the running 
$\alpha_{\mathrm{s}}$. Once selected, the $p_{\perp\mathrm{evol}}^2$
is translated into an actual $Q^2$ by the inverse relation
$Q^2 = p_{\perp\mathrm{evol}}^2/(1-z)$, with trivial Jacobian:
$\mathrm{d}Q^2/Q^2 \; \mathrm{d}z = \mathrm{d}%
p_{\perp\mathrm{evol}}^2/p_{\perp\mathrm{evol}}^2 \; \mathrm{d}z$. From 
$Q^2$ the correct $p_{\perp}^2$, including the $z Q^4/\hat{s}$
term, can be constructed.

Emissions on the two incoming sides are interspersed to form a single
falling $p_{\perp}$ sequence, $p_{\perp\mathrm{max}} > 
p_{\perp 1} > p_{\perp 2} > \ldots > p_{\perp\mathrm{min}}$. 
That is, the $p_{\perp}$ of the latest branching considered sets 
the starting scale of the downwards evolution on both sides, 
with the next branching occurring at the side that gives the 
largest such evolved $p_{\perp}$. 

In a branching $a \to b c$, the newly reconstructed mother $a$ is 
assumed to have vanishing mass --- a heavy quark would have to be 
virtual to exist inside a proton, so it makes no sense to put it on 
mass shell. The previous mother $b$, which used to be massless, now 
acquires the spacelike virtuality $Q^2$ and the correct $p_{\perp}$ 
previously mentioned, and kinematics has to be adjusted accordingly. 

In the old algorithm, the $b$ kinematics was not constructed until 
its spacelike virtuality had been set, and so four-momentum was 
explicitly conserved at each shower branching. In the new algorithm,
this is no longer the case. (A corresponding change occurs between 
the old and new timelike showers, as noted above.) Instead it is the 
set of partons produced by this mother $b$ and the current mother $d$ 
on the other side of the event that collectively acquire the 
$p_{\perp}$ of the new $a \to b c$ branching. Explicitly, when the 
$b$ is pushed off-shell, the $d$ four-momentum is modified accordingly, 
such that their invariant mass is retained. Thereafter a set of 
rotations and boosts of the whole $b+d$-produced system bring them 
to the frame where $b$ has the desired $p_{\perp}$ and $d$ is restored 
to its correct four-momentum. 

Matrix-element corrections can be applied to the first, i.e.\ hardest
in $p_{\perp}$, branching on both sides of the event, to improve the 
accuracy of the high-$p_{\perp}$ description. Also several other aspects 
are directly inherited from the old algorithm. 

Work on the algorithm is ongoing. In particular, an optimal description
of kinematics for massive quarks in the shower, i.e.\ $\mathrm{c}$ and 
$\mathrm{b}$ quarks, remains to be worked out.

Some first tests of the algorithm are reported elsewhere 
\cite{showerpthere}. In general, its behaviour appears rather similar
to that of the old algorithm.

\subsection{Outlook} 

The algorithms introduced above are still in a development stage. 
In particular, it remains to combine the two. One possibility would be to 
construct the spacelike shower first, thereby providing a list of 
emitted partons with their respective emission $p_{\perp}$ scales. 
This list would then be used as input for the timelike shower, where 
each emission $p_{\perp}$ sets the upper evolution scale of the 
respective parton. This is straightforward, but does not allow a fully 
factorized evolution, i.e.\ it is not feasible to stop the evolution 
at some $p_{\perp}$ value and continue downwards from there in a 
subsequent call. The alternative would be to intersperse spacelike 
and timelike branchings, in one common $p_{\perp}$-ordered sequence.

Obviously the finished algorithms have to be compared with data,
to understand how well they do. One should not expect any major 
upheavals, since checks show that they perform similarly 
to the old ones at current energies, but the hope is for a somewhat
improved and more consistent description. The step thereafter would 
be to study specific processes, such as $\mathrm{W} + n$ jets, to find 
how good a matching can be obtained between the different $n$-jet 
multiplicities, when initial parton configurations are classified by 
their $p_{\perp}$-clustering properties. The \texttt{PYCLUS} algorithm 
here needs to be extended to cluster also beam jets. Since one cannot 
expect a perfect match between generated and clustering-reconstructed 
shower histories, it may become necessary to allow trial showers and 
vetoed showers over some $p_{\perp}$ matching range, but hopefully then 
a rather small one. If successful, one may expect these new algorithms 
to become standard tools for LHC physics studies in the years to come. 



\section[Matching Matrix Elements and Parton Showers
with HERWIG and PYTHIA]{MATCHING MATRIX ELEMENTS AND PARTON SHOWERS
WITH HERWIG AND PYTHIA~\protect\footnote{Contributed by:
  {S.~Mrenna}}} 
\label{mrenna}

\newcommand{\CK}{{\sf CKKW}}
\newcommand{\PY}{{\sf PYTHIA}}
\newcommand{\HW}{{\sf HERWIG}}
\newcommand{\MAD}{{\sf MADGRAPH}}
\newcommand{\epem}{\rm{e}^+\rm{e}^-}




\subsection{Introduction}

  Parton-shower (PS) Monte Carlo event generators are  an 
  important tool in the
  experimental analyses of collider data. 
  These computational programs are based on the differential cross
  sections for simple
  scattering processes (usually $2\to2$ particle scatterings) together
  with a PS simulation of additional QCD radiation that
  naturally connects to a model of hadronization.
  As the PS algorithms are based
  on resummation of the leading soft and collinear logarithms, these
  programs may not reliably estimate 
  the radiation of hard jets, which, in turn, may bias
  experimental analyses.
 
  Improvements have been developed to correct the emission of the 
  hard partons in the PS.
  In \PY\
  \cite{Sjostrand:2000wi,Sjostrand:2001yu,Sjostrand:2003wg}, 
  corrections were added for $\rm{e}^+\rm{e}^-$ 
  annihilation \cite{Sjostrand:1987hx},
  deep inelastic scattering \cite{Bengtsson:1988rw},
  heavy particle decays \cite{Norrbin:2000uu}
  and vector boson production in hadron-hadron collisions \cite{Miu:1998ju}.
  Similarly, corrections were added to 
  \HW~\cite{Corcella:2000bw,Corcella:2002jc}
  in Refs. 
  \cite{Seymour:1992xa,Seymour:1994ti,Corcella:1998rs,Corcella:1999gs} 
  following the method described in \cite{Seymour:1995df}.
  
  The Catani-Krauss-Kuhn-Webber~(\CK)
  algorithm is a method for generalizing such corrections
  \cite{Catani:2001cc,Krauss:2002up}.
  Along with this development, computer programs have become  
  available~\cite{Maltoni:2002qb,Mangano:2002ea}
  which are capable of efficiently 
  generating multi-parton events in a format (the 
  Les Houches format\cite{Boos:2001cv}) that can be readily interfaced
  with \HW\ and \PY.
  Here, we report on how to use these programs combined with the \HW\
  and \PY\ Monte Carlo event generators to implement hard corrections to
  PS predictions.
  Several approaches
  are explored.  One adheres closely to the \CK\ algorithm, but
  uses \HW\ for adding an additional PS.  The second
  is more closely tuned to the specific PS generators themselves
  and calculates branching probabilities numerically (using exact
  conservation of energy and momentum) instead of
  analytically.  This is accomplished by generating pseudo-showers starting
  from the various stages of a PS history.  
  A comparison is also made with a much simpler method.

\subsection{Overview of the Correction Procedure}
\label{sec:overview}
PS's are used to relate the partons produced 
in a simple, hard 
interaction characterized by a large energy scale (large means
$\gg \Lambda_{QCD}$) to the partons at an energy scale near
$\Lambda_{QCD}$.  At this lower scale, a transition is made to
a non--perturbative description of hadronic physics, with physical,
long--lived particles as the final products.  This is possible, 
because the fragmentation functions for the highly-virtual partons
obey an evolution equation that can be solved analytically or
numerically.  This solution can be cast in the form of a Sudakov
form factor, which can be suitably normalized as a probability
distribution for {\it no} parton emission between two scales.
Using the Monte Carlo method, the evolution of a parton can
be determined probabilistically, consisting of steps when the
parton's scale decreases with no emission, followed by 
a branching into sub-partons, which themselves undergo the same
evolution, but with a smaller starting point for the scale.
The evolution is ended when the energy scale of parton  reaches the
hadronization scale $\sim \Lambda_{QCD}$.  Starting from the
initial (simple) hard process, a sampling of PS's generates
many topologies of many-parton final states, subject to certain
phase space and kinematic restrictions. 
However, the evolution equation (as commonly used)
only includes the soft and collinear fragmentation that 
is logarithmically enhanced, so that non--singular contributions
(in the limit of vanishing cut-offs) are ignored.  This means that
not enough gluons are emitted that are energetic and at a large angle from
the shower initiator, since there is no associated soft or collinear 
singularity.

In contrast, matrix element (ME) calculations give a description of a specific
parton topology, which is valid when the partons are energetic and 
well separated.  Furthermore, it includes interference between 
amplitudes with the same external partons but different internal structure.
However, for soft and collinear kinematics, the description in terms
of a fixed order of emissions is not valid, because it does not include
the interference between multiple gluon emissions which cannot be resolved.

The PS description of hard scattering would be improved if
information from the ME were included when calculating
emission probabilities.  A systematic method for this can be 
developed by comparing the PS and ME predictions
for a given fixed topology.  Consider 
a hard scattering ($\epem\to \gamma/Z\to
q\bar q$) followed by a PS off the outgoing $\rm q\bar q$
pair, with each branching $i$ characterized by
a variable $d_i$  The variables $d_i$ represent some virtuality or energy scales
that are evolved down to a cut-off $d_{\rm ini}$.  The PS
rate for this
given topology is a product of many factors: (1) the Born level cross
section for $\epem\to q\bar q$, (2) Sudakov form factors representing
the probability of no emission on each quark and gluon line, and
(3) the branching factors at each vertex (or splitting).
The ME prediction for this topology is somewhat more
complicated.  First, one needs to calculate the cross section for
the full initial- and final-state (here $\epem\to q\bar q g g q'\bar q'$).
Then, one needs to specify a particular topology.  There is no unique
way to do this, but a sensible method is to choose a clustering scheme
and construct a PS history.  Ideally, the clustering variable
would be the same as the virtuality $d_i$ used to generate the PS
in the usual way.  Having performed the clustering, one can
then make a quantitative comparison of the two predictions.

To facilitate the comparison, we first expand the PS 
prediction to the same fixed order in $\alpha_s$.  This is equivalent
to setting all the Sudakov form factors to unity.  In this limit,
we see that the PS product of the Born level cross section
and the vertex factors is an approximation to the exact ME
prediction.  As long as the values $d_i$ are all large, the
ME description is preferred theoretically, and the Sudakov
form factors are indeed near unity.  Therefore, the PS
description can be improved by using the exact clustered ME
prediction.  When the values $d_i$ are {\it not} all large, and
there is a strong ordering of the value ($d_1 \gg d_2 \cdots \gg d_{\rm ini}$)
then
the PS description is preferred theoretically.  
In this limit, the ME prediction reduces to the
product of Born level and vertex factors, provided that the argument
of $\alpha_s$ is chosen to match that used in the PS
(this should be related to $d_i$).  Therefore, the ME
prediction can be used to improve the PS description
in all kinematic regions provided that: (1) the correct argument
for $\alpha_s$ is used, and (2) the Sudakov form factors are inserted
on all of the quark and gluon lines.  This provides then an {\it interpolation}
scheme between the PS and the ME prediction.
As usual, there is a systematic uncertainty associated with how
one chooses to perform the interpolation.

This corrects the specific topology considered, 
but what of the rest of the topologies?  ME
calculations can be performed for those that are simple enough,
but technically there is a limitation.  Presently, $\epem\to 6$ parton
calculations can be performed using computational farms with appropriate
cuts.  A practical solution is to choose the cut-off $d_{\rm ini}$ large
enough that the ME calculations in hand saturate the
dominant part of the cross section.  Then, an ordinary PS
can be used to evolve the parton virtualities from $d_{\rm ini}$ 
down to the hadronization scale.  It has been shown that the
correct method for doing this consists of starting the PS's
at the scale where a parton was created in the clustered topology,
and then vetoing branchings with virtualities {\it larger} than $d_{\rm ini}$
\cite{Catani:2001cc}.

\subsection{Results}

\begin{figure}[hbtp]{
\subfigure{\includegraphics[width=.55\textwidth,angle=0]{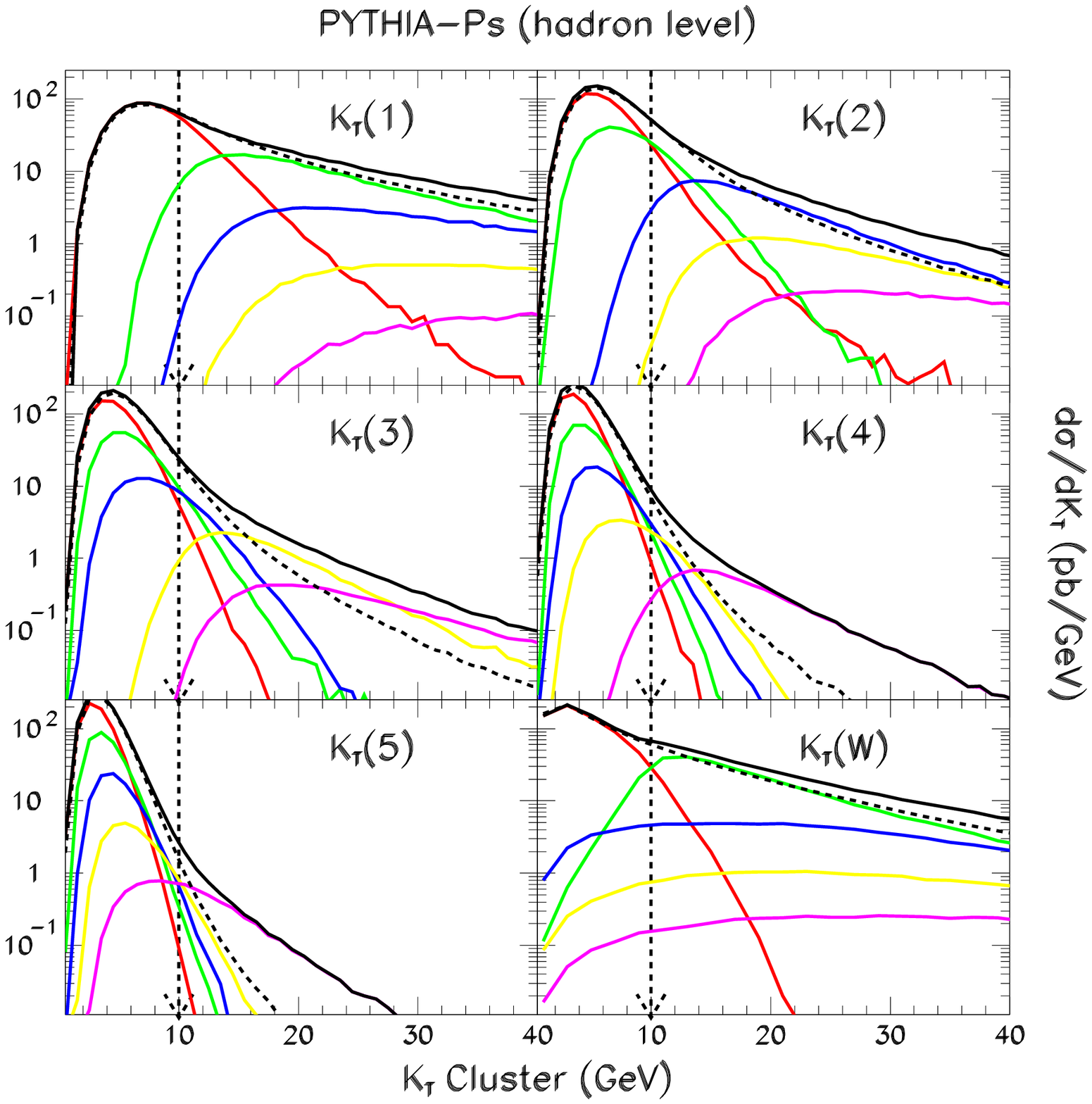}}
\subfigure{\includegraphics[width=.55\textwidth]{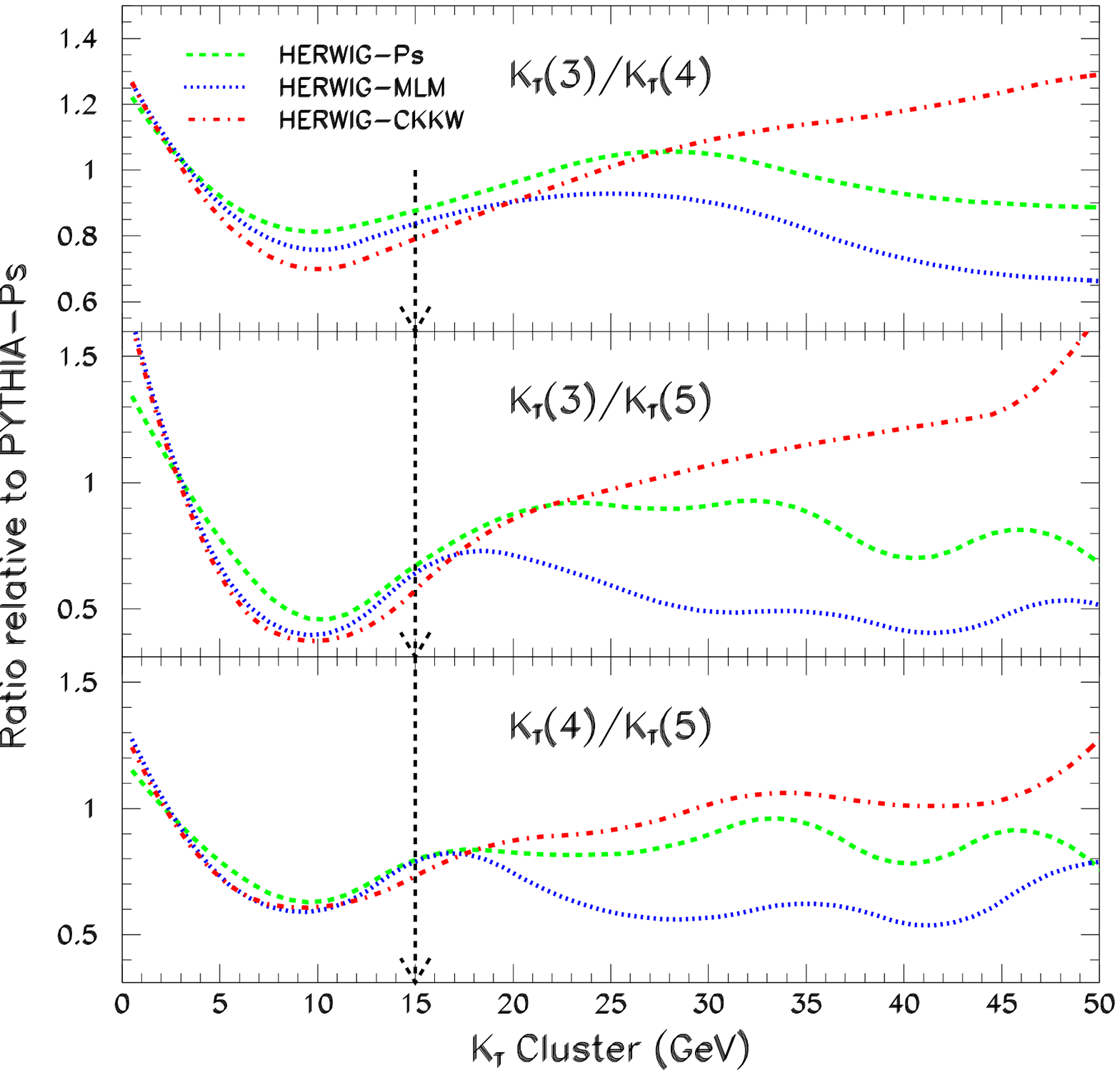}}
\caption{
(a) Differential $k_{Ti}$-cluster distributions 
  $d\sigma/d{k_{Ti}}$ 
         at the hadron level generated
         with the pseudo-shower procedure for
         $\rm{p}\bar{\rm{p}}\to W^{+}+X$ collisions at $\sqrt{s}=1.96$ TeV, 
        for $i=1-5$ and also showing the W$^+$ boson transverse momentum.
         The default result of \PY\ is shown as a dashed line, while
         the result of the pseudo-shower algorithm is shown as a 
         solid black line.
         The contribution to the pseudo-shower result 
          from the two (red), three (green), four (blue),
         five (yellow) and six (magenta) parton
         components is also shown.
         The matching scale 10 GeV is
         shown as a vertical arrow;
(b) Comparison of the ratio of various $k_T$-cluster
distributions from \HW\ and PY\ using the
pseudo-shower procedure, \HW\
 using the MLM procedure, and
\HW\ using the CKKW procedure
for a matching scale of 15 GeV.}
\label{fig:compare_ratio}
}
\end{figure}

\subsection{Discussion and Conclusions}
\label{sec:disc-concl}

We have compared three different procedures
of matching ME
predictions with PS's using a methodology close to the
\CK\ algorithm suggested in   \cite{Catani:2001cc,Krauss:2002up}: 
(1) a slightly expanded version of the \CK\ procedure using \HW\
as the PS generator (but not limited in principal
to \HW) and exploiting the freedom to choose scales and cut-offs; 
(2) a version of the \CK\ procedure relying on pseudo-showers
and matched closely to the scales and cut-offs of \PY\ and \HW;
and (3) a much simpler procedure based on the approach of
M. Mangano.  Results are summarized in Figs. 1 (a) and (b).

The \HW-\CK\ procedure uses all of the elements of the original
\CK\ procedure, but expands upon them.  Several choices of
scale were investigated as starting points for the vetoed PS, 
and a wide range of prefactors were explored as arguments
to the analytic NLL Sudakov form factor and $\alpha_S$.  The
variation of the results with these choices is shown in
the figures.  Optimized choices were settled upon based on
the smoothness of distributions, the agreement with \HW\ where
expected, and the apparent improvement over the default \HW\ 
predictions.   While this appears to be a tuning, the final
choices are easily justifiable.  Since \HW\ is an angular-ordered
shower, a variable such as $k_T$-cluster values is well suited as
a starting point for the \HW\ shower.  Because of the details 
of the \HW\ shower, a prefactor of $\frac{1}{2}$ for the scale
used in the Sudakov form factor is understandable, as well
as a prefactor of $\frac{1}{8}$ for the scale used in evaluating
$\alpha_S$.  The results presented are better at the parton level
than at the hadron level, which may require some tuning of the
\HW\ hadronization model, and are less important
at higher energies or
when the cut-offs are larger.

The pseudo-shower procedure uses the Sudakov form factors of
\HW\ and \PY\ to numerically calculate the Sudakov suppression.
Since the Sudakov form factor is a probability distribution for
no parton emissions, the suppression factor can be determined
by starting showers from different stages of the PS
history and discarding those events with emissions above 
a given cut-off.  Because of the nature of this approach, there
is less tuning of parameters.  To match the argument used in
$\alpha_S$ by default in \HW\ and \PY, a different clustering
scheme was used:  $p_T$ clustering or {\tt LUCLUS}-clustering.
Final results at the hadron level are shown in the figures.
In general, the hadron-level results are better than the
parton-level ones.  The use of {\tt LUCLUS} over {\tt KTCLUS}
was driven by the kinematics of the \PY\ shower.  We have
not checked whether {\tt KTCLUS} works as well or better for
the \HW\ results, and we leave this for future investigation.
We should also investigate the advantages of using the
{\it exact} clustering scheme of the individual generators:
invariant mass and angular ordering for \PY\ or just
angular ordering for \HW.  Also, since this work began, a
new model of final-state showering was developed for \PY\ which
is exactly of the {\tt LUCLUS} type.  This should also be
tested.

The MLM procedure is a logical extension of the procedure developed
by M. Mangano for adding PS's to $W$+multijet events.
It entails $k_T$-clustering the parton-level events, adding
a PS (with \HW\ in practice, but not limited to it), and rejecting
those events where the PS generates a harder emission
(in the $k_T$-measure) than the original events.  This
approach yields a matching which is almost as good as the more complicated
procedures based on the \CK\ procedures explored
in this work.  The reason is not a pure
numerical accident.  The MLM procedure rejects events (equivalently,
reweights them to zero weight) when the PS generates
an emission harder than the lowest $k_T$ value of the given
kinematic configuration.  This is equivalent to the first
step of the pseudo-shower procedure
in the calculation of the Sudakov suppression when applied to the highest
multiplicity ME.  The remaining difference
is in the treatment of the internal Sudakov form factors 
and the argument of $\alpha_S$.  The agreement between the
pseudo-shower and MLM procedures implies that the product of
Sudakov form factors on internal lines with the factors of
$\alpha_S$ evaluated at the clustering scale is numerically
equivalent to the product of $\alpha_S$ factors evaluated 
at the hard scale.  It is worth noting that, for
the process at hand, $q\bar q'\to W+X$, only two of the
cluster values can be very close to the cut-off, and thus
only two of the $\alpha_S(k_T)$ values can be very large.
Also, at the matching scales considered in this study,
$10-20$ GeV, with a factorization scale on the order 
of $M_W$, $Q_F=\sqrt{M_W^2+P_{TW}^2}$,
a fixed order expansion is of similar numerical
reliability as the ``all-orders'' expansion of a resummation
calculation.  In fact, the resummation (PS) 
expansion is ideally suited for $Q\ll M_W$, whereas the
fixed order expansion is best applied for $Q\sim M_W$.

Based on the study of these three procedures, we can make
several statements on the reliability of predicting
the shapes and rates of multijet processes at collider energies.
\begin{enumerate}
\item The three matching procedures studied here can be recommended.
They are robust to variation of the cut-off scale.
\item The relative distributions in $k_T$, for example, are
reliably predicted.  
\item The variation in the relative distributions from the three
procedures depends on the variable.  For variables within the
range of the ME's calculated, the variation is 20\%.
For variables outside this range, which depend on the
truncation of the ME calculation, the variation is
larger 50\%.  Of course, it is important to study the experimental
observables to correctly judge the senstivity to the cut-off and
methodology of matching.

\item More study is needed to determine the best method for
 treating the highest multiplicity ME contributions.

\item The subject of matching is far from exhausted.  The procedures
presented here yield an improvement over previous matching prescriptions.
However, these methodologies are an {\it interpolation} procedure.
\end{enumerate}

\subsection*{Acknowledgments}
This work was initiated at 
the Durham Monte Carlo workshop 
and has been performed in collaboration with P. Richardson.
I thank T. Sj\"{o}strand 
and J. Huston for discussion, encouragement
and for asking the difficult questions.
I also thank the Stephen Wolbers and the Fermilab Computing Division
for access to the Fixed Target Computing Farms.




\section[$W$ Boson, Direct Photon and Top Quark Production: 
Soft-Gluon Corrections]{$W$ BOSON, DIRECT PHOTON AND TOP QUARK PRODUCTION: 
SOFT-GLUON CORRECTIONS~\protect\footnote{Contributed by:
  {N.~Kidonakis}}}
\label{kid_softgluon}


\subsection{Introduction}

$W$-boson, direct photon, and top quark production are all processes
of considerable interest, useful in testing  the Standard 
Model and searching for new physics. The hadroproduction
cross sections  for these processes have been calculated fully through
next-to-leading order (NLO). Threshold corrections
are known to be important in current hadron colliders and 
attempts have been made to calculate these soft-gluon corrections
at next-to-next-to-leading order (NNLO) and beyond.
Here I present results from the latest calculations.

In general, at each order in perturbation theory
the partonic cross section $\hat{\sigma}$  for a hard-scattering
process includes ``plus'' distributions with respect to a kinematical variable,
denoted say as $s_2$, that measures distances from partonic threshold.
At $n$-th order in the strong coupling $\alpha_s$ (beyond the leading order)
these distributions are of the type
\begin{equation}
\left[\frac{\ln^{l}(s_2/M^2)}{s_2} \right]_+, \hspace{10mm} l\le 2n-1\, ,
\label{zplus}
\end{equation}
where $M$ is a hard scale, such as a  mass
or transverse momentum, relevant to the process at hand.
These logarithmic terms are the soft-gluon corrections
and they arise from incomplete cancellations near partonic
threshold between graphs with real emission and virtual graphs. 
This is due to the limited phase space available for real gluon emission
near partonic threshold. These threshold corrections, calculated
in the eikonal approximation, can be formally shown to
exponentiate  \cite{Kidonakis:1996aq,Kidonakis:1997gm,
Kidonakis:1998nf,Kidonakis:1999ze} as a result of the factorization 
properties of the cross section.
The logarithms with $l=2n-1$ are denoted as leading (LL), 
with $l=2n-2$ as next-to-leading (NLL), with $l=2n-3$ as 
next-to-next-to-leading (NNLL), 
and with $l=2n-4$ as next-to-next-to-next-to-leading (NNNLL).
We note that the virtual corrections appear in $\delta(s_2)$ terms.
A unified approach and master formulas for the calculation
of these corrections for any process at NNLO have recently been presented
in Ref. \cite{Kidonakis:2003tx}. 
For the processes discussed in the next three sections, the LL, NLL,
and NNLL terms have been calculated fully.
In the NNNLL terms we have not included some process-dependent two-loop
contributions \cite{Kidonakis:2003tx} which, however, we expect to be
small.

\subsection{$W$ Boson Production}

The production of $W$ bosons  in hadron colliders  
is a process of relevance in testing the Standard Model, 
calculating backgrounds to new physics such as
associated Higgs boson production, and luminosity monitoring.

The calculation of the complete NLO
cross section for $W$ hadroproduction at large transverse momentum was 
presented in Refs. \cite{Arnold:1989dp,Gonsalves:1989ar,Gonsalves:1990ae}.
The NLO results displayed an enhancement of the 
differential distributions in transverse momentum $Q_T$ of the $W$ boson.
The $Q_T$ distribution falls rapidly with increasing $Q_T$, spanning five
orders of magnitude in the 30 GeV $< Q_T <$ 190 GeV region 
at the Tevatron. 

$W$-boson production at high transverse momentum receives corrections 
from the emission of soft gluons from the partons in the process.
The resummation and NNLO-NNLL corrections were studied in 
Ref. \cite{Kidonakis:1999ur}. More recently the NNLO-NNNLL corrections
were studied in Ref. \cite{Kidonakis:2003xm}.
These threshold corrections further enhance the cross section and
reduce the scale dependence \cite{Kidonakis:2003xm}.

For the hadronic production of a high-$Q_T$ $W$ boson, with mass $m_W$,
the lowest-order partonic subprocesses are  
$q(p_a)+g(p_b) \longrightarrow W(Q) + q(p_c)$ and
$q(p_a)+{\bar q}(p_b) \longrightarrow W(Q) + g(p_c)$. 
The partonic kinematical invariants in the process are 
$s=(p_a+p_b)^2$, $t=(p_a-Q)^2$, $u=(p_b-Q)^2$, which satisfy 
$s_2 \equiv s + t + u - Q^2=0$ at partonic threshold.
Here $s_2=(p_a+p_b-Q)^2$ is the invariant mass of the system recoiling 
against the $W$ boson and it parametrizes the inelasticity of 
the parton scattering.
The partonic cross section $\hat{\sigma}$ 
includes distributions with respect to $s_2$ of the type
$[\ln^{l}(s_2/Q_T^2)/s_2]_+$.

\begin{figure}
\begin{center}
\includegraphics[width=7.95cm]{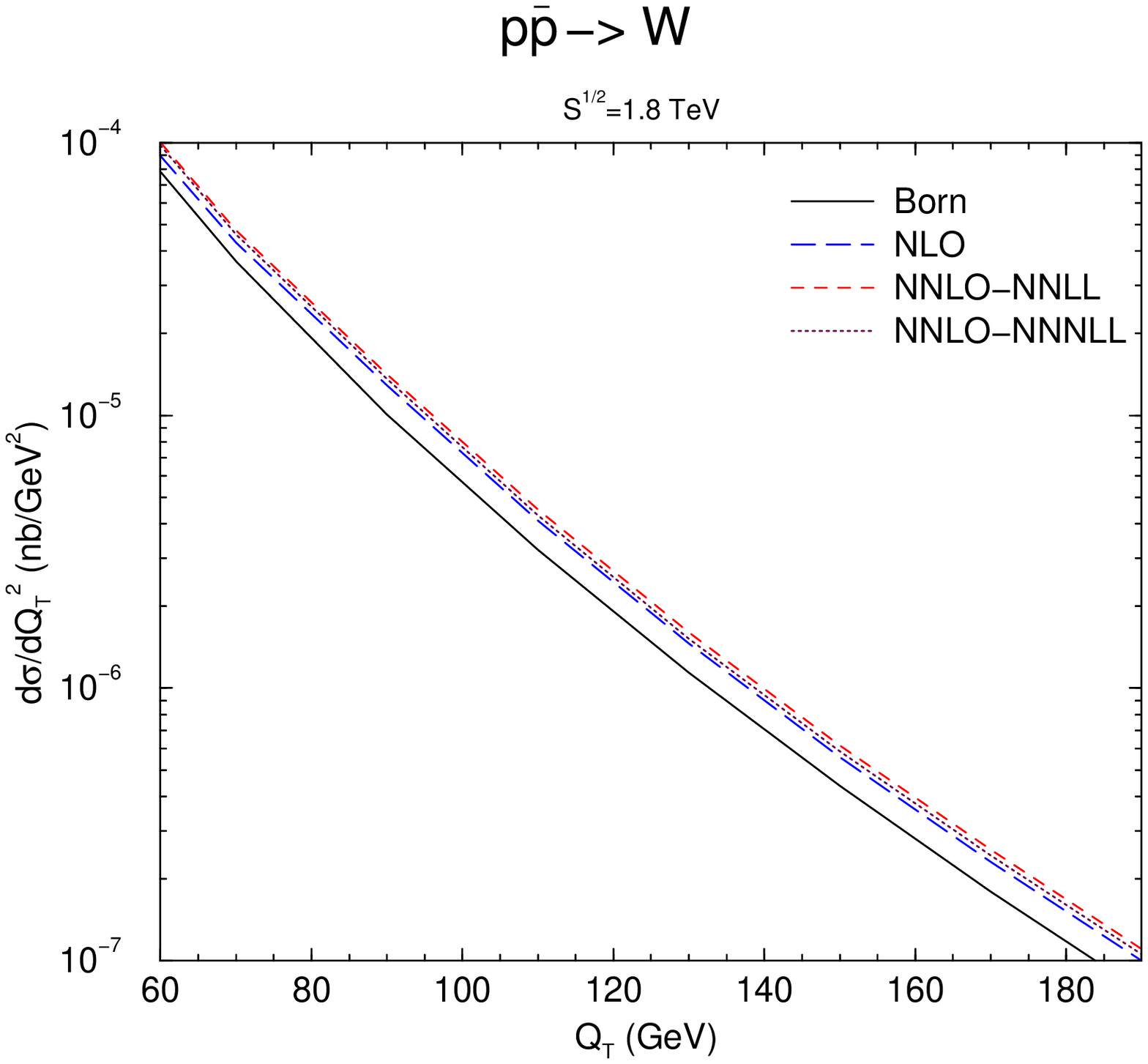} 
\includegraphics[width=7.95cm]{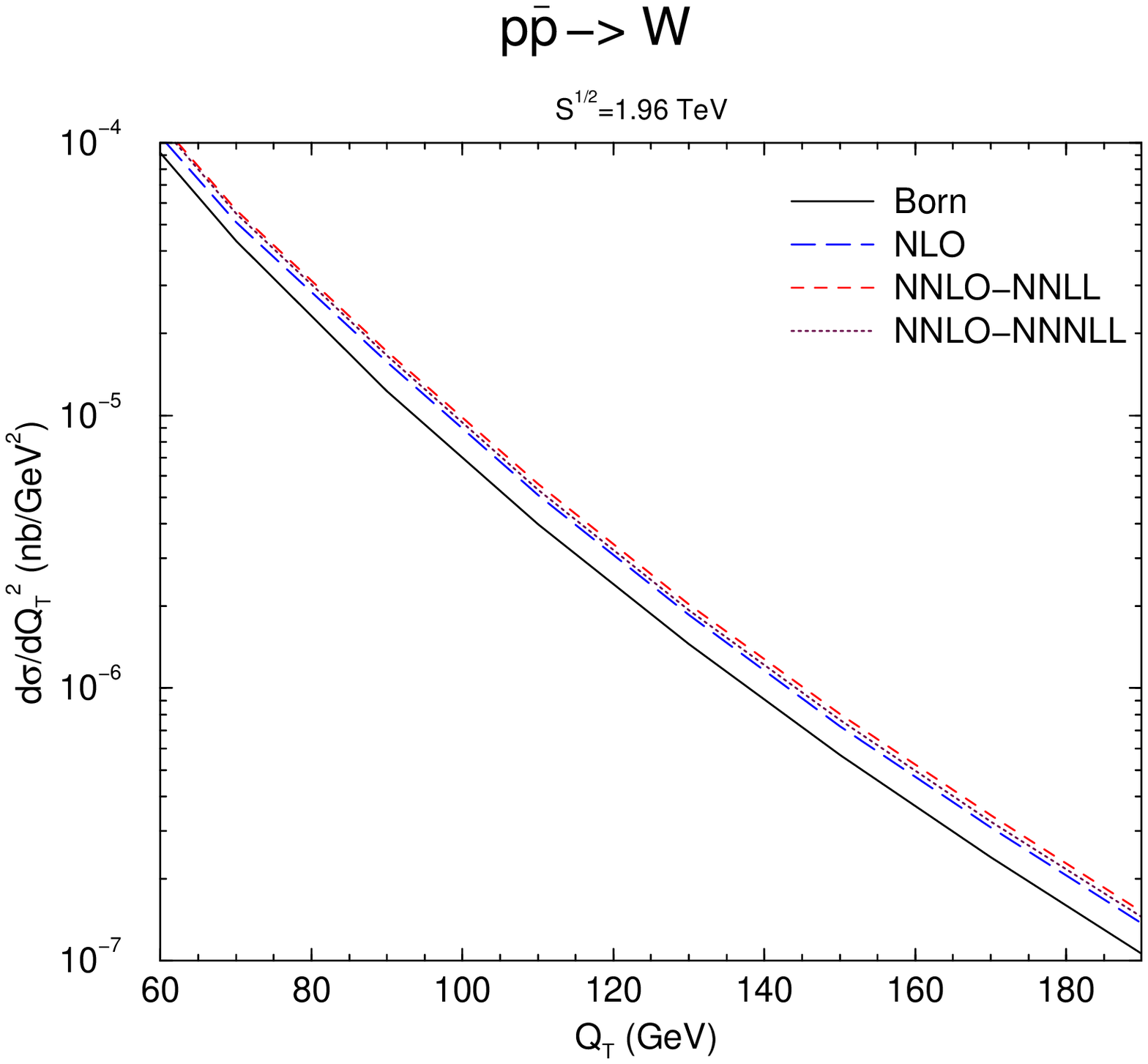} 
\caption{The differential cross section,
$d\sigma/dQ_T^2$, for $W$ hadroproduction in $p \bar p$ collisions
at the Tevatron at (left) $\sqrt{S}=1.8$ TeV and (right) 1.96 TeV,
with $\mu_F=\mu_R=Q_T$.
Shown are the Born, NLO, NNLO-NNLL, and NNLO-NNNLL results.}
\label{WQTplots}
\end{center}
\end{figure}

In Fig.~\ref{WQTplots} we plot the transverse momentum distribution,
$d\sigma/dQ_T^2$, for $W$ hadroproduction at the Tevatron Run I 
with $\sqrt{S}=1.8$ TeV and Run II with $\sqrt{S}=1.96$ TeV.
We use the MRST2002 NNLO parton densities \cite{Martin:2002aw}.
We set the factorization scale $\mu_F$ and the renormalization scale
$\mu_R$ equal to $Q_T$.
We focus on the high-$Q_T$ region where the soft-gluon approximation holds
well and the corrections are important.
We see that the NLO corrections provide a significant enhancement of the Born
cross section. The NNLO-NNLL corrections provide a further
modest enhancement of the $Q_T$ distribution. 
If we increase the accuracy by including the NNNLL contributions,
which are negative,
then we find that the NNLO-NNNLL cross section lies between
the NLO and NNLO-NNLL results.

\begin{figure}
\begin{center}
\includegraphics[width=7.95cm]{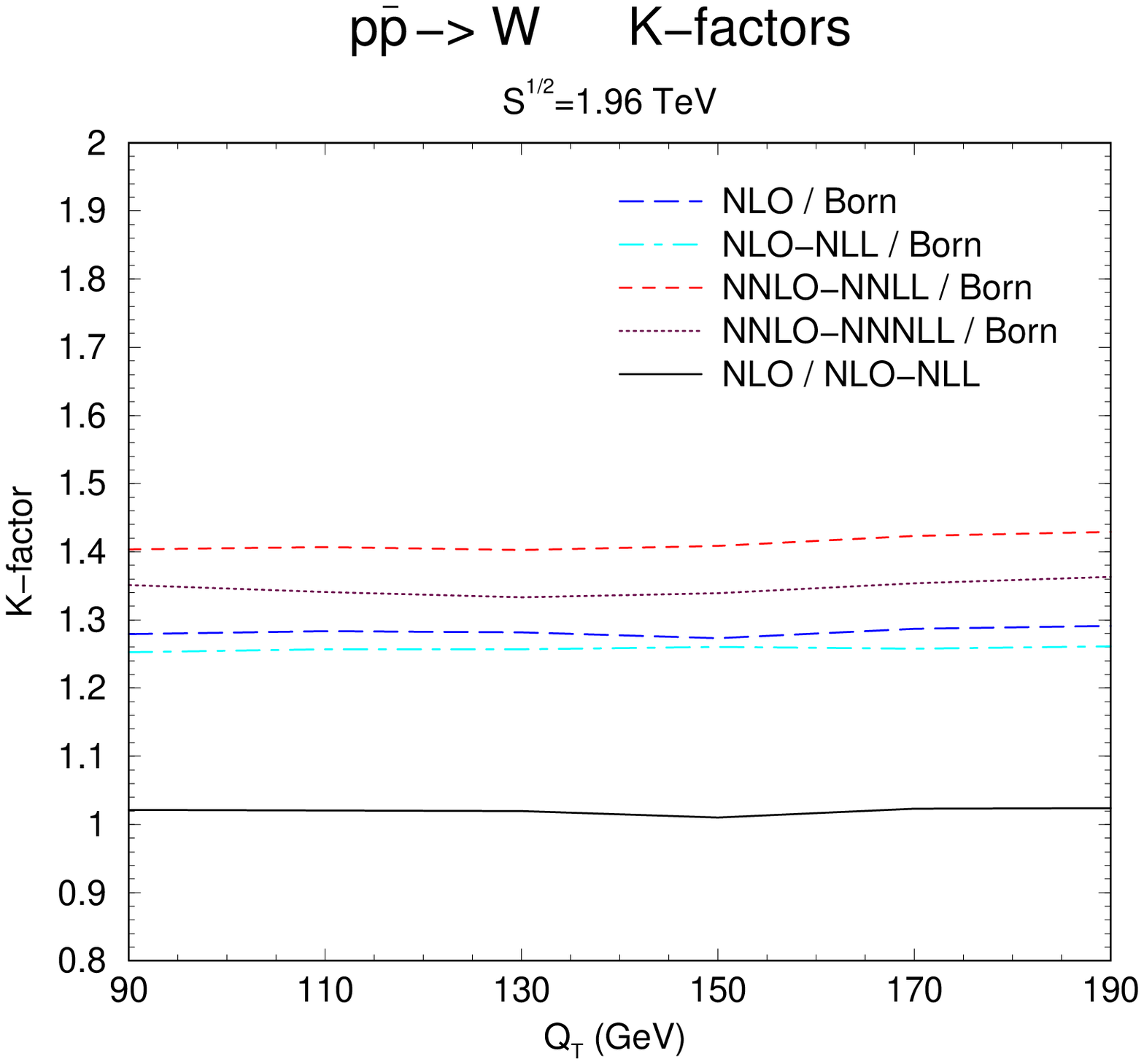} 
\includegraphics[width=7.95cm]{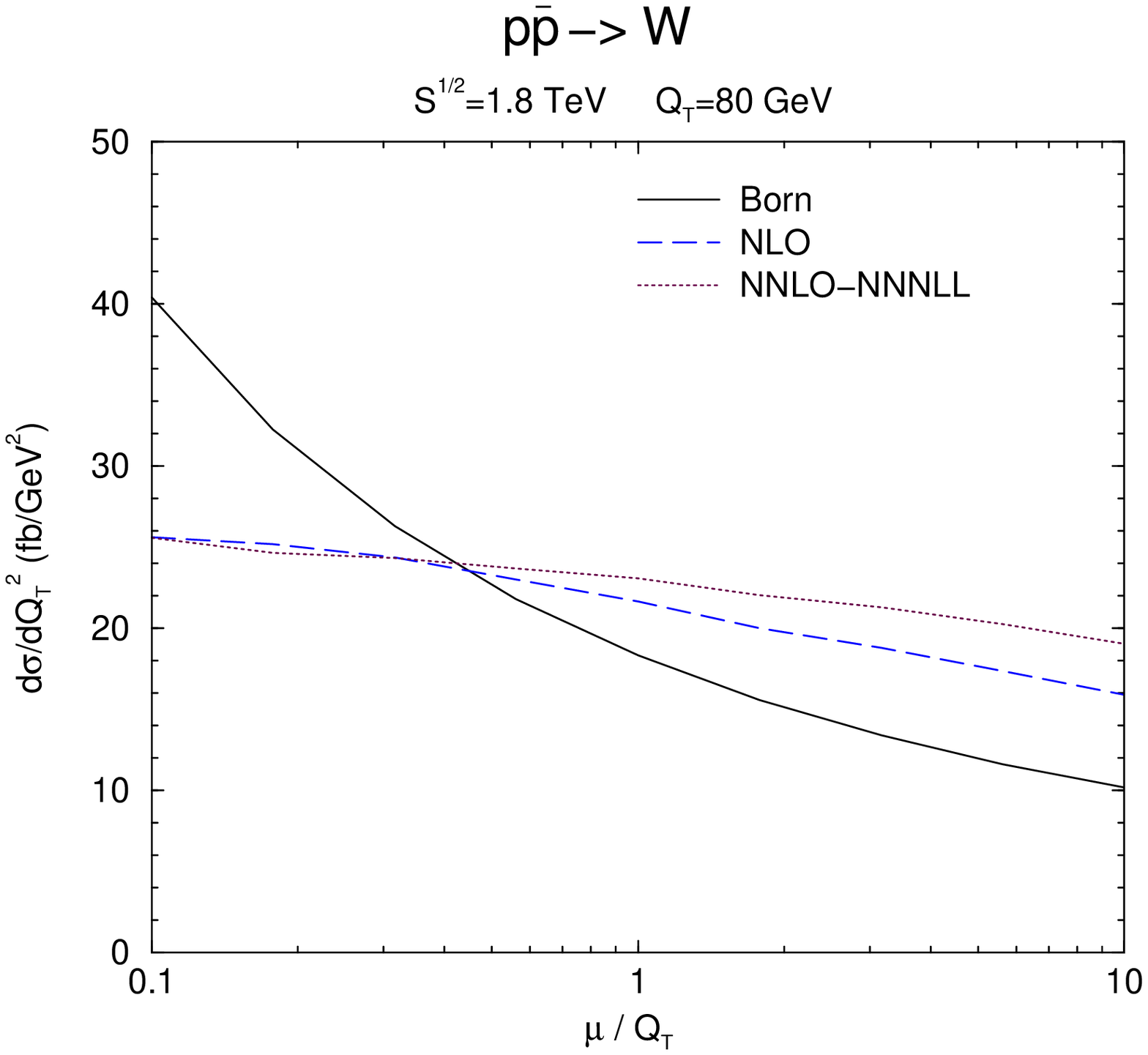} 
\caption{
Left: The $K$-factors for the differential cross section,
$d\sigma/dQ_T^2$,  for $W$ hadroproduction 
at the Tevatron with $\sqrt{S}=1.96$ TeV 
and $\mu_F=\mu_R=Q_T$.
Right: The $\mu$ dependence of the differential cross section,
$d\sigma/dQ_T^2$, for $W$ hadroproduction 
at the Tevatron with $\sqrt{S}=1.8$ TeV and $Q_T=80$ GeV.  
Here $\mu \equiv \mu_F=\mu_R$.
}
\label{WmuKplots}
\end{center}
\end{figure}

The $K$-factors are shown at $\sqrt{S}=1.96$ TeV in the left frame
of Fig.~\ref{WmuKplots}.  We see that the $K$-factors 
are moderate, and nearly constant over the $Q_T$ range shown even though
the distributions themselves span two orders of magnitude in this range.
It is also easy to see from the NLO/NLO-NLL curve 
that in the high $Q_T$ region
the soft-gluon approximation holds very well, as the NLO-NLL cross
section is almost identical to the full NLO result. 

The scale dependence of the differential cross section
is shown on the right frame of Fig.~\ref{WmuKplots}
for $Q_T=80$ GeV and $\sqrt{S}=1.8$ TeV. We plot $d\sigma/dQ_T^2$
versus $\mu/Q_T$ over two
orders of magnitude: $0.1 < \mu/Q_T < 10$. We note the good
stabilization of the cross section when the NLO corrections are
included, and the further improvement when the NNLO-NNNLL corrections
(which include all the soft and virtual NNLO scale-dependent terms) are added.

\subsection{Direct Photon Production}

Direct photon production is an important process 
for determinations of the gluon distribution function.
The NLO cross section for direct photon production 
has been given in Refs. \cite{Aurenche:1984ws,Aurenche:1988fs,Gordon:1993qc}. 
The role of higher-order soft-gluon
corrections has also been addressed more recently.
Threshold resummation and NNLO corrections for direct photon production
have been presented in Refs. \cite{Kidonakis:1999hq,Kidonakis:2003bh}. 

At lowest order, the parton-parton scattering  subprocesses are 
$ q(p_a)+g(p_b) \rightarrow  \gamma(p_{\gamma}) + q(p_J)$
and $q(p_a)+{\bar q}(p_b) \rightarrow \gamma(p_{\gamma}) + g(p_J)$. 
We define the Mandelstam invariants 
$s=(p_a+p_b)^2$, $t=(p_a-p_{\gamma})^2$, and 
$u=(p_b-p_{\gamma})^2$,
which satisfy $s_4 \equiv s+t+u=0$ at threshold.
Note that the photon transverse momentum is
$p_T=(tu/s)^{1/2}$. Here we calculate the 
cross section $E_{\gamma} \; d^3\sigma/d^3 p_{\gamma}$
in the $\overline {\rm MS}$ scheme.
The soft corrections to the cross section appear in the form of 
plus distributions
$[\ln^l(s_4/p_T^2)/s_4]_+$.

In order to show the effect of including the NNLO threshold terms, 
we start with a complete NLO calculation of the appropriate cross section 
using a program \cite{Baer:1990ra} which employs the phase-space 
slicing technique \cite{Harris:2001sx}. 
The original NLO calculation has been extended 
to include a complete NLO treatment of the bremsstrahlung contribution. 
The Set 2 fragmentation functions of \cite{Bourhis:2000gs} are used along 
with the CTEQ6M parton distribution functions \cite{Pumplin:2002vw}. 
In all cases the 
factorization and renormalization scales have been set equal to a common 
scale $\mu$.
Once the NLO results are obtained, the approximate NNLO contributions 
are then added to them. 

\begin{figure}
\begin{center}
\includegraphics[width=7.95cm]{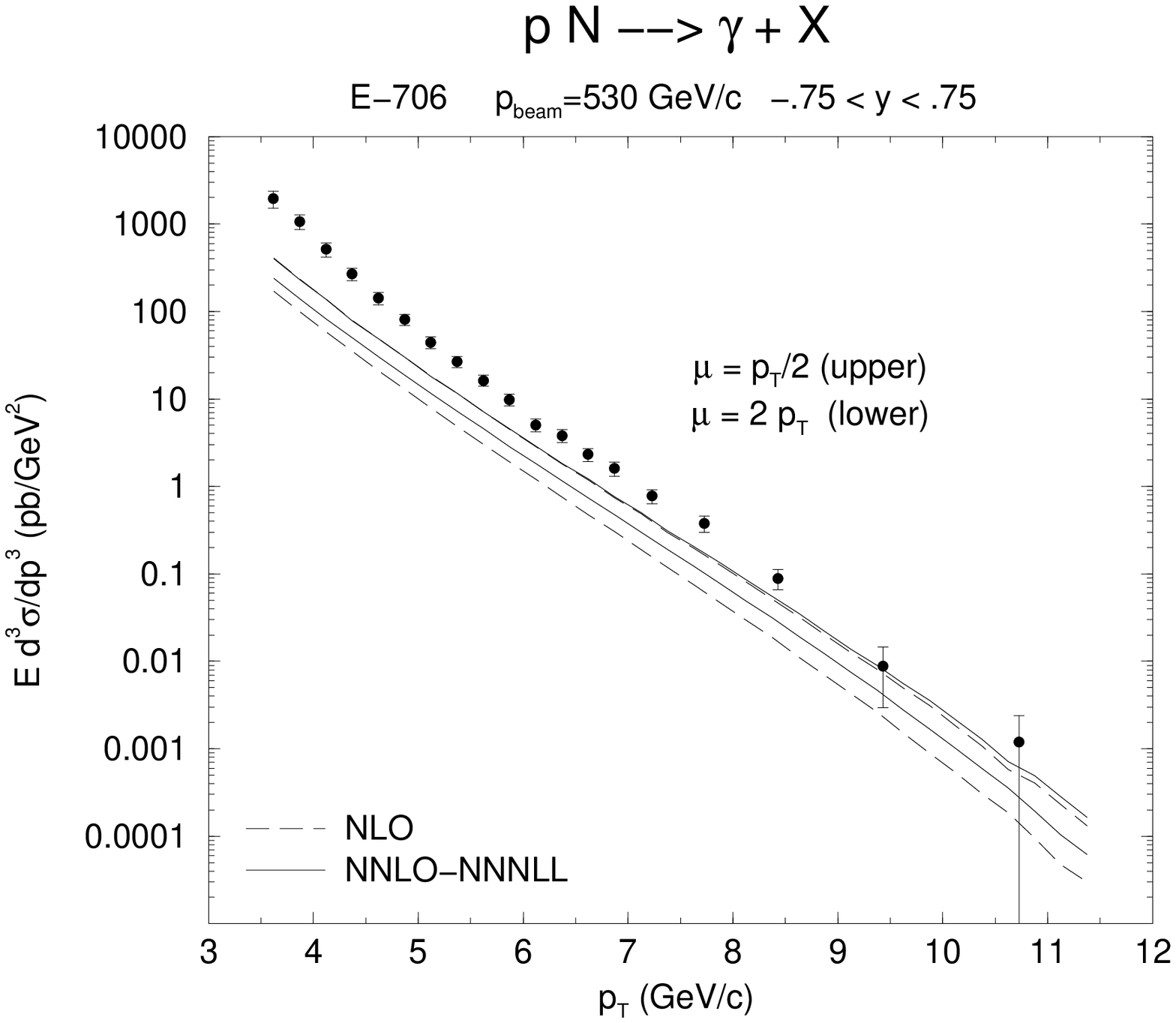} 
\includegraphics[width=7.95cm]{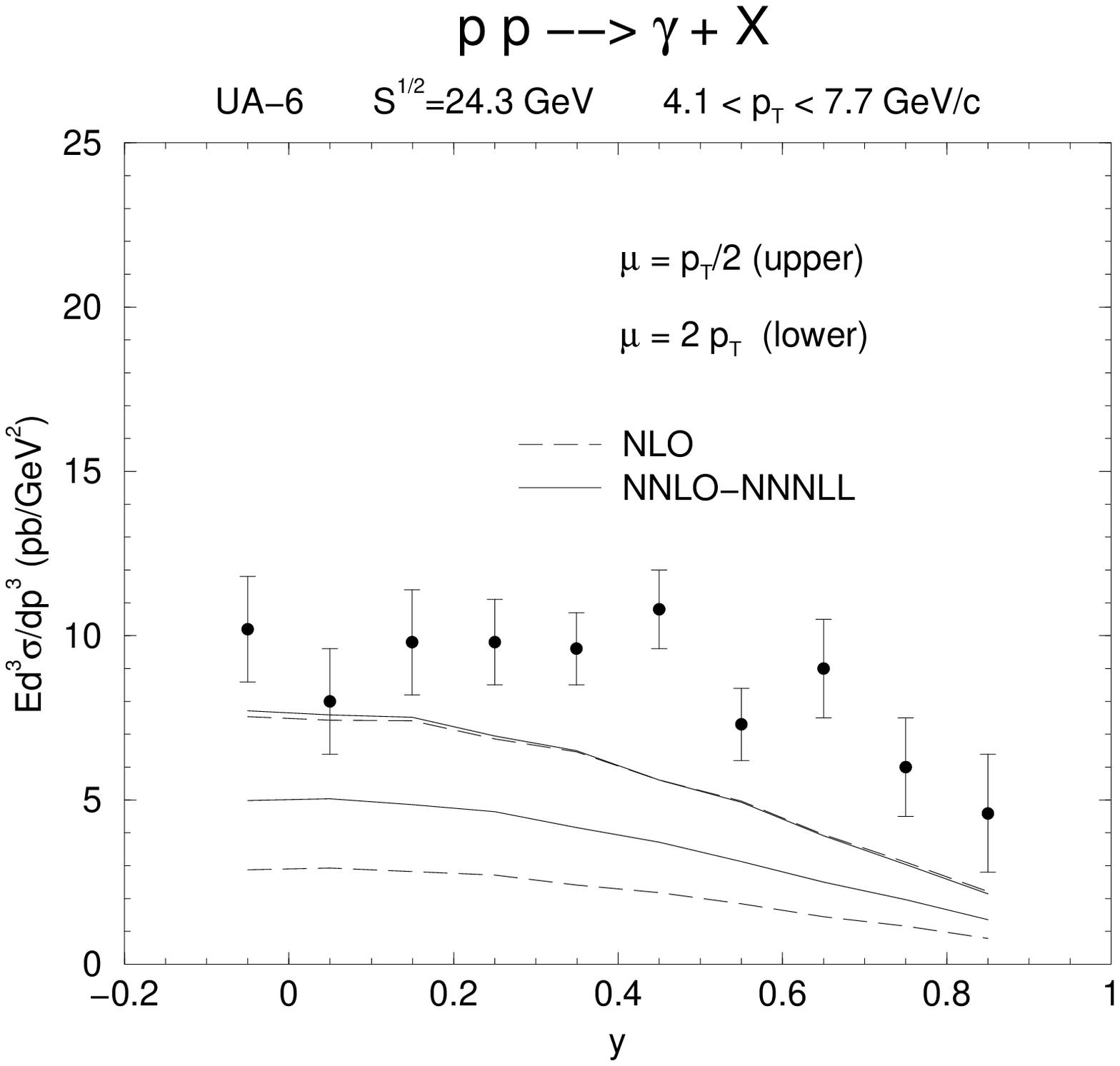} 
\caption{
NLO and NNLO-NNNLL results for direct photon production
in hadronic collisions. 
Left: theory compared to data from the E-706 Collaboration 
\cite{Apanasevich:1998hm} at p$_{\rm beam} = 530\ $ GeV/$c$.
Right: theory compared to $pp$ data for the rapidity distribution 
from the UA-6 Collaboration \cite{Ballocchi:1998au} 
at $\sqrt S = 24.3\ $ GeV.}
\label{dirphotplot}
\end{center}
\end{figure}

In the left frame of Fig.~\ref{dirphotplot} a  comparison 
is made to data from the 
E-706 Collaboration \cite{Apanasevich:1998hm}. 
The NNLO-NNNLL curve at $\mu=p_T/2$ is practically indistinguishable
from the NLO except at high $p_T$. However, the $\mu=2p_T$ NNLO result
is much higher than NLO, and as a result the scale dependence 
at NNLO is considerably reduced.
The theoretical band lies  
below the data at the lower end of the range covered by the data. 
In the right frame of Fig.~\ref{dirphotplot} the rapidity dependence 
is shown for the UA-6 proton proton data \cite{Ballocchi:1998au}. 
Again, the NNLO-NNNLL terms give a negligible 
contribution for the choice $\mu=p_T/2$ and the overall scale dependence 
is greatly reduced when the NNLO terms are added. 

\subsection{Top Quark Production}

Recent calculations for top hadroproduction 
include NNLO soft-gluon 
corrections to the double differential cross section 
\cite{Kidonakis:2000ui,Kidonakis:2001nj,Kidonakis:2003qe} 
from threshold resummation techniques. 
The latest calculation \cite{Kidonakis:2003qe}
includes NNNLL and some virtual $\zeta$ terms (defined in 
Ref.~\cite{Kidonakis:2003qe}) 
at NNLO. When not all terms are known there is some difference
between single-particle-inclusive (1PI) and pair-invariant-mass
(PIM) kinematics. When NNNLL terms are included,
the kinematics dependence of the cross section vanishes
near threshold and is reduced away from it relative to NNLL accuracy. 
The factorization and renormalization
scale dependence of the cross section is also greatly reduced.

We study the partonic process $ij \rightarrow t {\overline t}$ 
with $ij = q {\bar q}$ and $gg$.  
In 1PI kinematics, a single top quark is identified,
$i(p_a) + j(p_b) \longrightarrow t(p_1) + X[{\overline t}](p_2)$
where $t$ is the identified top quark of mass $m$
and $X[{\overline t}]$ is the remaining final state that contains
the ${\overline t}$.
We define the kinematical invariants 
$s=(p_a+p_b)^2$, $t_1=(p_b-p_1)^2-m^2$, $u_1=(p_a-p_1)^2-m^2$
and $s_4=s+t_1+u_1$. At threshold, $s_4 \rightarrow 0$,
and the soft corrections appear as $[\ln^l(s_4/m^2)/s_4]_+$.
In PIM kinematics, we have instead
$i(p_a) + j(p_b) \longrightarrow t{\overline t}(p) + X(k)$.
At partonic threshold, $s=M^2$, with $M^2$ the pair mass squared.
The soft corrections appear as $[\ln^l(1-z)/(1-z)]_+$,
with $z=M^2/s \rightarrow 1$ at threshold.

\begin{figure}
\begin{center}
\includegraphics[width=15.5cm]{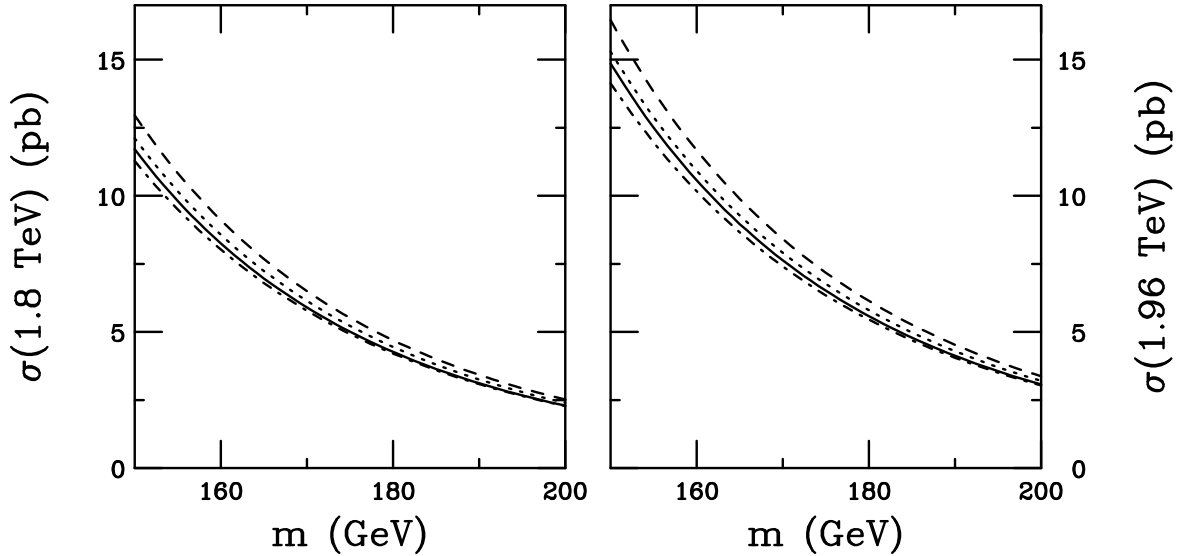} 
\caption{The $t \overline t$ total cross sections in $p \overline p$
collisions at $\sqrt{S} = 1.8$ TeV
(left frame) and 1.96 TeV (right frame) are shown as functions of $m$ for
$\mu = m$.  The NLO (solid), and approximate NNLO 1PI (dashed), 
PIM (dot-dashed) and average (dotted) results are plotted.}
\label{ttbarcs}
\end{center}
\end{figure}

In Fig.~\ref{ttbarcs} we present the NLO and 
approximate NNLO $t \overline t$
cross sections at $\sqrt{S} = 1.8$ TeV (left frame) and 1.96 TeV
(right frame) for $\mu=\mu_F=\mu_R=m$.  We use the MRST2002 NNLO parton
densities \cite{Martin:2002aw}. The NNLO
results include the soft NNNLL and virtual $\zeta$ terms
in 1PI and PIM kinematics.  
We also show the average of the two kinematics results
which may perhaps be closer to the full NNLO result.  

\begin{figure}
\begin{center}
\includegraphics[width=7.95cm]{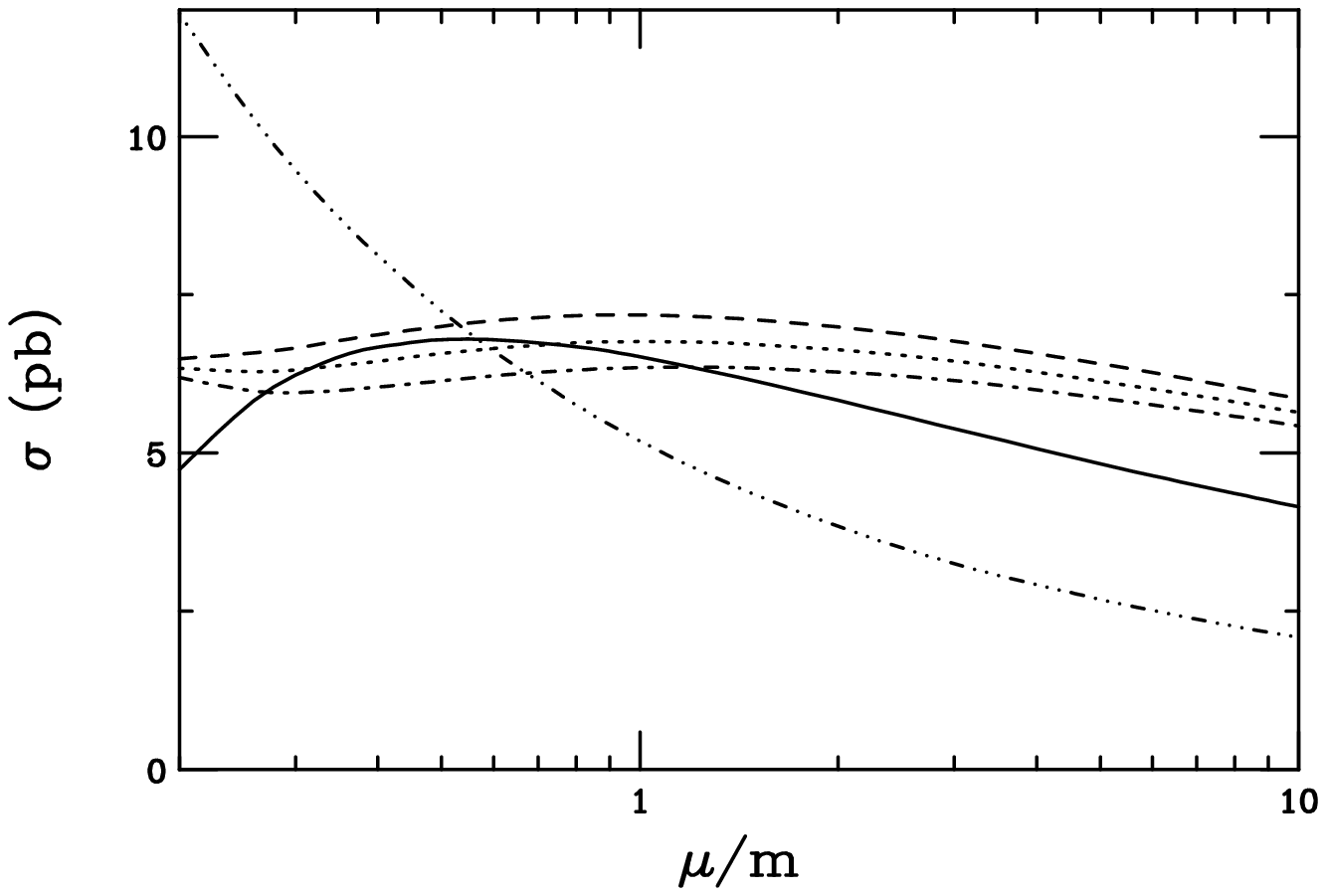} 
\includegraphics[width=7.95cm]{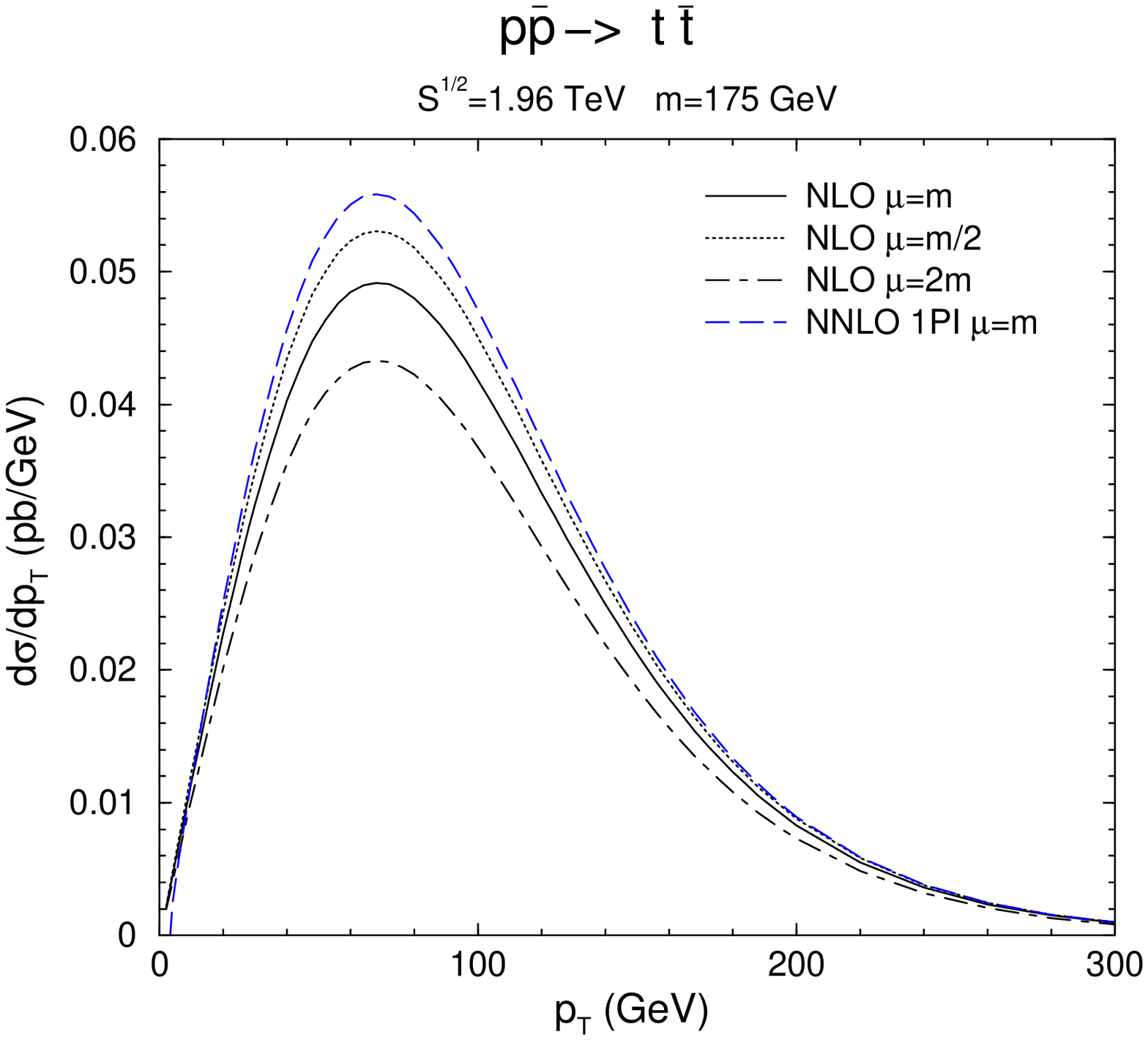} 
\caption{$t {\bar t}$ production in $p \overline p$ collisions
at $\sqrt{S} = 1.96$ TeV with $m = 175$ GeV and MRST20002 
NNLO pdf's \cite{Martin:2002aw}.
Left: The scale dependence of the total cross section
as a function of $\mu/m$. The LO (dot-dot-dot-dashed),
NLO (solid), and approximate NNLO 1PI (dashed), PIM (dot-dashed) 
and average (dotted) results are shown.
Right: The NLO and approximate NNLO top quark $p_T$ distributions are shown.}  
\label{mupt}
\end{center}
\end{figure}

In the left frame of Fig.~\ref{mupt} we show the scale dependence of 
the cross section, in the region 
$0.2 < \mu/m < 10$, at $\sqrt{S} = 1.96$ TeV.  
The NLO cross section has a milder dependence on scale than the LO result.
The NNLO cross section exhibits even less dependence on $\mu/m$, 
approaching the independence of scale corresponding to a true physical 
cross section. The change in the NNLO cross section in the
range $m/2 < \mu < 2m$, normally displayed as a measure of uncertainty
from scale variation, is less than 3\%.
In the right frame of Fig.~\ref{mupt} we show the top quark transverse 
momentum distributions at $\sqrt{S} = 1.96$ TeV. 
At NNLO we observe an enhancement of the NLO distribution
with no significant change in shape.

Finally we note that recently \cite{Kidonakis:2003sc} 
NNLO threshold corrections
were calculated for top quark production via flavor-changing 
neutral-current (FCNC) processes at the Tevatron and HERA colliders.

\vskip1cm
\noindent

\subsection*{Acknowledgements}

I would like to thank Jeff Owens, Agustin Sabio Vera, and Ramona Vogt for
fruitful collaborations.
The author's research has been supported by a Marie Curie Fellowship of 
the European Community programme ``Improving Human Research Potential'' 
under contract number HPMF-CT-2001-01221.


\section[Extending threshold exponentiation beyond logarithms for DIS 
and Drell-Yan]{EXTENDING THRESHOLD EXPONENTIATION BEYOND LOGARITHMS FOR DIS 
AND DRELL-YAN~\protect\footnote{Contributed by:
  {T.O.~Eynck, E.~Laenen, L.~Magnea}}}
\label{sec:thresh-exp}

\def\ifm{\ifmmode}
\def\msb{\ifm \overline{\rm MS}\,\, \else $\overline{\rm MS}\,\, $\fi}
\def\msbns{\ifm \overline{\rm MS}\, \else $\overline{\rm MS}\, $\fi}
\def\e{\epsilon}
\def\a{\alpha}
\def\as{\alpha_s}
\def\gtil{{\widetilde {G}}}









\subsection{Introduction}
\label{elmintr:resum;qcdsm}

Threshold resummation~\cite{Sterman:1987aj,Catani:1989ne}
sums terms in cross sections that 
grow as a production threshold is approached. There is however
empirical~\cite{Kramer:1998iq}, theoretical~\cite{Parisi:1980xd,Sterman:1987aj,Magnea:1990zb,Contopanagos:1997nh,Akhoury:1998gs,Akhoury:2003fw},
as well as speculative~\cite{YuLDprico} evidence of the fact that the
formalism enabling threshold resummation 
could be extended to
include classes of terms that are either constant or decrease
upon the approach of threshold. The resummations of
such terms, once put on a theoretically sound basis, would have
significant phenomenological consequences, as shown in
Ref.~\cite{Kramer:1998iq} for the case of Higgs production at hadron
colliders. 

The first evidence that the dominant non-logarithmic perturbative
contributions could be exponentiated goes back
to~\cite{Parisi:1980xd}, where it was shown that the partonic
Drell-Yan cross section in the DIS factorization scheme contains the
ratio of the timelike to the spacelike Sudakov form factor: large
perturbative contributions are left over in the exponentiated form of
this ratio after the cancellation of IR divergences.  This observation
was made more precise in Ref.~\cite{Sterman:1987aj}.  There, the
resummation of threshold logarithms for the Drell-Yan process was
proven to all logarithmic orders, making use of a procedure of
refactorization: the Mellin transform of the cross section is
expressed near threshold, approached by letting the Mellin variable $N$ grow very large,
as a product of functions, each
organizing a class of infrared and collinear enhancements; the
refactorization is valid up to corrections which are suppressed by
powers of $N$ at large $N$, so that terms
independent of $N$ can be treated by the same methods used to resum
logarithms of $N$.

It has also been clear for some time~\cite{Contopanagos:1997nh} that
at least a subset of the terms enhanced by logarithms but suppressed
by a power of $N$ can be resummed: in the \msb scheme, for example,
they arise from the exponentiation of the \msb quark distribution.
More recently, a factorization analysis of these terms was
performed~\cite{Akhoury:1998gs,Akhoury:2003fw}, for the case of the
longitudinal structure function in DIS, where however $(\log N)/N$
terms are the leading ones. It would be of considerable interest to
extend this analysis to other processes. Since these logarithms arise
at one loop from finite remainders of collinear singularities, which
are suppressed by infrared power counting, it is conceivable that
joint resummation~\cite{Laenen:2000ij} might provide a framework for
an all-order treatment.

Here we report on recent work~\cite{Eynck:2003fn}, where the results
of Refs.~\cite{Sterman:1987aj,Magnea:1990zb} were exploited to show
that for processes which are electroweak at tree level (such as DIS
and vector boson production through electroweak annihilation at
colliders) the exponentation of $N$-independent terms is in fact
complete. The generalization of this result to processes with
nontrivial color exchange is not straightforward, but it would of
phenomenological interest for many processes to be studied at the LHC,
including prompt photon, heavy quark and jet production.

\subsection{Extended exponentiation for electroweak annihilation}
\label{elmexpo:resum;qcdsm}

Consider the Drell-Yan partonic cross section in the DIS scheme. The 
refactorization procedure described in Ref.~\cite{Sterman:1987aj} leads
to the expression
\begin{equation}
\widehat{\omega}_{\mathrm{DIS}} (N) =
\frac{(\Gamma(Q^2, \e))^2}{\left|\Gamma(-Q^2, \e)\right|^4}
\left(\frac{\psi_r(N, \e)}{\chi_r(N, \e)} \right)^2
\frac{U_r(N,\e)}{V_r^2(N,\e)} \frac{1}{J_r^2(N,\e)} + {\cal O}(1/N)~.
\label{elmsumdis:resum;qcdsm}
\end{equation}
The presence of the final-state jet function $J_r(N,\epsilon)$ e.g. is 
direct consequence of dividing the unfactorized Drell-Yan cross section
by the deep-inelastic structure function.
A similar expression emerges in the \msb scheme,
\begin{equation}
 \label{elmomms:resum;qcdsm}
 \widehat{\omega}_{\msb}(N) =
 \left(\frac{|\Gamma(Q^2, \e)|^2}{\phi_v(\e)^2} \right) \:
 \left(\frac{\psi_r(N, \e)^2 \, U_r(N, \e)}{\phi_r(N,
 \e)^2} \right) + {\cal O}(1/N)~.
\end{equation}
Notice the absence of $J_r$, which does not occur in the \msb density.
Here $\Gamma$ is the quark form factor; $\phi$, $\psi$ and $\chi$ are
different quark distributions, with the same collinear singularities
but different finite contributions: $\phi$, in particular, is the \msb
distribution, consisting only of collinear poles; $U$ and $V$ are
eikonal functions describing soft radiation at wide angles; finally
$J$ is a jet function responsible for soft and collinear contributions
from the final state DIS current jet. All functions exponentiate, and
in each case real ($r$) and virtual ($v$) contributions have been
separated.  Finite virtual contributions in both cases can be expressed
just in terms of the quark form factor. Note that although each
function is divergent and thus depends on $\e = 2 - d/2$, the partonic
cross sections are finite.

All functions involved in Eqs.~(\ref{elmsumdis:resum;qcdsm}) and
(\ref{elmomms:resum;qcdsm}) can be precisely defined in terms of quark
fields and eikonal lines. Renormalization group analysis and explicit
evaluation lead to expressions that have the familiar exponential
form~\cite{Catani:1989ne}, with corrections involving $N$-independent
terms. For the DIS scheme one finds
\begin{eqnarray}
\label{elmfindis:resum;qcdsm}
 \widehat{\omega}_{\mathrm{DIS}} (N) & = &  
 \left|\frac{\Gamma(Q^2, \e)}{\Gamma(-Q^2, \e)} \right|^2
 \exp \Big[ \mathcal{F}_{\mathrm{DIS}}(\a_s) \Big] ~\exp \Bigg[ \int_0^1 \! dz \,
 \frac{z^{N - 1} - 1}{1 - z} \\ 
 & \times & \Bigg\{ 2 \int_{(1 - z)Q^2}^{(1 - 
 z)^2 Q^2} \! \frac{d \xi^2}{\xi^2} \, A \left( \a_s( \xi^2) \right) 
 - 2 B \left(\a_s \left((1 - z) Q^2 \right) \right) + 
 D \left( \a_s \left((1 - z)^2 Q^2 \right) \right) 
 \Bigg\} \Bigg] \nonumber .
\end{eqnarray}
Similarly, for the \msb scheme one has the expressioon
\begin{eqnarray}
 \widehat{\omega}_{\msb}(N) & = & \left|\frac{\Gamma(Q^2,
 \e)}{\Gamma(- Q^2, \e)} \right|^2
 \left( \frac{\Gamma(-Q^2,\e)}{ \phi_v(Q^2,\e)} \right)^2
 \exp \Big[ \mathcal{F}_{\msbns} (\a_s) \Big] ~\exp \Bigg[ \int_0^1 \! dz \,
 \frac{z^{N - 1} - 1}{1 - z} 
 \nonumber \\ & \times &
 \Bigg\{ 2 \, \int_{Q^2}^{(1 - z)^2 Q^2} 
 \frac{d \xi^2}{\xi^2} \, A \left(\a_s(\xi^2) \right)
 + D \left(\a_s \left((1 - z)^2 Q^2 \right) \right) \Bigg\} 
 \Bigg]~.
\label{elmfinms:resum;qcdsm}
\end{eqnarray}
The functions $\mathcal{F}_{\mathrm{DIS}}$ and $\mathcal{F}_{\msbns}$
are given to order $\alpha_s$ below in Eqs.~(\ref{elmrattwo:resum;qcdsm}) and (\ref{elmmsexp:resum;qcdsm}).
As is well-known~\cite{Sterman:1987aj,Catani:1989ne}, the function $A$ is 
responsible for leading logarithms of $N$ to all orders. To achieve NLL
accuracy one must compute it to two loops, obtaining
\begin{equation}
 A^{(1)} = C_F~, \qquad
 A^{(2)} = \frac{1}{2} \left[C_A C_F \left(\frac{67}{18} - \zeta(2) \right) -
 n_f C_F \left(\frac{5}{9} \right)\right]~.
\label{elmcoeffa:resum;qcdsm}
\end{equation}
To NLL accuracy, one also needs the functions $B$ and $D$ to one loop, 
which are given by
\begin{equation}
 B^{(1)} = - \frac{3}{4} C_F~, \qquad D^{(1)} =  0~.
\label{elmcoeffb:resum;qcdsm}
\end{equation}
The remaining ingredients, collecting $N$-independent terms, are easily 
computed at one loop. For the DIS scheme one needs
\begin{equation}
 \left| \frac{\Gamma(Q^2, \e)}{\Gamma(-Q^2, \e)} \right|^2 = 
 \exp \left[ \frac{\a_s(Q)}{\pi} C_F \left(3 \zeta(2) \right) \right]~,
 \qquad
 \mathcal{F}_{DIS} \left( \a_s \right) = \frac{\a_s}{\pi} C_F \left( \frac{1}{2} + 
 \zeta(2) \right)~.
 \label{elmrattwo:resum;qcdsm}
\end{equation}
For the \msb scheme, one further needs
\begin{equation}
 \frac{\Gamma(-Q^2,\e)}{ \phi_v(Q^2,\e)} = \exp \left[
 \frac{\a_s}{\pi} C_F \left(\frac{\zeta(2)}{4} - 2 \right)\right]~,
 \qquad \mathcal{F}_{\msbns} \left( \a_s \right) = \frac{\a_s}{\pi} C_F
 \left(- \frac{3}{2} \zeta(2) \right)~. 
 \label{elmmsexp:resum;qcdsm}
\end{equation}
Notice finally that by taking the ratio of
Eq.~(\ref{elmfinms:resum;qcdsm}) and Eq.~(\ref{elmfindis:resum;qcdsm})
one finds directly the (square of) the DIS structure function
$F_{2,(\msbns)}(N)$, factorized in the \msb scheme, which can
then be computed to the same accuracy without introducing other
information. The ratio of form factors drops out from
$F_{2,(\msbns)}(N)$, as does the function
$D$~\cite{Gardi:2002xm}, except for a contribution to the running
between the physical scales $Q^2$ and $(1 - x) Q^2$, which can be
systematically reexpressed as a modification of the function $B$.

All the functions appearing in Eq.~(\ref{elmfindis:resum;qcdsm}) and
Eq.~(\ref{elmfinms:resum;qcdsm}) can be explicitly evaluated at two
loops by matching with the complete two-loop calculation of
Ref.~\cite{vanNeerven:1992gh}.  
An alternative, and often simpler method to determine
the two-loop coefficients in these expressions uses equations
derived from the fact that real and virtual contributions in
the factorized expressions 
(\ref{elmmsexp:resum;qcdsm}) and (\ref{elmomms:resum;qcdsm}) are separately
finite \cite{Eynck:2003fn}.

It is also important to keep in mind
that the exponentiation of $N$--independent terms does not have the
predictive power of the standard resummation of threshold
logarithms. In that case, typically, an entire tower of logarithms can
be exactly predicted to all orders by performing just a low order
calculation. Here, on the other hand, functions such as $\mathcal{F}_{DIS}$ and
$\mathcal{F}_{\msbns}$ receive new nontrivial contribution at each perturbative
order. The exponentiation pattern is nonetheless nontrivial, and
higher order terms predicted by the exponentiation can be considered
representative of the size of the complete higher order
correction~\cite{Eynck:2003fn}.




 \section[Joint resummation for top quark production]{JOINT
   RESUMMATION FOR TOP QUARK PRODUCTION~\protect\footnote{Contributed by: {A.~Banfi, E.~Laenen}}}
\label{sec:joint-resumm-top}


The formalism of joint resummation \cite{Li:1998is,Laenen:2000ij} 
for hadronic cross sections of
distributions singular at partonic threshold and at zero recoil
has so far been applied to only a few
processes. Recent studies involve processes
that proceed at lowest order through a 
$2 \rightarrow 1$ electroweak ($Z/W$ production \cite{Kulesza:2002rh})
or Yukawa interaction (Higgs production \cite{Kulesza:2003wn}).
For these cases, the observables are the production cross sections 
at fixed mass $M$ and measured $Q_T$. Partonic threshold is 
then defined by $z\equiv M^2/\hat{s} = 1$, where $\hat{s}$ is the
partonic center of mass (cms) energy squared,
and zero recoil by $Q_T=0$. At any finite order, 
the distributions take the form of
plus-distributions $\left[\ln^{k}(1-z)/(1-z)\right]_+$ and
$\left[\ln^{k}(M/Q_T)/Q_T\right]_+$. Note that in these 
observables the latter distributions enter the physical
cross sections, whereas the former are defined, after factorization,
in the context of a perturbative analysis of the hard scattering.

In Ref.~\cite{Laenen:2000de} the case of the
prompt photon hadroproduction cross section at measured $p_T$
was analyzed, and a preliminary numerical study performed. 
In this case, a single-particle inclusive process proceeding
through a $2 \rightarrow 2$ reaction at lowest order,
the identification of the
recoil variable $Q_T$ can only be made in the context of 
a refactorization analysis,
just like the threshold variable $z$.
 Through such an analysis, it 
is possible to identify a reduced hard scattering with cms energy 
squared $\tilde{s}$ and at transverse momentum $\vec{Q_T}$ with
respect to the hadronic cms system. Note that this transverse momentum
is invariant w.r.t. longitudinal boosts. In this way, the hard
scattering need produce a photon with transverse momentum
$\vec{p_T}' = \vec{p_T} - \vec{Q_T}/2$. It still
remains to implement a procedure that 
consistently matches the joint resummed cross section to finite
order calculations.
In Refs.~\cite{Laenen:2000ij,Laenen:2000de} a simple
cut-off $\bar{\mu}$ was used in the integration over $Q_T$, with $\bar{\mu}$
smaller that $2p_T$. For the prompt photon case, the hard
scattering is singular at $Q_T = 2p_T$.

In this brief report we present a preliminary study of the application
of joint resummation to another prominent single-particle
inclusive cross section, the $p_T$ distribution of 
top quarks produced in hadronic collisions. Our motivation
is to see what effect joint resummation has on a
distribution in a TeV collider process that is nevertheless
near threshold.
Two key differences with the prompt photon case are \textit{(i)}
the heavy quark mass $m$, preventing a singularity
in the reduced hard scattering function when $Q_T = 2 p_T$, and
\textit{(ii)} the possibility of 
multiple  colored states for the produced top quark pair.

The jointly resummed expression for this observable can be written as
\begin{equation}
    \frac{d\sigma_{AB\to t\bar{t}+X}}{dp_T}
 = \int d^2 Q_T \, \theta(\bar{\mu}-Q_T)
\frac{d\sigma_{AB\to  t\bar{t}+X}}{dp_T d^2 \vec Q_T}\,,
\label{sec:resum;qcdsm:bl1}
\end{equation}
where
\begin{equation}
\label{sec:resum;qcdsm:bl2}
  \begin{split}
    &\frac{d\sigma_{AB\to t\bar{t}+X}}{dp_T d^2 \vec Q_T}= 
    p_T \int\frac{d^2 b}{(2\pi)^2}e^{i \vec b \cdot  \vec Q_T}
    \int\frac{dN}{2\pi i} \phi_{a/A}(N,\mu)\>\phi_{b/B}(N,\mu)\>
    e^{E_{ab}(N, b)}\\
    &\times\frac{e^{-2\>C_F \>t(N)\>(\mathrm{Re} L_\beta+1)}}{4\pi S^2}\left(
      \tilde M^2_{\bf 1}(N)+\tilde M^2_{\bf 8}(N)
    e^{2 \>t(N)\> \mathrm{Re} \Gamma_{\bf 8}}
    \right) 
        \left(\frac{S}{4(m^2+|\vec p_T-\vec Q_T/2|^2)}\right)^{N+1}
    \!\!\!\!\>.  
  \end{split}
\end{equation}
where the impact vector $\vec b$ is Fourier conjugate to $\vec Q_T$,
and the variable $N$ is Laplace conjugate to the 
variable
\begin{equation}
\label{sec:resum;qcdsm:bl3}
1-  x_T^2=1- \frac{4 m_T^2}{S}\>
\end{equation}
where $x_T^2=1$ defines hadronic threshold. Furthermore 
\begin{equation}
\label{sec:resum;qcdsm:bl4}
\begin{split}
&t(N)=\int_Q^{Q/N}\frac{dk_t}{k_t}\frac{\alpha_s(k_t)}{\pi}\>,\quad
 \mathrm{Re} L_\beta=\frac{1+\beta^2}{2\beta}
  \left(\ln\frac{1-\beta}{1+\beta}\right)\>, \quad
 \mathrm{Re} \Gamma_{\bf 8} = 
  \frac{C_A}{2}\left(\ln\frac{m_T^2}{m^2}+ \mathrm{Re} L_\beta\right)\>.
\end{split}  
\end{equation}
where $ \beta=\sqrt{1-4m^2/s}$.
The exponentional function $E_{ab}$ is given in Ref.\cite{Kulesza:2002rh}
and $M^2_{\bf 1}(N), M^2_{\bf 8}(N)$ are the 
Laplace moments of the lowest order heavy quark production
matrix elements for either the $q\bar{q}$ or $gg$ channel,
the index labelling the color-state of the heavy quark pair.

In two figures we illustrate the effect of joint resummation over threshold
resummation for the top quark ($m=175$ GeV) $p_T$ spectrum
for at Run II Tevatron, and only  for the dominant $q\bar{q}$ 
channel. It produces the top quark pair at lowest order in an octet state. 
We use $\alpha_s=0.1$ and toy densities
$\phi_a(x) = x(1-x)^3$. 
 \begin{figure}[htbp]
   \begin{center}
     \includegraphics[width=\textwidth]{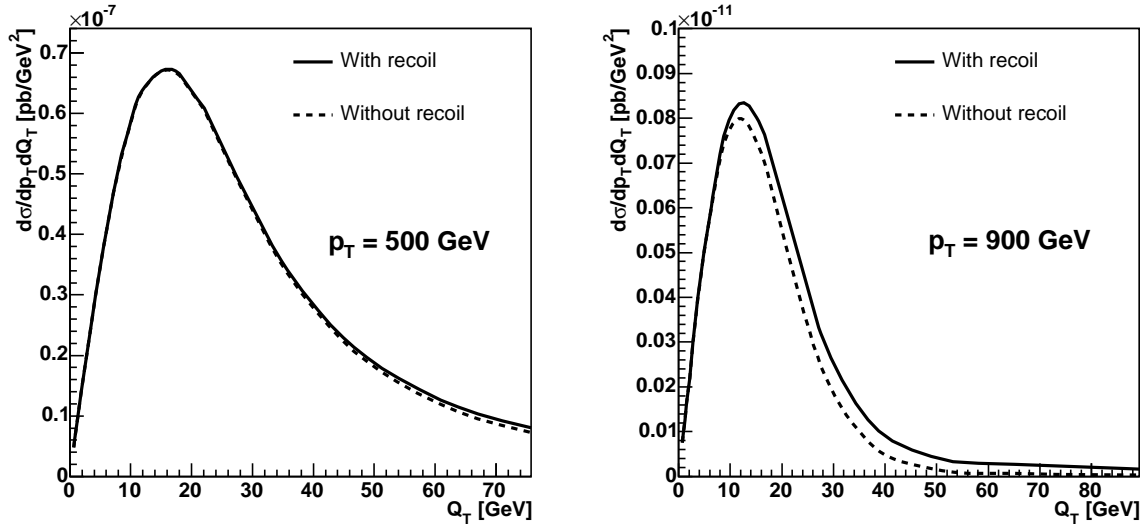}
     \caption{The $Q_T$ profile for two $p_T$ values.}
\label{fig:resum;qcdsm:profilettb}
\end{center}
 \end{figure}
In Fig.~\ref{fig:resum;qcdsm:profilettb} we exhibit 
the $Q_T$ profile of Eq.~\eqref{sec:resum;qcdsm:bl2} for two rather large $p_T$ values,
in analogy to Fig.~1 in Ref.\cite{Laenen:2000de}. Note that these profiles
are only of theoretical relevance, only their integral over $\vec Q_T$ can be measured. 
Recoil effects can be removed by neglecting $\vec Q_T$ in the 
last factor in Eq.~\eqref{sec:resum;qcdsm:bl2}. We observe a small enhancement
over the threshold-resummed result, in particular at very large $p_T$.
Although not shown in the
figure, there is in our case 
no divergence when $Q_T$ approaches $2 p_T$.

To keep the recoil soft with respect to the hard scale, we choose
the cut-off $\bar{\mu} = 0.2 m_T$.
We can now show, in Fig.~\ref{fig:resum;qcdsm:ptspectrattb} 
the effect of joint resummation
on the $p_T$ distribution, via Eq.~\eqref{sec:resum;qcdsm:bl1}.
We observe only a small enhancement at very large 
$p_T$ values. The suppression at
lower $p_T$ values is a consequence of the $Q_T$ cut $\bar{\mu}$.
A proper matching procedure should resolve this issue.
Actually we note that this suppression is 
absent if we choose $\bar{\mu}$ around 200 GeV.
 \begin{figure}[htbp]
   \begin{center}
     \includegraphics[width=0.8\textwidth]{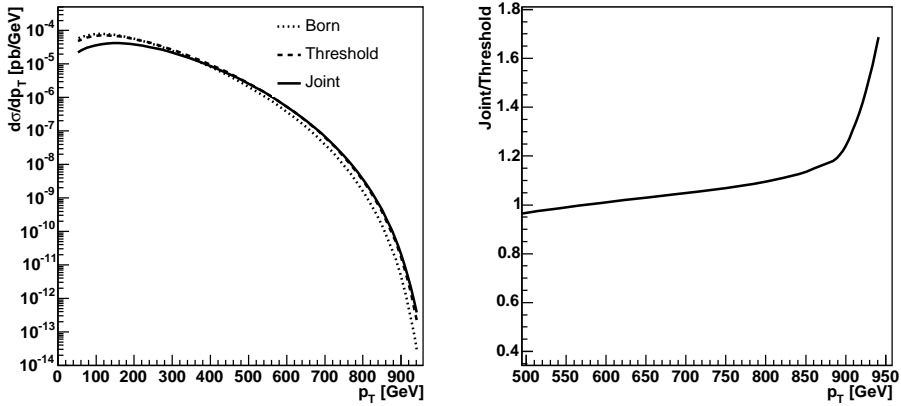}
     \caption{The $p_T$ spectra for top
       quark producation at Tevatron run II, using toy PDF's
       $x(1-x)^3$, in the $q\bar{q}$ channel, and their
ratio. }\label{fig:resum;qcdsm:ptspectrattb}
\end{center}
 \end{figure}
The enhancement at large $p_T$ is a consequence of what is effectively a 
smearing of the recoilless $p_T$ spectrum induced by perturbative radiation.

A more extenstive study including
the $gg$ channel and use of realistic parton distribution functions is 
forthcoming \cite{banfi04}.

\vskip1cm
\noindent

\subsection*{Acknowledgements}

This work is supported by the Foundation for Fundamental 
Research of Matter (FOM) and the National Organization for Scientific 
Research (NWO).


\section[A comparison of predictions for SM Higgs boson production at
the LHC]{A COMPARISON OF PREDICTIONS FOR SM HIGGS BOSON PRODUCTION AT
  THE LHC~\protect\footnote{Contributed by: {C.~Balazs, M.~Grazzini,
  J.~Huston, A.~Kulesza, I.~Puljak}}}\label{huston}


\def\ltap{\raisebox{-.4ex}{\rlap{$\,\sim\,$}} \raisebox{.4ex}{$\,<\,$}} 
\def\gtap{\raisebox{-.4ex}{\rlap{$\,\sim\,$}} \raisebox{.4ex}{$\,>\,$}} 












\subsection{Introduction}

The dominant mechanism for the production of a SM
Higgs boson at the
LHC is gluon-gluon fusion through a heavy (top) quark loop.
For this reason this channel
has attracted a large amount of theoretical attention~\cite{Cavalli:2002vs}.
Recently, the total cross section has been calculated to NNLO 
in the strong coupling constant $\alpha_s$
(i.e. at order ${\cal O}(\alpha_s^4)$)
\cite{Catani:2001ic,Harlander:2001is,Harlander:2002wh,Anastasiou:2002yz,Ravindran:2003um}
and also contributions from multiple soft gluon emission have been
consistently included to NNLL accuracy \cite{Catani:2003zt}.
In addition to the size of the total
rate, a knowledge of the shape of the Higgs boson $p_T$ distribution
is essential for any search and analysis strategies at the LHC. In
particular, the $p_T$ distribution for the Higgs boson is expected
to be harder than the one of its corresponding backgrounds.
The Higgs boson $p_T$ distribution has been computed with LL parton shower
Monte Carlos (HERWIG~\cite{Corcella:2002jc}
and PYTHIA~\cite{Sjostrand:2003wg}),
and through
various resummed calculations. The latter techniques are the more powerful
ones, but it is primarily the former that experimentalists at the LHC
have to rely upon, because of their flexibility in allowing to test
the effects of the various kinematic cuts which may optimize search 
strategies.

In the kinematic region $p_T^2\ll m_H^2$, where most of the events
are expected,
large logarithmic corrections
appear of the form $\alpha_s^n \ln^{m}m_H^2/p_T^2$ that spoil the validity
of the fixed order perturbative expansion.
The $p_T$ distribution can be written as
\begin{eqnarray}
\label{deco}
\frac{d\sigma}{dp_T^2} = \frac{d\sigma^{\rm res.}}{dp_T^2} +
\frac{d\sigma^{\rm fin.}}{dp_T^2}\, .
\end{eqnarray}
The first term contains all
logarithmically-enhanced contributions and requires their resummation to all
orders. The second term is free from logarithmically-enhanced contributions
and can be evaluated at fixed order in perturbation theory.
The method to perform the all-order resummation is well known:
to correctly take into account momentum conservation
the resummation must be performed in the
impact parameter ($b$) space \cite{Parisi:1979se,Dokshitzer:1980hw}.
The large logarithmic contributions are exponentiated in the
Sudakov form factor, which
in the CSS \cite{Collins:1985kg}
approach takes the form
\begin{eqnarray}
S_c =
\int_{b_0^2/b^2}^{m_H^2} \frac{d \mu^2}{\mu^2}
\left[
A_c \left( \alpha_s(\mu) \right) \ln \left( \frac{m_H^2}{\mu^2} \right) +
B_c \left( \alpha_s(\mu) \right)
\right] ,
\label{Eq:Sudakov}
\end{eqnarray}
where $b_0=2e^{-\gamma}$ and $c=q,g$.
The $A_c$ and $B_c$ functions
are free of large logarithmic corrections
and can be computed as expansions in the strong
coupling constant $\alpha_s$:
\begin{eqnarray}
A_c(\alpha_s) = &&
\sum_{n=1}^\infty
\left( \frac{\alpha_s}\pi \right)^n A_c^{(n)}, \\
B_c(\alpha_s) = &&
\sum_{n=1}^\infty
\left( \frac{\alpha_s}\pi \right)^n B_c^{(n)}.
\end{eqnarray}
The functions $A_c$ and $B_c$ control soft and flavour-conserving collinear radiation
at scales $1/b \ltap \mu \ltap m_H$.
Purely soft radiation at a very low scales $\mu \ltap 1/b$ cancels out because
the cross section is infrared safe and
only purely collinear radiation
up a scale $\mu \sim 1/b$ remains, which is taken into account by
the coefficients
\begin{equation}
C_{ab}(\alpha_s,z) =
\sum_{n=1}^\infty
\left( \frac{\alpha_S}\pi \right)^n C_{ab}^{(n)}(\alpha_s,z).
\end{equation}
Beyond NLL accuracy, to preserve the process independence of
the resummation formula,
an additional (process dependent)
coefficient $H$ is needed \cite{Catani:2000vq},
which accounts for
hard virtual corrections and has an expansion
\begin{equation}
H_c(\alpha_s)=1+\sum_{n=1}^{\infty}
\left( \frac{\alpha_s}\pi \right)^n H_c^{(n)}.
\end{equation}
In the case of Higgs boson production through $gg$ fusion, the relevant
coefficients $A_g^{(1)}$, $A_g^{(2)}$ and $B_g^{(1)}$
are known \cite{Catani:1988vd}
and control the resummation up to NLL accuracy
\footnote{There are two different classification schemes of the LL, NLL, NNLL, etc
terms and their corresponding B contents. Here we use the most popular
scheme. Another is discussed in Ref.~\cite{Giele:2002hx}.}.
The NNLL coefficients $C^{(1)}_{ab}$ and $H_g^{(1)}$ are also known
\cite{Kauffman:1992cx,Catani:2000vq}.
The NNLL coefficient $B_g^{(2)}$
has been computed in Refs.~\cite{deFlorian:2000pr,deFlorian:2001zd},
whereas $A_g^{(3)}$ is not yet known exactly.
In the following we assume that its value
is the same that appears in threshold resummation \cite{Vogt:2000ci}.

\subsection{Predictions for $p_t$ spectra and comparisons}

In the 1999 Les Houches workshop,
a comparison~\cite{Catani:2000zg,Balazs:2000sz} of the
HERWIG and PYTHIA (2 versions) predictions for the Higgs boson $p_T$
distribution with those of a $p_T$ resummation program
(ResBos~\cite{Balazs:1997xd,Balazs:2000wv}) was carried out. This comparison was
continued in the 2001 workshop and examined the impact
of the $B^{(2)}$ coefficient~
\cite{Cavalli:2002vs}.
In the meantime, a
number of new theoretical predictions have become available, both from
resummation and from the interface of NLO calculations 
with parton shower Monte Carlos. For these proceedings,
we have carried out a comparison of most of the
available predictions for the Higgs boson $p_T$ distribution at the LHC. We
have used a Higgs boson mass of 125 GeV
and either  the MRST2001 or the CTEQ5M pdf's.
The difference between the two pdf's for the production of a 125 GeV
mass Higgs boson is  of the order of a few percent.
Before comparing the different predictions, we comment on the various approaches in turn.

Parton shower MC programs such as HERWIG, which implements
angular ordering exactly, implicitly
include the $A^{(1)}$, $A^{(2)}$ and $B^{(1)}$ coefficients
and thus correctly sum the LL
and part of the N$^k$LL contributions. However, in the most straightforward
implementations, MC cannot correctly treat hard radiation. 
By contrast, the PYTHIA MC,
which does not provide an exact implementation
of angular ordering, has
a hard matrix element correction
\footnote{Very recently hard matrix element corrections for Higgs productions
have been implemented in HERWIG as well \cite{Corcella:2004fr}.}.
Recently, an approach to match NLO calculations to
parton showers generators, MC@NLO \cite{Frixione:2002ik,Frixione:2003ei},
has been proposed, and applied, amongst the other, to Higgs production.
This method joins the virtues of
NLO parton level generators
(correct treatment of hard radiation, exact NLO normalization)
to the ones of MC. It thus can be compared
to a resummed calculation at NLL+NLO accuracy.

As far as resummed calculations are concerned,
we first consider two implementation of the CSS approach.
The ResBos code
includes the $A_c^{(1,2,3)}$, $B_c^{(1,2)}$ and $C_{ab}^{(1)}$
coefficients
in the low-$p_T$ region and matches this to the NLO distribution at high $p_T$.
NNLO effects at high $p_T$ are approximately taken into account by
scaling the second term in Eq.~(\ref{deco}) with a K-factor.
The matching is performed through a switching procedure whose uncertainty
will be considered in the following.
The calculation of Berger and Qiu~\cite{Berger:2002ut} also performs a $p_T$
resummation in $b$ space and is accurate to NLL. The coefficient
$B^{(2)}$ is included but the matching is still to NLO.
Note that in both these approaches the integral of the spectrum
is affected by higher-order contributions
included in a non-systematic manner
whose effect is not negligible for Higgs production.

The prediction by Bozzi, Catani, de Florian and Grazzini~\cite{Bozzi:2003jy} (labeled Grazzini et al. in the following)
is based on an implementation of the $b$-space
formalism described in \cite{Catani:2000vq,Bozzi:2003jy}. The calculation has
the highest nominal accuracy since it matches NNLL resummation at small
$p_T$ to the NNLO result at high $p_T$\cite{deFlorian:1999zd}.
This approach includes the coefficients
$C_{ab}^{(2)}$ and $H_g^{(2)}$ in approximated form.
The main differences with respect to the standard CSS approach
are the following.
A unitarity constraint is imposed, such that the total cross section at the
nominal (NNLO) accuracy is exactly recovered upon integration.
A study of uncertainties from missing higher order contributions
can be performed as it is normally done in fixed order calculations, that is,
by varying renormalization and factorization scales around the central value,
that is chosen to be $m_H$.

Finally, we discuss the $p_T$ distribution of
Ref.~\cite{Kulesza:2003wn} (Kulesza et al.).
This is obtained using a
joint resummation formalism, by which
both threshold and low-$p_T$
logarithmic contributions are resummed to all orders. This approach
has been formally developed to NLL accuracy, but the NNLL coefficients
$A^{(3)}, B^{(2)}, C^{(1)}$ and $H^{(1)}$ can also be incorporated.
The matching is still performed to NLO.
Even though a low mass Higgs boson
at the LHC is produced with relatively low $x$ partons, threshold effects
can still be significant due to the large color charge in the
$gg$ initial state as well as steep $x$ dependence of the gluon
distribution functions at low $x$. This leads to an increased sensitivity
to Sudakov logarithms associated with partonic threshold for gluon-induced
processes, as shown in Ref.~\cite{Catani:2003zt}. 

It is known that the low-$p_T$ region is sensitive to
non-perturbative effects. These are expected to be less
important in the gluon channel due to the larger colour charge
of the $gg$ initial state \cite{Balazs:2000sz}.
Different treatments of non-perturbative effects are included in
the ResBos, Berger et al. and Kulesza et al calculation,
whereas Grazzini et al. prediction is purely perturbative.

 \begin{figure}[htbp]
   \begin{center}
     \includegraphics[width=0.5\textwidth]{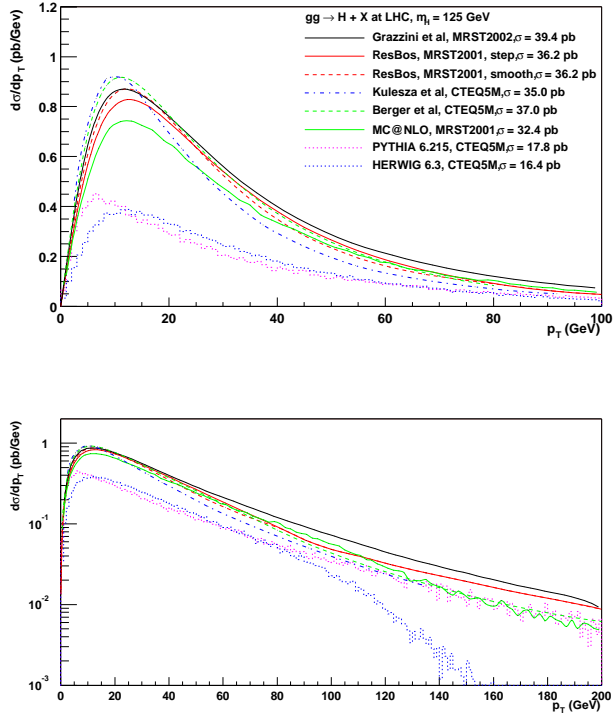}
     \caption{The absolute predictions for the production of a 125 GeV mass  Higgs boson at the LHC.}
\label{fig:higgs_abs_norm}
\end{center}
 \end{figure}

The absolute predictions for the cross sections
are shown in Figure~\ref{fig:higgs_abs_norm}.
All curves are obtained in the $m_{\rm top}\to\infty$ limit.
HERWIG and PYTHIA cross sections are significantly smaller than
the other predictions, their normalization being fixed to LO.
In the high-$p_T$ region, the HERWIG prediction drops quickly due to the
lack of hard matrix element corrections. PYTHIA, in
contrast, features the hard matrix element corrections.
We also note that PYTHIA prediction is significantly
softer than all the other curves, and thus
its overall shape is fairly different
from all the other predictions.

The MC@NLO cross section, about 32.4 pb, is roughly
twice that of the HERWIG and PYTHIA predictions, being fixed to the NLO total
cross section.

Two predictions ({\em step, smooth}) are shown for ResBos 
which differ in the manner in which the matching at high $p_T$
is performed. Their difference can be considered as
an estimate of the ambiguity in the switching procedure.
The two curves correspond to the same total cross section of about 36.2 pb,
which is about 8 \% higher than the NLO cross section.
This is the effect of the higher-order terms
that enter the prediction for the total rate in
the context of the CSS approach.
A slightly softer curve is obtained by Berger and Qiu.
The predicted cross section (37 pb) is close to that of ResBos. 

The Grazzini et al. prediction has an integral of about 39.4 pb, which
corresponds to the total cross section at NNLO. Contrary to what is done
in Ref.~\cite{Bozzi:2003jy}, here the curve is obtained with
MRST2002 NNLO partons and three-loop $\alpha_s$.
The difference with the result obtained with MRST2001 NNLO PDFs
is completely negligible.

Concerning the Kulesza et al curve,
the subleading terms associated with low $x$ emission
(i.e. in the limit opposite to partonic threshold) and of which only a
subset is included in the joint resummation formalism, play an important role
numerically. As a result, the total cross section turns out to be 35 pb,
about $10\%$ lower than the pure threshold result, which is 39.4 pb
 \cite{Kulesza:2003wn}.

We now want to examine in more detail the relative shapes of the
predictions plotted in Figure.~\ref{fig:higgs_abs_norm}.
In Figure.~\ref{fig:higgs_no_norm} all the predictions are normalized
to the Grazzini et al. cross section of 39.4 pb.

 \begin{figure}[htbp]
   \begin{center}
     \includegraphics[width=0.5\textwidth]{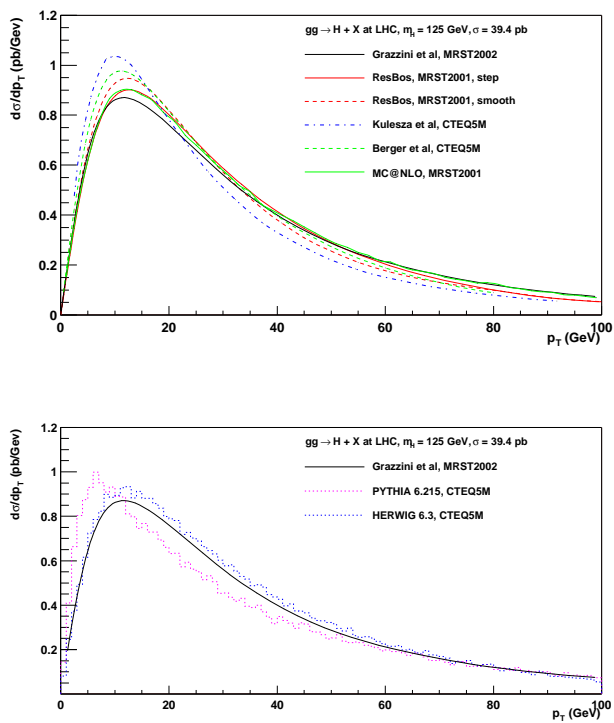}
     \caption{The predictions for the production of a 125 GeV mass  Higgs boson 
at the LHC, all normalized to the same cross section for better shape comparison.}
\label{fig:higgs_no_norm}
\end{center}
 \end{figure}

In the region of small and moderate $p_T$ (say, smaller than 100 GeV) 
all of the predictions are basically consistent with each other, with
the notable exception of PYTHIA, which predicts a much softer spectrum.
The curve of Kulesza et al. is also softer than the others.

For larger $p_T$, HERWIG gives unreliable predictions, since the
transverse momentum is generated solely by means of the parton shower,
and therefore it lacks hard matrix element effects.
The Grazzini et al. and ResBos curves are harder than
MC@NLO for large $p_T$. There are two reasons for this. Grazzini 
et al. implement  the NNLO matrix elements exactly, corresponding to the
emission of two real partons accompanying
the Higgs in the final state \cite{deFlorian:1999zd};
ResBos mimics these contributions, by multiplying the NLO matrix elements
by the K factor. MC@NLO, on the other hand, contains only NLO matrix
elements (one real parton in the final state). Secondly, Grazzini et al. 
and ResBos choose the renormalization and factorization scales equal to $m_H$,
whereas in MC@NLO these scales are set equal to the transverse mass of the
Higgs, $\sqrt{m_H^2+p_T^2}$. The difference is small at the level of
total rates, but it is not negligible in the tail of the $p_T$ distribution.

\subsection{Conclusions}

Up to now, the ATLAS  and CMS  experiments have relied
primarily on the predictions from HERWIG and PYTHIA in designing both
their experiments as well as defining their search and analysis
strategies. In the last few years, a number of tools for and predictions 
of the Higgs boson cross section at the LHC have become available, with 
the inclusion of beyond-the-leading-order effects at different level of 
accuracy. In the case of total rates, NNLO results have recently become
available; their consistent inclusion in experimental analysis will
allow to further decrease the estimated lower bound on the integrated 
luminosity to be collected for discovery.

In this contribution, we primarily 
focused on the predictions for the $p_T$ spectrum, comparing the results
of Monte Carlos with those obtained with analytically-resummed calculations.
In contrast to the situation in 1999, all of the predictions, with the
exception of PYTHIA, result in the same general features, most notably in
the position of the peak. However, differences do arise, because of 
different treatments of the higher orders. It is an interesting question
beyond the scope of this review that of whether these differences are
resolvable at the experimental level, which may lead to modify the strategy
for searches. In order to answer this, studies including realistic 
experimental cuts must be performed with the newly available tools.


\section[Matrix-element corrections to $gg/q\bar q\to$ Higgs in HERWIG]
{MATRIX-ELEMENT CORRECTIONS TO $gg/q\bar q\to$ HIGGS IN HERWIG~\protect\footnote{Contributed
  by: {G.~Corcella, S.~Moretti}}}
\def\s#1{{\small#1}}
\def\lsim{\:\raisebox{-0.5ex}{$\stackrel{\textstyle<}{\sim}$}\:}
\def\gsim{\:\raisebox{-0.5ex}{$\stackrel{\textstyle>}{\sim}$}\:}
\def\PD{\s{PDG}}
\def\TA{{\small TAUOLA}}
\def\HW{\s{HERWIG}}
\def\JS{\s{JETSET}}
\def\PY{\s{PYTHIA}}
\def\IS{\s{ISAJET}}
\def\IW{\s{ISAWIG}}
\def\SM{\s{SM}}
\def\MSSM{{MSSM}}
\def\SY{\s{SUSY}}
\def\QCD{\s{QCD}}
\def\QED{\s{QED}}
\def\DIS{\s{DIS}}
\def\LEP{\s{LEP}}
\def\LHC{\s{LHC}}
\def\OPAL{\s{OPAL}}
\def\PDF{\s{PDFLIB}}
\def\CERN{\s{CERN}}
\def\RPV{\rlap{/}{R}$_{\mbox{\scriptsize p}}$}
\def\BNV{\rlap{/}{B}}
\def\Ord{\buildrel{\scriptscriptstyle <}\over{\scriptscriptstyle\sim}}
\def\OOrd{\buildrel{\scriptscriptstyle >}\over{\scriptscriptstyle\sim}}
\def\gh{\Gamma_{\scriptscriptstyle \rm H}}
\def\gtap{\raisebox{-.4ex}{\rlap{$\sim$}} \raisebox{.4ex}{$>$}}
\def\ltap{\raisebox{-.4ex}{\rlap{$\sim$}} \raisebox{.4ex}{$<$}}
\def\ycut{$y_{\mbox{\tiny cut}}$}
\def\mw{m_{\scriptscriptstyle \rm W}}
\def\lms{\Lambda_{\overline{\rm MS}}}
\def\half{\mbox{\small $\frac{1}{2}$}}
\def\thlf{\mbox{\small $\frac{3}{2}$}}
\def\as{\alpha_{\mbox{\tiny S}}}
\def\ee{e^+e^-}
\def\MC{Monte Carlo}
\def\VEV#1{\langle{#1}\rangle}
\def\qbar{\bar{q}}
\def\Qbar{\bar{Q}}
\def\dbar{\bar{d}}
\def\ubar{\bar{u}}
\def\sbar{\bar{s}}
\def\cbar{\bar{c}}
\def\bbar{\bar{b}}
\def\tbar{\bar{t}}
\def\pbar{\bar{p}}
\def\B0bar{\overline{B^0}}
\def\lbar{\bar{\l}}
\def\l{\ell}

\subsection{The Higgs transverse momentum}

In order to investigate Higgs boson production
via $gg\to$ Higgs (see Ref.~\cite{Kunszt:1997yp}), 
one needs to account for multi-parton radiation
for the sake of performing trustworthy phenomenological analyses
\cite{Cavalli:2002vs,Djouadi:2000gu,Balazs:2000sz}. 
Standard Monte Carlo (MC) algorithms 
\cite{Corcella:2000bw,Corcella:2002jc,Sjostrand:2003wg} describe parton
radiation in the soft and/or collinear approximation
of the parton shower (PS), but can have regions
of phase space, so-called `dead zones', where no radiation is allowed.
Here, one can however 
rely on higher-order tree-level results, as in this region
the radiation is neither softly nor collinearly enhanced. 
Several methods have been recently suggested in order to match PS
and fixed-order matrix elements (MEs) \cite{Seymour:1995df,Norrbin:2000uu}, 
also including the virtual one-loop terms 
\cite{Frixione:2002ik,Frixione:2003ei,Dobbs:2001gb}. 

\subsection{The HERWIG implementation}

In this note, we briefly mention that the
same strategy which has already been used to
implement real ME corrections to 
$e^+e^-$ annihilation into quark pairs \cite{Seymour:1992xa}, 
Deep Inelastic Scattering (DIS) \cite{Seymour:1994ti},
top quark decay \cite{Corcella:1998rs} and vector boson hadro-production 
\cite{Corcella:1999gs} has now also been
adopted for the case of Higgs hadro-production via gluon-gluon fusion,
in the context of the HERWIG event generator 
\cite{Corcella:2000bw,Corcella:2002jc}. That is, the dead zone is here
populated by using the exact next-to-leading order (NLO)
tree-level ME result and the
PS in the already-populated region is corrected using the exact
amplitude any time an emission is capable of being the
{\sl hardest so far}.

\subsection{Numerical results and comparisons}

The MEs squared for the real 
corrections to $gg\to H$ that we have used can be found in
\cite{Baur:1990cm}, where top mass effects are fully included.
The real NLO corrections to $q\bar q\to H$ are instead rather straightforward:
the formulae we used can be read from Eq.~(3.62) of \cite{Moretti:2002eu}
with appropriate Yukawa couplings and crossing.
In the new HERWIG default version, in line with \cite{Corcella:1999gs}, 
ME corrections use
the Higgs transverse mass $m_T^2=q_T^2+m_H^2$ as the scale for 
$\alpha_S$ and for the Parton Distribution Functions (PDFs)
while the $gg,q\bar q\to H$ contributions use $m_H^2$.
We shall also assume that the intrinsic transverse momentum of the
initial-state partons is equal to
$q_{T,\mathrm{int}}=0$, the HERWIG default value.

\begin{figure}
\hspace*{2.5truecm}
{\epsfig{file=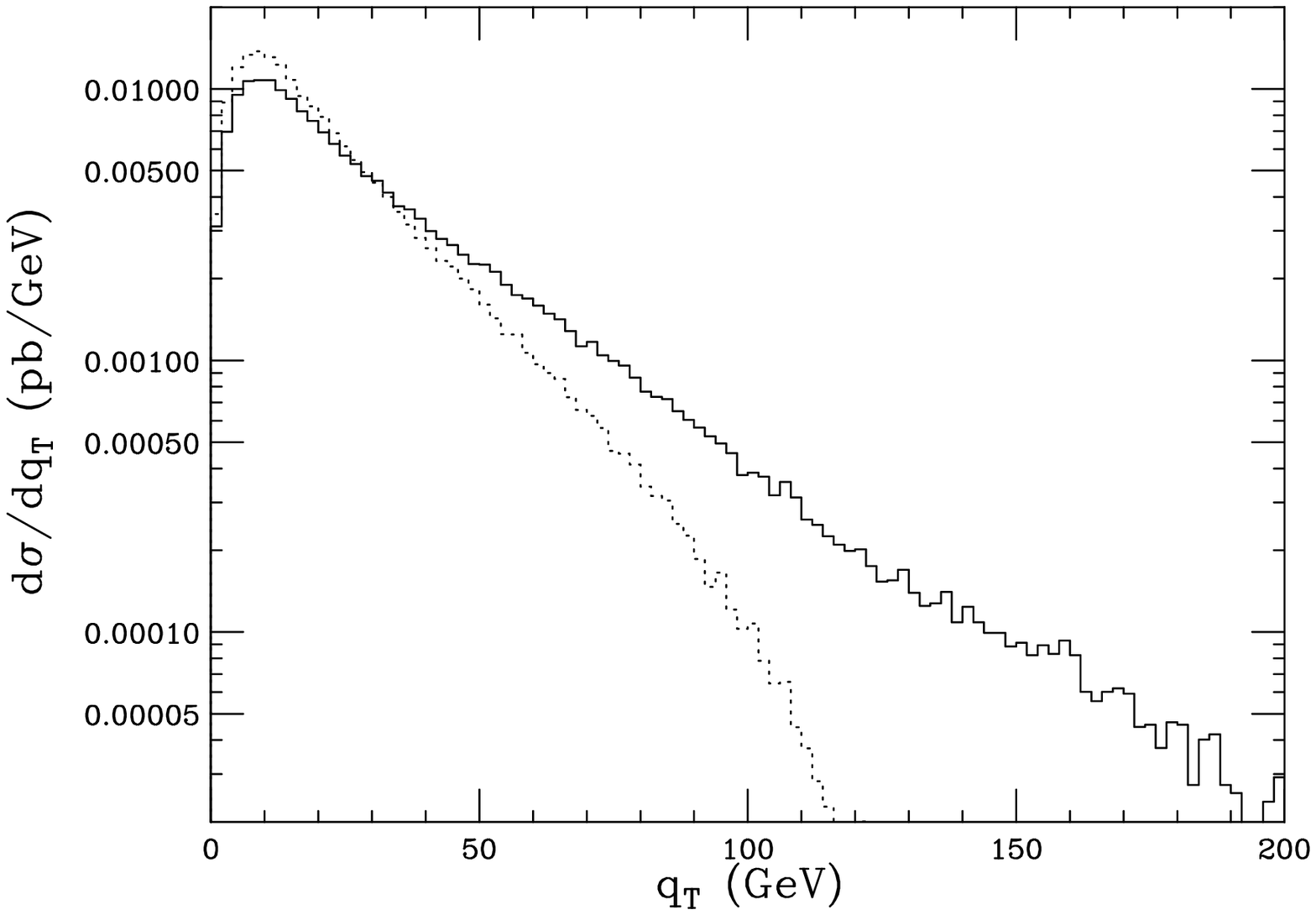,width=5.0cm,angle=0}}
{\epsfig{file=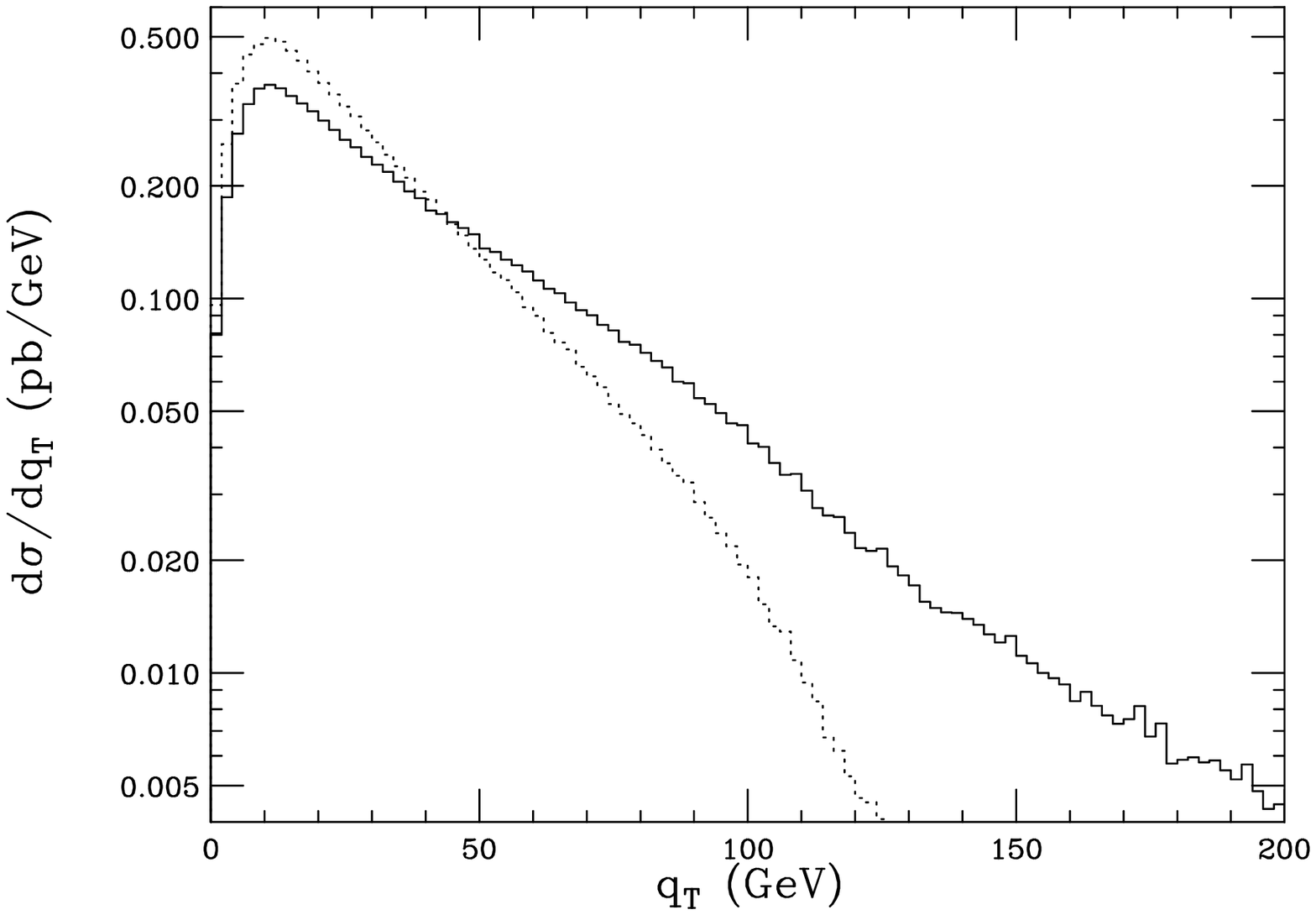,width=5.0cm,angle=0}}
\caption{\small Higgs transverse momentum distribution according to
HERWIG with (solid) and without (dotted) ME corrections,
at Tevatron (left, $\sqrt s_{p\bar p}=2$ TeV) and LHC
 (right, $\sqrt s_{pp}=14$ TeV).
We have set the Higgs mass to $m_H=115$~GeV.}
\label{HW}
\end{figure}

\begin{figure}
{\epsfig{file=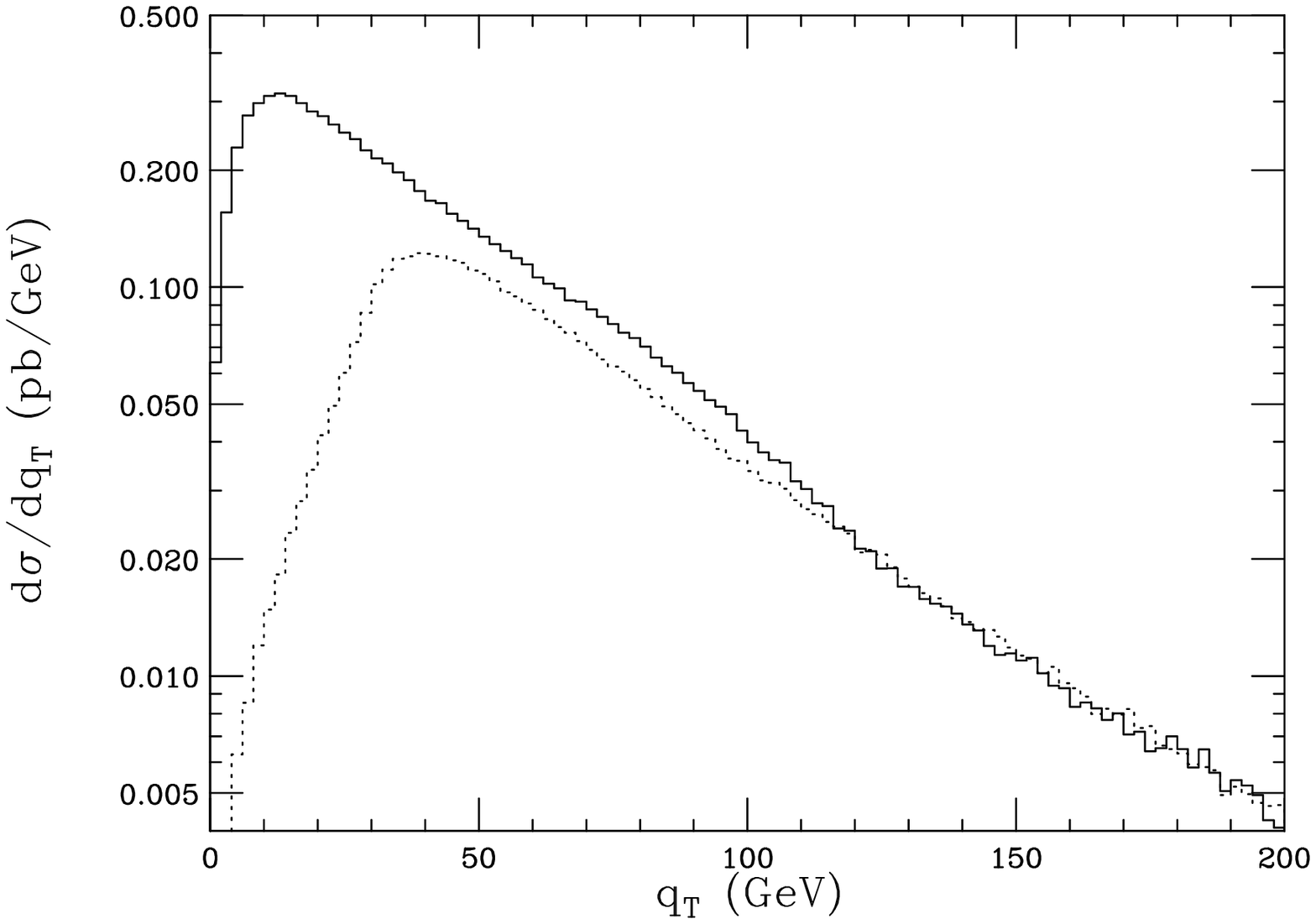,width=5.0cm,angle=0}}
{\epsfig{file=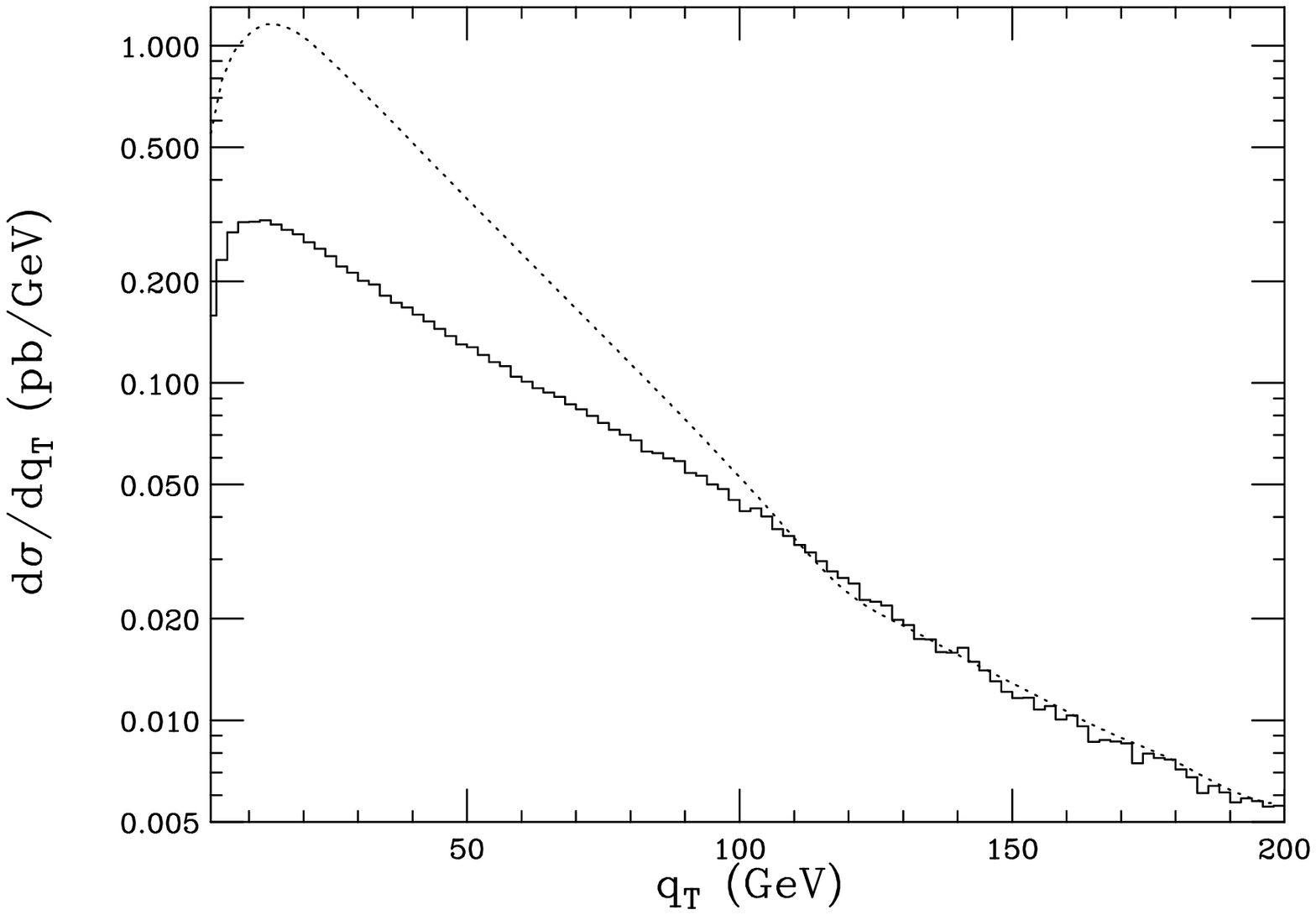,width=5.0cm,angle=0}}
{\epsfig{file=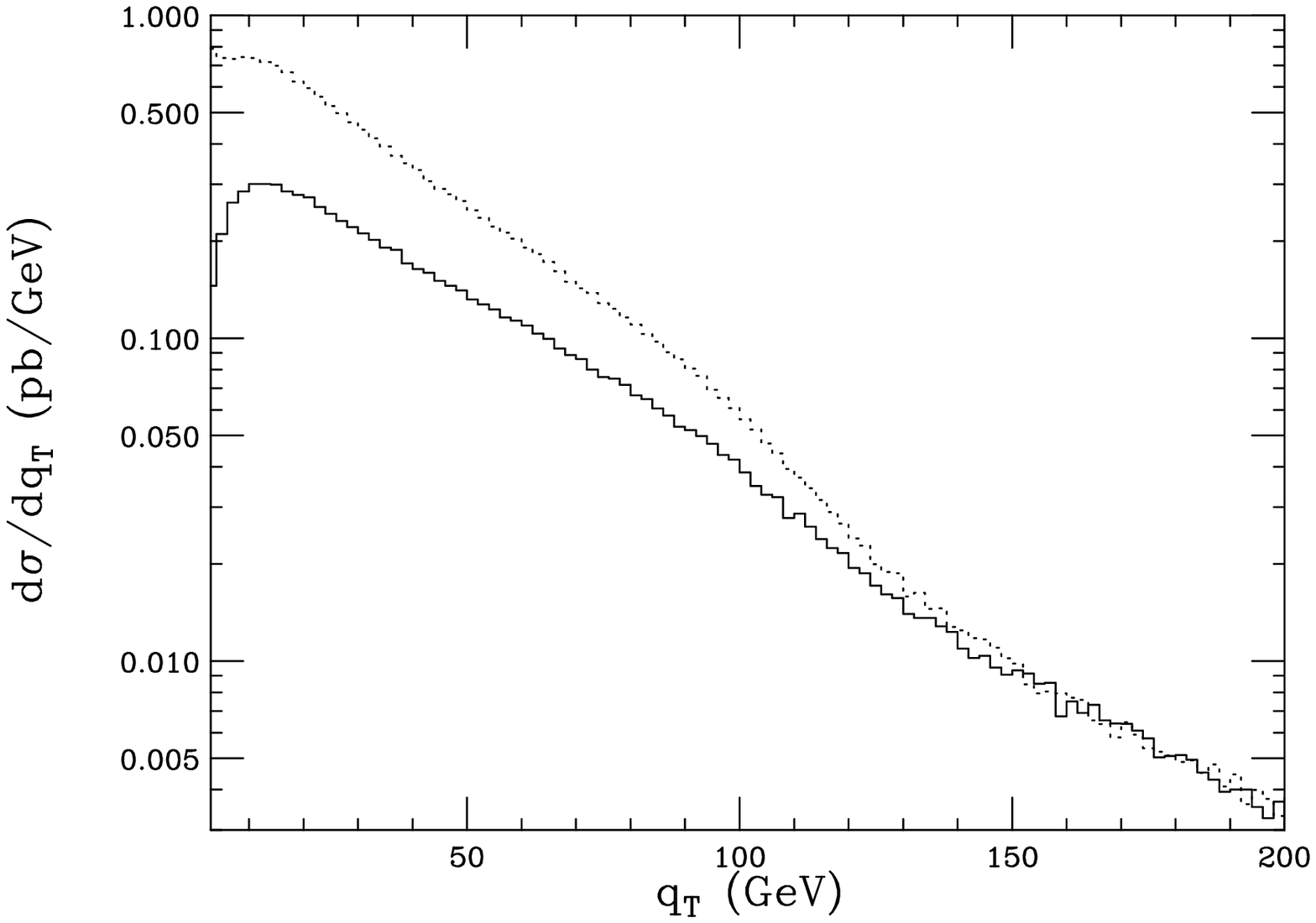,width=5.0cm,angle=0}}
\caption{\small Left: comparison of ME-corrected HERWIG
predictions (solid) to the `$H$ + jets' result from \cite{Baur:1990cm}
(dotted). Centre: comparison of ME-corrected HERWIG
predictions (solid) to the NLO and resummed calculation of \cite{Bozzi:2003jy}
(dotted).
Right: comparison of ME-corrected HERWIG predictions
(solid) to the MC@NLO results from the code described
in Ref.~\cite{Frixione:2003vm} (dotted). Here,
$q\bar q\to H$ processes have been turned off.}
\label{comparisons}
\end{figure}

By adopting the HERWIG defaults,
we first consider Higgs production at the Tevatron and the LHC within the
MC itself, by plotting the $q_T$ distribution  with (solid histogram) and 
without (dotted) ME corrections: see Fig.~\ref{HW}. 
Beyond $q_T\simeq m_H/2$ the ME-corrected 
version allows for many more events. In fact, one can prove 
that, within the standard algorithm, $q_T$ is constrained to be
$q_T<m_H$. At small $q_T$ the prediction which includes ME
corrections displays a suppression. By default, after the
latter are put in place,
the total normalization still equals
the LO rates. Hence, it is obvious that the enhancement
at large $q_T$ implies a reduction of the number of
events which are generated at small $q_T$ values.

In Fig.~\ref{comparisons} (left plot) 
we present the improved HERWIG spectrum (solid) for the LHC,
along with the result obtained running the so-called
`$H$ + jets' process (dotted), where the hard process is always 
one of the corrections to $gg\to H$. 
In order to perform such a comparison, we have turned 
the $q\bar q\to H$ hard process off, as `$H$ + jets' in HERWIG
does not currently
implement the corrections to quark-antiquark annihilation.
Furthermore, we have chosen $q_{T\mathrm{min}}=30$~GeV for the 
`$H$ + jets' generation.
As expected, at small $q_T$ the two predictions are
fairly different but at large transverse momentum they agree well.

In Fig.~\ref{comparisons} (centre plot) 
we compare the new HERWIG version with the
resummed calculation of Ref.~\cite{Bozzi:2003jy}. For the
sake of comparison with HERWIG, which includes leading logarithms and
only some subleading terms, we use the results
of \cite{Bozzi:2003jy} in the NLL approximation (rather than the default
NNLL one), matched to the NLO prediction. In order for such a comparison to be
trustworthy, we have to make parameter choices similar to \cite{Bozzi:2003jy}:
namely, we adopt a top quark with infinite mass in the
loop and $m_H=125$~GeV, with $\alpha_S$ and PDFs (both from HERWIG
defaults) evaluated at $m_H^2$. While
the normalization (LO in HERWIG, NLO in Ref.~\cite{Bozzi:2003jy}) and
the small-$q_T$ behaviour of the two curves are clearly different, 
the large transverse momentum predictions are in  good agreement, as
in both approaches it is the real NLO ME that
dominates the event generation at large $q_T$.

Finally, in Fig.~\ref{comparisons} (right plot), 
we compare the results of standard HERWIG after ME
corrections with the MC@NLO event generator 
(version 2.2) of Ref. \cite{Frixione:2003vm}, the latter 
implementing both real
and virtual corrections to the hard-scattering process, in such
a way that predicted observables (including normalization) 
are correct to NLO accuracy. As version 2.2 of the MC@NLO
includes only the corrections
to Higgs production in the gluon-fusion channel, we  again have
turned the quark-annihilation process off in our routines. 
As observed in the comparison with the resummed calculation, the two spectra 
differ in normalization and at small $q_T$, but agree in the 
large-transverse-momentum region.

\subsection{Conclusions}

Between the described implementation and the one available within the 
MC@NLO option,
we believe that HERWIG is presently a reliable event generator for (direct)
Higgs production from parton fusion at hadron colliders both at small and 
large transverse
momentum. In fact, all currently available
ME corrections will play an important role to perform any
analysis on Higgs searches at present and future colliders. In particular,
the option described here may be the most convenient choice for when the
phase space is limited to transverse momentum values such that $q_T\gsim m_H$.

\subsection*{Acknowledgements}

SM would like to thank the 2003 Les Houches workshop organisers
for their kind invitation and the Royal Society (London, UK) for financial
support.




\section[CAESAR: automating final-state resummations]{CAESAR:
  AUTOMATING FINAL-STATE RESUMMATIONS~\protect\footnote{Contributed
  by: {A.~Banfi, G.P.~Salam, G.~Zanderighi}}} 
\label{sec:caes-autom-final}


Event shapes and jet resolution parameters (final-state `observables')
measure the extent to which the energy-flow of the final state departs
from that of a Born event.  Their study has been fundamental for
measurements of the strong coupling~\cite{Bethke:2000ai,Jones:2003yv} as
well as the QCD colour factors~\cite{Kluth:2000km}; final states
also provide valuable information on the yet poorly understood
transition from parton to hadron level (see~\cite{Dasgupta:2003iq} for
a recent review).
In the region where an observable's value $v$ is small, one should resum
logarithmically enhanced contributions that arise at all
orders in the perturbative series. For a number of observables such a
resummation has been carried out manually at next-to-leading
logarithmic (NLL) 
accuracy~\cite{Catani:1993ua}. But achieving NLL accuracy requires a
detailed analysis of the observable's properties, and is often
technically involved.
We have instead recently proposed~\cite{Banfi:2003je} a new
approach based on a general NLL resummed master formula valid
for a large class of final-state observables (which includes many of
the final-state observables for which a NLL resummation already
exists). We have also formulated the formal requirements that an observable
should satisfy so as to be within the scope of the master formula.

To illustrate these requirements we consider a Born event consisting
of $n$ hard partons or `legs'.
For an observable
(a function $V$ of all final-state momenta) to be resummed in the
$n$-jet limit it should:
\begin{enumerate}
\item vanish smoothly after addition of an extra 
  soft particle collinear to any leg $\ell$, with 
  the following behaviour:
  \begin{equation}
    \label{eq:single}
       V(\{{\tilde p}\}, k) \simeq
    d_{\ell}\left(\frac{k_t}{Q}\right)^{a_\ell}e^{-b_\ell\eta}\, 
    g_\ell(\phi)\>.
  \end{equation}
Here $Q$ is a hard scale of the problem; $\{{\tilde p}\}$ represents
the Born (hard) momenta after recoil from the emission, which is defined in
terms of its transverse momentum $k_t$ and rapidity $\eta$ with
respect to leg $\ell$, and where relevant, by an azimuthal angle
$\phi$ relative to a Born event plane.

\item be \emph{recursively} infrared and collinear (rIRC) safe:
  meaning roughly that, given an ensemble of emissions, the addition
  of a relatively much softer or more collinear emission does not
  significantly alter the value of the observable, no matter how soft
  and/or collinear the other emissions are. This condition 
  is necessary to ensure exponentiation of leading
  logarithms, and is not satisfied for example by the JADE algorithm
  three-jet resolution parameter~\cite{Brown:1990nm}.
  
\item be continuously global, meaning that the observable is sensitive
  to emissions in the whole of the phase space (`global') in a
  continuous way, the formal requirement being $\partial_\eta
  \partial_{\ln k_t} \ln V(\{{\tilde p}\}, k) = 0$ and $\partial_\phi
  \partial_{\ln k_t} \ln V(\{{\tilde p}\}, k) = 0$
  ($k_t$ being defined with respect to the nearest leg).
  For non-continuously-global observables one must account for
  non-global logarithms (known only for large $N_c$)
  \cite{Dasgupta:2001sh}.  Their evaluation for a general observable
  would necessitate a (quite non-trivial) determination of the
  phase-space boundaries associated with the observable.
\end{enumerate}

Given the above conditions, the NLL resummation for the observable's
distribution (the probability $\Sigma(v)$ that the observable's value
is less than $v$) for a fixed Born configuration is given by the
`master' formula~\cite{Banfi:2003je}:
\begin{equation}
  \label{eq:Master}
  \Sigma(v) = 
  e^{-R(v)} {\cal F}(R'(v))
  \>,\qquad R'(v) = -v\frac{dR(v)}{dv}\>.
\end{equation}
The function $R(v)$ is a \emph{Sudakov} exponent that contains all
leading (double) logarithms and all NLL (single-log) terms that can be
taken into account by exponentiating the contribution to $\Sigma(v)$
from a single emission.  This function depends parametrically on
$a_\ell, b_\ell, d_\ell$ and on the azimuthal average $\left\langle
  \ln g_\ell \right\rangle$; its full expression is reported in
\cite{Banfi:2003je}.
Multiple emission effects, for example the fact that even if all
$V(\{\tilde p\},k_i) < v$, one might nevertheless have $V(\{\tilde
p\},k_1,\ldots,k_n) > v$, are accounted for by the NLL
function ${\cal F}$, which can be computed via a Monte Carlo
procedure~\cite{Banfi:2001bz}.

The advantage of having introduced a master formula is that the
resummation of the observable can be performed entirely automatically.
The master formula and applicability conditions are encoded in a
computer program (CAESAR,
Computer Automated Expert Semi-Analytical Resummer), which
given only the observable's definition in the form of a computer
routine, returns the observable's distribution $\Sigma(v)$ at NLL
accuracy (where possible).

As an example we present explicit results for the specific
case of the (global) transverse thrust in hadronic dijet production,
defined as
\begin{equation}
  \label{eq:Ttg}
  \tau_{\perp} \equiv 1-\max_{\vec n_\perp} \frac{\sum_i |{\vec
      p}_{\perp i}\cdot {\vec
      n_\perp}|}{\sum_i p_{\perp i}}\,,
\end{equation}
where the sums run over all particles in the final state, the ${\vec
  p}_{\perp i}$
are the particle transverse momenta (with respect to the beam direction)
and $\vec
n_\perp$ is 
a unit transverse vector.  The program, probing the observable with
randomly chosen soft and collinear emissions, is able to verify that
the applicability conditions hold and to determine the parameters
$a_\ell, b_\ell, d_\ell$, as well as the function $g_\ell(\phi)$.  It
then applies the Monte Carlo procedure introduced in
\cite{Banfi:2001bz} to compute the function ${\cal F}$. The results from
this analysis
are then plugged into the master formula~(\ref{eq:Master}) to
compute $\Sigma(\tau_\perp)$ at NLL accuracy. The resulting
differential distribution $D(\tau_\perp)\equiv
d\Sigma(\tau_\perp)/d\ln\tau_\perp$ (integrated over a range of Born
configurations, with the cuts given below) is shown in
figure~\ref{fig:othr} for the most 
relevant partonic subprocesses at the Tevatron run II c.o.m. energy
$\sqrt s = 1.96$TeV. 
 We select events with two outgoing jets with
$E_\perp > 50$GeV and $|\eta|<1.0$, use the CTEQ6M parton density
set~\cite{Pumplin:2002vw}, corresponding to $\alpha_s(M_Z) = 0.118$,
and set both the renormalisation and factorisation scale at the Born
partonic c.o.m. energy.  The curves in figure~\ref{fig:othr} show a
degree of separation between the various partonic channels --- this
information could perhaps be exploited in fits of parton
distributions.
\begin{figure}[htbp]
  \begin{center}
    \includegraphics[width=.5\textwidth]{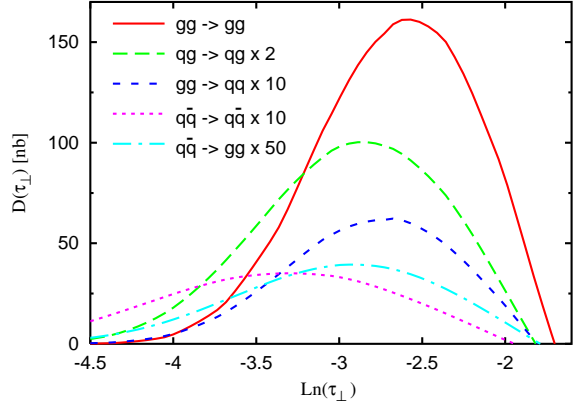} 
  \end{center}
\caption{The resummed differential distribution at NLL accuracy for the
  global transverse thrust.}
\label{fig:othr}
\end{figure}

To conclude we remark that since the only input to CAESAR is a
computer subroutine for an observable, it offers for the first time
the possibility for non-experts to easily obtain rigorous NLL resummed
predictions.
Work remains to be done both to release the first version of
CEASAR and to implement automated matching of NLL resummation with
fixed order results. In particular, addressing this last issue will
open the way for a vast amount of phenomenological analyses.



\section[Combined Effect of QCD Resummation and QED Radiative 
Correction to $W$ Boson Mass Measurement at the LHC]{COMBINED EFFECT OF QCD RESUMMATION AND QED RADIATIVE 
CORRECTION TO $W$ BOSON MASS MEASUREMENT AT THE LHC~\protect\footnote{Contributed
  by: {Q.-H.~Cao, C.-P.~Yuan}}}
\label{paper_lhc}
\def\D0{D\O~}



\subsection{Introduction}
As a fundamental parameter of the Standard Model (SM), the mass of the
$W$-boson ($M_{W}$) is of particular importance. Aside from being
an important test of the SM itself, a precision measurement of $M_{W}$,
together with an improved measurement of top quark mass ($M_{t}$),
provides severe indirect bounds on the mass of Higgs boson
($M_{H}$). With a precision of 15 MeV for $M_{W}$~\cite{Haywood:1999qg} and 2 GeV for
$M_{t}$ at the LHC~\cite{unknown:1999fr},  $M_{H}$ in the SM
can be predicted with an uncertainty of about 30\%~\cite{Haywood:1999qg}.
Comparison of these indirect constraints on $M_{H}$ with the results
from direct Higgs boson searches, at the 
LEP2, the Tevatron and the CERN
Large Hadron Collider (LHC), will be an important test of the SM. In order
to have a precision measurement of $M_{W}$, the theoretical uncertainties,
dominantly coming from the transverse momentum of the $W$-boson ($P_{T}^{W}$),
the uncertainty in parton distribution function (PDF) and the
electroweak (EW) radiative
corrections to the $W$ boson decay, must be controlled
to a better accuracy~\cite{Baur:2000bi,Haywood:1999qg}.

At the LHC, most $W$ bosons are produced in the small transverse momentum region.
When $P_{T}^{W}$ is much smaller than $M_{W}$,
every soft-gluon emission will induce a large
logarithmic contribution to the $P_{T}^{W}$ distribution so that an
order-by-order perturbative calculation
in the theory of Quantum chromodynamics (QCD)
cannot accurately describe the
$P_{T}^{W}$ spectrum
and the contribution from multiple soft-gluon emission,
which contributes to all orders in the expansion of the strong coupling
constant $\alpha_s$, needs to be summed to all orders.
It has been shown that by applying a renormalization group
analysis, the multiple soft-gluon radiation effects can be resummed
to all orders to predict the $P_{T}^{W}$ distribution which agrees
with experimental data~\cite{Balazs:1995nz, Balazs:1997xd}. RESBOS,
a Monte Carlo (MC) program ~\cite{Balazs:1997xd}
resumming the initial-state soft-gluon radiations
of the hadronically produced lepton pairs through EW vector boson
production and decay at hadron colliders
$p\bar{p}/pp\rightarrow V(\rightarrow\ell_{1}\bar{\ell_{2}})X$,
has been used by the CDF and \D0 Collaborations at the Tevatron to
compare with their data in order to determine $M_W$.
However, RESBOS does not include any
higher order EW corrections to describe the vector boson decay. The EW
radiative correction, in particular the final-state QED
correction, is crucial for precision measurement of $W$ boson mass
at the Tevatron, because photon emission from the final-state charged
lepton can significantly modify the lepton momentum which is used in
the determination of $M_{W}$. In the CDF Run Ib $W$ mass measurement,
the mass shifts due to radiative effects were estimated to be $-65\pm20$
MeV and $-168\pm10$ MeV for the electron and muon channels, respectively~\cite{Affolder:2000bp}.
The full next-to-leading order (NLO) $O(\alpha)$ EW
corrections have been calculated~\cite{Dittmaier:2001ay, Baur:1998kt}
and resulted in WGRAD~\cite{Baur:1998kt}, a MC program
for calculating $O(\alpha)$ EW radiative corrections to the process
$p\bar{p}\rightarrow\nu_{\ell}\ell(\gamma)$. However, WGRAD does not include
the dominant correction originated from the initial-state multiple soft-gluon
emission. The inclusion of both the initial-state QCD and final-state
QED corrections into a parton level MC program is urgently required
in order to reduce the theoretical uncertainties in interpreting the experimental
data at the Tevatron. It was shown in Refs. \cite{Dittmaier:2001ay, Baur:1998kt} that at the NLO,
the EW radiative correction in $p\bar{p}\rightarrow\ell\nu_{l}(\gamma)$ is
dominated by the final-state QED (FQED) correction. Hence, in this
paper we present a consistent calculation which includes both the
initial-state multiple soft-gluon QCD resummation and the final-state
NLO QED corrections, and develop an upgraded version of
the RESBOS program,
called RESBOS-A~\footnote{A Fortran code that implements the theoretical calculation is presented in this work.},
to simulate the signal events. Here, we only present
the phenomenological impacts on a few experimental observables, the transverse mass of $W$ boson ($M_T^W$) and
the transverse momentum of charged lepton ($p_T^{\ell}$), that are most sensitive to the
measurement of $M_W$. We focus our attention on
the electron only, though our analysis procedure also applies
to the muon. The detailed formula, the SM inputting parameters and the kinematics cuts
are given in Ref.~\cite{resbosa}.

\subsection{Precision Measurement of W Mass}
In Fig.~\ref{fig:tm}, we show various theory predictions on
the $M_{T}^{W}$ distribution.
The legend of the figure is defined as follows:
\begin{itemize}
\item LO : including only the Born level initial-state contribution,
\item RES : including the initial-state multiple soft-gluon corrections
via QCD resummation,
\item LO QED : including only the Born level final-state contribution,
\item NLO QED : including the final-state NLO QED corrections.
\end{itemize}
For example, the solid curve (labelled as RES+NLO QED) in
Fig.~\ref{fig:tm}(a) is the prediction from our combined calculation, 
given by Eqs.~(1) and (2) of
Ref~\cite{resbosa}.
\begin{figure}
\begin{center}
\includegraphics[width=10cm]{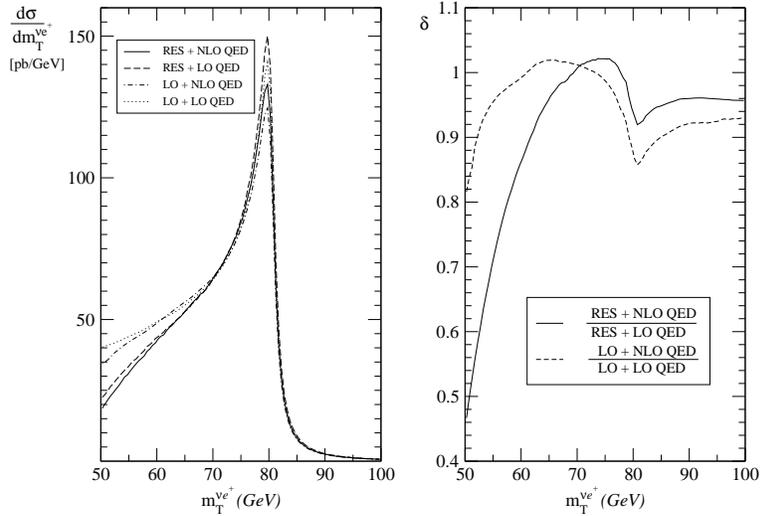}
\caption{Transverse mass distribution of $W^+$ boson \label{fig:tm}}
\end{center}
\end{figure}

As shown in Fig.~\ref{fig:tm}(a), compared to the lowest order cross
section (dotted curve), the initial state QCD resummation effects (dashed
curve) increase the cross section at the peak of the $M_{T}^{W}$ distribution
by about $6\%$, and the final state NLO QED corrections (dot-dashed
curve) decrease it by about $-12\%$, while the combined
contributions (solid curve)
of the QCD resummation and FQED corrections reduce it by $6\%$.
In addition to the change in magnitude,
the line-shape of the $M_{T}^{W}$ distribution
is significantly modified by the effects of QCD resummation
and FQED corrections. 
To illustrate this point, we plot the ratio of the (RES+NLO QED)
differential cross sections to the LO ones
as the solid curve in Fig.~\ref{fig:tm}(b).
The dashed curve is for the ratio of (LO+NLO QED) to LO.
As shown in the figure,
the QCD resummation effect dominates the shape of
$M_{T}^{W}$ distribution for
$65\,{\rm GeV\leq M_{W}\leq95\,{\rm GeV}}$, while
the FQED correction reaches its maximal effect around the Jacobian
peak ($M_{T}^{W}\simeq M_{W}$).
Hence, both corrections must be included to accurately predict
the distribution of $M_{T}^{W}$ around the Jacobian region to determine
$M_W$. We note that after including the effect due to the finite
resolution of the detector (for identifying an isolated electron or
muon), the size of the FQED correction is largely 
reduced~\cite{Dittmaier:2001ay, Baur:1998kt}.
\begin{figure}
\begin{center}
\includegraphics[width=10cm]{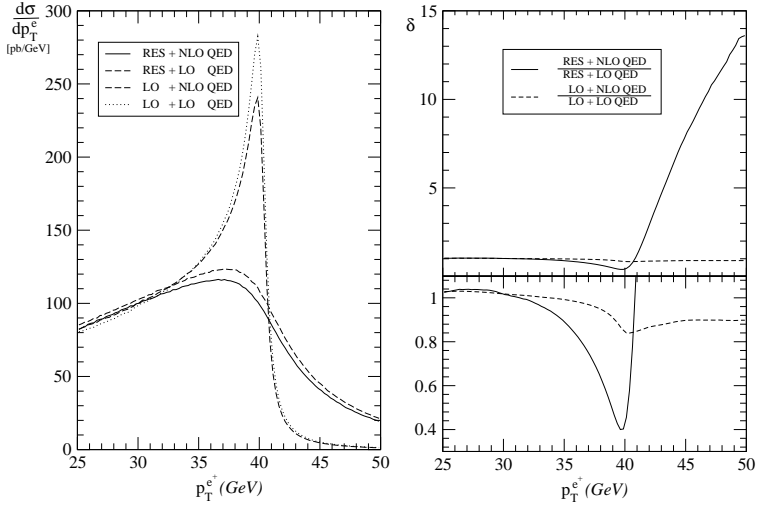}
\caption{Transverse momentum distributions of $e^+$\label{fig:pte}}
\end{center}
\end{figure}

Although the $M_{T}^{W}$ distribution is an optimal observable
for determining $M_{W}$ at the LHC with a low luminosity, it requires
 an accurate measurement of the missing transverse momentum
direction which becomes more difficult to control with a high luminosity 
option (when multiple scattering becomes important).
On the other hand, the transverse momentum
of the decay charged lepton ($p_{T}^{e}$) is less sensitive to
the detector resolution, so that it can be used to measure $M_W$
and provide important cross-check on the result derived from
the $M_{T}^{W}$ distribution,
for they have different systematic uncertainties. 
Another important feature of this observable is that
$p_{T}^{e}$ distribution is more sensitive to the transverse momentum
of $W$ boson. Hence, the QCD soft-gluon resummation effects,
the major source of $p_{T}^{W}$, must be included to reduce
the theoretical uncertainty of this method.
In Fig.~\ref{fig:pte}(a), we show the $p_{T}^{e}$
distributions predicted by various theory calculations, 
and in Fig.~\ref{fig:pte}(b), the ratios
of the higher order to lowest order cross sections as a function
of $p_{T}^{e}$. The lowest order distribution
(dotted curve) shows a clear and sharp Jacobian peak at 
$p_{T}^{e}\simeq M_{W}/2$,
and the distribution with the 
NLO final-state QED correction (dot-dashed curve) also
exhibits the similar Jacobian peak with the peak magnitude
reduced by about $15\%$. But the clear and sharp Jacobian peak of
the lowest order and NLO FQED distributions 
(in which $p_{T}^{W}=0$)
are strongly smeared by
the finite transverse momentum of the $W$ boson induced by multiple
soft-gluon radiation, as clearly demonstrated by the 
QCD resummation distribution
(dashed curve) and the combined contributions of the QCD resummation and
FQED corrections (solid curve).
Similar to the $M_{T}^{W}$ distribution, the QCD resummation effect
dominates the whole $p_{T}^{e}$ range, while the FQED correction
reaches it maximum around the Jacobian peak (half of $M_W$).
The combined contribution of the QCD resummation
and FQED corrections reaches the order of $45\%$ near 
the Jacobian peak. Hence, these lead us to conclude that the
QCD resummation effects are crucial in the measurement of $M_{W}$
from fitting the Jacobian kinematical edge of the $p_{T}^{e}$
distribution. 

As shown in Ref.~\cite{Balazs:1997xd},
the effect from the initial state QCD gluon resummation to the W transverse
mass distribution is dominated by the perturbative Sudkov contribution and
is not very sensitive to the non-perturbative parameters ($g_1, g_2$ and $g_3$)
of the CSS resummation formalism. On the other hand, as shown in Fig.~\ref{fig:tm},
the final state QED correction can largely modify the W transverse mass
distribution, though a definite conclusion can only be drawn after including
the effect of detector resolution.

In our calculation we have included the contributions from the final
state QED correction together with the initial state QCD resummation to
predict the production and decay of W bosons produced at the LHC.
Since the exact matrix elements have been used in the calculation, the spin
correlations among the initial state partons and final state leptons are
correctly implemented. Hence, the kinematic distributions of the final
state leptons, and the corresponding experimental observables, can be
reliably predicted.

In order to study the impact of the presented calculation to the determination
of the $W$ boson mass, the effect due to the finite resolution of
the detector should be included, which will be presented elsewhere.

We thank P. Nadolsky and J.W. Qiu for helpful discussions. This work
was supported in part by NSF under grand No. PHY-0244919 and PHY-0100677.


\section[Resummation for the Tevatron and LHC electroweak boson production at
small $x$]{RESUMMATION FOR THE TEVATRON AND LHC ELECTROWEAK BOSON PRODUCTION AT
  SMALL $x$~\protect\footnote{Contributed
  by: {S.~Berge, P.~Nadolsky, F.~Olness, C.-P.~Yuan}}} 
\label{sec:resumm-tevatr-lhc}
 \def\alphas{\alpha_s}

In the production of electroweak bosons, precise knowledge of the 
transverse mass $M_T$ and transverse momentum
$q_T$ provides detailed information about the 
production process, including the mass of the boson and associated radiative 
corrections.  At the Tevatron, $q_T$ distributions of $Z^0$ 
bosons offer insight into soft gluon radiation, and this information 
is then used for precision extraction of the $W$ boson mass. 
At the LHC, good knowledge of the transverse distribution 
of Higgs bosons $H^0$ will be needed to efficiently separate Higgs boson 
candidates from the large QCD background. Accurate predictions 
for the small-$q_T$ region are obtained via resummation of large
logarithms $\ln^n(q_T/Q)$ arising from unsuppressed soft and
collinear radiation in higher orders of perturbation theory.

As we move from the 2 TeV Tevatron to the 14 TeV LHC, 
typical values of partonic momentum fractions  $x$ 
for producing $W$, $Z^0$, and $H^0$ bosons  become 
smaller,  thus enhancing $\ln(1/x)$ terms in higher orders of
$\alphas$. It is not entirely known how these terms 
(not included in a fixed-order cross section or conventional 
$q_T$ resummation) will affect $W$, $Z^0$, and $H^0$ production 
at the LHC energies, in part because no Drell-Yan $q_T$ data
is available yet in the relevant region of $x$ 
of a few $10^{-3}$ or less.

Studies \cite{Nadolsky:1999kb,Nadolsky:2000ky} in the crossed channel
of semi-inclusive deep-inelastic scattering (SIDIS) suggest that
hadronic $q_T$ distributions at small $x$ cannot be 
straightforwardly described within the Collins-Soper-Sterman (CSS) resummation 
framework~\cite{Collins:1985kg}, if the nonperturbative Sudakov function
behaves like its large-$x$ counterpart from the Drell-Yan process. 
A $q_T$ distribution in SIDIS at $x < 10^{-2}$ is
substantially broader than the conventional CSS prediction. 
The broadening effect can be modeled by including an extra
$x$-dependent term in the Sudakov exponent. To describe the data, 
the extra term must grow quickly as $x \rightarrow 0$. 
It noticeably contributes to the
resummed form factor at intermediate impact parameters 
($b \sim 1/q_T < 1\mbox{~GeV}^{-1}$),
 which hints at its origin 
from perturbative physics. 
A possible interpretation of this term is that it mimics 
higher-order contributions  of the form $\alphas^m \ln^n (1/x)$, 
which are not included in the resummed cross section. Due to the
two-scale nature of the $q_T$ resummation problem, the non-resummed
$\ln(1/x)$ terms may affect the $q_T$ distribution even when 
they leave no discernible trace in inclusive DIS structure functions. 
The DIS structure functions 
depend on one hard scale 
(of order $Q$),  while the CSS resummation formula (cf. Eq.~(\ref{eq:w})) 
also includes contributions from large impact parameters $b$ (small momentum
scales).  As $b$ becomes large, the series 
$\alphas^m(1/b) \ln^n (1/x)$ in the CSS formula may begin to diverge
at a larger value of $x$ than the series 
$\alphas^m(Q) \ln^n (1/x)$ in the inclusive structure functions. 
For this reason, transition to $k_T$-unordered (BFKL-like \cite{Kuraev:1976ge,Balitsky:1978ic}) 
physics may happen at larger $x$ in $q_T$ distributions than in
inclusive (one-scale) observables.

The $q_T$ broadening discussed above was observed in semi-inclusive
DIS processes. In this study, we explore its possible 
implications for the (crossed) Drell-Yan process. We begin by
examining the resummed transverse momentum distribution for the Drell-Yan 
process \cite{Collins:1985kg}, following notations from 
Ref.~\cite{Landry:2002ix}:
\begin{equation}
\frac{d\sigma}{dy d q_T^2} 
=
\frac{\sigma_0}{S} \ 
\int
\frac{d^2 b}{(2\pi)^2} \ 
e^{-i \vec{q}_T \cdot \vec{b}} \ 
\widetilde{W}(b,Q,x_A,x_B) +Y(q_T,Q,x_A,x_B).
\label{eq:w}
\end{equation}
Here $x_{A,B} \equiv Q e^{\pm y}/\sqrt{S}$, the integral is  the 
Fourier transform of a resummed form factor $\widetilde W$ given in impact 
parameter ($b$) space, and $Y$ is a regular (finite at $q_T \rightarrow 0$) 
part of the fixed-order cross section. In the small-$b$ limit, 
the form factor $\widetilde{W}$ is given by a product 
of a perturbative Sudakov exponent $e^{-{\cal S}_P}$ and 
generalized parton distributions ${\cal \overline{P}}(x,b)$:
\begin{equation}
\left.\widetilde{W}(b,Q,x_A,x_B)\right|_{b^2 \ll \Lambda_{QCD}^{-2}} 
=
e^{-{\cal S}_P(b,Q)} 
\ {\cal \overline{P}}(x_A, b)  \ {\cal \overline{P}}(x_B,b).  \ 
\label{eq:p}
\end{equation}
At moderately small $x$, where the representation (\ref{eq:p}) for 
$\widetilde W$  holds, we write these generalized parton distributions 
in the form
\begin{equation}
\left.{\cal \overline{P}}(x,b)\right|_{b^2 \ll \Lambda_{QCD}^{-2}} \simeq 
({\cal C}\otimes f)(x,b_0/b) \  e^{- \rho(x) \, b^2},
\label{eq:Psmallx}
\end{equation}
where  ${\cal C}(x,b_0/b)$ are coefficient functions, 
 $f(x,\mu)$ are  conventional parton distributions, 
and $b_0=2e^{-\gamma_E}=1.12...$ is a commonly appearing constant factor. 

The term $e^{- \rho(x) \, b^2}$ in $\overline{\cal P}(x,b)$
will provide an additional $q_T$ broadening,
with an $x$ dependence specified by $\rho(x)$. For example, it may 
approximate $x$-dependent higher-order contributions 
that are not included in the finite-order expression for $({\cal C}\otimes f)$.
We parametrize $\rho(x)$ in the following functional form: 
\begin{equation}
\rho(x) = 
c_0 \left( \sqrt{\frac{1}{x^2}+ \frac{1}{x_0^2}}
-\frac{1}{x_0}\right),
\label{eq:ax}
\end{equation}
such that $\rho(x) \sim c_0/x$ for $x \ll x_0$, and
$\rho(x) \sim 0$ for $x \gg x_0$.
This parameterization ensures that the formalism reduces to the usual 
CSS form  for large $x$ ($x \gg x_0$) and introduces an additional source of
$q_T$ broadening (growing as $1/x$) at small $x$ ($x \ll x_0$). 
The parameter $c_0$ determines 
the magnitude of the broadening for a given $x$, while $x_0$ specifies
the value of $x$ below which the broadening effects become important.
In principle, $c_0$ and $x_0$ may depend on the hard scale $Q$;
in this first study, we neglect this dependence. Based on the observed 
dependence 
$\rho(x) \sim 0.013/x$ at $x \lesssim 10^{-2}$ in SIDIS energy flow
data~\cite{Nadolsky:2000ky}, we choose $c_0=0.013$ and $x_0=0.005$ 
as a representative choice for our plots.

As $x\rightarrow 0$, the additional broadening term 
in Eq.~(\ref{eq:Psmallx}) 
affects the form factor $\widetilde{W}$ both at perturbative 
($b \lesssim 1\mbox{~GeV}^{-1}$) and nonperturbative ($b \gtrsim
1\mbox{~GeV}^{-1}$) impact parameters. In addition, the resummed cross
section contains conventional non-perturbative contributions
from power corrections, which become important at large impact parameters 
($b \gtrsim
1\mbox{~GeV}^{-1}$). 
We introduce these corrections by replacing the impact parameter $b$ in 
functions ${\cal S}_P$ and $({\cal C}\otimes f)$ with a variable 
$b_* = b/\sqrt{1+b^2/(0.25\mbox{~GeV}^{-2})}$~\cite{Collins:1985kg} 
and including 
a nonperturbative Sudakov exponent 
$\exp{\{-{\cal S}_{NP}(b,Q)\}}$.
The function ${\cal S}_{NP}(b,Q)$ is parametrized by a 3-parameter 
Gaussian form from a recent global fit to low-energy
Drell-Yan and Tevatron Run-1 $Z^0$ data \cite{Landry:2002ix}.
Combining all the terms, we have:
\begin{eqnarray}
\frac{d\sigma}{dy d q_T^2} 
& = &
\frac{\sigma_0}{S} \ 
\int
\frac{d^2 b}{(2\pi)^2} \ 
e^{-i \vec{q}_T \cdot \vec{b}} \, 
({\cal C}\otimes f)(x_A,b_0/b_*) \,
({\cal C}\otimes f)(x_B,b_0/b_*) \nonumber \\
& \times & 
e^{- {\cal S}_P(b_*,Q) - {\cal S}_{NP}(b,Q) - b^2 \rho(x_A)- b^2 \rho(x_B)} 
+ Y.  
\label{eq:full}
\end{eqnarray}


\begin{figure}[ht] 
\begin{center}
\leavevmode
\vbox{
  \hbox{
     \includegraphics[width=0.45 \hsize]{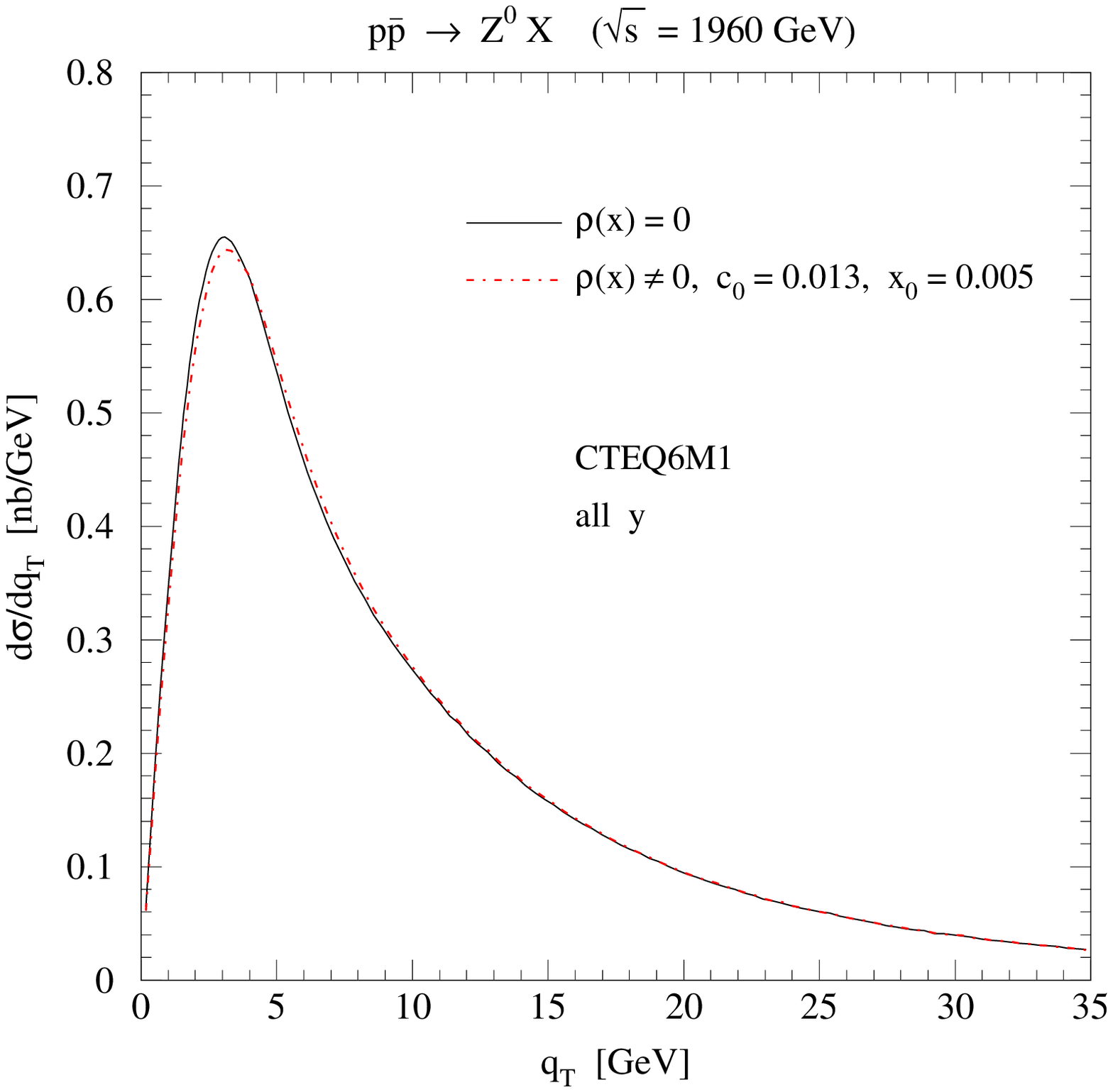} 
     \includegraphics[width=0.45 \hsize]{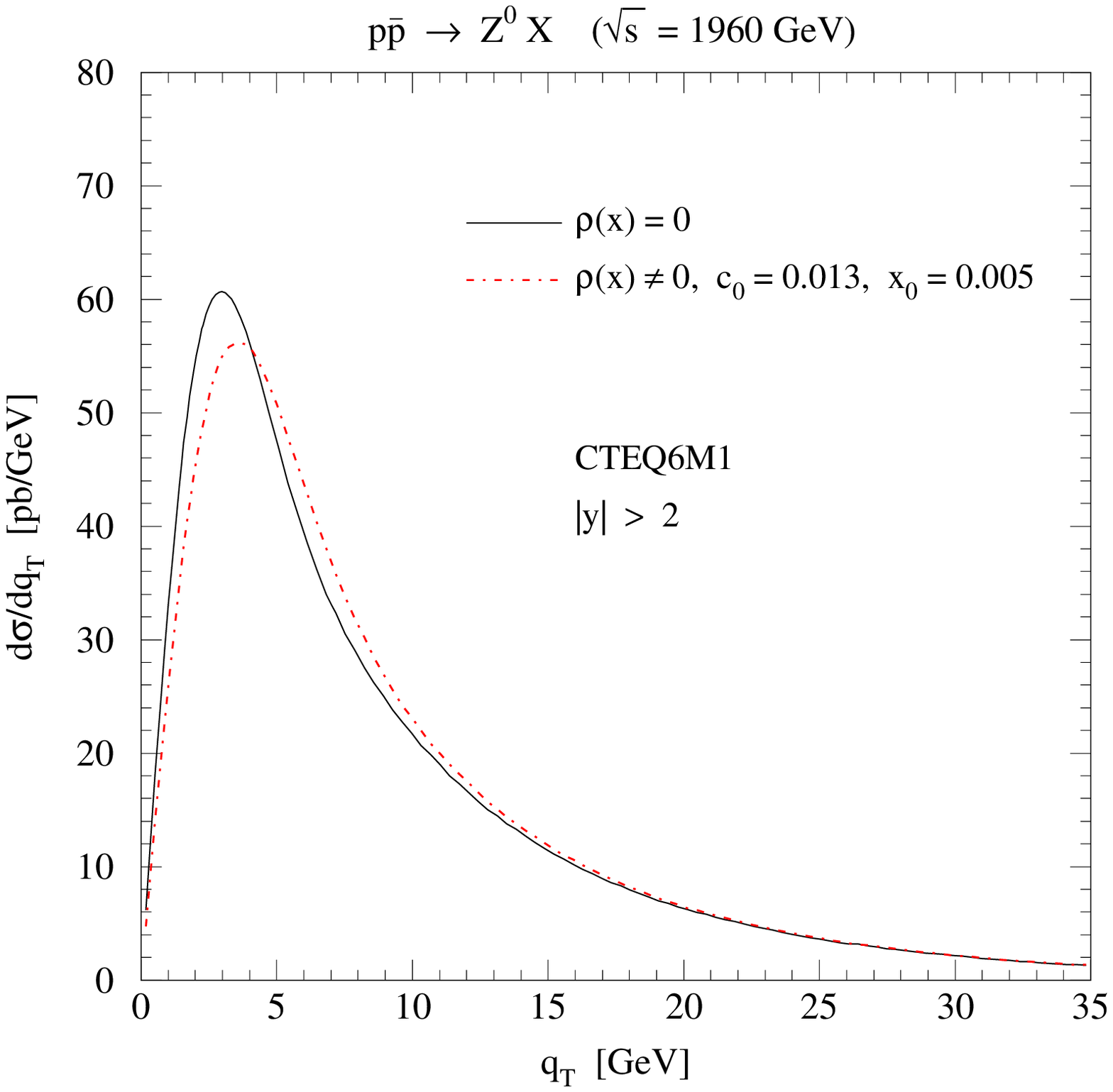} 
  }
}
\\\quad\quad(a) \hspace{6.5cm} (b)
\vskip -00pt
\caption{$q_T$ distributions of $Z^0$ bosons in the  Tevatron Run-2;
(a) integrated over the full range of $Z$ boson rapidities; (b)~integrated
over the forward regions $|y| > 2$.  The solid curve is a standard 
CSS cross section, calculated using the 3-parameter Gaussian 
parametrization~\protect\cite{Landry:2002ix} of the nonperturbative 
Sudakov factor. 
The dashed curve includes additional terms responsible for 
the $q_T$ broadening in the small-$x$ region (cf.~Eq.~(\ref{eq:full})). 
\label{fig:z} 
}
\vskip -00pt
\end{center}
\end{figure}

\begin{figure}[ht] 
\begin{center}
\leavevmode
\vbox{
  \hbox{
     \includegraphics[width=0.45 \hsize]{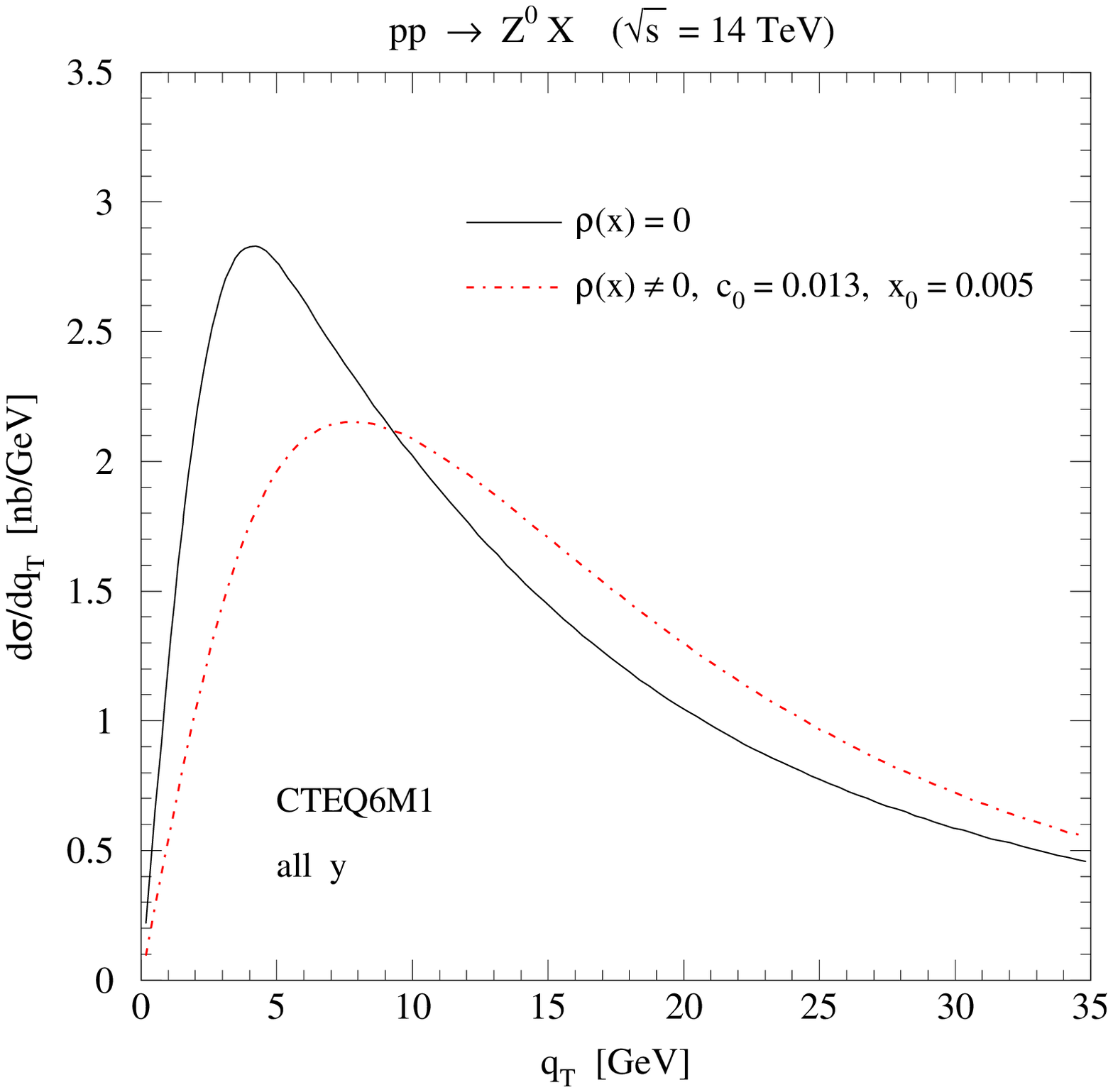}
     \includegraphics[width=0.45 \hsize]{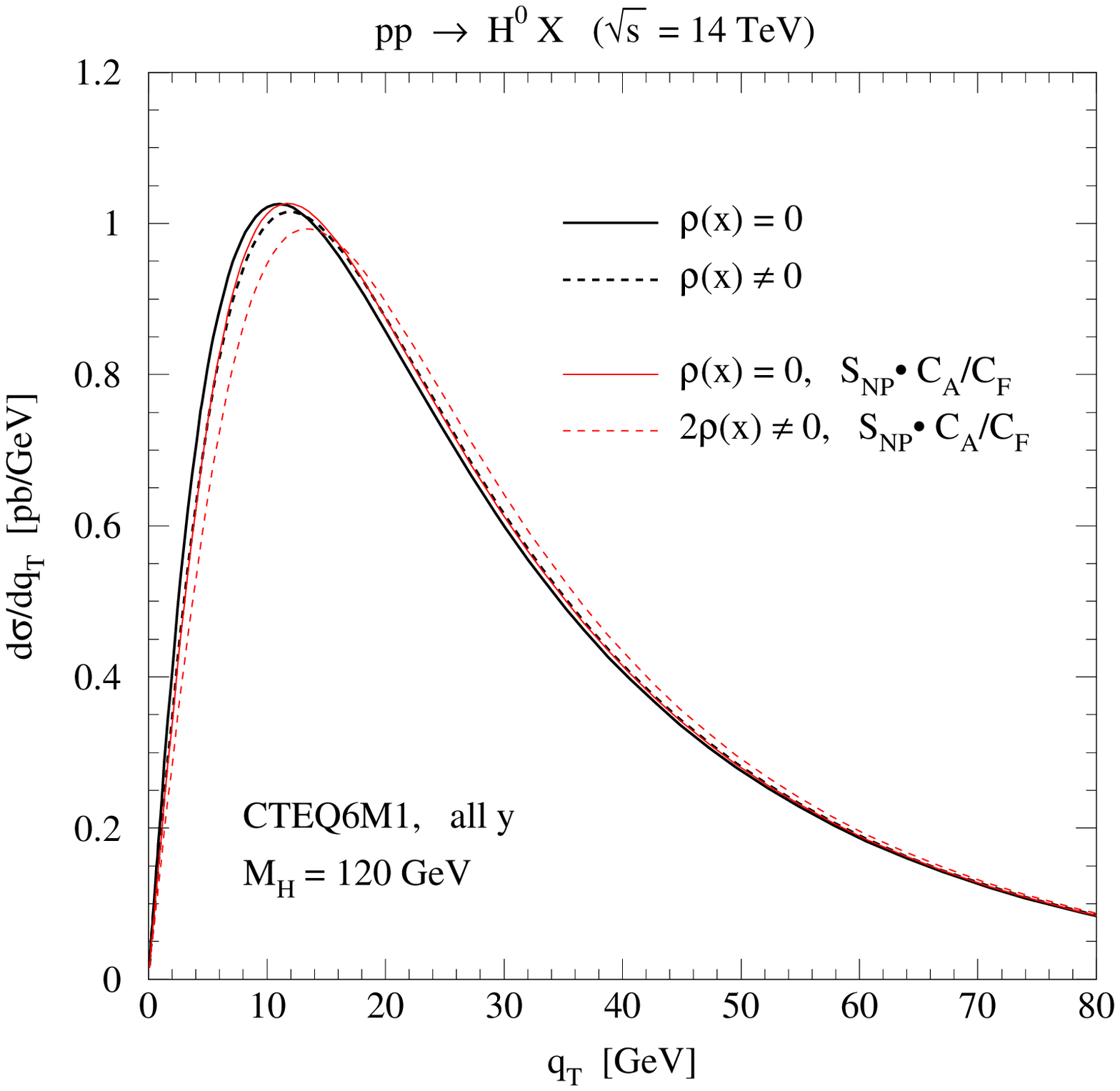}
  }
}
\\\quad\quad(a) \hspace{6.5cm} (b)
\vskip -00pt
\caption{
$q_T$ distributions of (a) $Z^0$ bosons and (b) Standard Model Higgs bosons
at the Large Hadron Collider, integrated over the full range of 
boson rapidities.
\label{fig:h} 
}
\vskip -00pt
\end{center}
\end{figure}

Figs.~\ref{fig:z} and \ref{fig:h} show the comparison of the 
resummed cross section
(\ref{eq:full}) with the additional broadening term 
($\rho(x) \neq 0$) to the resummed cross
section without such a term ($\rho(x) = 0$). We consider cross sections for the
production of $Z^0$ and $H^0$ bosons, calculated according to the procedures
in Refs.~\cite{Balazs:1997xd} and \cite{Balazs:2000wv}, respectively. 
The numerical  calculation was realized using the programs Legacy and ResBos 
\cite{Landry:2002ix,Balazs:1997xd}, and with the CTEQ6M1 parton distribution 
functions~\cite{Stump:2003yu}. The perturbative Sudakov factor was 
included up to ${\cal O}(\alphas^2)$, and the functions 
$({\cal C}\otimes f)$ up to ${\cal O}(\alphas)$. The relevant perturbative
coefficients can be found in Refs.~\cite{Collins:1985kg,Kodaira:1982nh,Davies:1984hs,Catani:1988vd,Kauffman:1991jt,Yuan:1992we,deFlorian:2000pr,Catani:2000vq}.

Fig.~\ref{fig:z}(a) shows the differential distribution 
$d\sigma/d q_T$ for $Z$ boson production in the Tevatron Run-2,
integrated over the rapidity $y$ of the $Z$ bosons. We observe that 
the distribution (\ref{eq:full}) with the additional small-$x$ term 
(the dashed curve) essentially coincides with the
standard CSS distribution (the solid curve). When $y$ is integrated
over the full range, both resummed cross sections 
are dominated by contributions from $x \sim 0.05 \gg x_0$, where 
the additional broadening (given by the function $\rho(x)$) is negligible. 
For this reason, the Tevatron distributions that are inclusive in $y$ 
(e.g., the Run-1 $Z^0$ boson data) 
will not be able to distinguish the small-$x$ broadening effects from
uncertainties in the nonperturbative Sudakov function ${\cal S}_{NP}$.

In contrast, the small-$x$ broadening does lead to observable differences
in the $q_T$ distributions in the forward rapidity region, where one of 
the initial-state partons carries a smaller momentum fraction than
in the central region. Fig.~\ref{fig:z}(b)
shows the cross section $d\sigma/dq_T$ for $Z$ bosons satisfying $|y|>2$.
The peak of the curve with $\rho(x) \neq 0$ is lower and shifted toward 
higher $q_T$. While this difference 
was not large enough to be observed in the Tevatron Run-1, 
it seems to be measurable in the Run-2 given the improved 
acceptance and higher luminosity of the upgraded Tevatron
collider. The small-$x$ broadening is more pronounced in $W$ boson production 
due to the smaller mass of the $W$ boson.

We now turn to the LHC, where the small-$x$ broadening may be observed in the
whole rapidity range due to the increased center-of-mass energy. 
Fig.~\ref{fig:h}(a) displays the distribution $d\sigma/dq_T$ 
for $Z^0$ production with and without the small-$x$ terms. Here, 
the difference is striking even if $y$ is integrated out. 
Effects of a similar magnitude are present in $W$ boson production,
and they are further enhanced in the forward regions.

The small-$x$ broadening is less spectacular, but visible, 
in the production of light Standard Model Higgs bosons via the effective 
$ggH$ vertex in the limit of a heavy top quark mass. 
Fig.~\ref{fig:h}(b) displays the resummed cross sections for production 
of Higgs bosons with a mass $M_H = 120$ GeV for several choices of 
${\cal S}_{NP}$ and the broadening term. We first compare cross
sections for $\rho(x) = 0$ and $\rho(x) \neq 0$  (thick lines), 
where the functions ${\rho}(x)$ and ${\cal S}_{NP}(b,Q)$ are taken 
to be the same as in $Z^0$ boson production. The difference between 
the two cross sections is not large, due to a harder $q_T$ spectrum 
in the Higgs boson case. The peaking in the $gg$-dominated
$H^0$ distribution occurs at $q_T = 10-20$ GeV, i.e., 
beyond the region where the function $\rho(x)b^2$ play its 
dominant role. This is different from the $q\bar q$-dominated $Z^0$
distribution, where the peak is located at $q_T \sim 5-10$ GeV and
is strongly affected by $\exp{\{-\rho(x)b^2\}}$.
Hence, for the same function $\rho(x)$ as in the $Z^0$ boson case, 
the difference between the curves with and without $\rho(x)$ is 
minimal. 

The harder $q_T$ spectrum in the Higgs boson case is induced by a larger
leading-logarithm coefficient ($C_A$) in $gg$ channels, as compared
to the leading-logarithm coefficient $C_F$ in $q\bar q$ channels. This suggests 
that the $Q$-dependent part (and possibly other terms) 
of the nonperturbative Sudakov function ${\cal S}_{NP}$ 
in Higgs boson production
is also multiplied by a larger color factor than in the Drell-Yan process. 
We estimate this effect by multiplying ${\cal S}_{NP}$ 
by the ratio of the leading color factors 
in Higgs and $Z^0$ boson production processes, $C_A/C_F=9/4$ 
(the thin solid line). 
The resulting change turns out to be small because of the
reduced sensitivity of the Higgs boson cross section to nonperturbative 
contributions.
 
The $\ln(1/x)$ terms may be enhanced in the case of the
Higgs bosons as well, due to the direct coupling of the Higgs bosons 
to gluon ladders. At present, we do not have a reliable estimate
of the small-$x$ broadening in gluon-dominated channels. However, this
broadening would have to be quite large to affect $q_T$ of 10-20 GeV or 
more, i.e., in the region where selection cuts on $q_T$ of the Higgs boson candidates
will be imposed. For example, increasing the function $\rho(x)$ by a factor 
of two as compared to the $Z^0$ boson case would lead to a distribution 
shown by the thin dashed line. While at $q_T \gtrsim 20$~GeV this effect 
is relatively small as compared to other theory uncertainties 
(e.g., higher-order corrections), it may affect precision calculations of
$q_T$ distributions needed to separate the Higgs boson signal from 
the background in the $\gamma \gamma$ mode.

Additional constraints on the small-$x$ behavior of the resummed 
cross sections in the $gg$ channel could be obtained from examination 
of photon pair production
away from the Higgs signal region. As the mass of the photon 
pair decreases, $\gamma\gamma$ production in the gluon fusion channel via a quark box diagram becomes 
increasingly important. For instance, the subprocess 
$gg \rightarrow \gamma \gamma$ contributes up to 40\% of the total 
cross section at $Q=80$ GeV \cite{Bern:2002jx}. By comparing $q_T$ 
distributions in $p p \to \gamma \gamma$ and  $p p \to Z$ in the same region 
of $Q$, one may be able to separate the $q \bar{q}$ and $gg$ components
of the resummed cross section and learn about the $x$ dependence in
the $gg$ channel.

To summarize, we argue that a measurement of transverse momentum 
distributions of forward $Z$  bosons at the Tevatron will
provide important clues about the physics of QCD factorization and 
possibly discover broadening of $q_T$ distributions associated 
with the transition to small-$x$ hadronic dynamics. 
Based upon the analysis of $q_T$ broadening 
effects observed in semi-inclusive DIS, we have estimated similar effects 
in the (crossed) processes of electroweak boson production at hadron-hadron
colliders. While the estimated impact on the Higgs boson cross section 
$d\sigma/dq_T$ at high $q_T$ was found to be minimal, 
much larger effects may occur 
in $W$ and $Z$ boson production in the forward region at the Tevatron Run-2, 
and at the LHC throughout the full rapidity range. If present,
the small-$x$ broadening will have to be taken into consideration 
in precision studies of electroweak boson production.
Additionally, its observation will provide insights about the
transition to $k_T$-unordered (BFKL-like) dynamics in
multi-scale distributions at hadron-hadron colliders.

\vskip.1cm
\noindent
\subsection*{Acknowledgments}
\indent

 We thank C.~R.~Schmidt for  valuable discussions.
 This work was supported by 
the U.S. Department of Energy under grant  DE-FG03-95ER40908,
the National Science Foundation under grant PHY-0244919, 
and the Lightner-Sams Foundation.




\section[The High Energy Limit of QCD and the BFKL Equation]{THE HIGH
  ENERGY LIMIT OF QCD AND THE BFKL EQUATION~\protect\footnote{Contributed
  by: {J.~R.~Andersen, V.~Del Duca,A.~De Roeck, A.~Sabio Vera  }}} 
\label{LH_Smallx}

The Balitsky--Fadin--Kuraev--Lipatov (BFKL)
\cite{Lipatov:1976zz,Fadin:1975cb,Kuraev:1976ge,Kuraev:1977fs,Balitsky:1978ic,Balitsky:1979ns}
formalism resums a class of large logarithms dominant in the Regge limit of
scattering amplitudes, where the center of mass energy $\sqrt{s}$ is large
and the momentum transfer $\sqrt{-t}$ fixed. The cross--section for a general
process $A+B \rightarrow A'+B'$ within this approach in the high energy limit
can be written in the factorised form
\begin{eqnarray}
\label{cross--section1}
\sigma(s) &=&\int 
\frac{d^2 {\bf k}_a}{2 \pi{\bf k}_a^2}
\int \frac{d^2 {\bf k}_b}{2 \pi {\bf k}_b^2} ~\Phi_A({\bf k}_a) ~\Phi_B({\bf k}_b)
~f \left({\bf k}_a,{\bf k}_b, \Delta = \ln{\frac{s}{s_0}}\right),
\end{eqnarray}
where $\Phi_{A,B}$ are the impact factors characteristic of the particular
scattering process and $f\left({\bf k}_a,{\bf k}_b,\Delta\right)$ is the
gluon Green's function describing the interaction between two Reggeised
gluons exchanged in the $t$--channel with transverse momenta ${\bf k}_{a,b}$.
When these two transverse momenta are large and of similar magnitude the
Regge scale $s_0 = \left|{\bf k}_a\right|\left|{\bf k}_b\right|$ is chosen as
the scaling factor in the large logarithms. The implicit $s_0$--dependence of
the NLL impact factors cancels that of the BFKL gluon Green's function so as
to render the cross section independent of $s_0$ to this accuracy. In the
leading logarithmic (LL) approximation terms of the form $\left(\alpha_s
  \Delta \right)^n$ are resummed, while in the next--to--leading logarithmic
(NLL) approximation \cite{Fadin:1998py,Ciafaloni:1998gs} contributions of the
type $\alpha_s \left(\alpha_s\Delta \right)^n$ are also taken into account.

The gluon Green's function is obtained as the solution of an integral
equation, the so called BFKL equation, where radiative corrections enter
through its kernel. At LL it is possible to construct the complete
eigenfunctions of this kernel and, consequently, calculate the solution
analytically. At NLL this is only possible up to terms directly related to
the running of the strong coupling; this means that solving the BFKL equation
at NLL with the full kernel is a very challenging problem. Good progress has
been made in the last few years in this field: Studies of the stability of
the perturbative expansion were performed in
\cite{Ross:1998xw,Levin:1998pk,Salam:1998tj,Ciafaloni:1998iv,Brodsky:1998kn,Schmidt:1999mz,Forshaw:1999xm,Ciafaloni:1999yw,Ball:1999sh,Altarelli:1999vw,Altarelli:2000mh}
and of running coupling effects in
\cite{Kovchegov:1998ae,Armesto:1998gt,Thorne:1999rb,Thorne:1999sg,Forshaw:2000hv,Thorne:2001nr,Ciafaloni:2001db,Altarelli:2001ji,Ciafaloni:2002xf,Ciafaloni:2002xk}.
Among the most recent work studying the gluon Green's function an analysis
based on a new renormalisation group improved small x resummation scheme was
proposed in Ref.~\cite{Ciafaloni:2003ek}; in Ref.~\cite{Altarelli:2003hk} an
anomalous dimension including running couplings effects was constructed; and
in Ref.~\cite{Ciafaloni:2003rd,Ciafaloni:2003kd} the Green's function and
gluon splitting function were studied including a particular resummation
scheme.

In this contribution we report on the progress made in the last year to solve
the NLL BFKL equation exactly in a novel way: Using the numerical
implementation of an iterative solution presented in
Ref.~\cite{Andersen:2003an,Andersen:2003wy} found by explicitly separating
the virtual contributions to the kernel from the real emissions in transverse
momentum space. This is achieved by introducing a phase space slicing
parameter in dimensional regularisation.  How to obtain this solution, which
includes the angular correlations present in the kernel and the running
coupling effects, is presented in the next section.

\subsection[A novel solution to the NLL BFKL
  equation]{A novel solution to the NLL BFKL
  equation~\protect\footnote{Contributing authors: J.~R.~Andersen,
    A.~Sabio Vera}}
\label{BFKL@NLLA}

It is convenient to introduce the Mellin transform in $\Delta$ space 
\begin{eqnarray}
\label{Mellin}
f \left({\bf k}_a,{\bf k}_b, \Delta\right) 
&=& \frac{1}{2 \pi i}
\int_{a-i \infty}^{a+i \infty} d\omega ~ e^{\omega \Delta} f_{\omega} 
\left({\bf k}_a ,{\bf k}_b\right)
\end{eqnarray}
in order to write the NLL BFKL equation in dimensional regularisation as 
\begin{eqnarray}
\omega f_\omega \left({\bf k}_a,{\bf k}_b\right) &=& \delta^{(2+2\epsilon)} 
\left({\bf k}_a-{\bf k}_b\right) + \int d^{2+2\epsilon}{\bf k} ~
\mathcal{K}\left({\bf k}_a,{\bf k}+{\bf k}_a\right)f_\omega \left({\bf k}+{\bf k}_a,{\bf k}_b \right),
\end{eqnarray}
with the kernel $\mathcal{K}\left({\bf k}_a,{\bf k}\right) = 2
\,\omega^{(\epsilon)}\left({\bf k}_a\right)
\,\delta^{(2+2\epsilon)}\left({\bf k}_a-{\bf k}\right) +
\mathcal{K}_r\left({\bf k}_a,{\bf k}\right)$ depending on the gluon Regge
trajectory, which includes the virtual contributions, and a real emission
component \cite{Fadin:1998py}. The delta function in the driving term of the
integral equation corresponds to the limit of two gluon exchange.

The phase space slicing parameter, $\lambda$, is introduced 
through the approximation
\begin{eqnarray}
f_\omega \left({\bf k}+{\bf k}_a,{\bf k}_b \right)
&=& f_\omega \left({\bf k}+{\bf k}_a,{\bf k}_b \right)
\left(\theta\left({\bf k}^2-\lambda^2\right)+
\theta\left(\lambda^2-{\bf k}^2\right)\right) \nonumber\\
&\simeq& f_\omega \left({\bf k}+{\bf k}_a,{\bf k}_b \right)
\theta\left({\bf k}^2-\lambda^2\right)+
f_\omega \left({\bf k}_a,{\bf k}_b \right)
\theta\left(\lambda^2-{\bf k}^2\right),
\end{eqnarray}
which is a valid one for small values of the infrared 
parameter $\lambda$. With this separation it is possible to 
show that the $\epsilon$ poles cancel when the real emission 
below the cut off is combined with the virtual 
contributions; i.e.
\begin{eqnarray}
\omega_0 \left({\bf q},\lambda\right) &\equiv&\lim_{\epsilon \to 0} \left\{ 2\, \omega^{(\epsilon)}\left({\bf q}\right) + \int d^{2+2\epsilon}{\bf k} \,
\mathcal{K}_r^{(\epsilon)} \left({\bf q},{\bf q}+{\bf k}\right) 
\theta \left(\lambda^2-{\bf k}^2\right) \right\} \nonumber\\
&&\hspace{-2.4cm}= - \bar{\alpha}_s \left\{\ln{\frac{{\bf q}^2}{\lambda^2}}
+ \frac{\bar{\alpha}_s}{4}\left[\frac{\beta_0}{2 N_c}\ln{\frac{{\bf q}^2}{\lambda^2}}\ln{\frac{\mu^4}{{\bf q}^2 \lambda^2}}+\left(\frac{4}{3}-\frac{\pi^2}{3}+\frac{5}{3}\frac{\beta_0}{N_c}\right)\ln{\frac{{\bf q}^2}{\lambda^2}}-6 \zeta(3)\right]\right\}.
\end{eqnarray}
Using the notation 
\begin{eqnarray}
\omega_0 \left({\bf q},\lambda \right) \equiv - \xi\left(\left|{\bf q}\right|\lambda\right) \ln{\frac{{\bf q}^2}{\lambda^2}} + {\bar{\alpha}_s}^2 \frac{3}{2} \zeta (3)
\end{eqnarray}
 and 
\begin{eqnarray}
\xi \left({\rm X}\right) \equiv \bar{\alpha}_s +  
\frac{{\bar{\alpha}_s}^2}{4}\left[\frac{4}{3}-\frac{\pi^2}{3}+\frac{5}{3}\frac{\beta_0}{N_c}-\frac{\beta_0}{N_c}\ln{\frac{{\rm X}}{\mu^2}}\right]
\end{eqnarray}
the NLL BFKL equation takes the simple form
\begin{eqnarray}
\label{nll}
\left(\omega - \omega_0\left({\bf k}_a,\lambda\right)\right) f_\omega \left({\bf k}_a,{\bf k}_b\right) &=& \delta^{(2)} \left({\bf k}_a-{\bf k}_b\right)\nonumber\\
&&\hspace{-4cm}+ \int d^2 {\bf k} \left(\frac{1}{\pi {\bf k}^2} \xi \left({\bf k}^2\right) \theta\left({\bf k}^2-\lambda^2\right)+\widetilde{\mathcal{K}}_r \left({\bf k}_a,{\bf k}_a+{\bf k}\right)\right)f_\omega \left({\bf k}_a+{\bf k},{\bf k}_b\right),
\end{eqnarray}
where $\widetilde{\mathcal{K}}_r \left({\bf q},{\bf q}'\right)$ can be found in Ref. \cite{Andersen:2003an}. 

Eq. (\ref{nll}) can now be solved using an iterative procedure in the
$\omega$ plane similar to the one in
\cite{Kwiecinski:1996fm,Schmidt:1997fg,Orr:1997im} for the LL approximation.
The final solution is obtained after Mellin transform back to energy space,
$\Delta$.  The expression for the gluon Green's function then reads (using
the notation $y_0 \equiv \Delta$):
\begin{eqnarray}
\label{ours}
&&\hspace{-1cm}f({\bf k}_a ,{\bf k}_b, \Delta) 
~=~ \exp{\left(\omega_0 \left({\bf k}_a,{\lambda}\right) \Delta \right)}
\left\{\frac{}{}\delta^{(2)} ({\bf k}_a - {\bf k}_b) \right. \\
&&\hspace{-1cm}+ \sum_{n=1}^{\infty} \prod_{i=1}^{n} 
\int d^2 {\bf k}_i \left[\frac{\theta\left({\bf k}_i^2 - \lambda^2\right)}{\pi {\bf k}_i^2} \xi\left({\bf k}_i^2\right) +
\widetilde{\mathcal{K}}_r \left({\bf k}_a+\sum_{l=0}^{i-1}{\bf k}_l,
{\bf k}_a+\sum_{l=1}^{i}{\bf k}_l\right)\frac{}{}\right]\nonumber\\
&& \hspace{-1cm} \times \left. 
\int_0^{y_{i-1}} d y_i ~ {\rm exp}\left[\left(
\omega_0\left({\bf k}_a+\sum_{l=1}^i {\bf k}_l,\lambda\right)-\omega_0\left({\bf k}_a+\sum_{l=1}^{i-1} {\bf k}_l,
{\lambda}\right)\right) y_i\right] \delta^{(2)} \left(\sum_{l=1}^{n}{\bf k}_l 
+ {\bf k}_a - {\bf k}_b \right)\right\}, \nonumber
\end{eqnarray}
where $n$ corresponds to the number of emissions, or, alternatively, to the
number of iterations of the kernel.

This solution has been implemented in a Monte Carlo integration routine to
study the behaviour of the gluon Green's function. In
Ref.~\cite{Andersen:2003an,Andersen:2003wy} it was shown how, for a fixed
value of $\Delta$, only a finite number of iterations contributes to the
final value of the solution. As the available energy in the scattering
process increases more terms in the expansion in Eq.~(\ref{ours}) are needed.
Independence on the $\lambda$ scale is achieved when its value is small
compared to the initial transverse momenta ${\bf k}_{a,b}$.

As an example of the potential of this approach we reproduce here some
results. In Fig.~\ref{fig:scale_scan} the value of the modulus of ${\bf k}_b$
is fixed and the dependence on the modulus of ${\bf k}_a$ is studied.  At LL
there is complete agreement with the analytic solution, while the NLL result
is always lower. This plot is calculated for a particular low value of the
energy scale. The discontinuity present in this figure has its origin in the
initial condition of the integral equation and its effect diminishes as the
available energy in the scattering is larger. In Fig.~\ref{fig:Delta_scan}
the rise of the gluon Green's function with energy is calculated. The slower
rise at NLL compared to LL is a well known feature of the NLL corrections.
The central lines for both the LL and NLL results are obtained by choosing
the renormalisation scale $\mu=k_b$.  The coloured bands correspond to a
variation of the renormalisation scale from $k_b/2$ to $2\,k_b$. An advantage
of this numerical method of solution is that it is possible to study the
angular dependences in the BFKL ladder. As an example we show
Fig.~\ref{fig:angle_scan} where the contribution to the NLL solution from
different angles between the two dimensional vectors ${\bf k}_{a,b}$ is
plotted. This analysis shows how the emissions are less correlated when the
energy is larger.

Future work using this approach will include the study of the gluon Green's
function in the non forward case; the calculation of the solution to the
$N=4$ Supersymmetric equation; an analysis of the effect of resumming the
strong coupling to all orders; and the investigation of the gluon
distribution at small x including all the scale invariant and running
coupling NLL effects.
\begin{figure}[tb]
  \centering
  \includegraphics[width=10cm]{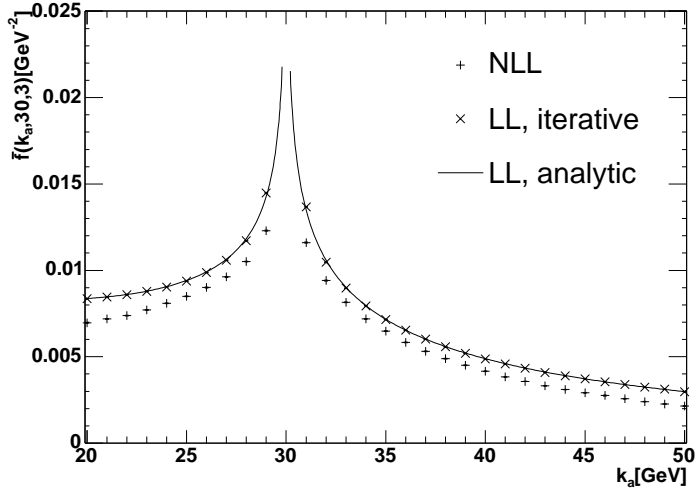}
\caption[*]{$k_a$ dependence of the LL and NLL gluon Green's function at $\mu=k_b=30$~GeV for two values of $\Delta$.}
\label{fig:scale_scan}
\end{figure}
\begin{figure}[tb]
  \centering
  \includegraphics[width=10cm]{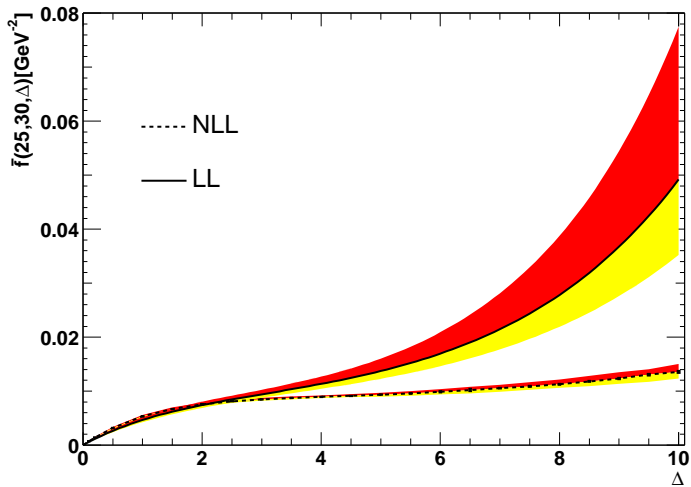}
\caption[*]{$\Delta$--dependence of the NLL gluon Green's function evaluated
    for $k_a=25$~GeV and $k_b=30$~GeV.}
\label{fig:Delta_scan}
\end{figure}
\begin{figure}[tb]
  \centering
  \includegraphics[width=10cm]{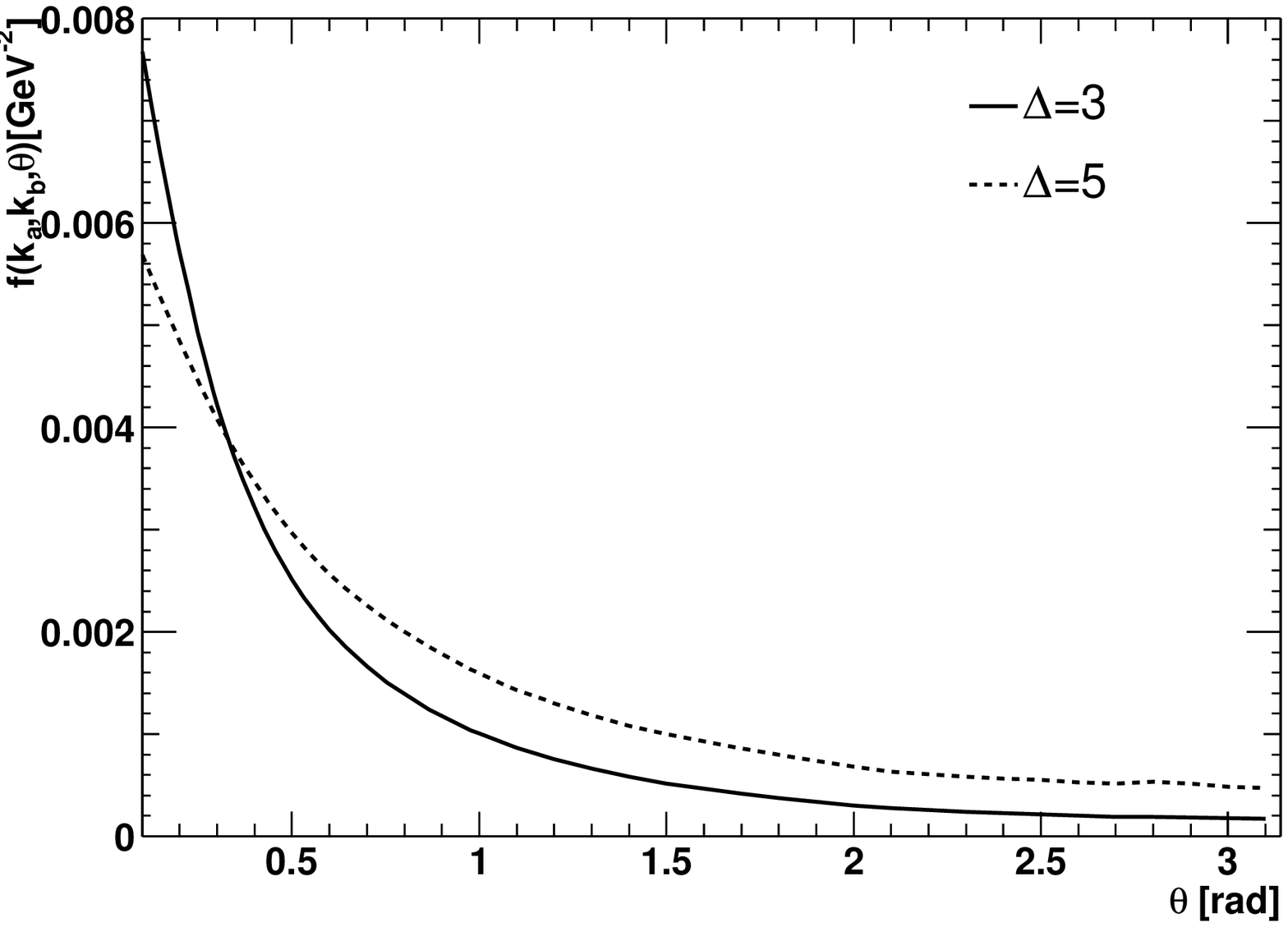}
\caption{The dependence of the gluon Green's function at NLL on the angle
    between $\mathbf{k}_a$ ($k_a=25$~GeV) and $\mathbf{k}_b$ ($k_b=30$~GeV)
    for the choice of $\Delta=3$ and $\Delta=5$. The renormalisation point is
    chosen at $\mu=k_b$.}
\label{fig:angle_scan}
\end{figure}

\subsection[BFKL Phenomenology at Colliders]{BFKL Phenomenology at
    Colliders~\protect\footnote{Contributing authors: J.~R.~Andersen,
    V.~Del Duca}} 
\label{sec:BFKLphen}
When confronting BFKL predictions with data, several points are worth
observing. First of all, present day colliders do not operate at ``asymptotic
energies'' where the high energy exponent dominates the BFKL prediction,
leading to a prediction of an exponential rise in cross section with an
intercept of $\bar\alpha_s4\ln 2$. The logarithms resummed are kinematically
generated, and in the derivation of the standard analytic solution to the
BFKL equation, the transverse momentum of the gluons emitted from the BFKL
evolution has been integrated to infinity.  It is therefore apparent that any
limits on the phase space probed in an experiment can have a crucial impact
on the theoretical prediction. Such limits can either be the cuts implemented
in the measurement or the overall limit on the available energy at a
collider.  Taking hadronic dijet production as an example, the energy
constraint will obviously not just limit the possible rapidity separation of
the leading dijets, but also the amount of possible radiation from the BFKL
evolution, especially so when the leading dijets are close to the kinematical
boundary.  For a multi--particle final state described by two leading dijets
with transverse momentum and rapidity $(p_{a/b\perp},y_{a/b})$ and $n$ gluons
described by $(k_{i\perp},y_{i})$, the square of the total energy of the
event is given by $\hat s=x_a x_b s$ where $\sqrt{s}$ is the energy of the
hadron collider and
\begin{equation}
  x_{a(b)}=\frac{p_{a\perp}}{\sqrt s}e^{(-)y_a}+\frac{p_{b\perp}} {\sqrt
  s}e^{(-)y_b} + 
  \sum_{i=1}^{n}\frac{k_{i\perp}}{\sqrt s}e^{(-)y_i}\label{eq:xs},
\end{equation}
where the minus sign in the exponentials of the right--hand side applies to
the subscript $b$ on the left--hand side. Since in the analytic BFKL approach
the contribution to $x_{a(b)}$ from the BFKL radiation is inaccessible, this
approach systematically underestimate the exact value of the $x$'s, and can
thus grossly overestimate the parton luminosities.

The iterative approach of Refs.\cite{Schmidt:1997fg,Orr:1997im} to solving
the LL BFKL equation solves this problem. The results of this approach
coincides with the LL limit of the solution of the NLL BFKL equation of
Sec.~\ref{BFKL@NLLA}. At LL the change in the $t$--channel momentum at each
step in the iteration corresponds to the momentum carried by one emission.
The method therefore allows for the reconstruction of the full final state
configurations contributing to the BFKL evolution, and it is possible to
study quantities such as multiplicities and distribution in transverse
momentum of the emitted gluons~\cite{Andersen:2003gs}. Only this
reconstruction of the full final state allows for the observation of energy
and momentum conservation. The effects of energy and momentum conservation
have been studied in several
processes~\cite{Orr:1998hc,Andersen:2001kt,Andersen:2001ja}. When no phase
space constraints are imposed, the iterative solution reproduces the known
analytic solution to the LL BFKL equation.

In Ref.~\cite{Andersen:2001kt} we have, in light of the recent D0
measurement, re--analysed the Mueller--Navelet proposal of the study of dijets
at hadron colliders in search of BFKL signatures. The main result of this
study is that the difference between D0 and the Mueller--Navelet analysis in
the reconstruction of the parton momentum fractions, the presence of an upper
bound on the momentum transfer, and the contribution of the BFKL gluon
radiation to the parton momentum fractions (at Tevatron energies) lower the
parton flux in such a way as to approximately cancel the rise in the
subprocess cross section with increasing dijet rapidity separation
($\hat\sigma_{jj} \sim \exp(\lambda\Delta y$)) predicted from the standard
BFKL approach. This strong pdf suppression is due to the dijet production
being driven by the gluon pdf, which is very steeply falling in $x$ for the
region in $x$ of interest. This means that even the slightest change in $x$
has a dramatic impact on the parton flux and thus the prediction for the
cross section.  Also, the experimental cuts implemented have been shown to
invalidate the Mueller--Navelet analysis and extraction of a 'BFKL
intercept'.

\begin{figure}[tbp]
  \centering
  \includegraphics[width=10cm]{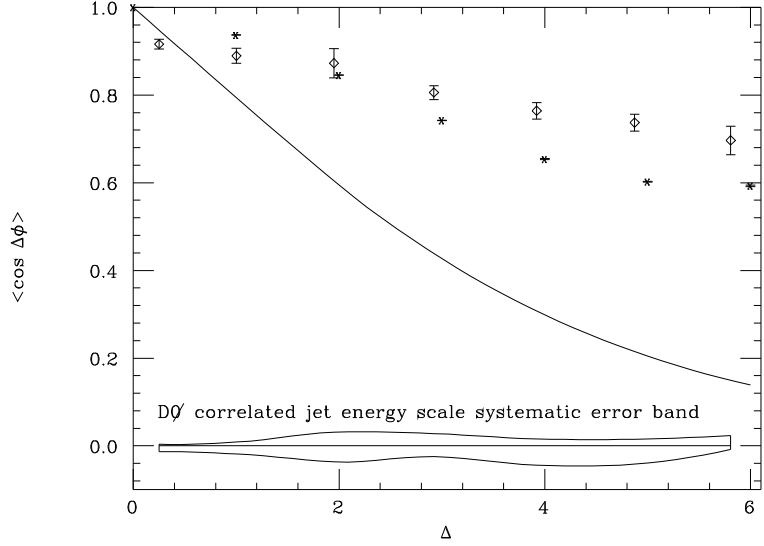}
  \caption{Dijet angular correlation at the Tevatron. D0 data are marked by
    the diamonds, BFKL prediction (LL supplemented by running coupling and
    including energy/momentum conservation) given by the stars. The LL BFKL
    prediction without energy/momentum conservation is marked by the solid
    line.  Figure taken from Ref.~\cite{Orr:1997im}.}
  \label{fig:dijet_tevatron}
\end{figure}
As shown by this analysis, the expected rise in cross section over the LO QCD
results does not survive once energy and momentum conservation is taken into
account at the LHC or the Tevatron. However, other BFKL signatures should
still be present; among those studied the most is the azimuthal decorrelation
of dijets\cite{DelDuca:1995fx,Stirling:1994zs}. In
Fig.~\ref{fig:dijet_tevatron} we have plotted the data and prediction for the
angular correlation between the leading dijets as a function of the inter--jet
rapidity separation at the Tevatron. It should be noted that no detailed jet
definition was applied in the BFKL prediction. It is, however, believed that
a proper implementation of a jet finding algorithm will not change the
partonic LL BFKL prediction significantly, since the gluons emitted from the
BFKL evolution are typically well separated in rapidity. It is seen from
Fig.~\ref{fig:dijet_tevatron} that the BFKL prediction respecting energy and
momentum conservation is predicting slightly too much azimuthal
decorrelation, but far less than an estimate based on a naive LL BFKL
analysis ignoring the energy taken up by the BFKL radiation. On the other
hand, a fixed NLO analysis shows too little decorrelation, while the
prediction from \texttt{HERWIG} is in agreement with data. This shows that
the decorrelation is dominated by soft gluon effects.

\subsection[Experimental Opportunities at the LHC]{Experimental
  Opportunities at the LHC~\protect\footnote{Contributing authors:
    A.~De Roeck}} 
\label{sec:forward;exp}
Studies of low-$x$ and BFKL dynamics at colliders typically require
experiments with a large acceptance.  Presently five experiments are planned
at the LHC.  Two of these, CMS~\cite{cms} and ATLAS~\cite{atlas} are general
purpose experiments with an acceptance in pseudorapidity $\eta$ of roughly
$|\eta| < 2.5$ for tracking based measurements and $|\eta| < 5$ for
calorimeteric based measurements.  Here $\eta$ is defined as $
-\ln\tan\theta/2$, with $\theta$ the polar angle of the particle.  Hadronic
jets can be detected and measured up to approximately $\eta =4.5$ while muons
and electrons can be identified up to about $\eta = 2.5$.  Extensions of the
detector range are being investigated, as discussed below.

The TOTEM experiment~\cite{totem} will use the same interaction point as the
CMS experiment. TOTEM is an experiment to measure the elastic and total cross
section and will use roman pot detectors to measure scattered protons in
elastic and diffractive $pp$ interactions, and charged particle detectors for
tagging inelastic events in the regions $3 < |\eta| < 5$ (T1) and $5.3 <
|\eta| < 6.5$ (T2). CMS is studying to add a calorimeter, called CASTOR, in
the forward region directly behind T2. CASTOR will have an electromagnetic
and hadronic readout section and an acceptance in the range $5.4 <|\eta|
<6.7$. The TOTEM trigger and data readout will be such that these can be
included in the CMS datastream, such that both experiments can run as one
experiment.  That way, a full detector with good coverage for jets and
electrons/photons with rapidities of up to 7 will be available. The position
of the T1, T2 trackers and the forward calorimeter, along the beamline and
integrated with CMS, is shown in Fig.~\ref{fig:castor}.

Roman pot detectors are presently also considered by ATLAS~\cite{rijsenbeek}.
ATLAS further plans for a cerenkov/quartz fiber detector, called LUCID, for
luminosity monitoring, which will cover the region $5.2 < \eta< 6.2$.  This
detector could possibly be used for the tagging of rapidity gaps in an event.

\begin{figure}[htb] 
\centering
  \includegraphics[height=6cm]{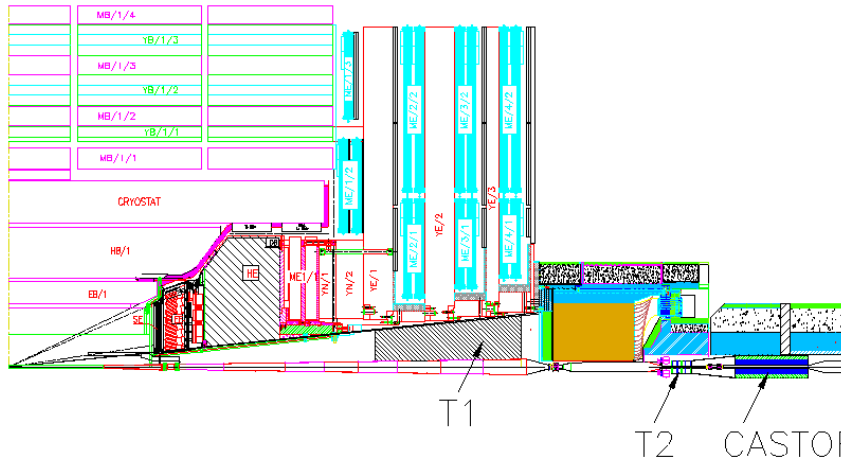}
\caption{Position of the inelastic event tagging detectors of 
TOTEM, T1, T2 and CASTOR integrated with CMS}
\label{fig:castor}
\end{figure}
\par

\begin{figure}[t] 
\centering
  \includegraphics[height=7cm]{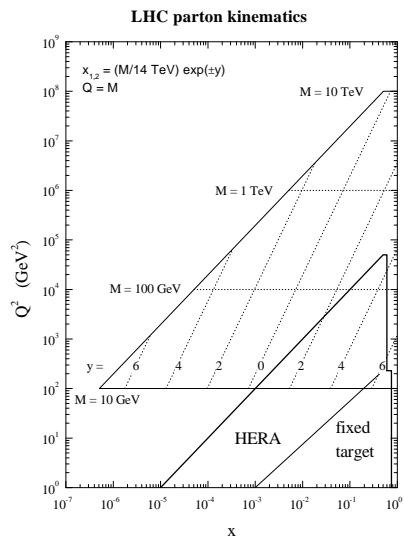}
\caption{
The kinematic plane $(x,Q^2)$ and the reach of the LHC, together with 
that of the existing data (HERA, fixed target). Lines of constant
pseudo--rapidity
are shown to indicate the kinematics of the produced objects in the LHC 
centre of mass frame\cite{Martin:1999ww}.}
\label{lhckin}
\end{figure}
\par

The extended coverage of the detectors will allow to probe processes at
low-$x$ via Drell--Yan, jets, W and direct photon production.  E.g. the
Drell--Yan process $q\overline{q} \rightarrow e^+e^-$ has a simple
experimental signature. The $x_{1,2}$ values of the two incoming quarks
relate to the invariant mass of the two electron system $M_{ee}$ as $x_1\cdot
x_2\cdot s \simeq M^2_{ee}$, hence when one of the $x_{1,2}$ values is large
(say $x> 0.1$), very low-$x$ can be probed with low mass Drell--Yan pairs.
From Fig.~\ref{lhckin}~\cite{Martin:1999ww} we observe that in order to reach
small masses (small scales) and low-$x$, will require to probe large values
of $\eta$.  Thus the resulting electrons will dominantly go in the very
forward direction.

If low mass Drell--Yan pairs, prompt photons or jets (which have the same
kinematics) can be measured in the CMS/TOTEM forward detectors, the parton
distributions can be probed down to values of $10^{-6}-10^{-7}$, i.e. at
lower values than for any other collider. Depending on the low-$x$ dynamics,
the predictions for jet, Drell--Yan and other cross sections can differ by a
factor two or more in this region~\cite{Kimber:2001nm}. Experimental
challenges on extracting the low-$x$ signals with respect to machine and QCD
backgrounds are still under study.

The extended detector capabilities also allow for larger rapidity distances
(i.e longer gluon ladders) between two measured hard jets, one of the golden
BFKL measurements at a hadron collider (see Sec.~\ref{sec:BFKLphen}).  Note
that the experiments will also measure all activity in between these two
jets.  Since BFKL does not have strong $k_T$ ordering the number and energy
spectrum of the mini--jets produced along the ladder may be a significant
footprint for BFKL.  The main challenge will be to define a suitable trigger
for this two jet channel. The presently foreseen trigger thresholds in ATLAS
and CMS are around 150-200 GeV for one and two jet events, while for BFKL
studies particularly jets in the energy range of 20-100 GeV are more of
interest, since the distributions of these will be less distorted by
kinematical constraints.

In principle it is also possible to run the LHC at lower energies, e.g.  at 3
and 8 TeV, with significant luminosities ($> 10^{32}$cm$^{-2}$s$^{-1}$),
allowing to make parton density independent ratios of di--jet production at
different energies, over a large range of rapidity.

For luminosities up to $10^{33}$cm$^{-2}$s$^{-1}$ or below, the number of
overlap events in one bunch crossing is small enough such that rapidity gaps
can be used to detect interactions with colour singlet
exchange~\cite{DeRoeck:2002uc}.  The extended range covered in $\eta$ will
allow to study events with multi--gaps. Hard diffractive phenomena, often
linked to small-$x$ dynamics, can be studied further in detail via the
tagging of the scattered protons in the roman pot detectors along the
beam--line.  Together with the measurement of the hard probes in the central
LHC detectors, these events will allow for a detailed study of diffractive
phenomena in $pp$ collisions at the highest energies.


\clearpage


\section[Pion Pair Production at the LHC: Comparing QCD@NLO with
PYTHIA]{PION PAIR PRODUCTION AT THE LHC: COMPARING QCD@NLO WITH
  PYTHIA~\protect\footnote{Contributed
  by: {T.~Binoth, K.~Lassila-Perini}}} 
\label{2pi}


\subsection{Introduction}

The indirect mass bounds for the Higgs boson from precision 
measurements indicate 
that the favoured range is below 200 GeV. 
A prominent search channel for neutral  Higgs bosons in the mass
window 80~GeV~$< m_H <$~140~GeV is the rare decay
$H \to \gamma\gamma$, as electromagnetic signals are preferable
compared to hadronic ones due to huge backgrounds for the latter.
The important backgrounds for this channel are
prompt di-photon production and processes with one real photon
and a jet with an isolated neutral meson. In addition, the abundant
two jet rate at the LHC leads
to a large reducible background where a meson in both jets
produces an electromagnetic signature in the detector. This
background needs to be rejected very efficiently.
It is thus an important question to
predict the respective rates as precisely as possible
to allow for reliable simulations. 

Experimental studies dominantly use PYTHIA \cite{Sjostrand:2000wi} 
to predict the reaction rates for signal and background processes.
PYTHIA is based on leading order $2 \to 2$ matrix elements
combined with initial and final state parton showers coupled
to a hadronization model. One can expect that calculations
in next-to-leading order in the QCD coupling lead to an improvement
of the predictions in two respects. 
Firstly, distributions which are sensitive to
hard radiation of an extra parton  will not be described properly
by a parton shower which is on the other hand expected 
to describe the collinear regions in phase space more reliably
than a fixed order calculation. Secondly, inclusion of loop 
effects typically reduces the dependence on unphysical scales
and thus leads to more reliable predictions. 

The experimental studies show that the reducible pion+photon and di-pion 
contribution can be reduced by isolation cuts to a level which is acceptable for
Higgs searches in the two photon channel \cite{unknown:1999fq,unknown:2002zz}.
It is the aim of this study to cross check if this statement holds also at 
next-to-leading order (NLO) precision. 
We compare in the following production rates
of pion pairs at high $p_T$ calculated with PYTHIA and
the DIPHOX code \cite{Binoth:1999qq}. The latter contains 
partonic matrix elements at NLO for the production of photon pairs,
photon+hadron and hadron pairs and is flexible enough
to account for diverse experimental cuts which can be modeled
on the partonic level. For the description of 
hadrons in DIPHOX the  model of collinear fragmentation 
is used. Partonic matrix elements are folded with fragmentation
functions, $D_{p\to h}(x,Q)$ which stand for the probability to
find a hadron $h$ with an energy fraction $z$ relative to the jet
formed by the initial parton $p=g,q$. This probability is
scale dependent. The scale dependence is governed by Altarelli-Parisi evolution. 
Initial distributions are fitted to experimental data
in a certain $x$ range and extrapolated to the  high (low) 
x range by some standard parameterizations. 
We note at this place that in the case of severe isolation cuts
typically the tails of the fragmentation functions become 
numerically important which are not experimentally  well restricted \cite{Binoth:2002wa}.   

DIPHOX has been confronted with many different data sets from fixed target 
experiments to Tevatron data and showed an excellent agreement between
di-pion or di-photon observables, whenever the condition for fixed 
order perturbative calculations were fullfilled \cite{Binoth:1999qq,Binoth:2002ym,Binoth:2000zt}. 
We are thus in a situation to make a NLO prediction for photon pair, photon+pion
and pion pair production at the LHC \cite{Binoth:2002wa,Binoth:2000is,Binoth:2001jd}.
We want to mention that there is another calculation for di-hadron production at NLO available 
\cite{Owens:2001rr}\footnote{In \cite{Bern:2002jx} corrections
for two photon production via a quark loop at the two-loop
level are taken into account.}.  

A similar study
for photon+pion production was presented by us in a previous
Les Houches report \cite{Giele:2002hx}. We found good agreement between 
PYTHIA and DIPHOX for loose isolation cuts whereas for severe isolation
cuts the PYTHIA rates  were above the NLO predictions. 
Qualitatively we observed that isolation cuts are more efficient 
for the NLO calculation than in the PYTHIA result. We were lead
to the conclusion that the experimental studies are conservative for the photon pion 
background.   

\subsection{Comparison: PYTHIA VS. DIPHOX}


The basic underlying reaction for the production of pion pairs
at leading order is two jet production with numerous
different subprocesses like $q +\bar q \to q' +\bar q'$, 
$q +\bar q' \to q +\bar q'$, $q +\bar q \to g+g$, 
$q + g \to p+g$, $g + g \to g+g$. It is a priori not clear,
if there is a dominant process for different kinematical
regions and experimental cuts.  All subprocesses have to be 
considered. At NLO virtual corrections to the $2\to 2$ processes
have to be combined with numerous $2\to 3$ matrix elements with one 
unresolved particle. Details of the calculation can be found in \cite{Binoth:1999qq}.

We compared two distributions relevant for
Higgs boson searches, first the invariant mass distribution
of the pion pair, $d\sigma/dM_{\pi\pi}$ with 
$M_{\pi\pi}=(p_{\pi_1}+p_{\pi_1})^2$ and second the transverse
momentum distribution of the pion pair, $d\sigma/dq_T$ with
$q_T =  | \vec p_{T\;\pi_1} + \vec p_{T\;\pi_2} |$.

PYTHIA \cite{Sjostrand:2000wi} version 6.205 was used to generate the
events. High $p_T$ QCD processes were simulated and the
events with two neutral pions were accepted. The pions were
required to have  $p_T >$ 25 GeV and to be in the pseudo-rapidity
range of $|\eta| <$ 2.5.
The isolation was implemented by setting a threshold to the sum of $E_T$
of all particles (excluding the pion itself) in the isolation
cone $\Delta R = \sqrt{(\Delta \eta)^2 + \Delta \phi)^2}$.
                                                                                
The default parameters were chosen, apart from the longitudinal
fragmentation function, where a hybrid scheme was used and
for the multiple parton interactions, where a set of parameters
tuned with the collider data \cite{Moraes:2003zz} were used.
The pile-up effects from collisions within the same bunch
crossing were not taken into account, as we are here basically 
interested in a comparison of event rates of partonic interactions.
We note that the multiple interactions belong to the
regime of soft QCD, which is modeled in PYTHIA. The relevance of
multiple interactions is restricted to  $p_T$ values 
which are much below the scales  used in our study. 
We are not in a regime where these effects are expected 
to become numerically sizable.

As already mentioned DIPHOX contains all matrix elements 
indicated above together with  their next-to-leading order
virtual and real emission corrections. Using the same experimental cuts 
defined above, the same distributions as with PYTHIA
were produced by varying the renormalization ($\mu$), factorization ($M$)
and fragmentation ($M_f$) scales. We took $\mu=M=M_f= c\,M_{\pi\pi}/2$
with $c=1/2,1,2$. For the parton distribution functions 
we used \cite{Martin:2001es}, for the pion fragmentation function
we used a recent parametrization of Kramer, Kniehl and P\"otter (KKP) \cite{Kniehl:2000fe}
and an older one from Binnewies, Kramer and Kniehl (BKK) \cite{Binnewies:1995ju}.
This was done to get an idea about the uncertainties due to fragmentation functions,
although the KKP set should be more reliable, as it contains newer data.

\begin{figure}
\unitlength=1mm
\begin{picture}(190,85)
\put(1,6){$-$}
\put(62,6){$+$}
\put(62,27){$-$}
\put( 0,0){\epsfig{file=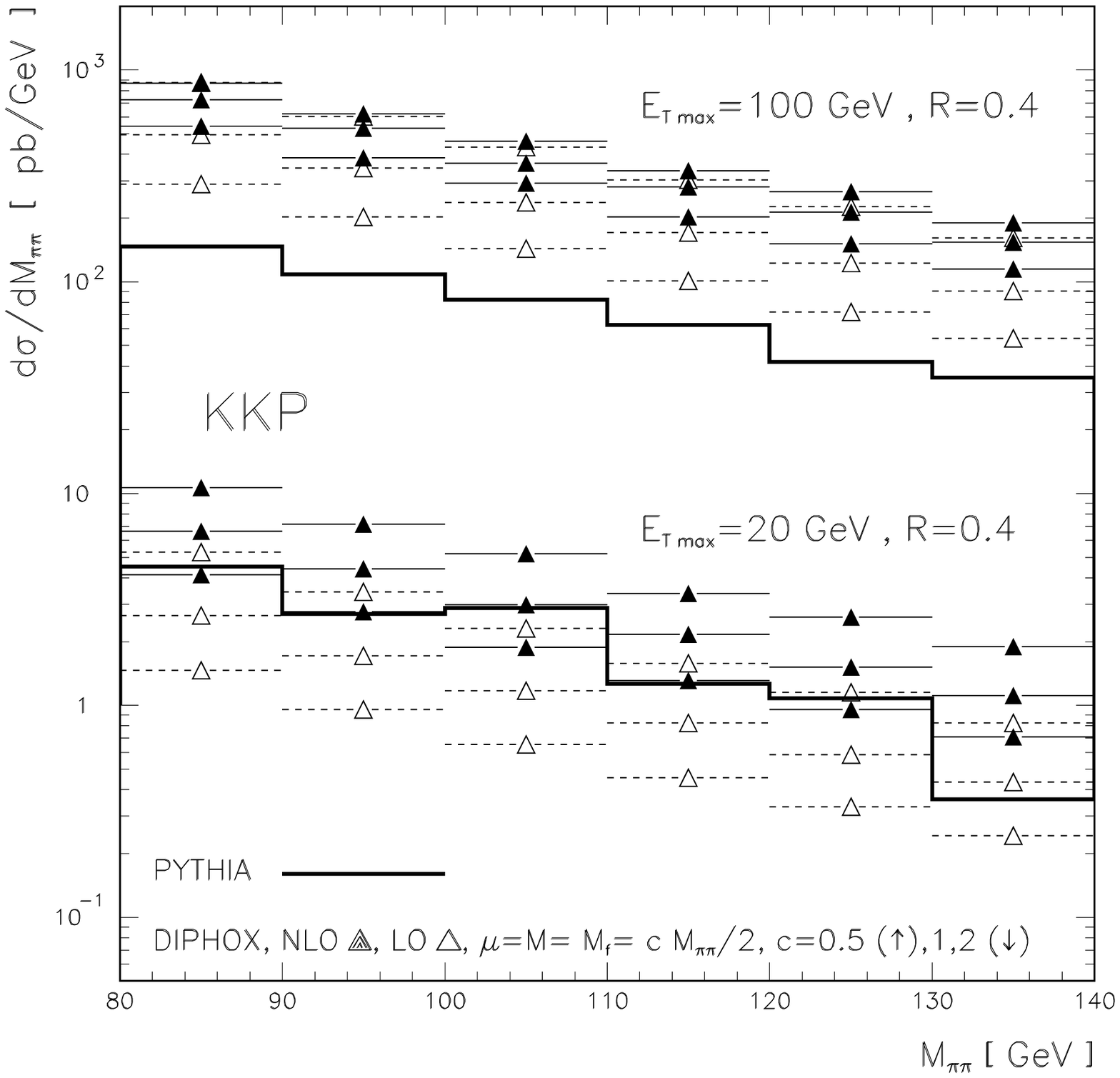,height=8.3cm}}
\put(80,0){\epsfig{file=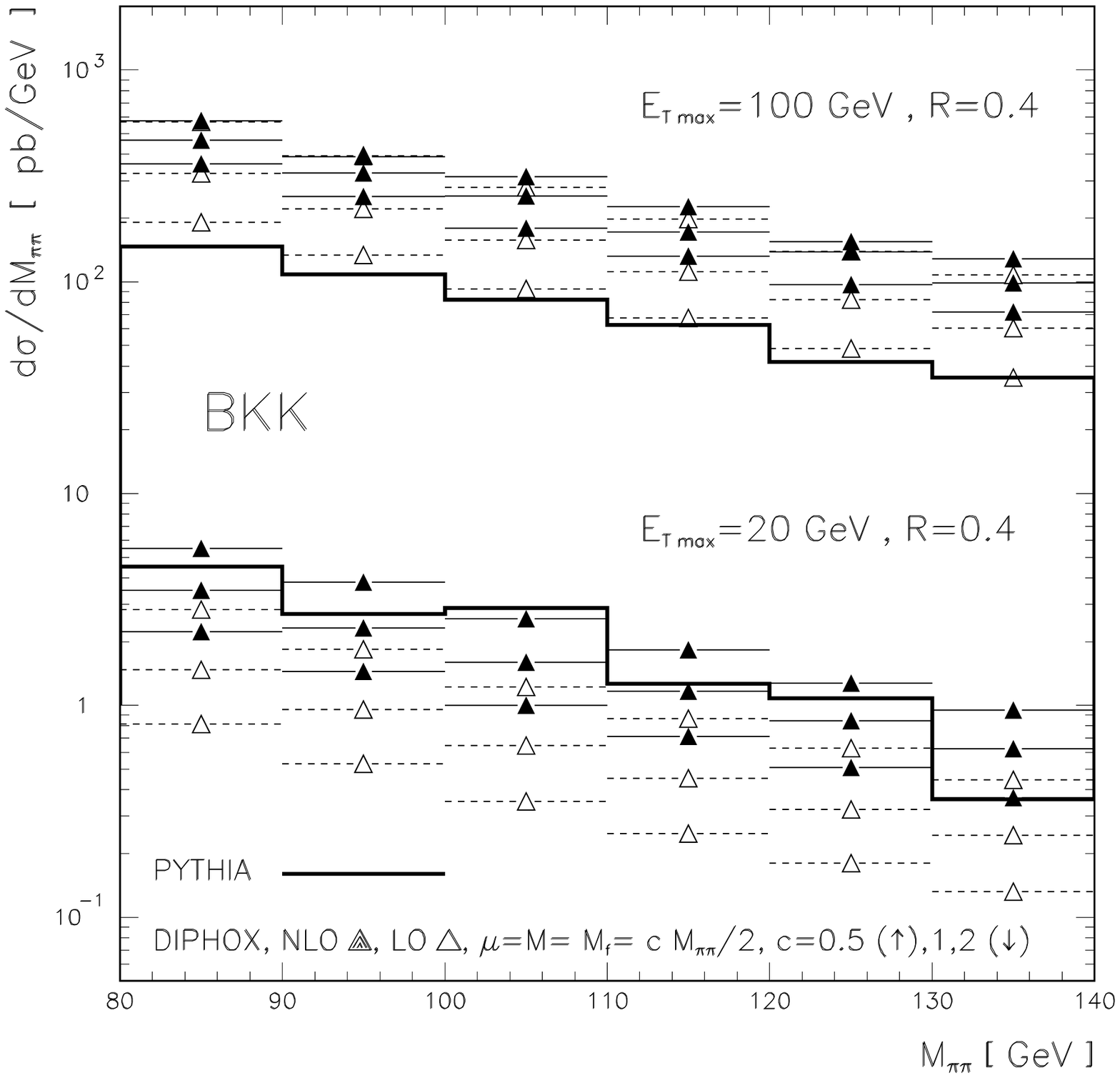,height=8.3cm}}
\end{picture}
\caption{The invariant mass distribution of pion pairs at LHC for loose ($E_{T max}=100$ GeV, $\Delta R<0.4$)
and a more realistic isolation ($E_{T max}=20$ GeV, $\Delta R<0.4$) criterion. The full line is the PYTHIA
prediction. For DIPHOX NLO (full triangle) and LO (open triangle) matrix elements were used  for 3 different scale choices
$\mu=M=M_f=c\; M_{\pi\pi}/2$, $c=0.5$ (upper), $c=1$ (middle), $c=2$ (lower). The left plot uses 
the KKP, the right plot the BKK fragmentation functions for the neutral pions.}\label{fig1}
\end{figure}

In Fig.~\ref{fig1} we show the invariant mass distribution 
for two different isolation criteria, $E_{T max}=20, 100$ GeV
in a cone $\Delta R<0.4$.
LO and NLO matrix elements were used for the DIPHOX predictions for three different
choices of the scales. 
Higher scales mean lower cross sections in the plot. In the left plot KKP fragmentation
functions were used, on the right the BKK parametrization. 
The PYTHIA prediction is shown as a full line. One observes that for $E_{T max}=100$ 
PYTHIA undershoots the DIPHOX curves considerably. For the KKP fragmentation functions
there is  difference between a factor 2 to 6 for LO and 3 to 6 for NLO matrix elements. 
By using the BKK fragmentation functions,
LO matrix elements and high scales ($c=2$) the accordance is improved. 
For the harder isolation cut, $E_{T max}=20$, the PYTHIA prediction shows a fair agreement
with the DIPHOX result for both fragmentation functions. 
It is inside the scale variation of the NLO result apart for the last bin which is slightly below
in the KKP case.
One observes that the isolation cut acts more efficiently
in DIPHOX than in PYTHIA. This was observed already in a 
earlier study \cite{Giele:2002hx}. It has  to do with the different fragmentation models used.
For more severe isolation preliminary results indicate that the PYTHIA distributions start to fall below the 
DIPHOX curves \footnote{The curve for $E_{T max}=10$ GeV is not included because more statistics 
is needed to make a definite statement.}.

The inclusion of NLO
matrix elements increases the cross section. In Fig.\ref{fig2} the ratio of NLO and LO
result for KKP (left figure of Fig.\ref{fig1}) is shown.  
Evidently the K-factors defined as this ratio depend on the scale choice and the experimental cuts, they
vary between 1 and 2 for $E_{T max}=100$ GeV and 2 and 3 for $E_{T max}=20$ GeV, see Fig.~\ref{fig2}. 
The inclusion of higher order terms improves the stability
of the DIPHOX prediction under scale variations. This is much less pronounced for the harder
isolation cut with $E_{T max}=20$ GeV. Here compensations between LO and NLO terms 
are spoiled by the isolation cut. We note that asymmetric $p_T$ cuts, not applied here, 
lead to larger K-factors for the same reason \cite{Binoth:2001jd}.

\begin{figure}
\begin{center}
\includegraphics[width=10.5cm]{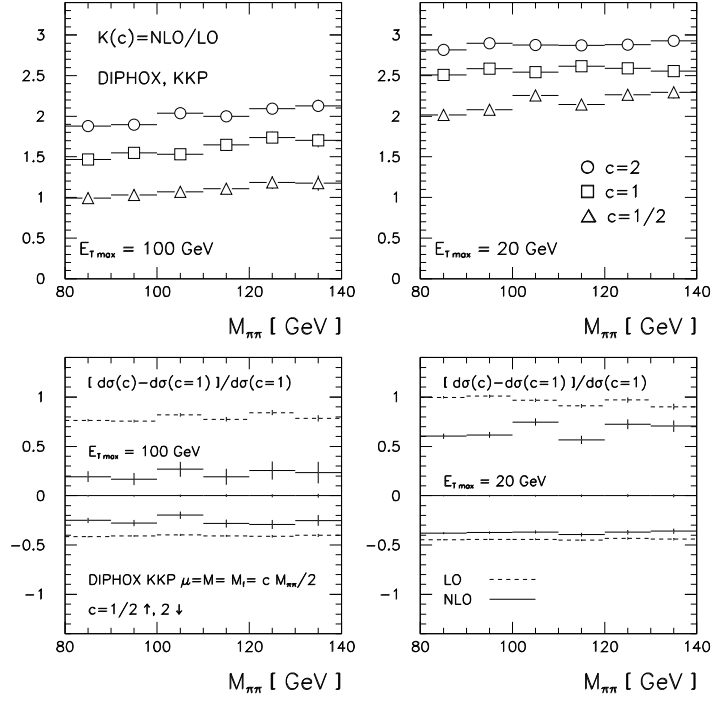} 
\caption{K factors (top) and relative scale variations (bottom) for the two isolation criteria. The scale variations 
are plotted for the LO (dashed) and the NLO prediction (full). The scales have been varied as
explained in the text. KKP fragmentation functions have been used.}\label{fig2}
\end{center}
\end{figure}

\begin{figure}
\begin{center}
\includegraphics[width=8.5cm]{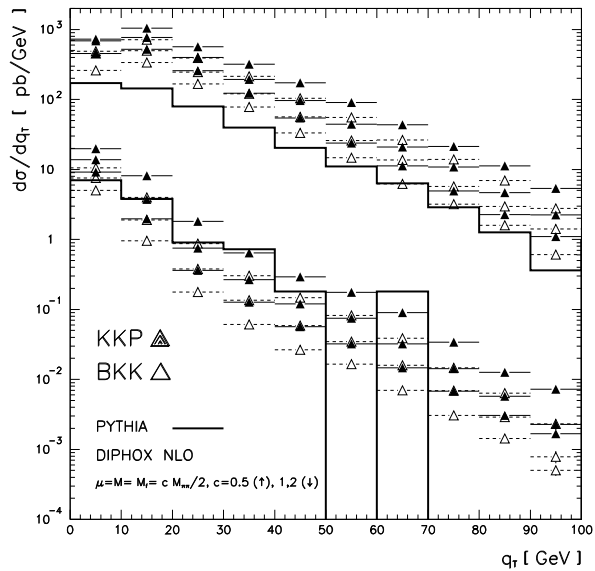} 
\caption{The $q_T$ distribution of the pion pair. PYTHIA vs. DIPHOX NLO. KKP (full triangle)
and BKK (open triangle) fragmentation functions are used.}\label{fig3}
\end{center}
\end{figure}
Finally we present the transverse momentum distribution of the pion pair in Fig.~\ref{fig3}.
Now, only the NLO prediction is shown as this distribution is sensitive
to $2 \to 3$ matrix elements not present in the LO calculation.
For a partonic $2\to 2$ reaction there is a kinematical restriction $p_T<E_{T max}$.
Distributions beyond this point are filled by the parton shower in the PYTHIA case and  
with  hard $2 \to 3$ matrix elements in DIPHOX which are
not present in PYTHIA. For $E_{T max}=100$ GeV this does not affect
the plotted range. 
Again, whereas  PYTHIA undershoots the NLO curves considerably for  $E_{T max}=100$ GeV,
the shape is very similar.
Fair agreement is found for $E_{T max}=20$ GeV for not too high values of $q_T$. The drop
in statistics of the PYTHIA curve can be understood by the missing $2 \to 3$  matrix elements
which are important for high $q_T$ values in the case of hard isolation.

\subsection{Conclusion}

We have presented a comparison between PYTHIA version 6.205 and DIPHOX version 1.2 for the pair production
of high $p_T$  pions at the LHC. This is an important reducible background for Higgs search in the
two photon channel.   
We observe that for loose isolation cuts the NLO result of DIPHOX is significantly higher 
than from PYTHIA which is due to
matrix elements not present in PYTHIA and different fragmentation models in both codes.
As isolation cuts act harder in DIPHOX than in PYTHIA, due to different fragmentation  models,
there is a fair agreement of both codes for a harder isolation cut. The trend is that
for more severe isolation cuts PYTHIA will even overshoot the NLO predictions.
We note that a further suppression of the dipion rate than shown in our plots is possible 
for example by using asymmetric $p_T$ cuts and pion/photon identification methods, but
this is beyond the scope of this study.
  
We arrive at the tentative conclusion  that experimental studies based on PYTHIA seem to 
be slightly below the more complete NLO calculation although the discrepancie is not dramatic. 
We have to stress that this statement depends strongly on the
behaviour of the fragmentation model.
Hard isolation means that fragmentation models are tested at high $x$ (the ratio between the pion energy and the jet energy).
Currently the fragmentation functions are not well constrained experimentally in the high $x$ tails.
Hopefully the situation improves soon by analyzing high $p_T$ pions at the Tevatron. Our study strongly
motivates a refitting of the fragmentation functions which eventually leads in turn to a good understanding
of the two photon rate at the LHC which is, apart from Higgs search, an interesting
observable concerning our understanding of QCD.    

\vskip1cm
\noindent

\section*{Acknowledgements}

We would like to thank Sasha Nikitenko for many useful discussions and 
also the conference organizers for the pleasant workshop.



 \section[QCD-induced spin phenomena in top quark
pair production at the LHC]{QCD-INDUCED SPIN PHENOMENA IN TOP QUARK
PAIR PRODUCTION AT THE LHC~\protect\footnote{Contributed
  by: {W.~Bernreuther, A.~Brandenburg, Z.G.~Si, P.~Uwer}}}
\label{sec:qcd-induced-spin}
 \newcommand{\one}{1\!\mbox{l}}

\subsection{Introduction}
The NLO QCD corrections to hadronic $t \bar t$ production have been
known for quite some time
\cite{Nason:1988xz,Nason:1989zy,Beenakker:1989bq,Beenakker:1991ma}. 
These results 
 were refined by resummation of soft gluon \cite{Laenen:1994xr}
and threshold \cite{Bonciani:1998vc,Kidonakis:2001nj}  logarithms. 
These predictions were made for the production of 
$t \bar t$ pairs averaged over their spins. 
 Because of the extremely short top-quark 
lifetime 
the spin properties of top quarks are transferred
to its decay products without being diluted by hadronization. Thus
quantities that involve the $t$ and/or $\bar t$ spin are also
``good'' observables in the sense that they can be reliably calculated
perturbatively, in particular within perturbative QCD. 
It is expected that 
such observables will  be very
useful  in exploring the interactions that are involved in top
quark production and decay. Besides standard model (SM) studies
they allow, for instance, the search for non-SM interactions, in
particular CP violating 
interactions of top quarks
\cite{Bernreuther:1993df,Bernreuther:1994hq,Bernreuther:1998qv}, 
or to pin down the nature of heavy
resonances that strongly couple to top quarks
\cite{Bernreuther:1998gs,Bernreuther:1998qv}, if such objects exist.

Needless to say, searches for new interactions with top spin observables
will require, on the theoretical side,  rather precise SM precictions. While SM interactions
induce only small  polarizations of $t$ and $\bar t$ quarks 
in  hadronic $t\bar t$ production, the QCD-induced correlation
between the  $t$ and $\bar t$ spins is large. They can be studied 
 both at the Tevatron 
and at the LHC by means of double differential angular distributions of
$t \bar t$  decay products. In this contribution we present results 
\cite{Bernreuther:2000yn,Bernreuther:2001bx,Bernreuther:2001rq,bbsu03}
for these distributions at NLO QCD.

\subsection{Theoretical framework}
\label{theory}
We consider  hadronic $t\bar t$ production and subsequent decay
into the following channels: 
\begin{equation}
  h_1 \, h_2\to t\bar{t}+X\to
  \left\{\begin{array}{lcc}
      \ell^+ \, \ell '^{-} &+& X\\ 
      \ell^{\pm} \, j_{\bar{t}(t)} &+& X\\ 
      j_t \, j_{\bar{t}}&+&X
    \end{array} \right.
  \label{reac1}
\end{equation}
Here $h_{1,2}=p,\bar{p}$, $\ell,\ell' =e,\mu,\tau$, and 
$j_t \ (j_{\bar{t}})$ denotes a  
jet originating from nonleptonic $t$ ($\bar{t}$) decay. 
A complete next-to-leading order
QCD analysis of (\ref{reac1})
involves the parton reactions
\begin{equation}
gg, q{\bar q} \ {\buildrel
 t{\bar t}\over \longrightarrow} \  b {\bar b} + 4 f,
\label{eq:ttrec1}
\end{equation}
\begin{equation}
gg, q{\bar q} \  {\buildrel
 t{\bar t}\over \longrightarrow} \  b {\bar b} + 4f  + g,
\label{eq:ttrec2}
\end{equation}
\begin{equation}
g + q ({\bar q})\  {\buildrel
 t{\bar t}\over \longrightarrow}\   b {\bar b} + 4f  + q ({\bar q}) .
\label{eq:ttrec3}
\end{equation}
where $f=q,\ell,\nu_{\ell}$. 
In view of the fact that the total width of the $t$ quark is much
smaller than its mass,  $\Gamma_t/m_t\approx 1\%$, it is adequate to
use the leading pole approximation (LPA) in calculating the NLO
differential cross sections of the above 
parton reactions. 
Within the LPA, the radiative corrections can be classified into
factorizable and non-factorizable  contributions.
The factorizable
 corrections were computed in 
\cite{Bernreuther:2000yn,Bernreuther:2001bx,Bernreuther:2001rq,bbsu03}
in the narrow width approximation $\Gamma_t/m_t\to 0$
for which the contributions to the
squared matrix element ${\cal M}$ of the respective
parton reaction are 
schematically of the form
\begin{equation}
\vert{\cal M}{\vert}^2 \propto {\rm Tr}\;[\rho
R{\bar{\rho}}]
 = \rho_{\alpha'\alpha}
R_{\alpha\alpha',\beta\beta'}{\bar{\rho}}_{\beta'\beta} .
\label{eq:trace}
\end{equation}
Here $R$ denotes the  density matrix which describes
 the production of on-shell
$t\bar t$ pairs in a specific spin configuration, and 
$\rho,{\bar{\rho}}$ are the density matrices describing the decay
of polarised $t$ and $\bar t$ quarks, respectively, into specific final states.
The subscripts in  (\ref{eq:trace}) denote the  $t$, $\bar t$ spin
indices. The spin-averaged production density matrices yield the 
NLO cross sections for $t\bar t$ being produced by $q\bar q$, $g g,$
$gq$, and $g \bar q$ fusion, which were  first determined by
\cite{Nason:1988xz,Nason:1989zy,Beenakker:1989bq,Beenakker:1991ma}. 
At NLO the  $q\bar q$ and  $g g$ fusion reactions
receive also non-factorizable corrections. They  were
computed by \cite{Beenakker:1999ya} in the semi-soft gluon 
approximation. While these 
contributions have an impact on, e.g.,  the $t,$ $\bar t,$ and $t\bar
t$ invariant mass distributions, they cancel in the integrated NLO
cross  sections
of the above reactions \cite{Fadin:1994dz,Melnikov:1994np}. Moreover, it can 
be shown \cite{bbsu03}  that these non-factorizable corrections
do not contribute either to the double differential distributions
which will be discussed below. Therefore we will not discuss them here
any further.

The factorizable contributions (\ref{eq:trace}) must be consistently
evaluated to order $\alpha_s^3$. This involves 
also the matrix elements
to order $\alpha_s$ of the main SM decay modes of the (anti)top 
quark in a given spin state, that is,  the semileptonic modes
$t \to  b \ell^+ \nu_{\ell},$ $b \ell^+ \nu_{\ell} g$  $ (\ell=e,\mu,\tau)$,
and the non-leptonic decays $t \to  b q {\bar q}',$ $
 b q {\bar q}' g$ where  $q {\bar q}'= u {\bar d}, c {\bar s}$ for the
dominant channels. For the computation of the double angular
distributions (\ref{dist}) the matrix elements of the 2-particle
inclusive parton reactions
$i \ {\buildrel
 t{\bar t}\over \longrightarrow} \ a_1 + a_2 + X $ are required. 
Here $a_1,a_2$ denotes a lepton or a jet. In the LPA this involves the
1-particle inclusive $t$ decay density matrix $2\rho^{t\to a_1}_{\alpha'\alpha}
= {\Gamma^{(1)}}(\one +{{\kappa}_1}\,{\bf{\tau}} \cdot
{\hat{\bf{q}}_1})_{\alpha'\alpha},$ where ${\hat{\bf{q}}_1}$ is the
direction of flight in the $t$ rest frame and $\Gamma^{(1)}$ is the
  partial width of the respective decay channel. An analogous  formula
  holds for $\bar t$ decay. 
The factor $\kappa_1$  is the $t$ spin analyzing power
of particle/jet $a_1.$  
It is clear that its value  is crucial for the 
experimental determination of top spin effects, in particular
of $t\bar{t}$ spin correlations. For the standard $V-A$ charged
current interactions these coefficients are known to order $\alpha_s$
for semileptonic  \cite{Czarnecki:1991pe} and non-leptonic 
\cite{Brandenburg:2002xr} modes. 
The charged lepton is  a perfect analyzer 
of the top quark spin, which is due to the fact that 
\begin{equation} 
\kappa_{\ell}=1-0.015\alpha_s \, .
\end{equation}
In the case of  hadronic top quark 
decays the spin analysing power of jets can be defined.
A detailed analysis was made
in  \cite{Brandenburg:2002xr}; we give here only two examples:
\begin{eqnarray} \label{kappa}
\kappa_b=-0.408\times(1-0.340\alpha_s)=-0.393 \, ,\\
\kappa_j=+0.510\times(1-0.654\alpha_s)=+0.474 \, ,
\end{eqnarray}   
where $\kappa_b$ is the analysing power of the $b$ jet
and  $\kappa_j$ refers to  the least energetic non-b-quark
jet defined by  the Durham algorithm. 
Using hadronic final states to analyze the spins of $t$
and/or $\bar t$ results in a loss  of analyzing power.
However, this is 
(over)compensated by the gain
in statistics and by the  efficiency with which the $t$ (${\bar t}$)
rest frames can be reconstructed. 

Let us now discuss double angular distributions 
$\sigma^{-1}d\sigma/{d\cos\theta_1d\cos\theta_2}$ for the 
channels  (\ref{reac1}) which are appropriate observables to
measure the (QCD induced) $t \bar t$ spin correlations.
Here $\theta_{1} (\theta_{2}) $ is the angle between the 
direction of flight of particle/jet $a_1 (a_2)$
in the $t(\bar{t})$ rest frame  with respect to
reference directions  $\hat{\bf a}$ ($\hat{\bf b}$) which will be specified below. 
We define the $t$ $(\bar t)$ rest frame by a rotation-free
boost from the $t\bar t$ zero-momentum frame, for reasons given below.
As mentioned above, non-factorizable corrections do not contribute, at
NLO, to
this type of distributions \cite{bbsu03}. Thus, to this order in the 
QCD coupling $\hat{\bf a}$ ($\hat{\bf b}$) can be interpreted as 
$t (\bar t)$ spin quantization axes, see below.  Integrating over the full phase
space and choosing $\hat{\bf a}$, $\hat{\bf b}$ as given  below
it can be shown \cite{bbsu03} that these distributions have the
generic form\footnote{QCD (or SM) generated  absorptive parts in the
parton scattering amplitudes induce a small $t$ and $\bar t$
polarization, which to order $\alpha_s^3$ is normal
to the $q{\bar q}, gg \to t{\bar t}$ scattering planes 
\cite{Bernreuther:1996cx,Dharmaratna:1996xd}.}
\begin{equation}\label{dist}
\frac{1}{\sigma}\frac{d\sigma}{d\cos\theta_1d\cos\theta_2}
=\frac{1}{4}(1-C\cos\theta_1\cos\theta_2).
\end{equation}
 The coefficient $C$
which signals the correlation between the $t\bar t$ spins depends,
for a given reaction, on the choice
of $\hat{\bf a}$ and $\hat{\bf b}.$  For the factorizable corrections the
exact (to all orders
in $\alpha_s$) formula $C = \kappa_1\kappa_2 D$ holds \cite{Bernreuther:2001rq}.
Here $D$  is the $t\bar{t}$ double spin asymmetry
\begin{equation}
\label{d}
D=\frac{N(\uparrow\uparrow)+N(\downarrow\downarrow)-N(\uparrow\downarrow)
-N(\downarrow\uparrow)}{N(\uparrow\uparrow)
+N(\downarrow\downarrow)+N(\uparrow\downarrow)+N(\downarrow\uparrow)},
\end{equation}
where $N(\uparrow\uparrow)$ denotes the number of $t\bar{t}$ pairs
with $t$ ($\bar{t}$) spin parallel to the reference axis $\hat{\bf a}$ 
($\hat{\bf b}$),  etc.
Thus $\hat{\bf a}$ and $\hat{\bf b}$  
can be identified with the
quantization axes of the $t$ and $\bar{t}$ spins, respectively, 
and $D$ directly 
reflects the strength of the correlation between the 
$t$ and $\bar{t}$ spins for the chosen axes. 

For $t\bar t$ production at the Tevatron it is well known that 
the so-called off-diagonal
basis \cite{Mahlon:1997uc}, which is defined by the requirement that 
$\hat{\sigma}(\uparrow\downarrow)$ $=\hat{\sigma}(\downarrow\uparrow)=0$
for the process $q\bar{q}\to t\bar{t}$ at tree level, yields 
a large coefficient $D$. It has been shown in \cite{Bernreuther:2001rq}
that the beam basis, where $\hat{\bf a}$ and $\hat{\bf b}$ are
identified with the hadronic beam axis, is practically as good as the
off-diagonal basis.  A further choice is the helicity basis, which is  suitable for
the LHC. 

An important theoretical issue of top quark spin physics beyond leading-order
QCD is the construction of infrared and collinear safe 
observables at the parton level.  In the case at
hand it boils down to the question  in which frame the
reference directions $\hat{\bf a}$ and $\hat{\bf b}$ are to be
defined. It has been  shown  \cite{bbsu03} that,
apart from the $t$
and $\bar t$ rest frames,  the $t\bar t$ zero
momentum frame (ZMF) is the appropriate frame for defining
collinear safe spin-momentum observables. 
The off-diagonal, beam, and helicity bases which we refer to in the
next section {\it are defined  in
the $t \bar t$ ZMF.}  Details will be given in  \cite{bbsu03}.

\subsection{Spin correlations at nlo: predictions for the Tevatron and the LHC}
\label{results}
We have computed the 2-particle  inclusive differential cross sections
for (\ref{reac1}) and, in particular, the double angular distributions (\ref{dist})
to NLO QCD, with  $\alpha_s$ and the top mass 
being defined  in the $\overline{{\mbox{MS}}}$ and in the on-shell scheme,
respectively. The 
mass factorization was performed in the $\overline{{\mbox{MS}}}$
scheme.) 

In Table~\ref{tab:Tevatron} we list our predictions \cite{bbsu03} 
for the spin correlation coefficient $C$ in the double 
differential distribution (\ref{dist}) at the Tevatron
for the three  different choices of references  axes $\hat{\bf a},$
$\hat{\bf b}$  discussed above. We use the CTEQ6L (LO) 
and CTEQ6.1M (NLO) parton distribution functions
(PDF) \cite{Stump:2003yu}.
Numbers are given for the dilepton, lepton$+$jet and all-hadronic
decay mode of the $t\bar{t}$ pair, where in the latter two cases 
the least energetic non-$b$-quark jet (defined by the Durham cluster
algorithm) was used as spin analyser. 
One sees that the spin correlations are largest in the
beam and off-diagonal basis. The QCD corrections reduce the LO results
for the coefficients $C$ by about 10\% to 30\%.

\begin{table}[htbp]\begin{center}

\begin{tabular}{|ccccc|}  
\hline
          &    & dilepton    &lepton+jet  & jet+jet  \\ \hline
$C_{\rm hel}$ &LO  & $-$0.471     & $-$0.240  & $-$0.123 \\
          &NLO & $-$0.352     & $-$0.168  & $-$0.080 \\ \hline
$C_{\rm beam}$ &LO &  $\phantom{-}$0.928    &  $\phantom{-}$0.474  
&  $\phantom{-}$0.242 \\ 
           &NLO&  $\phantom{-}$0.777    &  $\phantom{-}$0.370  &  
$\phantom{-}$0.176\\ \hline
$C_{\rm off}$   &LO& $\phantom{-}$0.937    &  $\phantom{-}$0.478 &  
$\phantom{-}$0.244 \\ 
           &NLO& $\phantom{-}$0.782   &  $\phantom{-}$0.372 &  
$\phantom{-}$0.177\\ 
\hline
\end{tabular}\end{center}
\caption{LO and NLO results for the spin correlation coefficient $C$ of 
the distribution (\ref{dist}) in the case of  
$p\bar{p}$ collisions at $\sqrt{s}=1.96$ TeV for different $t\bar{t}$
decay modes. The PDF CTEQ6L (LO) and CTEQ6.1M (NLO) of \cite{Stump:2003yu} 
were used, and $\mu=m_t=175$ GeV.}\label{tab:Tevatron}    
\end{table}

For the LHC it turns out that the spin correlations w.r.t. the 
beam and off-diagonal basis are quite small due to a cancellation 
of contributions from the $gg$ and $q\bar{q}$ initial states. 
Here, the helicity basis is
a good choice, and Table~\ref{tab:LHC} shows  our results for the $C$ coefficient
in that case. The QCD corrections are smaller  for the LHC
than for  the Tevatron; they  vary between 1 and 10$\%$.
For  both colliders the relative corrections 
$|(C_{\rm NLO}-C_{\rm LO})/C_{\rm LO}|$ 
are largest for the all-hadronic decay modes and smallest for the
dilepton decay modes.

\begin{table}[htbp]\begin{center}
\begin{tabular}{|ccccc|}   
\hline
   &    & dilepton    &lepton+jet  & jet+jet  \\ \hline
$C_{\rm hel}$ &LO  & 0.319    &  0.163 &  0.083 \\
          &NLO & 0.326    &  0.158 &  0.076 \\ 
\hline
\end{tabular}
\end{center}
\caption{Results for  $C_{\rm hel}$ 
for $pp$ collisions at $\sqrt{s}=14$ TeV using the same PDF and
parameters as in \protect{Table~\ref{tab:Tevatron}}. }\label{tab:LHC}

\end{table}
An interesting aspect of these double distributions is their high 
sensitivity to the quark and gluon
content of the proton \cite{Bernreuther:2001rq}; the reason being that
the contributions  to $C$ from $q\bar{q}$ and $gg$ initial
states have  opposite signs.
Table~\ref{tab:PDF} shows, for dilepton final states,  the dependence of the NLO results 
on the choice of the PDF.
While the results for the recent CTEQ6.1M and MRST2003 \cite{Martin:2003tt} PDF agree
at the percent level (this is not the case for previous versions
of the CTEQ and MRST PDF), the GRV98 \cite{Gluck:1998xa} PDF 
give significantly different results
at the Tevatron. This shows that 
measurements  of (\ref{dist})
may offer the possibility
to further constrain the quark and gluon content of the proton. 
%

\begin{table}[htbp]\begin{center} 

\begin{tabular}{|cccc|}  \hline
\multicolumn{4}{|c|}{Tevatron} \\ \hline
              &  CTEQ6.1M  &MRST2003      &GRV98\\ \hline
$C_{\rm hel}$
          & $-$0.352    &  $-$0.351    & $-$0.313  \\  \hline
$C_{\rm beam}$
           &  $\phantom{-}$0.777    &  $\phantom{-}$0.777 
& $\phantom{-}$0.732   \\ \hline
$C_{\rm off}$ &  
           $\phantom{-}$0.782    &  $\phantom{-}$0.782 & 
$\phantom{-}$0.736  \\ \hline \hline
 \multicolumn{4}{|c|}{LHC} \\ \hline
$C_{\rm hel}$
          & 0.326    &  0.327  & 0.339  \\ \hline
\end{tabular}
\end{center}
\caption{Spin correlation coefficients at NLO for different PDF for
 $p\bar{p}$ at $\sqrt{s}=1.96$ TeV (upper part) and 
 $pp$ at $\sqrt{s}=14$ TeV (lower part) for dilepton final states.}\label{tab:PDF}

\end{table}

All the 
 results above were obtained with  $\mu\equiv \mu_R=\mu_F=m_t=175$ GeV. A variation of
the scale $\mu$ between $m_t/2$ and $2m_t$  changes the central
results for $C$ at $\mu=m_t$ by roughly $\pm 5\%$.
Varying $m_t$ from $170$ to $180$ GeV changes the 
results for $C$ at the Tevatron by less than $5\%$, 
while for the LHC,  $C_{\rm hel}$ changes by less than a percent.

\subsection{Conclusion}
\label{concl}
Top-antitop spin correlations, which are predicted to be large
within the SM,  are expected to become 
a good tool for  analyzing in detail top quark
pair production and decay dynamics. They  can
be studied at the Tevatron and -- in view of the expected large $t\bar
t$ data samples -- especially at the LHC in the dilepton, single lepton and all-hadronic 
decay channels by measuring suitably defined 
double angular distributions.  
While the  NLO QCD corrections to these distributions are
of the order of 10 to 30$\%$ for the Tevatron, they are below 10$\%$
for the LHC. Work on soft gluon and threshold resummations will further
reduce the theoretical uncertainties.




\section[QCD Radiative Corrections to Prompt Diphoton Production in 
Association with a Jet at the LHC]{QCD RADIATIVE CORRECTIONS TO PROMPT DIPHOTON PRODUCTION IN 
ASSOCIATION WITH A JET AT THE LHC~\protect\footnote{Contributed
  by: {V.~Del Duca, F.~Maltoni, Z.~Nagy, Z.~Tr\'ocs\'anyi}}}
\label{sec:qcd-radi-corr}

\newcommand{\beq}{\begin{equation}}
\newcommand{\eeq}{\end{equation}}
\newcommand{\bea}{\begin{eqnarray}}
\newcommand{\eea}{\end{eqnarray}}
\newcommand{\nn}{\nonumber}
\newcommand\sss{\scriptscriptstyle}
\newcommand\asz{\alpha_{\sss S}}
\newcommand\mh{m_{\sss H}}
\newcommand\mgg{m_{\sss \gamma\gamma}}
\newcommand\mug{\mu_\gamma}
\newcommand\et{E_\perp}
\newcommand\ptz{p_\perp}
\newcommand\ptg{p_{\gamma\perp}}
\newcommand\ptgg{p_{\gamma\gamma\perp}}
\newcommand\ptjet{p_{{\rm jet}\perp}}
\newcommand\jet{+ {\rm jet}}
\newcommand\rd{ {\rm d}}

\newcommand\aem{\alpha_{\rm em}}
\newcommand\eps{\epsilon}
\newcommand\ord[1]{\rm O(#1)}

\def\eqn#1{Eq.~(\ref{#1})}
\def\eqns#1#2{Eqs.~(\ref{#1}) and~(\ref{#2})}
\def\eqnss#1#2{Eqs.~(\ref{#1})-(\ref{#2})}
\def\fig#1{Fig.~{\ref{#1}}}
\def\sec#1{Sect.~{\ref{#1}}}
\def\app#1{Appendix~\ref{#1}}
\def\tab#1{Table~\ref{#1}}
\def\bom#1{{\mbox{\boldmath $#1$}}}



\subsection{Introduction}
\label{sec:introduction}

Higgs production in association with a jet of high transverse energy
with a subsequent decay into two isolated photons, $pp\to H\jet \to
\gamma\gamma\jet$, is considered a very promising discovery channel for
a Higgs boson of intermediate mass (100\,GeV $\leq \mh \leq$
140\,GeV)~\cite{Dubinin:1997rq,Abdullin:1998er,DelDuca:2003uz}. The main
background to this signal is the $pp\to \gamma \gamma \jet$ channel,
where the photons are isolated.  In this channel the signal is a small
and narrow peak on a flat background ~\cite{DelDuca:2003uz}. Thus the QCD
prediction is not needed for predicting the background, which can be
well measured from the sidebands, but rather to optimize the selection
and isolation cuts for the experimental search.  For this purpose
usually a Monte Carlo program is used~\cite{Corcella:2002jc}, which however,
does not take into account the QCD radiative corrections. These
corrections are large and strongly dependent on the photon isolation
parameters~\cite{DelDuca:2003uz} and therefore cannot be ignored in the
analysis.

In perturbative QCD, the cross section for the production of a single
isolated photon in a collision of two hadrons $A$ and $B$ of momenta
$p_A$ and $p_B$, respectively, has the following general form:
\bea
&&
\!\!\!\!
\rd\sigma_{AB}(p_A,p_B;p_\gamma)=
\sum_{a,b}\int\!\rd x_a\,\rd x_b\,f_{a/A}(x_a,\mu_F) f_{b/B}(x_b,\mu_F) 
\,\rd\hat{\sigma}_{ab,\gamma}^{\rm isol}
 (x_a p_A,x_b p_B;p_\gamma;\mu_R,\mu_F,\mug)
\nn \\&&
+\sum_{a,b,c}
\int\!\rd x_a\,\rd x_b\,\rd z\,f_{a/A}(x_a,\mu_F) f_{b/B}(x_b,\mu_F)
\,\rd\hat{\sigma}_{ab,c}^{\rm isol}
\left(x_a p_A,x_b p_B;\frac{p_\gamma}{z};\mu_R,\mu_F,\mug\right) 
D_{\gamma/c} (z,\mug)\,.
\nn \\&&
\label{eq:fact}
\eea
The first term is called the direct component, where the subtracted
partonic cross sections \mbox{$d\hat{\sigma}_{ab,\gamma}^{\rm isol}$}
get contributions from all the diagrams with a photon leg. The second
term is called fragmentation component, where the subtracted partonic
cross sections \mbox{$d\hat{\sigma}_{ab,c}^{\rm isol}$} get contributions
from diagrams with only coloured external partons, with one of the partons
eventually fragmenting into a photon, in a way described by the 
(perturbatively uncalculable but universal) parton-to-photon
fragmentation function $D_{\gamma/c}$. If there are two isolated photons,
the equation contains four terms, a double direct, two
single-fragmentation and a double-fragmentation component~\cite{Binoth:1999qq}.
The direct and fragmentation components are not defined unambiguously
(except for the case of the `smooth' photon isolation to be used here),
finite terms can be shifted among the terms, only the sum is meaningful
physically.  The precise definition of the direct and fragmentation
terms, valid to all orders in perturbation theory, can be found for
instance in Ref.~\cite{Catani:2002ny}.

In perturbation theory beyond leading order, the isolated photon cross
section is not infrared safe. To define an infrared safe cross section,
one has to allow for some hadronic energy inside the photon isolation
cone. In a parton level calculation it means that soft partons up to a
predefined maximum energy are allowed inside the cone. This is also
natural in the experiment: complete isolation of the photon is not
possible due to the finite energy resolution of the detector.

There are two known ways to implement the photon isolation. The standard
way of defining an isolated prompt photon cross section, that matches
the usual experimental definition, is to allow for transverse hadronic
energy inside the photon isolation cone up to $E_{\perp,{\rm max}} =
\varepsilon p_{\gamma\perp}$, with typical values of $\varepsilon$
between 0.1 and 0.5\footnote{In experiments, often a fixed value of
$E_{\perp,{\rm max}}$ in the order of several GeV is used.}, and where
$p_{\gamma\perp}$ is taken either to be the photon transverse momentum
on an event-by-event basis or to correspond to the minimum value in the
$p_{\gamma\perp}$ range. Using this isolation prescription, both the
direct and the fragmentation terms contribute to \eqn{eq:fact}. In 
Ref.~\cite{Catani:2002ny} it was shown that a small isolation
cone for the photon leads to unphysical results in a fixed order
computation.  For a small cone radius $R_\gamma$, an all-order resummation of
$\asz \ln(1/R_\gamma^2)$ terms combined with a careful study of the
border line between perturbative and non-perturbative regions has to be
undertaken.

Even with resummation, the very narrow isolation cone is not favoured
from the practical point of view.  The ratio of the fragmentation
component to the direct one slowly increases with decreasing
$R_\gamma$. Since the fragmentation function is non-perturbative,
therefore, it has to be measured and the available functions have
relatively large errors. The fragmentation component is also strongly
dependent on $\varepsilon$, increasing rapidly with increasing
$\varepsilon$. Thus, the theoretical uncertainty related to the photon
fragmentation can be decreased with large cone sizes and small $\varepsilon$.

In order to avoid completely the uncertainties related to the
fragmentation component, Frixione introduced a `smooth' photon-isolation
when the fragmentation contribution is zero~\cite{Frixione:1998jh}.
This isolation means that the energy of the soft parton inside the
isolation cone has to converge to zero smoothly if the distance in the
$\eta-\phi$ plane between the photon and parton vanishes. Explicitly,
the amount of hadronic transverse energy  $E_\perp$ (which in our NLO
partonic computation is equal to the transverse momentum of the
possible single parton in the isolation cone) in all cones of radius $r
< R_\gamma$ must be less than 
\beq
E_{\perp,{\rm max}} = \varepsilon p_{\gamma\perp}
\left(\frac{1 - \cos r}{1 - \cos R_\gamma}\right)^n\:.
\label{eqn:frixione}
\eeq
The smooth isolation prescription can be viewed as if the singularity
in the quark-photon splitting were factorized into the fragmentation
component using phase-space cuts such that the finite remainder in the
fragmentation is shifted completely into the direct component in a
perturbatively computable way. 

The two prescriptions have their own advantages and less appealing
features. The standard isolation is easier to employ experimentally,
but has the following disadvantages from the theoretical point of
view:
(i)
it involves a non-perturbative component in its theoretical prediction
(fragmentation) with large errors due to pure experimental information;
(ii)
the theoretical prediction of the fragmentation is more difficult to
compute than that of the direct component.
The smooth isolation is easier to implement in NLO perturbation theory.
However, it is less favoured in the experiment because
(i)
it is difficult to employ it due to the finite granularity of the
detector;
(ii)
it was found to be less efficient for isolating photons from jets than
the cone isolation, approaching the efficiency of the latter with $n$
tending to zero \cite{wielers}.

Experimentally the standard cone isolation is recovered from the smooth
isolation when $n\to 0$. The smooth isolation leads to smaller cross
sections for the same values of the cone radius and $\eps$, because it
means a more severe cut into the phase space.  However, due to the
finite granularity of the detector, the two prescriptions yield equal
cross sections even for small, but non-zero values of $n$.  Small
values of $n$ can be chosen in perturbation theory, but $n=0$ cannot be
taken without including the fragmentation component.  If we are
interested only in the Higgs-boson search, we need a perturbative
prediction that is reliable for those values of the isolation
parameters which are found useful to increase the signal significance.
Studying the dependence of the perturbative prediction with the smooth
prescription on the isolation parameters, we can estimate the reliability
of the theoretical predictions.

In order to assess the dependence of the radiative corrections on the
isolation parameters, we use a partonic Monte Carlo program which
employs the dipole subtraction method~\cite{Catani:1997vz}, slightly
modified for better numerical control~\cite{Nagy:1998bb} as implemented
in the {\tt NLOJET++} package~\cite{Nagy:2001fj}, and use the smooth
isolation prescription at the parton level. We compute cross section
values at leading order and at NLO for a Large Hadron Collider
(LHC) running at 14\,TeV. The values shown at leading order were
obtained using the leading order parton distribution functions
(p.d.f.'s) and those at NLO accuracy were obtained using the NLO
p.d.f.'s of the CTEQ6 package~\cite{Pumplin:2002vw} (tables cteq6l1 and
cteq6m, respectively).  We used the two-loop running of the strong
coupling at NLO with $\Lambda_{\rm QCD} = 226$\,MeV and one-loop
running with $\Lambda_{\rm QCD} = 165$\,MeV at leading order.  The
renormalization and factorization scales are set to $\mu_R = \mu_F =
x_\mu \mu_0$, where for the reference value $\mu_0$ we use
$\mu_0^2 = (\mgg^2 + \ptjet^2)/4$
with $\mgg$ the invariant mass of the photon pair. This definition
reduces to the usual scale choice for inclusive photon pair production
if $\ptjet$ vanishes. Our prediction for the $\gamma\gamma\jet$
production cross section is intended for use in the detection of a
Higgs boson lighter than the top quark, therefore, we assume 5 massless
flavours and restrict all cross sections to the range of 80\,GeV $\le
\mgg \le$ 160\,GeV. The electromagnetic coupling is taken at the
Thomson limit, $\aem = 1/137$.  We use a jet reconstruction algorithm
and a set of event selection cuts, expected to be typical in Higgs
searches. In particular, in order to find the jet, we use the midpoint
cone algorithm~\cite{Blazey:2000qt} with a cone size of
$R = \sqrt{\Delta\eta^2 + \Delta\phi^2} = 1$, with $\Delta \eta$ the
rapidity interval and $\Delta \phi$ the azimuthal angle\footnote{In
our NLO computation the midpoint and seedless cone algorithms yield
identical cross sections.}.  Then, we require that both the jet and the
photons have $\ptz > 40\,$GeV and rapidity within $|\eta| < 2.5$. These
are the same selection cuts as used in Ref.~\cite{deFlorian:1999tp} for
computing the gluon initiated $\ord{\asz^3}$ corrections. Furthermore,
we isolate both photons from the partons in a cone of size $R_{j\gamma}$.

\subsection{Results}
\label{sec:results}

In \fig{fig:mgg}(a) we plot the invariant mass distribution of the
photon pair. Here we see the continuum background on which the Higgs
signal is expected to manifest itself as a narrow resonance in the
intermediate-mass range. The dotted (red) line is the leading order
prediction and the solid (red) one is the differential cross section at
NLO accuracy.  The striking feature of the plot is the rather large
correction.  The large corrections are partly due to the appearance of
new subprocesses at NLO as can also be read off the figure. The
gluon-gluon scattering subprocess begins to appear only at NLO accuracy,
and therefore it is effectively leading order. It is shown with a long
dashed-dotted (magenta) line: it yields a very small contribution. 
The bulk of the cross section comes from quark-gluon scattering both at
leading order and at NLO, shown with sparsely-dotted (blue) and
long-dashed (blue) lines. The quark-quark scattering receives rather
large corrections because the leading
order subprocess can only be a quark-antiquark annihilation process,
shown with short-dashed (green) line, while at NLO, shown with short
dashed-dotted (green) line, there are (anti-)quark-(anti-)quark
scattering subprocesses. Thus at NLO the parton luminosity is sizeably
larger. In addition, more dynamic processes are allowed, which include
$t$-channel gluon exchange. These contribute to enlarge the cross section
in phase space regions which are disfavoured at leading order.
\begin{figure}
\begin{center}
\includegraphics[width=.47\linewidth]{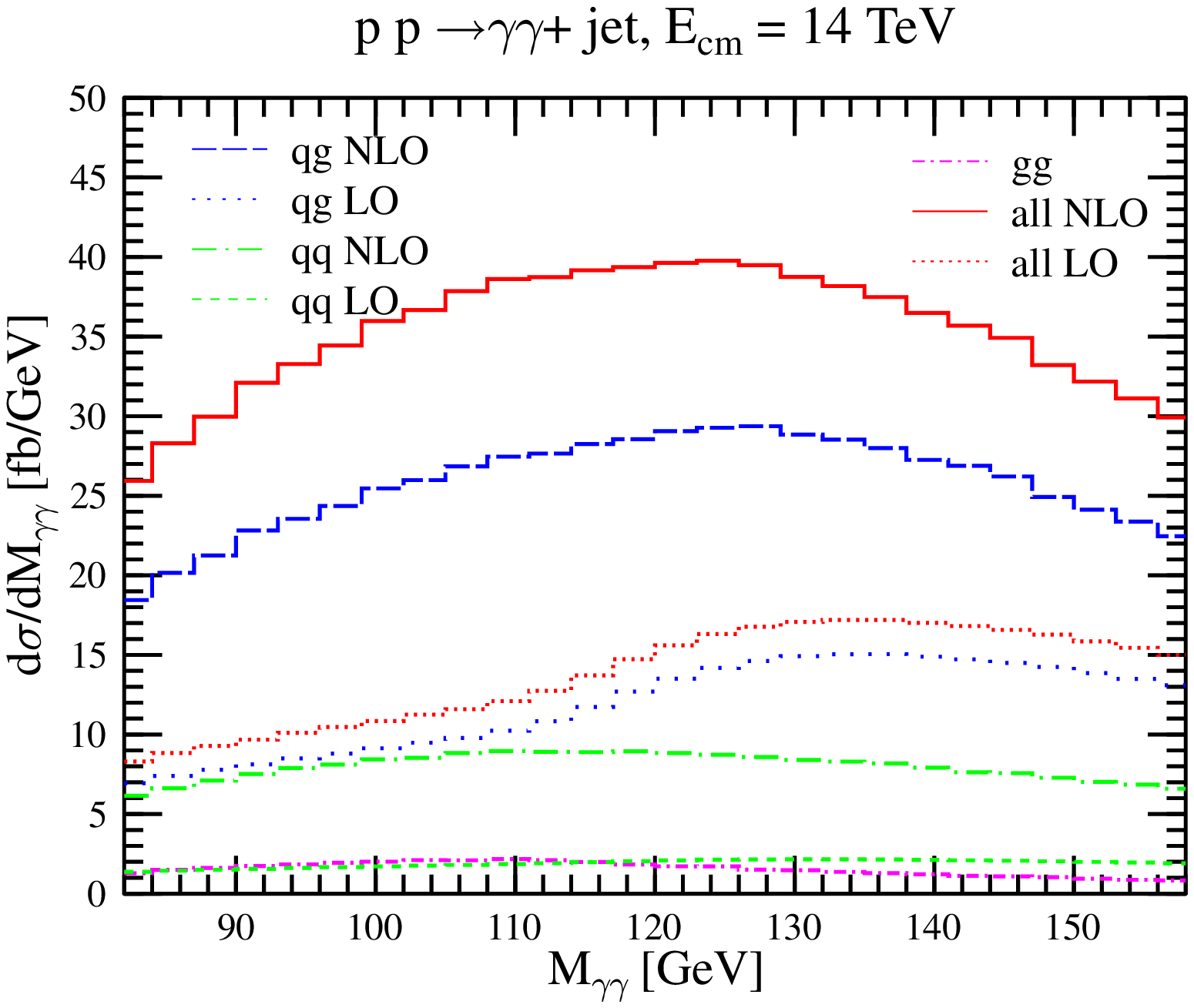}
\hfill
\includegraphics[width=.47\linewidth]{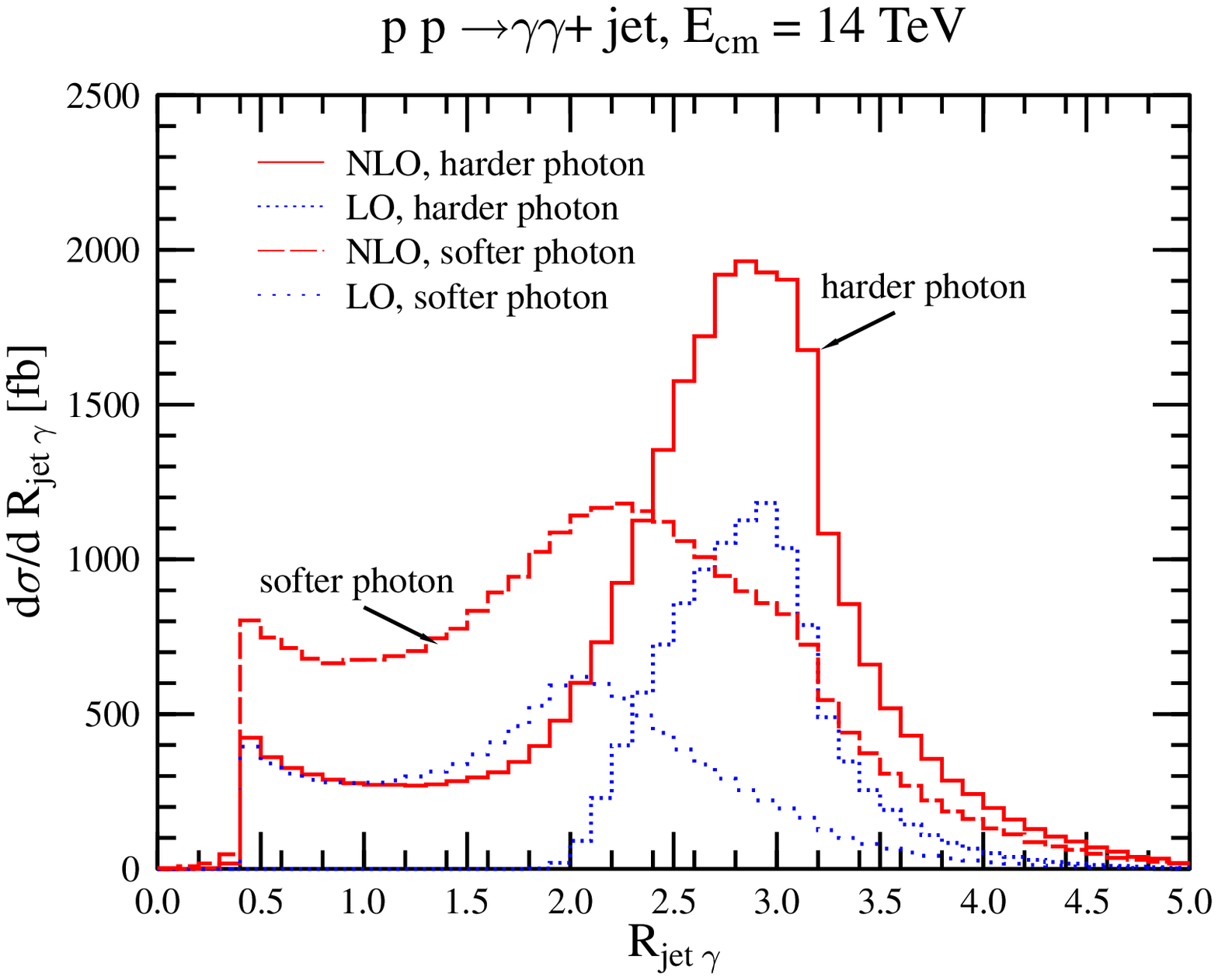}
\caption{
(a) The invariant mass distribution of the photon pair at LHC energy for
smooth isolation with $R_\gamma = 0.4$ and $\eps = 0.5$.
(b) The distributions in the distances between the jet and the photons in
the $\eta$-$\phi$ plane. 
The photons and the jet are required to have transverse momentum
$|p_\perp| \ge 40$~GeV and lie in the central rapidity region of
$|\eta| \le 2.5$. The jet is reconstructed according to the
midpoint algorithm.
\label{fig:mgg} }
\end{center}
\end{figure}

A part of the large radiative corrections is accounted for by the new
subprocesses; another part is due simply to the enlarged phase space,
as can be seen from of \fig{fig:mgg}(b), where the differential
distributions in the distance
$R_{j\gamma} = \sqrt{|\eta_j-\eta_\gamma|^2 + |\phi_j-\phi_\gamma|^2}$
between the jet and the photons in the $\eta$-$\phi$ plane are shown,
with a selection cut at $R_{j\gamma} \ge 0.4$. From the distribution
for the harder photon we see that the radiative corrections recieve
contributions from a larger part of the phase space than the prediction
at leading order.  In fact, on one hand a cut on
$R_{j\gamma_s} > 1.5$ affects the leading order and the NLO evaluation
in the same way because they have the same shape. Thus in that case the
cut does not reduce the correction to the $\mgg$ distribution. On the
other hand, a cut on $R_{j\gamma_h} > 1.5$ cuts the NLO correction, but
does not cut the leading order, so the correction is reduced.
Nevertheless, the reduction is less then 10\,\%, thus cutting on
$R_{j\gamma}$ at $R_{j\gamma} \ge 1.5$ reduces the NLO correction to
the invariant mass distribution of the photon pair, but marginally.
In the rest of this study we require $R_{j\gamma}\ge 1.5$ and $\ptgg
\ge 40$\,GeV, in addition to the same selection cuts as used in
\fig{fig:mgg}.

Next we consider the dependence of the isolated cross section on the
power $n$ in the definition of the smooth isolation \eqn{eqn:frixione}.
We show the dependence on $n$ for $R_\gamma = 0.4$, $\eps = 0.5$
and $R_\gamma = 0.7$, $\eps = 0.1$ in \fig{fig:ndep}(a). We see that the
radiative corrections are much more than 100\,\% and depend on $n$
strongly if the isolation cone is narrow and $\eps$ is large, therefore,
the fixed-order perturbative prediction at the NLO accuracy is not
reliable in this case.  On the other hand, the dependence on $n$ is
much milder and remains well below 100\,\% for $R_\gamma = 0.7$,
$\eps = 0.1$. Thus, the $n=0.1$ line can be considered a good
approximation to the prediction with standard photon isolation.

In order to assess the stability of the predictions against scale
variations, we show the cross section in a 3\,GeV bin around $\mgg =
120$\,GeV, that is the background for a hypothetical Higgs signal for
a Higgs particle of mass 120\,GeV. \fig{fig:ndep}(b) shows the cross section
for two sets of photon-isolation parameters. We show the scale variations
for varying the renormalization and factorization scales separately,
keeping the other scale fixed, as well as varying them simultaneously.
The lower three curves represent the leading order predictions.  At
leading order the dependence on the renormalization and factorization
scales is rather small, especially when the two scales are set equal
(densely dotted
line). Observing the predictions we conclude that the scale dependence at
leading order does not represent the uncertainty of the predictions due
to the unknown higher orders. The inclusion of the radiative corrections 
results mainly in the substantial increase of the cross section.
For the isolation parameters $R_\gamma = 0.4$ and $\epsilon = 0.5$
the NLO corrections at $x_\mu = 1$ are about 120\,\% of the leading
order prediction.  In addition, the scale dependence is not reduced by
the inclusion of the radiative corrections. If we require more
stringent photon isolation cuts, then we find smaller corrections and a
more stable prediction.  For instance, in \fig{fig:ndep}(b) we also show
the scale dependence of the cross section obtained with $R_\gamma = 1$
and $\eps = 0.1$. We find that the cross section at NLO is
about 40\,\% larger than the leading order prediction, and in this case
the scale dependence at NLO is reduced as compared to the one at
leading order accuracy.  The reduction of the scale dependence when the
stronger cuts ($R_\gamma = 1$ with $\epsilon = 0.1$) are used indicates
that the large correction obtained with the looser (default) cuts are
due mainly to real emission of soft and collinear partons.
\begin{figure}
\begin{center}
\includegraphics[width=.47\linewidth]{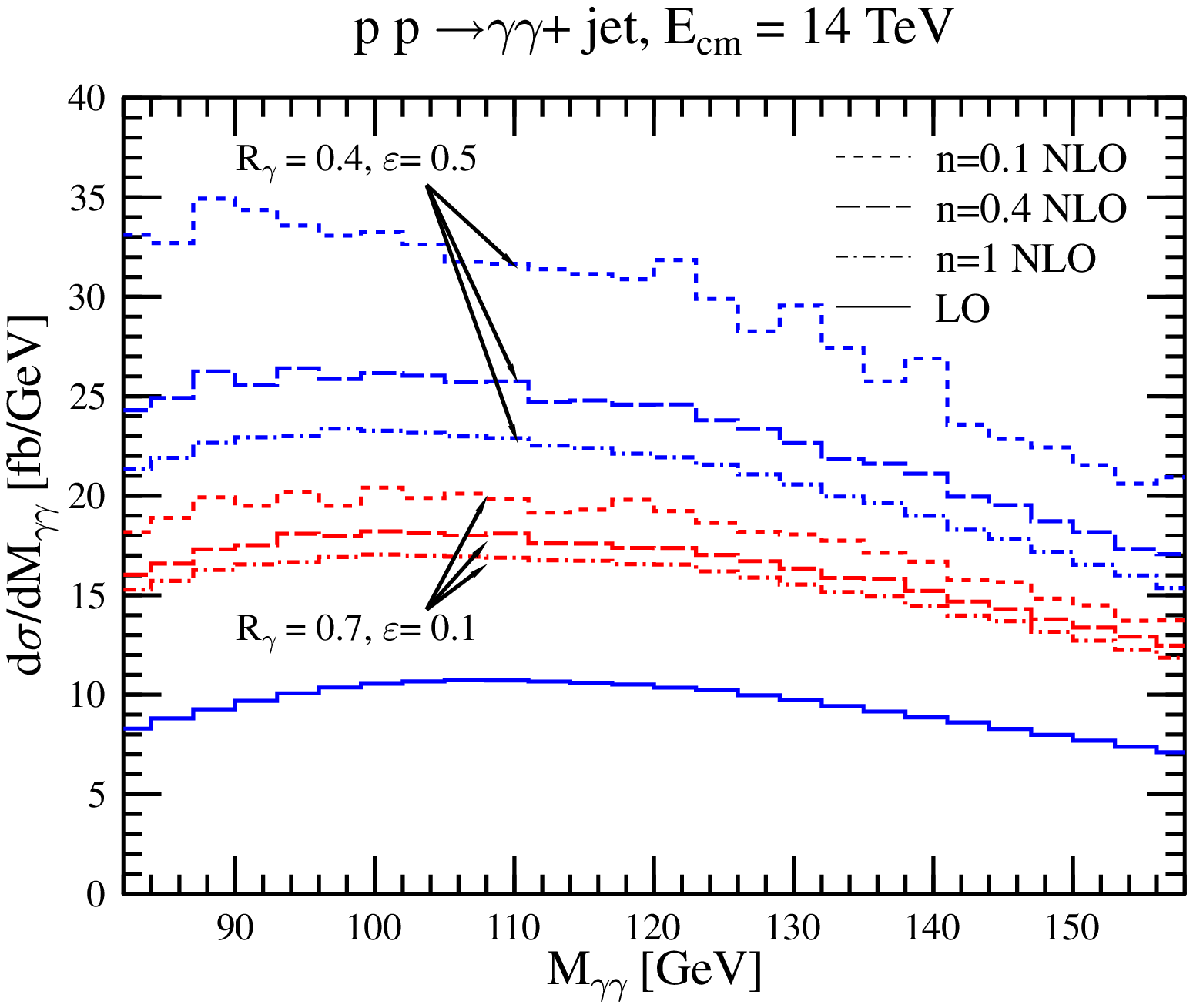}
\hfill
\includegraphics[width=.47\linewidth]{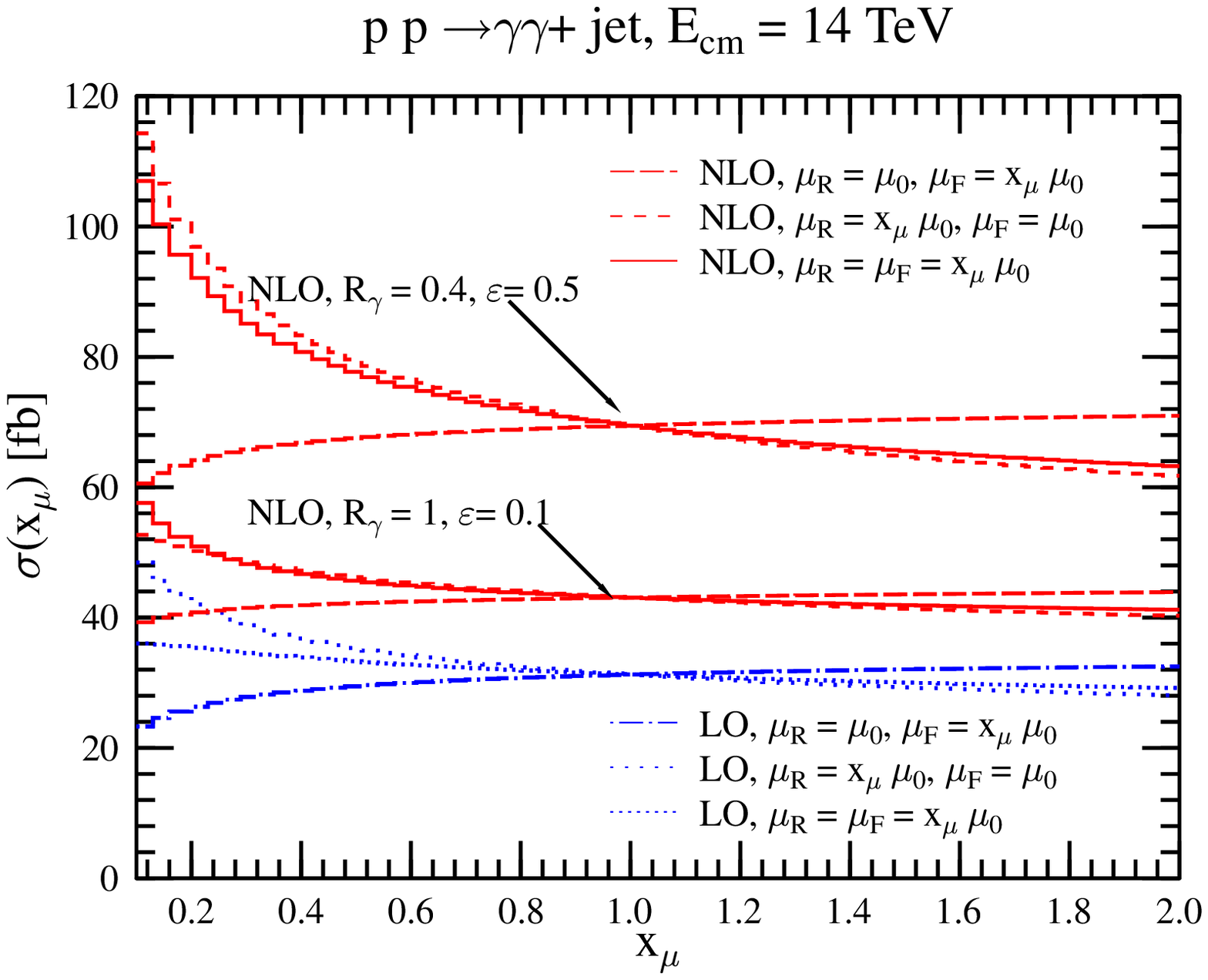}
\caption{
(a) Dependence of the invariant mass spectrum on the isolation parameter
$n$.
(b) Dependence on the renormalization and factorization scales of the cross
section in a bin of 118.5\,GeV $\le \mgg \le$ 121.5\,GeV at leading-order
and at NLO accuracy. 
\label{fig:ndep} }
\end{center}
\end{figure}

\subsection{Conclusions}
\label{sec:conclusions}

We conclude by summarizing our observations concerning the 
QCD radiative corrections to the $p p\to\gamma\gamma\jet$ process
that constitutes part of the irreducible background to the $p p\to
H\jet \to \gamma\gamma\jet$ discovery channel of an intermediate-mass
Higgs boson at the LHC.  We used a smooth photon isolation that is
infrared safe to all orders in perturbation theory and independent of
the photon fragmentation into hadrons.  We found large radiative
corrections, which are rather sensitive to the selection cuts and
photon isolation parameters. Choosing a mild photon isolation, i.e.~a
small isolation cone radius $R_\gamma = 0.4$ with relatively large
hadronic activity allowed in the cone results in more than 100\,\%
correction with as large residual scale dependence at NLO as at leading
order. In this case, the dependence on the isolation parameter $n$ is
also large. Since the radiative correction with standard photon isolation
would even be larger, we conclude that NLO perturbation theory does not
yield a reliable theoretical prediction for such a mild photon isolation.
Making the photon isolation more stringent, for instance increasing the
cone radius to $R_\gamma \ge 0.7$ and decreasing the hadronic activity
in the cone (using for instance $\eps = 0.1$) reduces both the
magnitude of the radiative corrections as well as the dependence on the
renormalization and factorization scales and that on the isolation
parameter $n$. The more stringent isolation is also useful for
decreasing the relative weight of the poorly known fragmentation
components in the cross section, and the prediction obtained with the
smooth isolation can be considered a good approximation (valid to the
extent of scale ambiguities) to those with the standard isolation. 
Our results show that a constant factor is not appropriate for taking
into account the radiative corrections to the irreducible background of
the $pp \to H\jet \to \gamma\gamma\jet$ discovery channel at the LHC.  

\subsection*{Acknowledgments}
This work was supported in part by US Department of Energy, contract
DE-FG0396ER40969 and by the Hungarian Scientific Research Fund grants
OTKA T-038240.

\section[Electroweak Radiative Corrections to Hadronic Precision Observables 
at TeV Energies]{ELECTROWEAK RADIATIVE CORRECTIONS TO HADRONIC PRECISION OBSERVABLES 
AT TEV ENERGIES~\protect\footnote{Contributed
  by: {E.~Maina, S.~Moretti, M.R.~Nolten, D.A.~Ross}}}
\label{SingleBosonLH}

\def\s#1{{\small#1}}
\def\lsim{\:\raisebox{-0.5ex}{$\stackrel{\textstyle<}{\sim}$}\:}
\def\gsim{\:\raisebox{-0.5ex}{$\stackrel{\textstyle>}{\sim}$}\:}
\def\PD{\s{PDG}}
\def\TA{{\small TAUOLA}}
\def\HW{\s{HERWIG}}
\def\JS{\s{JETSET}}
\def\PY{\s{PYTHIA}}
\def\IS{\s{ISAJET}}
\def\IW{\s{ISAWIG}}
\def\SM{\s{SM}}
\def\MSSM{{MSSM}}
\def\SY{\s{SUSY}}
\def\QCD{\s{QCD}}
\def\QED{\s{QED}}
\def\DIS{\s{DIS}}
\def\LEP{\s{LEP}}
\def\LHC{\s{LHC}}
\def\OPAL{\s{OPAL}}
\def\PDF{\s{PDFLIB}}
\def\CERN{\s{CERN}}
\def\RPV{\rlap{/}{R}$_{\mbox{\scriptsize p}}$}
\def\BNV{\rlap{/}{B}}
\def\Ord{\buildrel{\scriptscriptstyle <}\over{\scriptscriptstyle\sim}}
\def\OOrd{\buildrel{\scriptscriptstyle >}\over{\scriptscriptstyle\sim}}
\def\gh{\Gamma_{\scriptscriptstyle \rm H}}
\def\gtap{\raisebox{-.4ex}{\rlap{$\sim$}} \raisebox{.4ex}{$>$}}
\def\ltap{\raisebox{-.4ex}{\rlap{$\sim$}} \raisebox{.4ex}{$<$}}
\def\ycut{$y_{\mbox{\tiny cut}}$}
\def\mw{m_{\scriptscriptstyle \rm W}}
\def\mh{m_{\scriptscriptstyle \rm H}}
\def\lms{\Lambda_{\overline{\rm MS}}}
\def\half{\mbox{\small $\frac{1}{2}$}}
\def\thlf{\mbox{\small $\frac{3}{2}$}}
\def\as{\alpha_{\mbox{\tiny S}}}
\def\ee{e^+e^-}
\def\MC{Monte Carlo}
\def\VEV#1{\langle{#1}\rangle}
\def\qbar{\bar{q}}
\def\Qbar{\bar{Q}}
\def\dbar{\bar{d}}
\def\ubar{\bar{u}}
\def\sbar{\bar{s}}
\def\cbar{\bar{c}}
\def\bbar{\bar{b}}
\def\tbar{\bar{t}}
\def\pbar{\bar{p}}
\def\B0bar{\overline{B^0}}
\def\lbar{\bar{\l}}
\def\l{\ell}


\subsection{Weak corrections at TeV scales}

At TeV energy scales, next-to-leading order (NLO) 
Electro-Weak (EW) effects produce leading
corrections of the type $\alpha_{\rm{EW}}\log^2({\hat{s}}/M_W^2)$, where 
$\alpha_{\mathrm{EW}}\equiv \alpha_{\mathrm{EM}}\sin^2\theta_W$,
with $\alpha_{\mathrm{EM}}$ the Electro-Magnetic coupling
constant and $\theta_W$ the Weinberg angle. In fact,
for large enough $\hat s$ values, the centre-of-mass (CM)
energy at parton level, such EW effects may be competitive not
only with next-to-NLO (NNLO) (as $ \alpha_{\rm{EW}}\approx 
\alpha_{\rm{S}}^2$) but also with NLO QCD corrections (e.g., for
$\sqrt{\hat{s}}=3$ TeV, $\log^2({\hat{s}}/M_W^2)\approx10$).

These `double logs' are 
due to a lack of cancellation between virtual and real $W$-emission in
higher order contributions. This is in turn a consequence of the 
violation of the Bloch-Nordsieck theorem in non-Abelian theories
\cite{Ciafaloni:2000df}.
The problem is in principle present also in QCD. In practice, however, 
it has no observable consequences, because of the final averaging of the 
colour degrees of freedom of partons, forced by their confinement
into colourless hadrons. This does not occur in the EW case,
where the initial state has a non-Abelian charge,
as in an initial quark doublet in proton-(anti)proton scattering. 
Besides, these
logarithmic corrections are finite (unlike in
QCD), since $M_W$ provides a physical
cut-off for $W$-emission. Hence, for typical experimental
resolutions, softly and collinearly emitted weak bosons need not be included
in the production cross section and one can restrict oneself to the 
calculation
of weak effects originating from virtual corrections only. 
By doing so, similar
logarithmic effects, $\sim\alpha_{\rm{EW}}\log^2({\hat{s}}/M_Z^2)$, 
are generated also by $Z$-boson corrections.
Finally, all these purely weak contributions can  be
isolated in a gauge-invariant manner from EM effects which therefore may not
be included in the calculation. In fact, we have neglected the latter here. 

In view of all this,  it becomes of crucial importance to assess
the quantitative relevance of such weak corrections
affecting, in particular, key QCD processes studied at present and future 
hadron colliders. We show here results for the case of $b$-jet-,
prompt-photon and $Z$-production at Tevatron and LHC.

\subsection{Corrections to $\mathbf{b}$-jet-production}

In Fig.~\ref{fig:b} (left and right panels) we show the effects of the full  
${\cal O}(\alpha_{\rm S}^2\alpha_{\rm{EW}})$ contributions to
the $p\bar p\to b\bar b(g)$ and  $pp\to b\bar b(g)$ cross sections
at Tevatron and LHC, respectively. (For details of the calculation,
see Ref.~\cite{Maina:2003is}.) Results are shown for the total inclusive
$b$-jet production rate as a function of the jet transverse
energy. (Tree-level EW and NLO QCD effects are also given for comparison.)
At Tevatron, ${\cal O}(\alpha_{\rm S}^2\alpha_{\rm{EW}})$ terms
are typically negligible in the inclusive cross section, as the partonic
energy available is too small for the mentioned logarithms to be 
effective. At LHC, the contribution due to such terms grows
accordingly to the collider energy, reaching the --2\% level at 
transverse momenta of $\approx800$ GeV. 

Next, we study the forward-backward asymmetry at Tevatron, defined as
follows:
\begin{equation}\label{AFB}
A^b_{\rm{FB}}=
\frac{\sigma_+[p\bar p\to b\bar b(g)]-\sigma_-[p\bar p\to b\bar b(g)]}
     {\sigma_+[p\bar p\to b\bar b(g)]+\sigma_-[p\bar p\to b\bar b(g)]},
\end{equation}
where the subscript $+(-)$ iden\-ti\-fies events in which the $b$-jet
is produced with polar angle larger(smal\-ler) than 90 degrees respect to 
one of the two beam directions (hereafter, we use
the proton beam as positive $z$-axis).
The polar angle is defined in the CM
frame of the hard partonic scattering. In the center plot of 
Fig.~\ref{fig:b}, the solid curve
represents the sum of the tree-level contributions, that is, 
those of order $\alpha_{\rm{S}}^2$ and  
$\alpha_{\rm{EW}}^2$, whereas the dashed one also includes the
higher-order ones 
$\alpha_{\rm{S}}^2\alpha_{\rm{EW}}$. The effects of the one-loop weak corrections on 
this observable are rather large, indeed comparable to the effects
through order $\alpha_{\rm{S}}^3$ \cite{Kuhn:1998kw,Kuhn:1998jr}. 
In absolute terms, the asymmetry is of order $-4\%$ 
at the $W$, $Z$ resonance (i.e., for $p_T\approx M_W/2, M_Z/2$)
 and
fractions of percent elsewhere, hence it should be measurable
after the end of Run 2. We expect even larger effects at LHC, however,
some care is here necessary in order
to define the observable, which depends on
the configuration and efficiency of the experimental apparata 
(so we do not present the corresponding plot in this instance). 
The $\alpha_{\rm{S}}^3$ results presented here are 
from Ref.~\cite{Frixione:1997nh}.

\begin{figure}[htbpt]
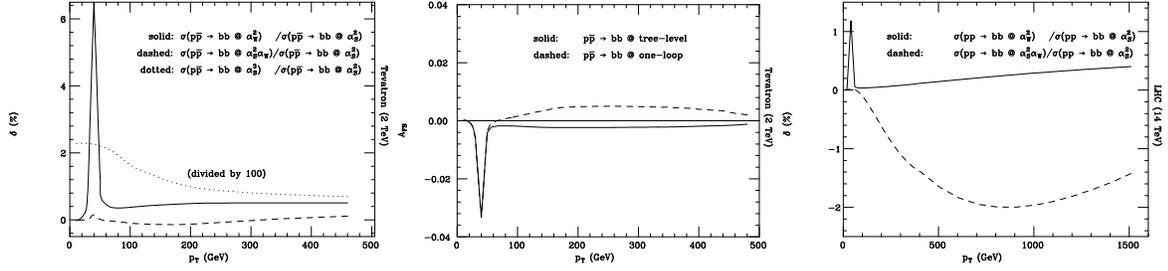

{\epsfig{file=moretti/ratiobb_Tev.ps,width=3.5cm,angle=90}}
{\epsfig{file=moretti/sigmabb_AFB_Tev.ps,width=3.5cm,angle=90}}
{\epsfig{file=moretti/ratiobb_LHC.ps,width=3.5cm,angle=90}}
\caption{The corrections (NLO-LO)/LO due to the 
$\alpha_{\rm{EW}}^2$, $\alpha_{\rm{S}}^2\alpha_{\rm{EW}}$ 
and $\alpha_{\rm{S}}^3$ terms
relative to the $\alpha_{\rm{S}}^2$ ones
vs. the transverse momentum
of the $b$-jet for $p\bar p\to b\bar b(g)$ and  $p p\to b\bar b(g)$ production at
Tevatron and LHC, left and right frame, respectively. 
(For LHC,
we do not show the corrections due to $\alpha_{\rm{S}}^3$ terms as results are
perturbatively unreliable.) In the middle frame, the forward-backward  
asymmetry vs. the transverse momentum
of the $b$-jet for $p\bar p\to b\bar b(g)$ events at
Tevatron, as obtained at tree-level ${\cal O}(\alpha_{\rm{EW}}^2)$ 
and one-loop ${\cal O}(\alpha_{\rm{S}}^2\alpha_{\rm{EW}})$ orders.}
\label{fig:b}
\vspace*{-0.5truecm}
\end{figure}

\subsection{Corrections to $\mathbf{\gamma}$- and $\mathbf{Z}$-production}

The neutral-current processes ($V=\gamma,Z$)
\begin{equation}\label{procs_neutral}
q\bar q \to g V\quad{\rm{and}}\quad q(\bar q) g\to q(\bar q) V
\end{equation}
are two of the cleanest probes of the partonic content of (anti)protons,
in particular of antiquark and gluon
densities. In order to measure the latter it is necessary to study
the vector boson $p_T$ spectrum. That is, to
compute $V$ production in association with a jet 
(originated by either a quark or a gluon).
We briefly report here on the full one-loop results for processes
(\ref{procs_neutral}) obtained through
${\cal O}(\alpha_{\rm{S}}\alpha_{\rm{EW}}^2)$. (For technical
details of the calculation, see Ref.~\cite{Maina:2002wz}.)

Fig.~\ref{fig:V} shows the effects of the 
${\cal O}(\alpha_{\rm{S}}\alpha_{\rm{EW}}^2)$
corrections relatively to the ${\cal O}(\alpha_{\rm{S}}\alpha_{\rm{EW}})$
Born results ($\alpha_{\rm{EM}}$ replaces $\alpha_{\rm{EW}}$ for photons),
as well as the absolute magnitude of the latter, as a function
of the transverse momentum. The corrections are found to be rather
large, both at Tevatron and LHC, particularly
for $Z$-production. In case of the latter,
such effects are of order --10\% at Tevatron 
and --15\% at LHC for $p_T\approx 500$ GeV. In general, above 
$p_T\approx100$ GeV,
they tend to (negatively) increase, more or less linearly, with $p_T$.
Such effects are indeed observable at both Tevatron and LHC. 
For example, at FNAL, for $Z$-production and decay into electrons and muons
with BR$(Z\rightarrow e,\mu)\approx 6.5\%$, assuming
$L= 2-20$ fb$^{-1}$ as integrated luminosity, in
a window of 10 GeV at $p_T = 100$ GeV, one finds
650--6500 $Z+j$ events through LO, hence a
$\delta\sigma/\sigma\approx -1.5\%$ EW NLO correction corresponds to 10--100 
fewer 
events. At CERN, for the same production and decay channel, assuming now 
$L= 30$ fb$^{-1}$, in a window of 40 GeV at $p_T = 450$ GeV, 
we expect about 1200 $Z+j$ events from LO, so that a 
$\delta\sigma/\sigma\approx -12\%$ EW NLO correction 
corresponds to 140 fewer events. In line with the normalisations seen
in the top frames of  Fig.~\ref{fig:V} and the size of the corrections
in the bottom ones, absolute rates for the photon are similar
to those for the massive gauge boson while ${\cal O}(\alpha_{\rm{S}}\alpha_{\rm{EW}}^2)$ corrections are about a factor of two 
smaller.

Altogether, these results point to the relevance of one-loop
${\cal O}(\alpha_{\rm{S}}\alpha_{\rm{EW}}^2)$ contributions for precision
analyses of prompt-photon and neutral Drell-Yan events at both Tevatron and
LHC, also
recalling that the residual scale dependence of the known
higher order QCD corrections
to processes of the type (\ref{procs_neutral}) is very small 
in comparison \cite{Arnold:1989ub,Arnold:1989dp,Giele:1993dj,Campbell:2003hd}. 
Another relevant aspect is
that such higher order weak terms introduce parity-violating
effects in hadronic observables \cite{Ellis:2001ba}, which can be observed 
at (polarised) RHIC-Spin \cite{Bunce:2000uv}.

\begin{figure}[htbp]
\begin{minipage}{\textwidth}
\includegraphics[width=.5\linewidth]{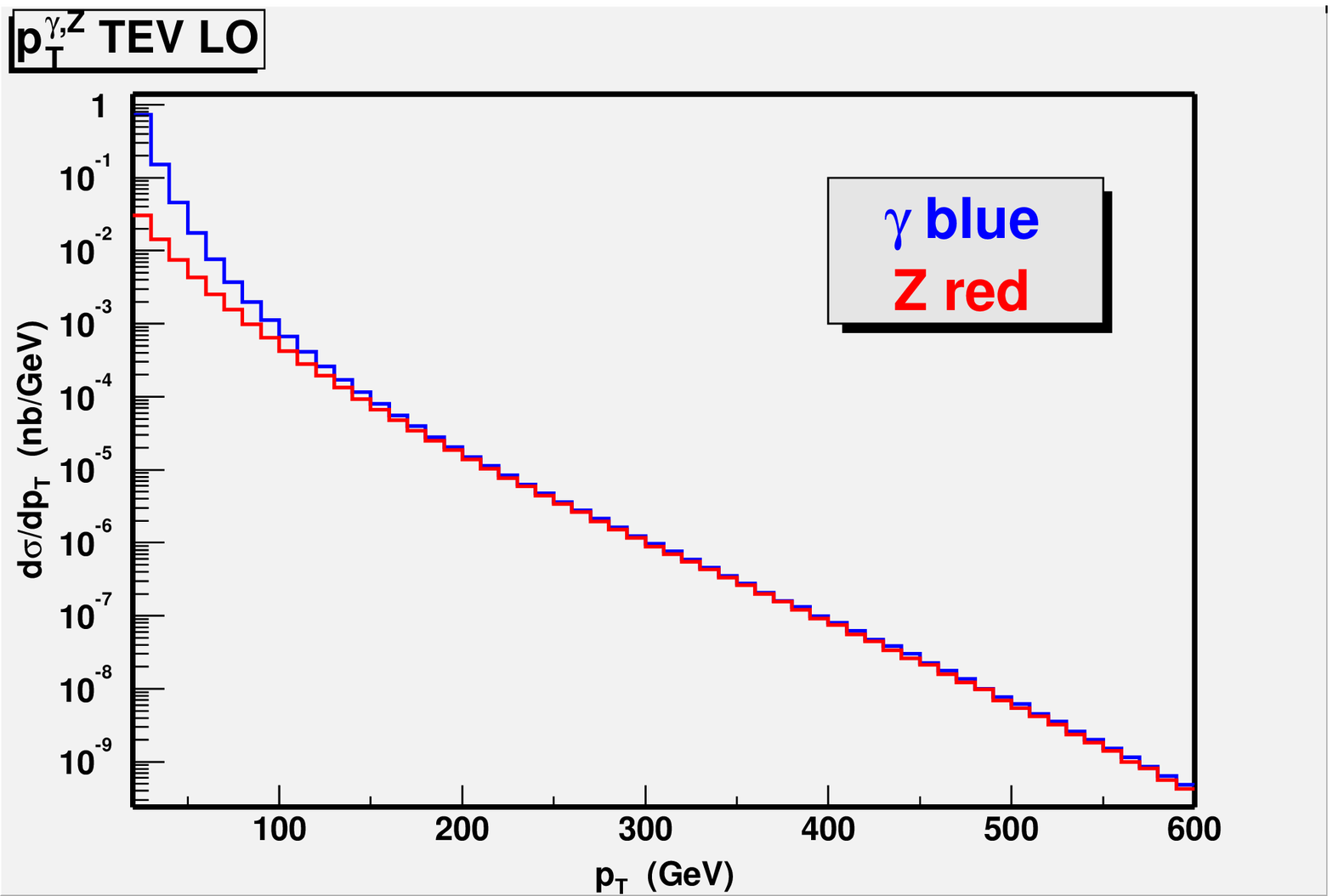}
\includegraphics[width=.5\linewidth]{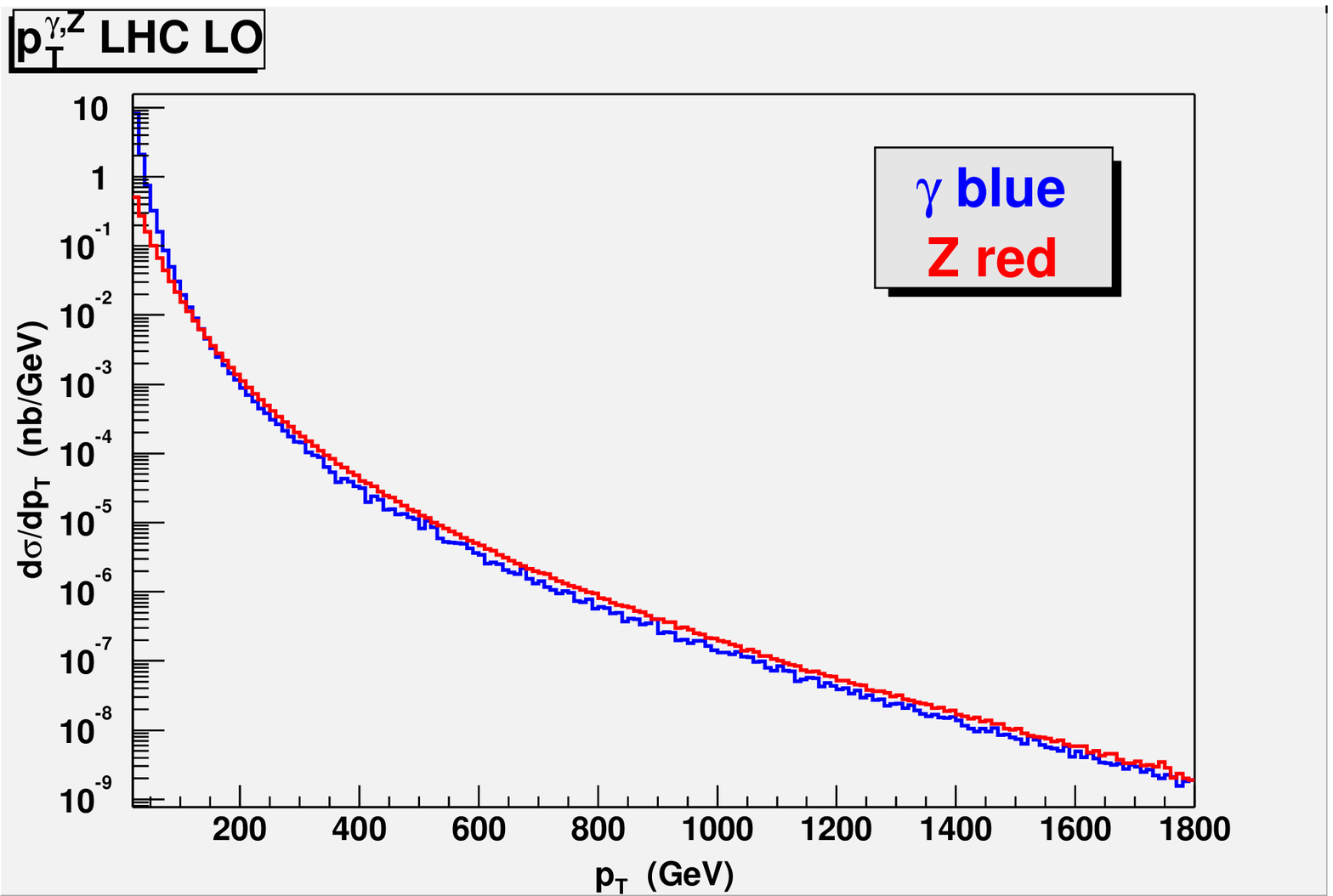}
\end{minipage}
\begin{minipage}{\textwidth}
\includegraphics[width=.5\linewidth]{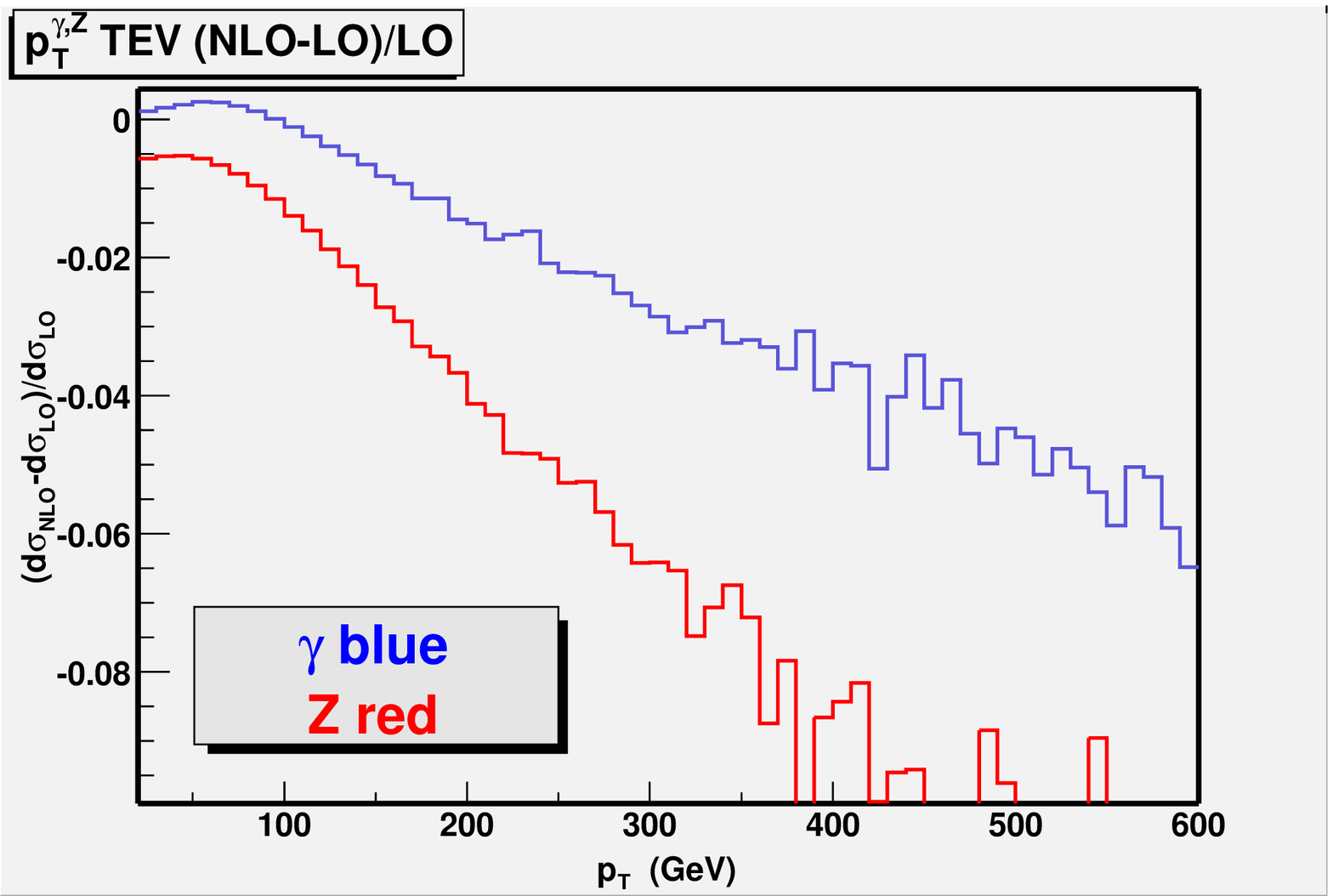}
\includegraphics[width=.5\linewidth]{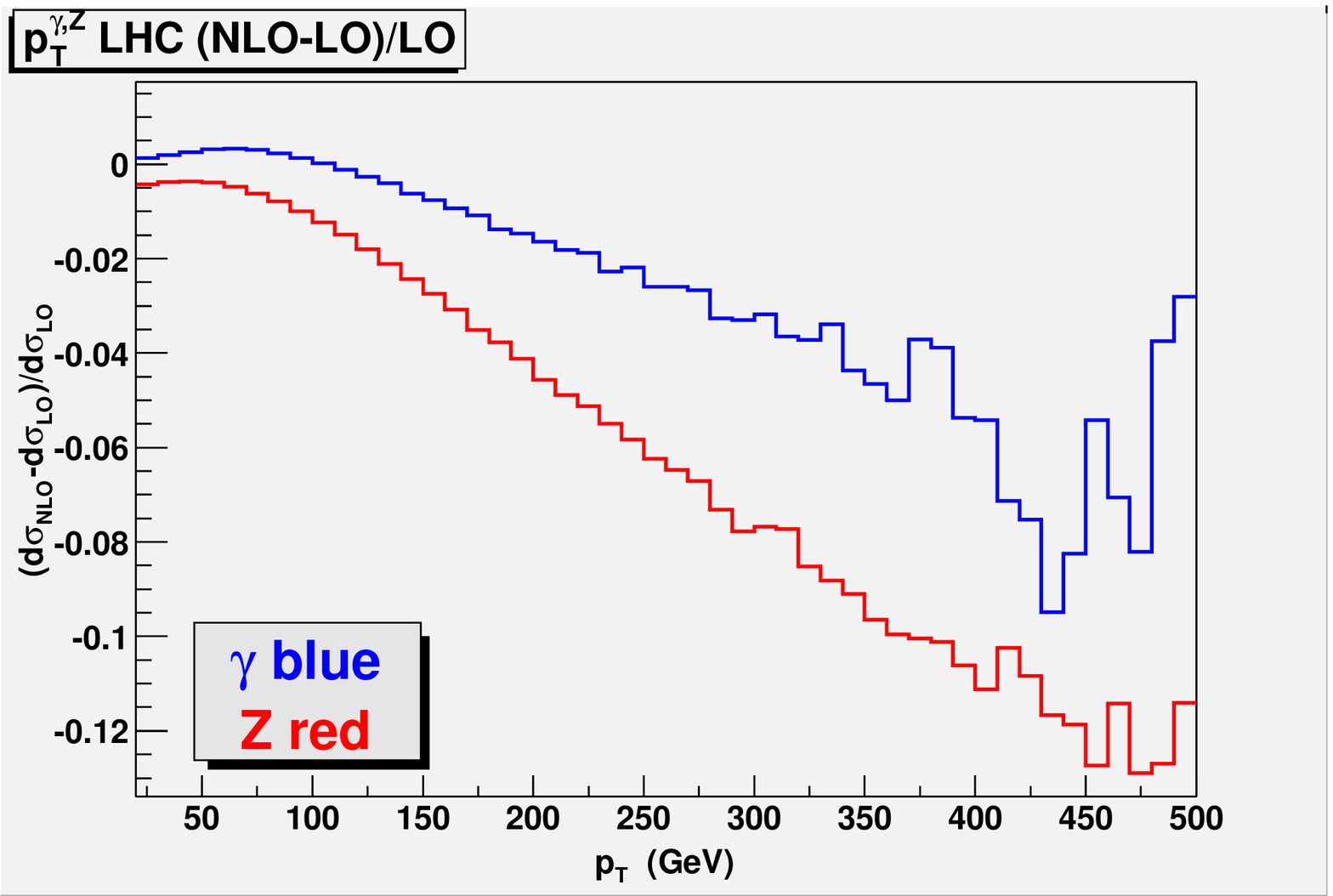}
\end{minipage}
\caption{The LO results through 
${\cal O}(\alpha_{\rm{S}}\alpha_{\rm{EW}})$
for the $\gamma$- and $Z$-production cross sections 
at Tevatron
and LHC, as a function of the transverse
momentum (top) as well as the size of the NLO 
corrections  through ${\cal O}(\alpha_{\rm{S}}\alpha_{\rm{EW}}^2)$
relatively to the former.} 
\label{fig:V}
\vspace*{-0.5truecm}
\end{figure}

\providecommand{\href}[2]{#2}



\section[Towards Automated One-Loop Calculations for Multi-Particle
Processes]{TOWARDS AUTOMATED ONE-LOOP CALCULATIONS FOR MULTI-PARTICLE
  PROCESSES~\protect\footnote{Contributed
  by: {T.~Binoth, J.Ph.~Guillet, G.~Heinrich, N.~Kauer, F.~Mahmoudi}}} 
\label{sec:towards-autom-one}





\subsection{Introduction}

In this decade experiments at hadron colliders explore the TeV scale.
The large center of mass energies lead generically to
multi-particle final states created  by QCD initial states.
The application of perturbative methods is justified if the scales 
of the problem are considerably larger than the proton mass.
Then a systematic separation
of long and short distance effects is possible and predictions for
cross sections can be made at the loop level, which is mandatory for 
a reliable estimate of production rates especially at hadron colliders.

The calculation of multi-particle production at the one-loop level
is very challenging due to the combinatorial complexity of 
the Feynman diagrammatic approach. 
Although the calculation of partonic $2 \to 2$ amplitudes at one-loop
is meanwhile standard, already the number of known $2 \to 3$  1-loop amplitudes is very restricted.
Up to now not a single Standard Model process which has generic  $2\to 4$ 
kinematics is computed at the one-loop level.
Needless to say  this is highly relevant
to many search channels for the Higgs boson at the LHC, like gluon fusion
and weak boson fusion, where additional jets have to be tagged to improve the
signal to background ratio. For signal reactions like 
$PP \to H + 0,1,2$ jets, with $ H \to \gamma\gamma,WW^*,\tau^+\tau^-$
which are available at one-loop level, not all amplitude calculations for the background exist. 
This is due to the fact that the signal case typically
contains only 5-point functions at 1-loop, whereas the background has
generic $2 \to 4$ kinematics. As an example for such reactions consider 
$PP \to b\bar b b\bar b + X$, $PP \to \gamma\gamma + 2\;\mbox{jets} + X$ or 
$PP \to ZZ + \gamma\gamma + X$. These require the evaluation of hexagon graphs like
\begin{figure}[htb]
\unitlength=1mm
\begin{picture}(150,25)
\put(20,-5){\epsfig{file = 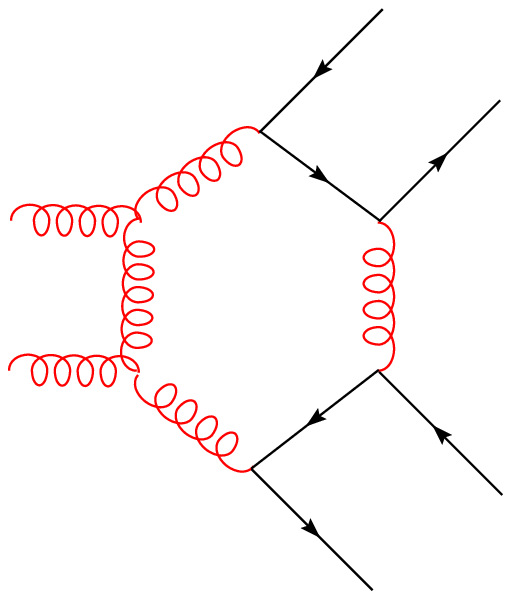,height=3.cm}}
\put(16,5){$g$}
\put(16,13){$g$}
\put(40,-5){$b$}
\put(47,0){$\bar b$}
\put(40,25){$b$}
\put(47,20){$\bar b$}
\put(70,-5){\epsfig{file = 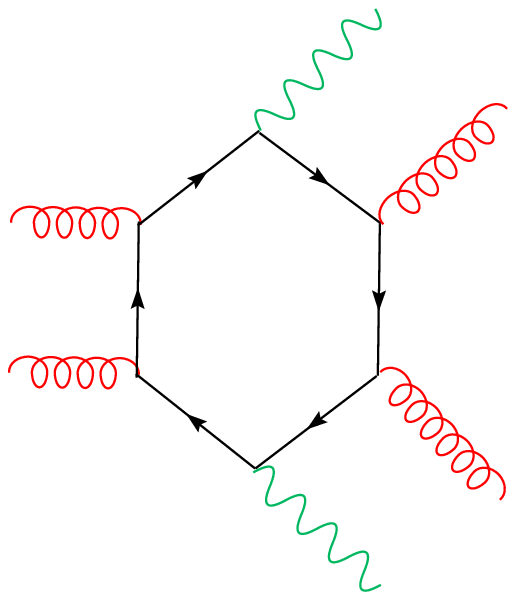,height=3.cm}}
\put(66,5){$g$}
\put(66,13){$g$}
\put(90,-3){$\gamma$}
\put(97,0){$g$}
\put(90,25){$\gamma$}
\put(97,20){$g$}
\put(120,-5){\epsfig{file = 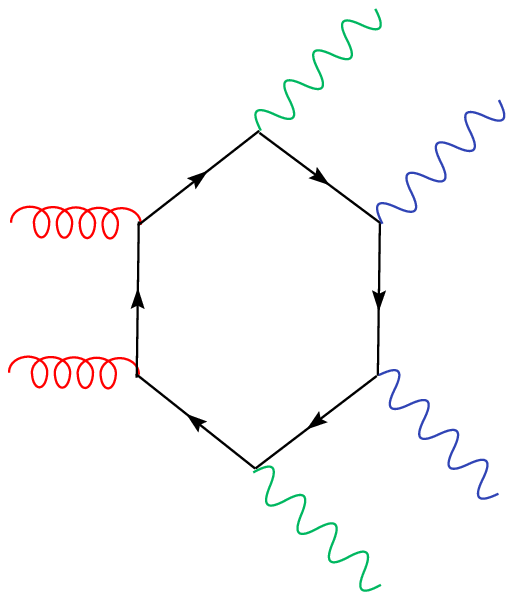,height=3.cm}}
\put(116,5){$g$}
\put(116,13){$g$}
\put(140,-5){$\gamma$}
\put(147,0){$Z$}
\put(140,25){$\gamma$}
\put(147,20){$Z$}
\put(132,14){$t$}
\end{picture}
\caption{Hexagon graphs for multi-particle production. The $t$ in the right graph
indicates that the fermion is a top quark.\label{FigHex}}
\end{figure}

The aim of our working group is to develop methods 
for the calculation of such hexagon amplitudes. The final goal is
a tool to tackle the computation of general  $2 \to 4$ processes at 
the one-loop level in a completely automated way. A basic ingredient
of such a tool are algebraic reduction formulas.
Our reduction formalism is described in the
next section. As an example for the efficiency of our methods 
we discuss the 5-point 1-loop amplitude $gg \to \gamma\gamma g$
in Section \ref{sec:analyt-calc-ggto}. 
The fully analytical  treatment
of hexagon one-loop amplitudes seems to be feasible analogously, if 
massless particles are considered. Examples of hexagon
amplitudes in the Yukawa model calculated with our approach 
can be found in  \cite{Binoth:2002qh,Binoth:2001vm}. Other, phenomenologically relevant 
examples are presently under study.     
For the massive case we suggest a numerical evaluation described in Section \ref{sec:numer-eval-multi}.

\subsection{Reduction Formalism}\label{sec:redu}
We will very briefly discuss the basic reduction formulas for  $N$-point
scalar and tensor integrals. More details can be found in \cite{Bern:1994kr,Binoth:1999sp}.  

\subsubsection{Tensor reduction}

Feynman diagrams correspond to combinations of tensor integrals. The momentum space
representation of an $N$-point tensor integral of rank $R$
in $D=4-2\epsilon$ dimensions is given by
\begin{equation}
I^{\mu_1\dots\mu_R}_N = \int \frac{d^Dk}{i\pi^{D/2}}\frac{k^{\mu_1}\dots k^{\mu_R}}{\prod_{j=1}^N (q_j^2 - m_j^2)}
\end{equation}    
Here $q_j$ is a linear combination of the loop momentum $k$ and external momenta $p_1,\dots p_N$.
If $N\ge 5$ standard reduction methods lead to a proliferation of terms with complicated
denominators. This has to be avoided, as otherwise a stable numerical evaluation of the amplitude
is hardly possible. By using  helicity or projection methods  loop momenta 
can  always be combined with external momenta such that they are expressed
by combinations of inverse propagators, e.g. with $s_j=p_j^2$, 
$s_{ij}=(p_i+p_j)^2$ , $q_1=k-p_1$ and $q_2=k-p_1-p_2$:
\begin{equation}
2 k\cdot p_2 = (q_1^2 -m_1^2)- (q_2^2 -m_2^2) + ( s_{12} - m_2^2 ) - ( s_{1} - m_1^2 )
\end{equation}
After expressing products of loop momenta with external vectors by propagators and canceling as many  propagators
as possible, one is left with tensor integrals which are irreducible. 
One finds that at most rank 1 $N$-point integrals have to be reduced for an $N$-point problem.
Explicit representations for the irreducible tensor integrals can be found in \cite{Binoth:1999sp}.
After complete tensor reduction one is left with a linear combination of
scalar integrals. 

\subsubsection{Scalar reduction}

To achieve as many analytic cancellations as possible the amplitude has
to be expressed by a basis of scalar integrals. To this end scalar $N$-point
integrals have to be reduced further. The scalar $N$-point function in momentum
and Feynman parameter space is given by
\begin{equation}
I_N^D = \int \frac{d^Dk}{i\pi^{D/2}}\frac{1}{\prod_{j=1}^N (q_j^2 - m_j^2)}
= (-1)^N \Gamma(N-D/2) \int\limits_{0}^1 d^Nz \frac{\delta(1-\sum_{j=1}^N z_j)}{\sum_{i,j=1}^N S_{ij}z_iz_j/2}
\end{equation}
where $S_{ij}=G_{ij} -G_{ii}/2-G_{jj}/2 +m_i^2+m_j^2$, $G_{ij}=2 \,r_i\cdot r_j$, $r_j=p_1+\dots +p_j$.
The basic reduction formula relates an $N$-point scalar integral to
$(N-1)$-point scalar integrals $I_{N-1,j}^D$,  where the $j$th propagator is omitted,
and $(D+2)$-dimensional remainder terms:
\begin{equation}
I_N^D = \sum\limits_{j=1}^N b_j I_{N-1,j}^D+
\left\{
\begin{array}{cl} 
-(1+2\epsilon) \frac{\det(G)}{\det(S)} \,I_N^{D+2} & \,N=4\\  
& \\
{\cal O}(\epsilon) &  \,N=5\\&\\
0 &\, N\ge 6
\end{array}\right.
\end{equation}
The reduction coefficients are $b_j=- \sum_{l=1}^N S_{jl}^{-1}$.
One finds by iteration that each $N$-point function and therefore each 
amplitude can be 
expressed by 2- and 3-point functions in $D$ dimensions and 4-point functions in 
$D+2$ dimensions only. These form the basic building blocks for 
an irreducible representation of the amplitude in terms of scalar functions.
The coefficients of these scalar integrals are expected to simplify to a 
large extent.  

\subsection{Analytic Calculation of $gg\to \gamma\gamma g$}
\label{sec:analyt-calc-ggto}
To give an example for our algebraic approach to $N$-point amplitudes 
we have reconsidered
the 5-point 1-loop  amplitude $gg\to \gamma\gamma g$. 
While this amplitude only had been extracted indirectly from the 
5-gluon amplitude \cite{Bern:1993mq} 
by replacing gluons by photons until recently, 
we present a direct calculation\,\cite{Binoth:2003xk}.

For convenience we define all particles as incoming.
\begin{equation}
\gamma(p_1,\lambda_1) + \gamma(p_2,\lambda_2) 
+ g( p_3,\lambda_3,c_3 ) + g( p_4,\lambda_4,c_4 ) + g( p_5,\lambda_5,c_5 ) \to 0
\end{equation}
Out-states can be obtained by crossing rules. 
In hadronic collisions this amplitude is relevant for the 
production of photon pairs
in association with a jet and as such a contribution of the background
to the Higgs boson search channel $H\to\gamma\gamma + \mbox{jet}$.
A phenomenological analysis has already been provided in \cite{deFlorian:1999tp,DelDuca:2003uz}.
The colour  structure of this amplitude is simple. It can be written  as
\begin{equation}
\Gamma^{\{\lambda_j\},\{c_j\}}[\gamma\gamma g g g \to 0] 
= \frac{Q_q^2 g_s^3}{i \pi^2} f^{c_3c_4c_5} {\cal A}^{\lambda_1\lambda_2\lambda_3\lambda_4\lambda_5}
\end{equation}
${\cal A}^{\lambda_1\lambda_2\lambda_3\lambda_4\lambda_5}$ are helicity dependent
linear combinations of scalar integrals and a constant term which is a remnant of
two-point functions with coefficients of order $(D-4)$.
Six independent helicity components exist: +++++,++++\,--,--\,++++,--\,--\,+++,
+++\,--\,--,\,--\,+++\,--.
As the amplitude is finite one expects that all 3-point functions which carry spurious
infrared poles cancel. The function basis of the problem is thus reduced to 2-point functions 
$I_2^D(s_{ij}) = \frac{\Gamma(1+\epsilon)\Gamma(1-\epsilon)^2}{\Gamma(2-2\epsilon)} \frac{(-s_{ij})^{-\epsilon}}{\epsilon}$,
4-point functions in $D+2$ dimensions  written as \cite{Binoth:2001vm}
\begin{equation}
F_1(s_{j_1j_2},s_{j_2j_3},s_{j_4j_5}) = 
\frac{1}{s_{j_4j_5}-s_{j_1j_2}-s_{j_2j_3}} \; I_4^6(p_{j_1},p_{j_2},p_{j_3},p_{j_4}+p_{j_5})
\end{equation}
and constant terms. 
From unitarity one expects that the  +++++\,,\,++++\,--\,,\,--\,++++ amplitudes should be polynomial.
The other helicity amplitudes will also contain non-polynomial functions
like logarithms and dilogarithms contained in  $I_2^D$ and $F_1$.

To give an example of each case  we show here the results for 
${\cal A}^{++++-}$ and ${\cal A}^{--+++}$
only, the remaining ones which
have also compact representations can be found in \cite{Binoth:2003xk}.
The result is expressed in terms of field strength 
tensors ${\cal F}^{\mu\nu}_j=p^\mu_j\epsilon^\nu_j-p^\nu_j\epsilon^\mu_j$
which satisfy the relations
\begin{eqnarray}
{\rm Tr}({\cal F}_i^\pm{\cal F}_j^\pm) &=&
2 \, p_i\cdot \epsilon_j^\pm p_j\cdot \epsilon_i^\pm - s_{ij}\, \epsilon_i^\pm\cdot \epsilon_j^\pm \nonumber\\
p_i\cdot {\cal F}_j^\pm \cdot p_k &=& ( s_{ij}\, p_k\cdot \epsilon_j^\pm - s_{jk}\, p_i\cdot \epsilon_j^\pm )/2
\end{eqnarray}
where $\epsilon_j^\pm$ are the polarization vectors of the gluons and photons.

For ${\cal A}^{++++-}$ which is polynomial we find the following result
\begin{eqnarray}
{\cal A}^{++++-} = \frac{\mbox{Tr}({\cal F}_1^+{\cal F}_2^+)\mbox{Tr}({\cal F}_3^+{\cal F}_4^+)}{s_{12}^2 s_{34}^2} 
\Bigl[ C^{++++-}  \; p_1\cdot {\cal F}_5^- \cdot p_3 - ( 3 \leftrightarrow 4 ) \Bigr]
\end{eqnarray} 
The coefficient is given by
\begin{eqnarray}
C^{++++-} &=& -\frac{s_{45}s_{13}s_{14}}{s_{35}s_{15}s_{24}} 
-\frac{s_{13}s_{45}}{s_{15}s_{35}}
+\frac{s_{45}^{2}}{s_{15}s_{24}}
-\frac{s_{12}^{2}+s_{45}^{2}-s_{12}s_{45}}{s_{35}s_{15}}
+\frac{s_{13}s_{15}}{s_{23}s_{45}}
+\frac{s_{13}-s_{34}}{s_{23}} \nonumber\\ &&
-\frac{s_{34}s_{45}}{s_{23}s_{15}}
+\frac{s_{15}-s_{25}}{s_{45}}
-\frac{s_{23}+s_{35}}{s_{13}}
-\frac{s_{23}s_{25}}{s_{13}s_{45}}
+\frac{s_{34}+s_{12}}{s_{15}}
\end{eqnarray}
We have checked that the corresponding amplitude has a $S_2\otimes S_2$ Bose symmetry when the photons
and the gluons with equal helicities are interchanged.

For ${\cal A}^{--+++}$ we split the result into three pieces with indices $F,B,1$, which belong to the  
 part proportional to 6-dimensional 
boxes $F_1$,  a part containing bubble graphs $I_2^D$,  and a constant term, respectively.   
\begin{eqnarray}
{\cal A}^{--+++} = {\cal A}^{--+++}_F + {\cal A}^{--+++}_B + {\cal A}^{--+++}_1 
\end{eqnarray} 
We find

\begin{eqnarray}
{\cal A}^{--+++}_F &=& \frac{\mbox{Tr}({\cal F}_1^-{\cal F}_2^-)\mbox{Tr}({\cal F}_3^+{\cal F}_4^+)}{s_{12}^2 s_{34}^2} 
\Bigl[ C^{--+++}_F\; p_1\cdot {\cal F}_5^+\cdot p_3 - (  3 \leftrightarrow 4 ) \Bigr] F_1(s_{13},s_{14},s_{25})\nonumber\\ &&
- (  4 \leftrightarrow 5 ) - (  5 \leftrightarrow 3 ) + (  1 \leftrightarrow 2 )  
       - (  1 \leftrightarrow 2, 4 \leftrightarrow 5 ) - (  1 \leftrightarrow 2, 5 \leftrightarrow 3 )\nonumber\\
{\cal A}^{--+++}_B &=& \frac{\mbox{Tr}({\cal F}_1^-{\cal F}_2^-)\mbox{Tr}({\cal F}_3^+{\cal F}_4^+)}{s_{12}^2 s_{34}^2} 
\Bigl[ C^{--+++}_B\; p_1\cdot {\cal F}_5^+\cdot p_3 - (  3 \leftrightarrow 4 ) \Bigr] I_2^D(s_{15})\nonumber\\ &&
- (  4 \leftrightarrow 5 ) - (  5 \leftrightarrow 3 ) + (  1 \leftrightarrow 2 )  
       - (  1 \leftrightarrow 2, 4 \leftrightarrow 5 ) - (  1 \leftrightarrow 2, 5 \leftrightarrow 3 )\nonumber\\       
{\cal A}^{--+++}_1 &=& \frac{\mbox{Tr}({\cal F}_1^-{\cal F}_2^-)\mbox{Tr}({\cal F}_3^+{\cal F}_4^+{\cal F}_5^+)}{2\,s_{34}s_{45}s_{35}}
\end{eqnarray} 
The indicated permutations have to be applied to the  indices of the field strength tensors, 
momenta and  Mandelstam variables. 
The coefficients are
\begin{eqnarray}
C^{--+++}_F &=& \frac{1}{2}\frac{s_{12}^2-2s_{13}s_{14}}{s_{35}s_{15}} 
                         - \frac{s_{14}}{s_{34}}- \frac{s_{14}}{s_{35}}\nonumber\\
C^{--+++}_B &=& \frac{s_{45}}{s_{15}}\left[\frac{s_{13}+s_{35}}{s_{14}+s_{45}}+\frac{s_{14}+s_{45}}{s_{13}+s_{35}}\right]
                         +\frac{s_{45}^2s_{13}}{s_{15} s_{35}^2}
             +\frac{s_{14}s_{35}}{s_{15} s_{45}}+2\frac{(s_{15}+s_{45})^2}{s_{35}^2}
             \nonumber\\ &&
                        -\frac{s_{13}+s_{35}}{s_{15}}
            -\frac{s_{14}s_{45}}{s_{15} s_{35}}
            -\frac{s_{45}^2}{s_{35} s_{15}}
            +\frac{s_{14}+s_{24}}{s_{45}}
            +\frac{s_{12}-s_{14}-s_{35}}{s_{14}+s_{45}}
            +2\frac{s_{14}(s_{15}+s_{45})}{s_{35}^2}
            \nonumber\\ &&
            +\frac{s_{23}^2s_{15}}{s_{35}^2(s_{13}+s_{35})}
            +\frac{2s_{45}+s_{15}}{s_{13}+s_{35}}
            -2\frac{(s_{15}+s_{45})s_{23}}{s_{35}(s_{13}+s_{35})}
            -\frac{(2s_{45}+s_{15})}{s_{35}}
            +\frac{s_{13}(2s_{45}+s_{15})}{s_{35}^2}
            \nonumber
\end{eqnarray} 
In the given expressions the  $S_2 \otimes S_3$ symmetry 
under exchange of the two photons and the three gluons is manifest 
after taking into account the omitted colour factor.

The result indicates that with our approach indeed a compact representation
of complicated loop amplitudes can be obtained. The application of our approach to
relevant 6-point amplitudes is presently under study. 

\subsection{Numerical Evaluation of Multi-Leg Integrals}
\label{sec:numer-eval-multi}
As already mentioned, our final aim is a complete automatisation
of one-loop calculations. 
The bottleneck for this goal is mainly given by the calculation of the 
virtual amplitudes: 
As the number of external legs increases, the growing number of invariants
renders the analytical expressions more and more complicated. If 
in addition massive particles are involved, the complexity 
of the resulting expressions rapidly approaches a limit where 
the analytical evaluation of the amplitude becomes unfeasible. 
Therefore, a numerical approach seems to be more appropriate 
to tackle different types of one-loop amplitudes  in a unified
and efficient way. 
Of course, it has to be stated what "numerical" means, as 
any method should finally lead to "numbers" to be compared 
to data. The important question is at what stage of the 
calculation the transition from analytical to numerical 
evaluation should be made. 

The conventional approach to calculate NLO cross sections
seeks to keep analytical expressions in the course 
of the calculation of a cross section as far as possible. 
Of course, there are good reasons to do so: 
If infrared (and/or ultraviolet) poles are present, 
they have to be isolated and canceled before any numerical evaluation
can be attempted. Further, analytic expressions are  flexible
in the sense that they can be used in "crossed" processes 
with different kinematics by analytic continuation.
On the other hand, the analytical approach 
may be troublesome  if the calculation of differential 
cross sections $d\sigma/d{\cal O}$ for some (infrared safe) 
observable ${\cal O}$  and/or the implementation of experimental cuts 
requires modifications of the analytic expressions.
In addition, concerning the virtual integrals, it is well known that 
even if a closed form exists, 
the implementation into a Monte Carlo  program may lead to 
numerical instabilities because the expressions are not 
appropriate for every phase space region.

These drawbacks of the "maximally analytical" approach  
suggest to make the transition analytical $\to$ numerical 
at an earlier stage of the calculation.   
A completely numerical approach has been suggested by 
D.~Soper \cite{Soper:1998ye,Soper:1999xk}, where the sum over 
cuts for a given graph 
is performed {\em before} the numerical integration over the loop momenta, 
in this way exploiting unitarity to cancel soft and collinear divergences. 
This method is very elegant, but choosing appropriate integration contours 
in the multi-dimensional parameter space is far from straightforward
and therefore might be hard to automate. 

In \cite{Nagy:2003qn}, a different approach has been suggested, 
where the calculation is split into "virtual" (loop) and 
"real" contributions as in the conventional approach, but 
a subtraction formalism for the UV and soft/collinear divergences of the 
one-loop graphs has been worked out, such that the subtracted integrals 
can be performed numerically in four dimensions. 
While the subtractions act on a graph by graph basis, 
the subtraction terms are added back (in analytically integrated form)  
{\em after} having been summed over the graphs, 
as only the summation leads to expressions which are simple 
enough to be integrated analytically. 

Another promising approach in this context is the one of \cite{Forde:2003jt},  
which tries to tackle the problem of infrared divergences 
by its very root: Starting from the observation that the 
clash between the long-range nature of the interactions in a 
massless gauge theory and the assumption of asymptotically free 
external states causes the appearance of IR singularities in the 
"conventional" amplitudes, they show that an appropriate 
redefinition of the external states, which includes the  long-range 
interactions, leads to S-matrix elements which are IR finite and 
apply it to the case $e^+e^-\to 2$\,jets at NLO.

The method we suggest here to calculate one-loop amplitudes 
is oriented at the aim of automatisation as a main guideline. 
It follows to a large extent the 
"analytical road" in treating virtual and real emission corrections separately.
This is feasible as the isolation of IR/UV divergences is straightforward if the 
discussed reduction formalism is used. UV renormalisation 
is well understood, and systematic methods for the combination 
of the IR divergences from the virtual corrections with 
their counterparts from the real emission contribution also exist
\cite{Giele:1992vf,Giele:1993dj,Keller:1998tf,Frixione:1997np,Catani:1997vz,Phaf:2001gc,Catani:2002hc}.
The main problem consists in calculating the remaining
finite terms of the $N$-point one-loop amplitudes, 
especially for $N=5,6$, and we will concentrate on this point in what follows.
By making the transition from analytical to numerical 
evaluation for these terms at an earlier stage than in the conventional 
approach, complicated cancellations between
numerous dilogarithms can be avoided.


We should mention that another approach to calculate loop integrals numerically is 
suggested by Passarino et al.\,\cite{Passarino:2001wv,Ferroglia:2002mz}, 
based on the Bernstein-Tkachov relation \cite{Tkachov:1997wh}. 

\subsubsection{Reduction to basic building blocks}

A one-loop $N$-point amplitude involving particles with 
arbitrary masses (including the case $m=0$)  will be 
reduced to basic building blocks using the method of 
\cite{Binoth:1999sp}, as outlined in Section \ref{sec:redu} 
As basic building blocks, we choose scalar 2-point functions $I^D_2$ and 
3-point functions $I^D_3$ in $D$ dimensions and $D+2$ dimensional 
box functions $I^{D+2}_4$.  The latter are infrared finite. 
Possible UV singularities are only contained in the 2-point 
functions and their subtraction is straightforward. 
The (soft and collinear) IR singularities are, as a result of the 
reduction, only contained in 2-point functions and 3-point functions
with one or two light-like legs. In this form, 
they are easy to isolate and to subtract from the amplitude. 
However, the resulting expression in general still has 
integrable (threshold) singularities which hinder a successful numerical
evaluation. For example, the  general 6-point integral 
(where all internal lines have different masses $m_i$ 
and all external legs are off-shell, $p_i^2=s_i\;, i=1\ldots 6$) 
depends on 21 
kinematic invariants with one non-linear constraint among them, 
and its analytic form contains hundreds of dilogarithms. 
A numerical evaluation of the latter 
leads to large cancellations in certain kinematic regions and 
such a representation is therefore inappropriate.

\subsubsection{Parameter representation of basic building blocks}

After reduction and separation of the divergent parts, 
we are left with finite integrals $I^D_3$ and $I^{D+2}_4$.
To evaluate them we first use standard Feynman parametrisation and 
then perform a sector decomposition\footnote{We define the 
step function $\Theta$ to be 1 if its argument is true, and 0 else.} 
\begin{equation}
\label{sectordeco}
1 = \Theta(x_1>x_2,\dots,x_N) +\Theta(x_2>x_1,\dots,x_N) 
+ \dots + \Theta(x_N>x_1,\dots,x_{N-1})
\end{equation}
for the integration over $N$ parameters ($N=3$ for the triangle, $N=4$ 
for the box).
Now, we carry out  {\it only one} parameter integration. 
We obtain
\begin{eqnarray}
&&I_3^{D}(s_1,s_2,s_3,m_1^2,m_2^2,m_3^2) = 
\nonumber\\&&\quad
\Bigl[  S_{Tri}^D(s_2,s_3,s_1,m_2^2,m_3^2,m_1^2) 
       +S_{Tri}^D(s_3,s_1,s_2,m_3^2,m_1^2,m_2^2) 
      + S_{Tri}^D(s_1,s_2,s_3,m_1^2,m_2^2,m_3^2) \Bigr]\nonumber
\end{eqnarray}
\begin{eqnarray}
&&S_{Tri}^{D=4}(s_1,s_2,s_3,m_1^2,m_2^2,m_3^2) 
=\int\limits_0^1 dt_1dt_2 
\frac{1}{(1+t_{1}+t_2)}\frac{1}{A t_2^2 + B t_2 + C - i \delta}\nonumber\\
&&= \int\limits_0^1 dt_1  \frac{2A}{\sqrt{R}} 
\left[ \frac{\log(2A+B-\sqrt{R}) - \log(B-\sqrt{R}) 
           - \log(2A+B+T)+\log(B+T)}{T+\sqrt{R}} \right.\nonumber\\
&&-
\left. \frac{\log(2A+B+\sqrt{R}) - \log(B+\sqrt{R}) 
           - \log(2A+B+T)+\log(B+T)}{T-\sqrt{R}}
\right] \label{Stri_num}\\
\mbox{with}&&\nonumber\\
A &=& m_2^2 \;,\quad
B = (m_1^2+m_2^2-s_2) t_1 + m_2^2 + m_3^2 -s_3 \nonumber\\
C &=& m_1^2 t_1^2  + ( m_1^2+m_3^2-s_1) t_1 + m_3^2 \nonumber\\
R &=& B^2 - 4 A C + i \delta \;,\quad
T = 2 A ( 1+t_1 ) - B\nonumber
\end{eqnarray}
We show the explicit expressions only for the triangle, 
the ones for the box are analogous and can be found in \cite{Binoth:2002xh}.
In the case of vanishing masses or invariants, as long as 
the functions remain IR finite, analogous 
expressions can be derived. 

\subsubsection{Singularity structure}
In order to analyse the singularity structure of the
integrands, we explicitly separate imaginary and real part.
One obtains
\begin{eqnarray}
&&S_{Tri}^{D=4}(s_1,s_2,s_3,m_1^2,m_2^2,m_3^2) 
= \int\limits_0^1 dt_1  \frac{4 A}{T^2-R} 
\Bigl\{
\Bigl[
\log(2A+B+T) - \log(B+T)
\Bigr] \nonumber\\&&
+\Theta(R<0)
\Bigl[ 
       \frac{\log(C) - \log(A+B+C)}{2}  \nonumber\\&&
+\frac{T}{\sqrt{-R}}
\Bigl(  
   \arctan\left(  \frac{\sqrt{-R}}{B} \right) 
  -\arctan\left(  \frac{\sqrt{-R}}{2 A+B}\right) + \pi\;\Theta(B < 0 < 2A+B)  
\Bigr)  
\Bigr] \nonumber\\&&
+\Theta(R>0)
\Bigl[ 
\frac{T - \sqrt{R}}{2\,\sqrt{R}}
\Bigl(
 \log\left( | 2 A + B - \sqrt{R} |\right) 
-\log\left( | B - \sqrt{R} |\right) -i \pi \Theta( B < \sqrt{R}< 2A+B)
\Bigr)\nonumber\\&&
- \frac{T + \sqrt{R}}{2\,\sqrt{R}}
\Bigl(
 \log\left( | 2 A + B + \sqrt{R} |\right) 
-\log\left( | B + \sqrt{R} |\right) +i \pi \Theta( B < -\sqrt{R}< 2A+B)
\Bigr)
\Bigr]
\Bigr\}\label{Stri_final}
\end{eqnarray} 
Three regions which lead to an imaginary part can be distinguished:
\begin{description}
\item[Region I:] $A+B+C>0, -2A<B<0, C>0 \Leftrightarrow 
                 ( B<\pm \sqrt{R}<2A+B )$.
\item[Region II:] $A+B+C>0, C<0 \Leftrightarrow 
               (B< \sqrt{R}<2A+B)\, \mbox{and not} \,(B<-\sqrt{R}<2A+B)$.
\item[Region III:]  $A+B+C<0, C>0 \Leftrightarrow  
               (B<-\sqrt{R}<2A+B)\, \mbox{and not} \,(B<\sqrt{R}<2A+B)$.
\end{description}  
Region I is an overlap region where the imaginary part 
has two contributions. In regions II and III only one of the 
$\Theta$--functions in (\ref{Stri_final})  contributes.
All critical regions are shown in Fig.~\ref{FigReg},
which illustrates the analytic structure of the integrand.
\begin{figure}[htb]
\unitlength=1mm
\begin{center}
\epsfig{file = 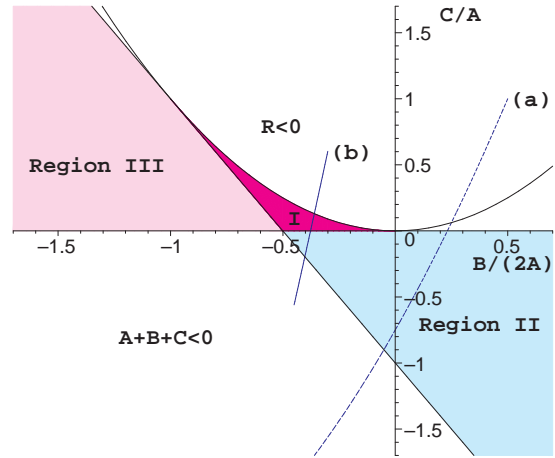,height=6.cm}
\end{center}
\caption{Analytic regions of the box and
triangle integrands. Within regions I,II and III the integrand
has an imaginary part. The integrable square-root and
logarithmic singularities are located at the boundaries of these 
regions.\label{FigReg}}
\end{figure}
Line segment (a) corresponds to
the integration region of a triangle function where only logarithmic 
singularities are present, while for line segment (b)  square-root
and logarithmic singularities are present. Note that  the 
box function $I_4^{D=6}$ has the same singularity structure. 
As $I^{D=4}_3$ and $I_4^{D=6}$ are the basic building blocks, 
this analysis of the singularity structure is done once and for all.


\subsubsection{Numerical integration}

We  now discuss the numerical evaluation of the analytic 
expressions derived above.\footnote{A more detailed description
can be found in Ref.~\protect\cite{Binoth:2002xh}.}
So far, we showed that any finite scalar
$N$-point function can be written as a linear combination
of the basic building block $S^{D=4}_{Tri}$ of Eq. (\ref{Stri_final})
and a similar 2-dimensional integral, $S^{D=6}_{Box}$,
with coefficients that are rational functions of the kinematical
invariants and masses.  To avoid numerical instabilities, the residual
dimensions of $S^{D=4}_{Tri}$ and $S^{D=6}_{Box}$ are then
integrated over numerically.  Since scalar function expressions
can contain dozens of building blocks and an amplitude has to be
evaluated many times to calculate a cross section, a fast method
to evaluate the basic building blocks is called for.  However,
the naive application of standard numerical integration techniques
is not sufficient to achieve this objective.  This is due to the
presence of integrable singularities and step discontinuities
in the integrands, which the detailed analysis above revealed.
It prevents the naive application of deterministic, integration-rule
based algorithms that are better suited for fast evaluation than the more
robust, but significantly slower Monte Carlo techniques commonly used
to evaluate multi-particle cross sections in high energy physics.

Several approaches can be pursued to achieve a sufficiently fast
and accurate numerical integration of the basic building
blocks.  A first direction are automatic methods, i.e.~methods
that do not require knowledge of the exact location and type
of the discontinuities.  The key to success here are adaptive
algorithms that iteratively divide the integration volume into
non-uniform subvolumes and apply basic numerical integration
methods to each subvolume until an optimal partition of the
integration volume minimizes the total error.  Using this
approach the 1-dimensional integral of $S^{D=4}_{Tri}$
can be integrated with negligible time requirements 
(fractions of a second).
The 2-dimensional integral of $S^{D=6}_{Box}$ is much
more challenging, but can be tackled in the same spirit
by combining deterministic and Monte Carlo integration techniques
(see Ref.~\cite{Kauer:2002hk} and references therein).
We note that the time required to 
integrate all $S^{D=6}_{Box}$ building blocks of the
scalar hexagon function using this approach
depends on the kinematical configuration and,
while sufficiently short at this stage, is no longer negligible.
If, at a subsequent stage, i.e.  
for the calculation of a certain cross section, 
the $S^{D=6}_{Box}$ building blocks
had to be computed in a time comparable to the one for the
$S^{D=4}_{Tri}$ functions, a second
direction could be pursued.  Since the location of all
singularities and step discontinuities is known analytically,
one can identify regions with continuous integrand and
in each region flatten the integrand either by transforming
integration variables or subtracting singular approximations.
The resulting bounded integrands could then be integrated
numerically with standard deterministic methods at a speed
that would facilitate millions of amplitude evaluations 
in a reasonable amount of time.

To demonstrate the practicality of our method to evaluate
multi-leg integrals, we show in
Fig.~\ref{hexagon-scan} a scan of the $2m_t = 350$ GeV
threshold of the 4-dimensional scalar hexagon function for a kinematical
configuration appropriate for the Feynman diagram to the right in Fig.~\ref{FigHex}.

\begin{figure}[htbp]
\begin{center}
\begin{minipage}[c]{.48\linewidth}
\flushright \includegraphics[width=5.2cm, angle=90]{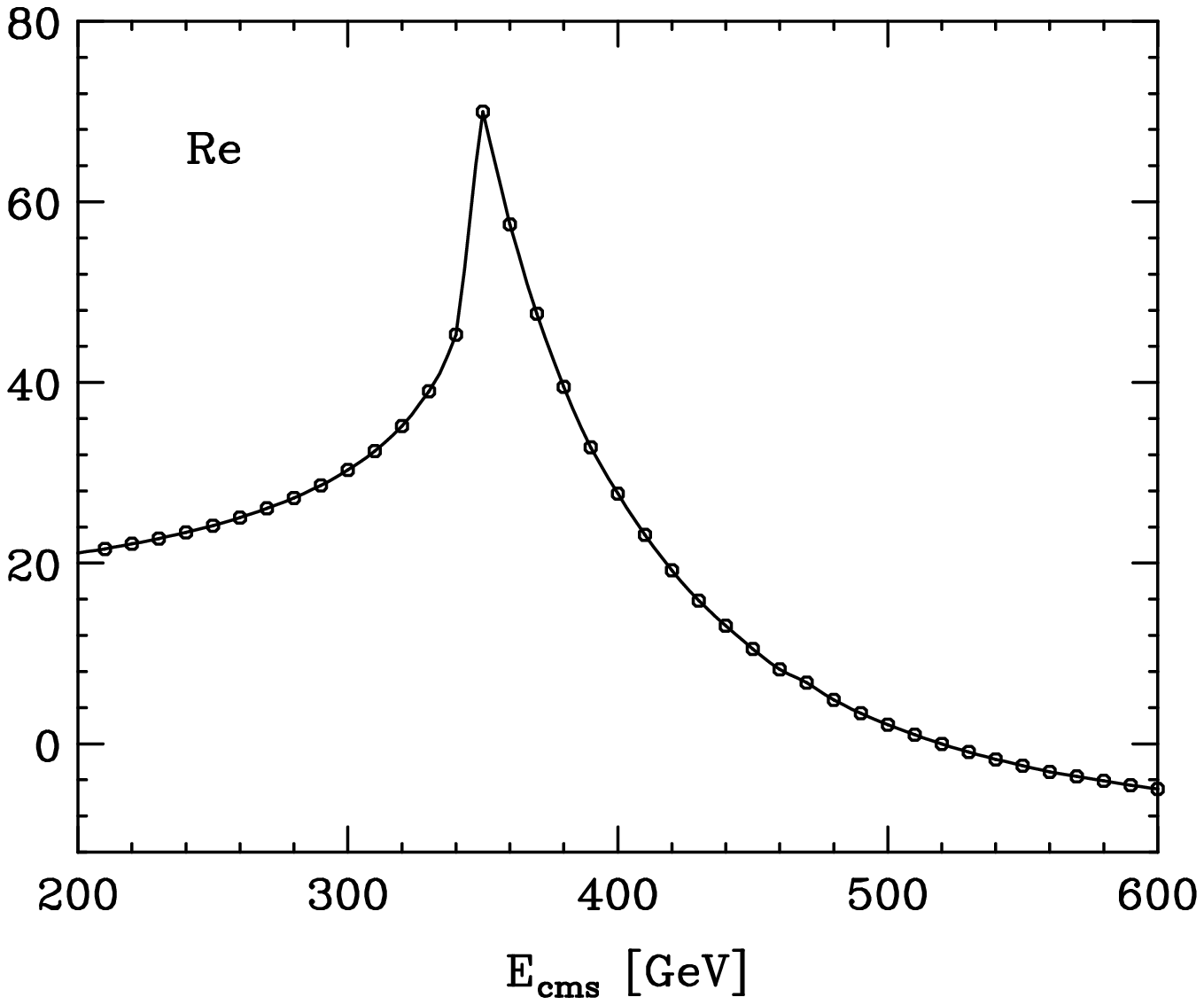} 
\end{minipage} \hfill
\begin{minipage}[c]{.48\linewidth}
\flushleft \includegraphics[width=5.2cm, angle=90]{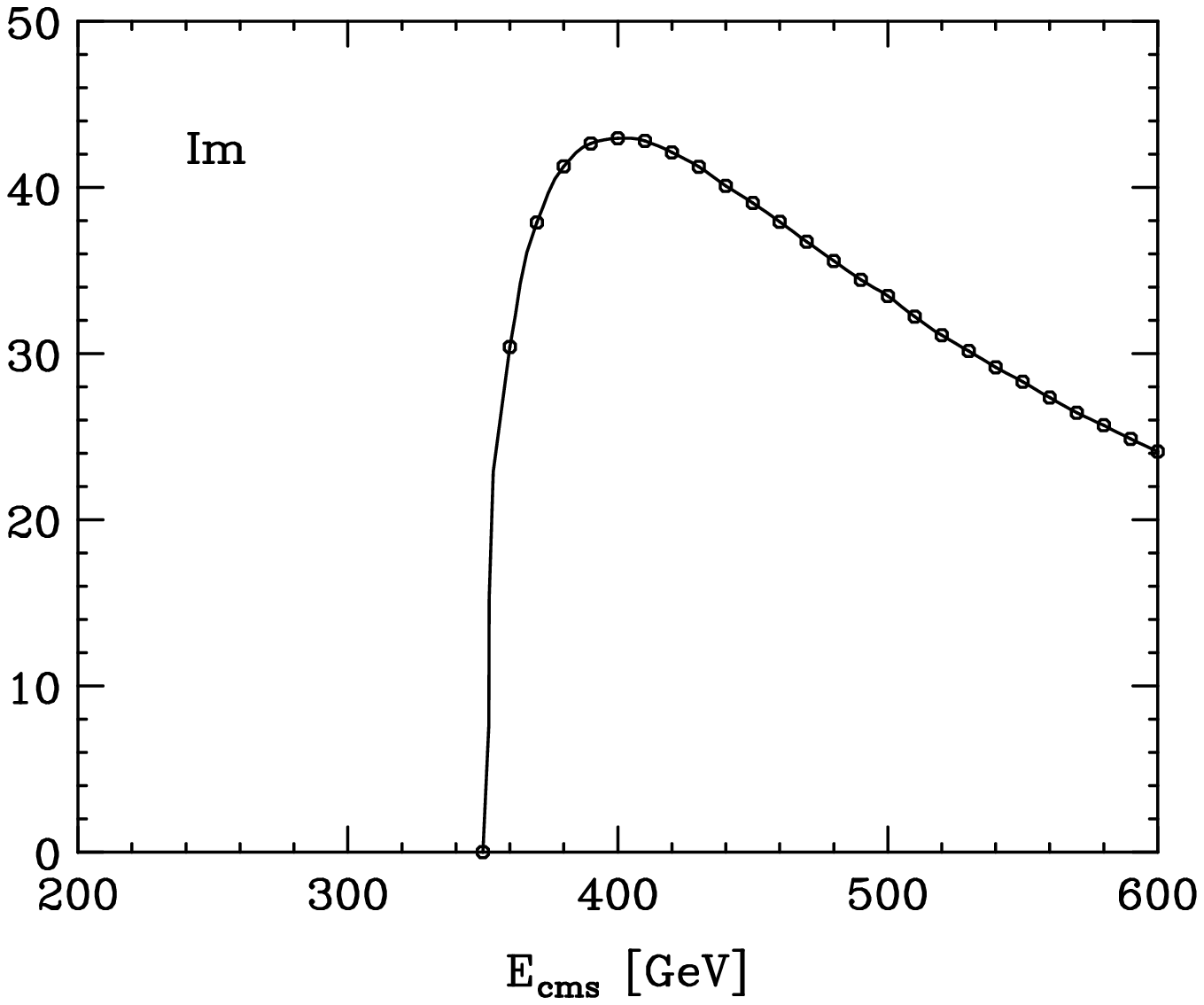}
\end{minipage}
\caption{Scan of the $2m_t = 350$ GeV threshold of the 4-dimensional scalar hexagon function 
which corresponds topologically and kinematically to the 
rightmost Feynman diagram of Fig.~\protect\ref{FigHex}.}
\label{hexagon-scan}
\end{center}
\end{figure}

\subsection{Conclusion}

We have outlined new algebraic/numerical approaches for 1-loop calculations. We have shown 
that our algebraic formalism leads to compact representations of 
complicated 1-loop amplitudes. Furthermore we have constructed numerical methods for 
general hexagon kinematics. 
The presented methods are parts of a project to describe multi-particle/jet
production at TeV colliders with $2 \to 4$ kinematics at the 1-loop level.    

\vskip1cm
\noindent

\section*{Acknowledgements}

We would like to thank the conference organizers!  
This research has been supported by the Bundesministerium f\"ur Bildung
und Forschung (BMBF, Bonn, Germany) under the contract under the contract number
05HT1WWA2.




\section[Infrared divergences at NNLO]{INFRARED DIVERGENCES AT NNLO~\protect\footnote{Contributed
  by: {G.~Heinrich, S.~Weinzierl}}}
\label{sec:infr-diverg-at}




\subsection{Introduction}

Fully differential NNLO calculations are needed to increase the accuracy of theoretical predictions
and are relevant to high-energy collider experiments, in particular for the Tevatron and the LHC.
They involve a variety of technical complications, such as the calculation of two-loop amplitudes,
a method for the cancellation of infrared divergences and stable and efficient numerical methods
for the final computer program.
There has been a significant progress in the calculation of two-loop amplitudes
in the past years
\cite{Bern:2000ie,Bern:2000dn,Anastasiou:2000kg,Anastasiou:2000ue,Anastasiou:2000mv,Anastasiou:2001sv,Glover:2001af,Bern:2001dg,Bern:2001df,Bern:2002tk,Garland:2001tf,Garland:2002ak,Moch:2002hm}.
Here we review the state of the art for the cancellation of infrared divergences
at NNLO.

Infrared divergences occur already at next-to-leading order.
At NLO real and virtual corrections contribute.
The virtual corrections contain the loop integrals and can have,
in addition to ultraviolet divergences, infrared divergences.
If loop amplitudes are calculated in dimensional regularisation,
the IR divergences manifest themselves as
explicit poles in the 
dimensional regularisation parameter $\varepsilon=2-D/2$.
These poles cancel with similar poles arising from
amplitudes with additional partons but less internal loops, when integrated over phase space regions where
two (or more) partons become ``close'' to each other.
In general, the Kinoshita-Lee-Nauenberg theorem
\cite{Kinoshita:1962ur,Lee:1964is}
guarantees that any infrared-safe observable, when summed over all 
states degenerate according to some resolution criteria, will be finite.
However, the cancellation occurs only after the integration over the unresolved phase space
has been performed and prevents thus a naive Monte Carlo approach for a fully exclusive
calculation.
It is therefore necessary to cancel first analytically all infrared divergences and to use
Monte Carlo methods only after this step has been performed.

At NLO, general methods to circumvent this problem are known.
This is possible due to the universality of the singular behaviour
of the amplitudes in soft and collinear limits.
Examples are the phase-space slicing method
\cite{Giele:1992vf,Giele:1993dj,Keller:1998tf}
and the subtraction method
\cite{Frixione:1996ms,Catani:1997vz,Dittmaier:1999mb,Phaf:2001gc,Catani:2002hc}.
It is worth to examine a simple NLO example in detail to understand the basic concepts
which are currently under discussion for an extension to NNLO.
We consider the NLO corrections to $\gamma^\ast \rightarrow 2 \; \mbox{jets}.$
The real corrections are given by the matrix element for 
$\gamma^\ast \rightarrow q g \bar{q}$ and read, 
up to colour and coupling factors
\begin{eqnarray}
 \left| {\cal A}_3 \right|^2 & = & 8 ( 1 - \varepsilon)  \left[
         \frac{2}{x_1 x_2} 
         - \frac{2}{x_1}
         - \frac{2}{x_2} 
         + (1-\varepsilon) \frac{x_2}{x_1}
         + (1-\varepsilon) \frac{x_1}{x_2} 
         - 2 \varepsilon 
        \right],
\end{eqnarray}
where $x_1=s_{12}/s_{123}$ and $x_2=s_{23}/s_{123}$.
This term is integrated over the three particle phase space
\begin{eqnarray}
\int d\phi_3 & = & \frac{4^{-4+3\varepsilon} \pi^{-5/2+2\varepsilon}}{\Gamma(1-\varepsilon) \Gamma(\frac{3}{2}-\varepsilon)}
     s_{123}^{1-2\varepsilon} 
     \int d^3 x \delta\left( 1 - \sum\limits_{i=1}^3 x_i \right) 
          x_1^{-\varepsilon} x_2^{-\varepsilon} x_3^{-\varepsilon}.
\end{eqnarray}
Singularities occur at the boundaries of the integration region at $x_1=0$ and $x_2=0$.
Historically, phase space slicing 
\cite{Giele:1992vf,Giele:1993dj,Keller:1998tf}
has been the first systematic method to treat the infrared singularities.
Here, one splits the integration region into different parts, shown in fig. \ref{fig:8}:
A soft region, given by $x_1<x_{min}$ and $x_2<x_{min}$, two collinear regions, corresponding to
$x_1<x_{min}, x_2>x_{min}$ and $x_1>x_{min}, x_2<x_{min}$ and a hard region region
$x_1>x_{min}, x_2>x_{min}$.
The hard region is free of singularities and the integration can be performed
numerically there.
In the remaining regions the matrix element is approximated by the soft or collinear factorisation formulae
and the integration
over a one-parton phase space can then be performed analytically.
Phase space slicing has the advantage, that different factorisation formulae may be used in different regions of phase space.
However, there are also some disadvantages: The method introduces a systematic error of order $x_{min}$,
it becomes rather intricate for colour-subleading terms
and it poses a numerical problem:
The hard region gives a contribution of the form
\begin{eqnarray}
a \ln^2 x_{min} + b \ln x_{min} + c.
\end{eqnarray}
The logarithms $\ln^2 x_{min}$ and $\ln x_{min}$ cancel against the contributions from the other regions,
but this cancelation happens only numerically.

Within the subtraction method 
\cite{Frixione:1996ms,Catani:1997vz,Dittmaier:1999mb,Phaf:2001gc,Catani:2002hc}
one subtracts a suitable approximation term $d\sigma^A$ 
from the real corrections $d\sigma^R$.
This approximation term must have the same singularity structure as the real corrections.
If in addition the approximation term is simple enough, such that it can be integrated analytically
over a one-parton subspace, then the result can be added back to the virtual corrections $d\sigma^V$.
\begin{eqnarray}
\sigma^{NLO} & = & \int\limits_{n+1} d\sigma^R + \int\limits_n d\sigma^V
= \int\limits_{n+1} \left( d\sigma^R - d\sigma^A \right) + \int\limits_n \left( d\sigma^V + \int\limits_1 
d\sigma^A \right).
\end{eqnarray}
Since by definition $d\sigma^A$ has the same singular behaviour as $d\sigma^R$, $d\sigma^A$
acts as a local counter-term and the combination $(d\sigma^R-d\sigma^A)$ is integrable
and can be evaluated numerically.
Secondly, the analytic integration of $d\sigma^A$ over the one-parton subspace will yield
the explicit poles in $\varepsilon$ needed to cancel the corresponding poles in $d\sigma^V$.
The subtraction method overcomes the short-comings of the slicing method, but there is a price to pay:
The approximation term is subtracted over the complete phase space and has to interpolate between
different singular regions.
At NLO this requires an interpolation between soft and collinear regions.
For the example discussed above the approximation term can be taken as a sum of two (dipole) subtraction
terms:
\begin{eqnarray}
d\sigma^A
 & = & 
\left| {\cal A}_2(p_1',p_3') \right|^2 \frac{1}{s_{123}}
       \left[
         \frac{2}{x_1 (x_1 + x_2)} - \frac{2}{x_1}
         + (1-\varepsilon) \frac{x_2}{x_1}
        \right]
 \nonumber \\
 & &
 +
\left| {\cal A}_2(p_1'',p_3'') \right|^2 \frac{1}{s_{123}}
 \left[
         \frac{2}{x_2 (x_1 + x_2)} - \frac{2}{x_2}
         + (1-\varepsilon) \frac{x_1}{x_2}
        \right] 
\end{eqnarray}
The momenta $p_1'$, $p_3'$, $p_1''$ and $p_3''$ are linear combinations of the original momenta $p_1$, $p_2$ and $p_3$.
The first term is an approximation for $x_1 \rightarrow 0$, whereas the second term is an approximation
for $x_2 \rightarrow 0$.
Note that the soft singularity is shared between the two dipole terms 
and that in general the Born amplitudes ${\cal A}_2$ are evaluated with different momenta.
The subtraction terms can be derived by working in the axial gauge. In this gauge only diagrams where the emission
occurs from external lines are relevant for the subtraction terms.
Alternatively, they can be obtained from off-shell currents.
Antenna factorisation 
\cite{Kosower:1998zr,Kosower:2002su,Kosower:2003cz,Kosower:2003bh}
allows to reduce the number of subtraction terms needed and 
interpolates smoothly between the $x_1 \rightarrow 0$ and
$x_2 \rightarrow 0$ regions.
\begin{figure}
\begin{center}
\resizebox{0.7\textwidth}{!}{
  \includegraphics[120pt,575pt][490pt,725pt]{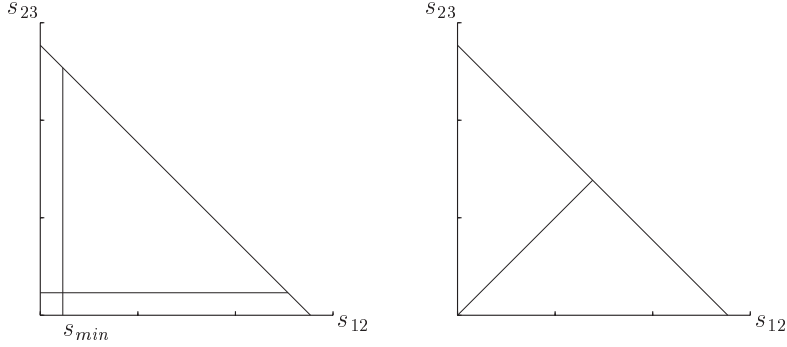}
}
\caption{Partitioning of the Dalitz plot for three final-state particles: (a) phase space slicing,
(b) sector decomposition.}
\label{fig:8}
\end{center}
\end{figure}
We briefly sketch how to obtain the antenna function from off-shell currents.
The amplitude for $\gamma^\ast \rightarrow q g \bar{q}$ is constructed from the quark and antiquark currents $J$ and $\bar{J}$ 
as follows:
\begin{eqnarray}
{\cal A}_3(p_1,p_2,p_3) & = & 
 \bar{J}(p_1,p_2) \; \varepsilon_\gamma \cdot \gamma \; J(p_3)
 +
 \bar{J}(p_1) \; \varepsilon_\gamma \cdot \gamma \; J(p_2,p_3).
\end{eqnarray}
$\varepsilon_\gamma$ is the polarisation vector of the photon, contracted into the Dirac matrix $\gamma_\mu$.
Singular contributions can only arise from the currents
$\bar{J}(p_1,p_2)$ and $J(p_2,p_3)$.
With a suitable pair of reconstruction functions $\hat{p_1}(p_1,p_2,p_3)$ and $\hat{p_2}(p_1,p_2,p_3)$ 
one achieves that the Born amplitude ${\cal A}_2$ is evaluated with the same set of momenta and one can
approximate the amplitude ${\cal A}_3$ by
\begin{eqnarray}
 \mbox{Ant}(p_1,p_2,p_3;\hat{p_1},\hat{p_2}) \;
 {\cal A}_2(\hat{p_1},\hat{p_2})
\end{eqnarray}
Note that in general the momenta $\hat{p_1}$ and $\hat{p_2}$ are non-linear functions of the original momenta
$p_1$, $p_2$ and $p_3$.

Once suitable subtraction terms are found, they have to be integrated over the unresolved phase space.
Here, one faces integrals with overlapping divergences, as one can already see from our simple example:
\begin{eqnarray}
     \int d^3 x \delta\left( 1 - \sum\limits_{i=1}^3 x_i \right) 
          x_1^{-\varepsilon} x_2^{-\varepsilon} x_3^{-\varepsilon}
       \left[
         \frac{2}{x_1 (x_1 + x_2)} - \frac{2}{x_1}
         + (1-\varepsilon) \frac{x_2}{x_1}
        \right]
\label{nlo-2}     
\end{eqnarray}
The term $1/(x_1+x_2)$ is an overlapping singularity. Sector decomposition 
\cite{Hepp:1966eg,Roth:1996pd,Binoth:2000ps}
is a convenient tool to disentangle
overlapping singularities. Here one splits the integration region into two sectors $x_1>x_2$ and $x_1<x_2$,
as shown in fig. \ref{fig:8}.
In the fist sector one rescales $x_2$ as $x_2'=x_2/x_1$, while in the second sector one rescales
$x_1'=x_1/x_2$.
Sector decomposition is discussed in detail in sect. \ref{sec:num}.

\subsection{The subtraction method at NNLO}
\label{sect:doubleunres}

The following terms contribute at NNLO:
\begin{eqnarray}
d\sigma_{n+2}^{(0)} & = & 
 \left( \left. {\cal A}_{n+2}^{(0)} \right.^\ast {\cal A}_{n+2}^{(0)} \right) 
d\phi_{n+2},  \nonumber \\
d\sigma_{n+1}^{(1)} & = & 
 \left( 
 \left. {\cal A}_{n+1}^{(0)} \right.^\ast {\cal A}_{n+1}^{(1)} 
 + \left. {\cal A}_{n+1}^{(1)} \right.^\ast {\cal A}_{n+1}^{(0)} \right)  
d\phi_{n+1}, \nonumber \\
d\sigma_n^{(2)} & = & 
 \left( 
 \left. {\cal A}_n^{(0)} \right.^\ast {\cal A}_n^{(2)} 
 + \left. {\cal A}_n^{(2)} \right.^\ast {\cal A}_n^{(0)}  
 + \left. {\cal A}_n^{(1)} \right.^\ast {\cal A}_n^{(1)} \right) d\phi_n,
\end{eqnarray}
where ${\cal A}_n^{(l)}$ denotes an amplitude with $n$ external partons and $l$ loops.
$d\phi_n$ is the phase space measure for $n$ partons.
Taken separately, each of these contributions is divergent.
Only the sum of all contributions is finite.
We would like to construct a numerical program for an arbitrary infrared safe observable ${\cal O}$.
Infrared safety implies that whenever a 
$n+l$ parton configuration $p_1$,...,$p_{n+l}$ becomes kinematically degenerate 
with a $n$ parton configuration $p_1'$,...,$p_{n}'$
we must have
\begin{eqnarray}
{\cal O}_{n+l}(p_1,...,p_{n+l}) & \rightarrow & {\cal O}_n(p_1',...,p_n').
\end{eqnarray}
To render the individual contributions finite, one adds and subtracts suitable
pieces
\cite{Weinzierl:2003fx,Weinzierl:2003ra}:
\begin{eqnarray}
\langle {\cal O} \rangle_n^{NNLO} & = &
 \int \left( {\cal O}_{n+2} \; d\sigma_{n+2}^{(0)} 
             - {\cal O}_{n+1} \circ d\alpha^{(0,1)}_{n+1}
             - {\cal O}_{n} \circ d\alpha^{(0,2)}_{n} 
      \right) \nonumber \\
& &
 + \int \left( {\cal O}_{n+1} \; d\sigma_{n+1}^{(1)} 
               + {\cal O}_{n+1} \circ d\alpha^{(0,1)}_{n+1}
               - {\cal O}_{n} \circ d\alpha^{(1,1)}_{n}
        \right) \nonumber \\
& & 
 + \int \left( {\cal O}_{n} \; d\sigma_n^{(2)} 
               + {\cal O}_{n} \circ d\alpha^{(0,2)}_{n}
               + {\cal O}_{n} \circ d\alpha^{(1,1)}_{n}
        \right).
\end{eqnarray}
Here $d\alpha_{n+1}^{(0,1)}$ is a subtraction term for single unresolved configurations
of Born amplitudes.
This term is already known from NLO calculations.
The term $d\alpha_n^{(0,2)}$ is a subtraction term 
for double unresolved configurations.
Finally, $d\alpha_n^{(1,1)}$ is a subtraction term
for single unresolved configurations involving one-loop amplitudes.

To construct these terms the universal factorisation properties of 
QCD amplitudes in unresolved limits are essential.
QCD amplitudes factorise if they are decomposed into primitive
amplitudes.
Primitive amplitudes are defined by
a fixed cyclic ordering of the QCD partons,
a definite routing of the external fermion lines through the diagram
and the particle content circulating in the loop.
One-loop amplitudes factorise in single unresolved limits as
\cite{Bern:1994zx,Bern:1998sc,Kosower:1999xi,Kosower:1999rx,Bern:1999ry,Catani:2000pi,Kosower:2003cz}
\begin{eqnarray}
\label{oneloopfactformula}
A^{(1)}_{n}
  & = &
  \mbox{Sing}^{(0,1)} 
  \cdot A^{(1)}_{n-1} +
  \mbox{Sing}^{(1,1)} \cdot A^{(0)}_{n-1}.
\end{eqnarray}
Tree amplitudes factorise in the double unresolved limits as
\cite{Berends:1989zn,Gehrmann-DeRidder:1998gf,Campbell:1998hg,Catani:1998nv,Catani:1999ss,DelDuca:1999ha,Kosower:2002su}
\begin{eqnarray}
\label{factsing}
A^{(0)}_{n}
  & = &
  \mbox{Sing}^{(0,2)} \cdot A^{(0)}_{n-2}.
\end{eqnarray}
To discuss the term $d\alpha_n^{(0,2)}$ let us consider as an example
the Born leading-colour contributions to $e^+ e^- \rightarrow q g g \bar{q}$,
which contribute to the NNLO corrections to
$e^+ e^- \rightarrow \mbox{2 jets}$.
The subtraction term has to match all double and single unresolved 
configurations.
It is convenient to construct $d\alpha_n^{(0,2)}$ as a sum over 
several pieces,
\begin{eqnarray}
d \alpha^{(0,2)}_{n} & = & 
 \sum\limits_{\mbox{\tiny topologies $T$}} {\cal D}_{n}^{(0,2)}(T).
\end{eqnarray}
Each piece is labelled by a splitting topology.
\begin{figure}
\begin{center}
\resizebox{!}{25mm}{
  \includegraphics[110pt,640pt][200pt,725pt]{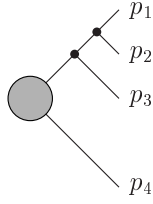}
}
\caption{Splitting topology.}
\label{fig:7}
\end{center}
\end{figure}
An example is shown in fig. \ref{fig:7}.
The term ${\cal D}_{n}^{(0,2)}(T)$ corresponding to the topology shown in
fig. \ref{fig:7} approximates singularities in $1/s_{12}$,
$1/(s_{12} s_{123})$ and part of the singularities in $1/s_{123}^2$.
Care has to be taken to disentangle correctly overlapping singularities
like $1/(s_{12}s_{23})$.
Details can be found in \cite{Weinzierl:2003fx}.

\subsubsection{One-loop amplitudes with one unresolved parton}
\label{sect:oneloop}

Apart from $d\alpha_n^{(0,2)}$ also the term
$d\alpha_n^{(1,1)}$, which approximates one-loop amplitudes 
with one unresolved parton, is needed at NNLO.
If we recall the factorisation formula (\ref{oneloopfactformula}),
this requires as a new feature 
the approximation of the one-loop singular function
$\mbox{Sing}^{(1,1)}$.
The corresponding subtraction term is proportional to the 
one-loop $1\rightarrow 2$ splitting function 
${\cal P}^{(1,1)}_{(1,0)\; a \rightarrow b c}$.
An example is the leading-colour part for the
splitting $q \rightarrow q g$ \cite{Weinzierl:2003ra}:
\begin{eqnarray}
\lefteqn{
{\cal P}^{(1,1)}_{(1,0)\; q \rightarrow q g, lc, corr} 
 =  
     - \frac{11}{6\varepsilon} {\cal P}^{(0,1)}_{q \rightarrow q g},
 +
 S_\varepsilon^{-1} c_\Gamma \left( \frac{-s_{ijk}}{\mu^2} \right)^{-\varepsilon} 
  y^{-\varepsilon}
 }
 \nonumber \\
 & &
     \left\{ 
         g_{1, corr}(y,z) \; {\cal P}^{(0,1)}_{q \rightarrow q g}
         + f_2 \frac{2}{s_{ijk}} \frac{1}{y} p\!\!\!/_{e} 
               \left[ 1 - \rho \varepsilon (1-y) (1-z) \right]
     \right\}.
\end{eqnarray}
This term depends on the correlations among the remaining hard partons.
If only two hard partons are correlated, $g_{1}$ is given by
\begin{eqnarray}
\lefteqn{
g_{1, intr}(y,z) 
 = 
} \nonumber \\ & &
  - \frac{1}{\varepsilon^2} 
 \left[ \Gamma(1+\varepsilon) \Gamma(1-\varepsilon) \left( \frac{z}{1-z} \right)^\varepsilon 
        + 1 
        - (1-y)^\varepsilon z^\varepsilon \; {}_2F_1\left( \varepsilon, \varepsilon, 1+\varepsilon; (1-y)(1-z) \right) \right].
\hspace{10mm}
\end{eqnarray}
Here, $y=s_{ij}/s_{ijk}$, $z=s_{ik}/(s_{ik}+s_{jk})$ and 
$f_2=(1-\rho\varepsilon)/2/(1-\varepsilon)/(1-2\varepsilon)$.
The parameter $\rho$ specifies the variant of dimensional regularisation:
$\rho  = 1$ in the conventional or 't Hooft-Veltman scheme and 
$\rho=0$ in a four-dimensional scheme.
For the integration of the subtraction terms 
over the unresolved phase space all occuring integrals are reduced to
standard integrals of the form
\begin{eqnarray}
\lefteqn{
\int\limits_0^1 dy \; y^a (1-y)^{1+c+d} \int\limits_0^1 dz \; z^c (1-z)^d \left[ 1 -z(1-y)\right]^e
  {}_2F_1\left( \varepsilon, \varepsilon; 1+\varepsilon; (1-y) z \right) =
} 
 \\ & &
 \frac{\Gamma(1+a) \Gamma(1+d) \Gamma(2+a+d+e) \Gamma(1+\varepsilon)}{\Gamma(2+a+d) \Gamma(\varepsilon) \Gamma(\varepsilon)}
 \sum\limits_{j=0}^\infty 
 \frac{\Gamma(j+\varepsilon) \Gamma(j+\varepsilon) \Gamma(j+1+c)}
      {\Gamma(j+1) \Gamma(j+1+\varepsilon) \Gamma(j+3+a+c+d+e)}. 
 \nonumber 
\end{eqnarray}
The result is proportional to  hyper-geometric functions 
${}_4F_3$ with unit argument and can be
expanded into a Laurent series in $\varepsilon$ 
with the techniques of \cite{Moch:2001zr,Weinzierl:2002hv}.
For the example discussed above one finds after integration
\cite{Weinzierl:2003ra}:
\begin{eqnarray}
\lefteqn{
{\cal V}^{(1,1)}_{(1,0)\; q \rightarrow q g, lc, intr} = 
 - \frac{1}{4\varepsilon^4}
 - \frac{31}{12 \varepsilon^3}
 + \left( -\frac{51}{8} - \frac{1}{4} \rho + \frac{5}{12} \pi^2 
          - \frac{11}{6} L
   \right) \frac{1}{\varepsilon^2} 
}
 \nonumber \\
 & & 
 + \left( - \frac{151}{6} - \frac{55}{24} \rho 
          + \frac{145}{72} \pi^2 + \frac{15}{2} \zeta_3
          - \frac{11}{4} L 
          - \frac{11}{12} L^2 
   \right) \frac{1}{\varepsilon}
 - \frac{1663}{16} - \frac{233}{24} \rho 
 + \frac{107}{16} \pi^2 + \frac{5}{12} \rho \pi^2 
 \nonumber \\ & & 
 + \frac{356}{9} \zeta_3 
 - \frac{1}{72} \pi^4
 - \frac{187}{24} L - \frac{11}{12} \rho L + \frac{55}{72} \pi^2 L
 - \frac{11}{8} L^2 - \frac{11}{36} L^3
 \nonumber \\
 & &
 + i \pi \left[
            - \frac{1}{4 \varepsilon^3}
            - \frac{3}{4 \varepsilon^2}
            + \left( - \frac{29}{8}
                     - \frac{1}{4} \rho + \frac{\pi^2}{3} \right) \frac{1}{\varepsilon}
            - \frac{139}{8} - \frac{11}{8} \rho 
            + \pi^2 + \frac{15}{2} \zeta_3 
     \right]
 + {\cal O}(\varepsilon),
\end{eqnarray}
where $L = \ln(s_{ijk}/\mu^2)$.


\subsection{Isolation of infrared poles by sector decomposition}\label{sec:num}

As is well known,  for ultraviolet divergences   
a general subtraction scheme to all orders can be 
established \cite{Bogoliubov:1957gp,Hepp:1966eg,Zimmermann:1973te}. 
For infrared poles, i.e. 
soft and collinear poles in Minkowski space, the situation is less settled. 
Of course, general cancellation theorems like the KLN 
theorem \cite{Kinoshita:1962ur,Lee:1964is} exist, 
but a local subtraction scheme acting on a graph by graph basis 
and being valid to all orders is not available. 
 
On the other hand, the use of dimensional regularisation in combination with 
Feynman (or alpha\,\cite{Smirnov:1991jn})  parameters allows in principle 
to isolate the infrared poles as powers in $1/\varepsilon$ for an arbitrary 
graph. The problem is that the corresponding parameter integrals 
get extremely complicated the more loops and scales are involved, 
in particular they exhibit an {\em overlapping} structure. 
A simple example of an overlapping singularity already has been given in eq.\,(\ref{nlo-2}).

To address this problem, 
an automated algorithm\footnote{The method of sector decomposition has been 
used first in \cite{Hepp:1966eg} for overlapping UV divergences, 
and applied in a different context in \cite{Roth:1996pd}.}
presented in \cite{Binoth:2000ps} and sketched below  
has been constructed which disentangles the overlapping regions 
in parameter space
by decomposing them iteratively into subsectors until the divergent 
contributions factorise. 
Arrived at the factorised form, subtractions can be implemented easily. 
The resulting parameter integrals for the pole coefficients are in general too 
complicated to be integrated analytically, but as they are finite, they can 
be integrated numerically. 

\subsubsection{Multi-loop integrals}\label{sec:numvirt}

The straightforward automatisation of the algorithm is one of its virtues, 
and this is in particular true when applied to virtual loop integrals, 
because the latter, after Feynman parametrisation, have a "standard form": 
An $L$-loop graph $G$ in $D$ dimensions with $N$ propagators is, after 
momentum integration,  of the form  
\begin{equation}
G = (-1)^N\Gamma(N-LD/2)\int
\limits_{0}^{\infty} d^Nx\, 
\delta(1-\sum_{l=1}^N x_l)\,
\frac{{\cal U}(\vec x)^{N-(L+1) D/2}}{{\cal F}(\vec x,\{s_{ij},m_i\})^{N-L
D/2}}\;.
\label{G}
\end{equation}
The functions ${\cal U}$ and ${\cal F}$ can be 
straightforwardly derived from the momentum representation, or 
they can be constructed from the topology of the corresponding 
Feynman graph\,\cite{Itzykson:1980rh,Smirnov:1991jn}.  
${\cal U}$ is a polynomial of degree $L$ in the Feynman parameters, 
${\cal F}$ is of degree $L+1$ and also depends on the kinematic 
invariants of the diagram. 
The sector decomposition uses representation (\ref{G}) as a starting point 
and proceeds as follows:
\begin{itemize}
\item The integration domain is split into $N$ parts, using the identity
\begin{equation}
\int_0^{\infty}d^N x =
\sum\limits_{l=1}^{N} \int_0^{\infty}d^N x
\prod\limits_{\stackrel{j=1}{j\ne l}}^{N}\theta(x_l- x_j)\;,
\end{equation} 
such that $G$ becomes a sum over $N$ integrals $G_l$, where in each 
"primary sector" $l$ the variable $x_l$ is the largest one. 
\item  The variables are transformed in each primary sector $l$ as follows:
\begin{eqnarray}
x_j = \left\{ \begin{array}{lll} x_l t_j     &  & j<l \nonumber\\
                                   x_l         &  & j=l \nonumber\\
                                   x_l t_{j-1} & & j>l \end{array}
                                   \right.
\label{trafo}                              
\end{eqnarray}
\item By construction, $x_l$ factorises from ${\cal U}$ and ${\cal F}$. 
We eliminate $x_l$ in each $G_l$  using 
$$\int dx_l/x_l\;\delta(1-x_l(1+\sum_{k=1}^{N-1}t_k ))=1\;.$$
\end{itemize}
By applying the sector decomposition iteratively, 
one finally arrives at a form where all singularities are factorised 
explicitly in terms of factors of Feynman parameters like 
$t_j^{-1-\kappa\varepsilon}$. Subtractions of the form 
\begin{eqnarray*}
\int_0^1 d t_j\,t_j^{-1-\kappa\varepsilon}\,{\cal F}(t_j,t_{i\not=j})=
-\frac{1}{\kappa\varepsilon}\,\,{\cal F}(0,t_{i\not=j})
+\int_0^1 d t_j\,t_j^{-1-\kappa\varepsilon}\,\bigg\{{\cal F}(t_j,t_{i\not=j})-
{\cal F}(0,t_{i\not=j})\bigg\}
\end{eqnarray*}
for each $j$, where $\lim_{t_{j}\to 0}{\cal F}(t_{j},t_{i\neq j})$
is finite by construction, and subsequent expansion in $\varepsilon$ 
leads to a representation of the graph $G$ as a Laurent series in $\varepsilon$:
\begin{equation}
G=\sum\limits_{k=-l}^{2L}\frac{C_k(\vec{x},\{s_{ij},m_i\})}{\varepsilon^k}\;.
\label{Laurent}
\end{equation}
The pole coefficients $C_k(\vec{x},\{s_{ij},m_i\})$ are sums of integrals 
over functions of Feynman parameters. Of course they also 
contain the kinematic invariants $\{s_{ij},m_i\}$ defined by the graph. 
For the numerical integration  of those functions, two cases can be 
distinguished: If the  diagrams depend only on a single scale, 
this scale can be factored out and the pole coefficients are just numbers 
which can be calculated once and for all. 
For diagrams depending on several scales, like for example the Mandelstam 
variables $s$ and $t$ in the case of the massless $L$-loop box, the 
kinematic invariants have to be fixed to certain values at which the diagram 
is evaluated. These have to be in the Euclidean region in order to 
avoid that thresholds spoil the numerical integration. 

As an example for a one-scale problem, we give the result for a five-loop
propagator diagram, shown in Fig.\,\ref{figTL5}.
\begin{figure}[htb]
\begin{picture}(100,30)(0,0)
\put(50,-20){\epsfig{file=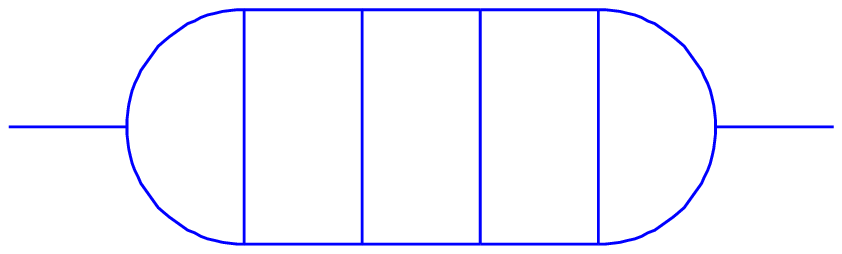,height=2.cm}}
\put(235,0){
$
G[5]=(-s)^{-4-5\varepsilon}\Gamma(4+5\varepsilon)\,\, 40.53
$
}
\end{picture}
\caption{\em A 5-loop propagator graph}\label{figTL5}
\end{figure}
Examples of diagrams depending on four scales are given in 
Fig.\,\ref{bhabha_graphs}, where the two straight lines 
flowing through the graphs denote massive propagators 
which can have different masses. 
\begin{figure}[htb]
\begin{picture}(100,20)(0,20)
\put(2,20){(a)}
\put(5,0){\epsfig{file=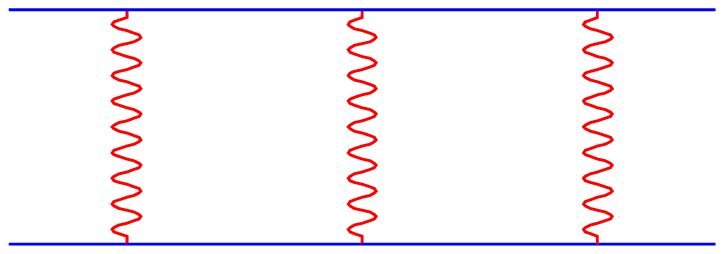,height=1.8cm}}
\put(155,20){(b)}
\put(160,0){\epsfig{file=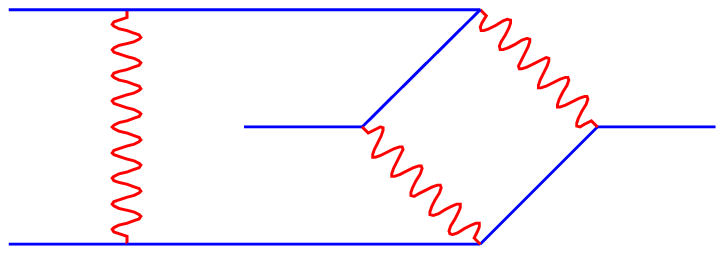,height=1.8cm}}
\put(305,20){(c)}
\put(310,0){\epsfig{file=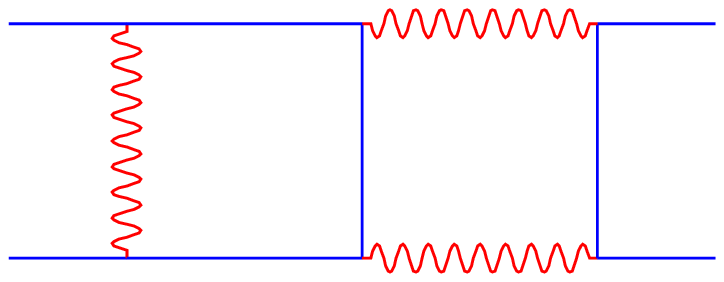,height=1.8cm}}
\end{picture}
\caption{\em The two-loop four point master topologies relevant for Bhabha scattering. The wavy lines
are massless and the straight lines are massive, with external 
legs on-shell.
The topologies from left to right are denoted by $G_a$, $G_b$, $G_c$.}
\label{bhabha_graphs}
\end{figure}
These graphs are the most complicated ones
occurring in the calculation of two-loop Bhabha scattering. 
Only topology $G_a$,  for the case of one single  mass scale,  
has been calculated analytically so far by Smirnov\,\cite{Smirnov:2001cm}.
A calculation of all three graphs only exists 
in the massless approximation \cite{Bern:2000ie}. 
Our numerical results at two different points for 
$G_{a,b,c}(s,t,u,m^2,M^2) = 
\Gamma^2(1+\varepsilon)\sum\limits_{i=0}^2 P_i/\varepsilon^i$
are given in Table \ref{tablebhabha}.
\small
\begin{table}[ht]
\begin{center}
\begin{tabular}{|c||c|c|c||c|c|c|}
\hline
&\multicolumn{3}{|c||}{}&\multicolumn{3}{|c|}{}\\
$(-s,-t,-u,m^2,M^2)$&\multicolumn{3}{|c||}{$(1/5,3/10,7/2,1,1)$}&
\multicolumn{3}{|c|}{$(5/3,4/3,5,1,3)$}\\
&\multicolumn{3}{|c||}{}&\multicolumn{3}{|c|}{}\\
\hline
&$G_a$&$G_b$&$G_c$&$G_a$&$G_b$&$G_c$\\
\hline
&&&&&&\\
$P_2$&-1.561   &-0.5255 &-1.152&  -0.08622& -0.03483 & -0.05832  \\
$P_1$&-5.335   &-0.2024 &-3.690&  -0.04195&  0.07556&   0.05389 \\
$P_0$& 1.421   &3.606   &1.555 &  0.7323& 0.1073 &     0.6847\\
&&&&&&\\
\hline
\end{tabular}
\end{center}
\caption{\em Results for the double box graphs for Bhabha scattering}
\label{tablebhabha}
\end{table}
\normalsize

\subsubsection{Phase space integrals}\label{sec:numreal}

As already explained in detail above, solving the problem of isolating and 
subtracting the infrared poles occurring in NNLO phase space integrals, 
and of integrating over the divergent subtraction terms, is a major 
step towards a (partonic) Monte Carlo program calculating 
processes like for example $e^+e^-\to 2$ or 3 jets at NNLO. 
The method of sector decomposition can also be very useful for this task, 
as it is a general method to isolate (overlapping) poles in parameter 
space. 
Its first application to phase space integrals can be found in 
\cite{Heinrich:2002rc}, further developments of the method will be 
sketched in the following. 

In order to be able to use the automated sector decomposition procedure
for phase space integrals, the phase space has to be cast into a 
"standard form", similar to (\ref{G}) for loop integrals. 
For example, a $1\to 4$ parton phase space is most conveniently written as 
\begin{eqnarray*}
\int d\, PS_4&=&K_{\Gamma}
\,(q^2)^{\frac{3D}{2}-4}
\int\Big\{\prod\limits_{j=1}^6 d x_j\,\Theta(x_j)\Big\}\,\delta(1-\sum\limits_{i=1}^6
x_i)\,
\Big[-\lambda(x_1x_6,x_2x_5,x_3x_4)\Big]^{\frac{D-5}{2}}
\Theta(-\lambda)\;,
\end{eqnarray*}
where the parameters $x_i$ are rescaled Mandelstam invariants, defined by
\begin{displaymath}
x_1=s_{12}/q^2, x_2=s_{13}/q^2, 
x_3=s_{23}/q^2,
x_4=s_{14}/q^2,x_5=s_{24}/q^2,
x_6=s_{34}/q^2
\end{displaymath}
and 
$K_{\Gamma}=(2\pi)^{4-3D}V(D-1)V(D-2)V(D-3)\,2^{1-2D}\;,\;
V(D) =2\pi^{D/2}/\Gamma(D/2).$\\
$\lambda(x_1x_6,x_2x_5,x_3x_4)$ is the K\"allen function,
$\lambda(x,y,z)=x^2+y^2+z^2-2(xy+xz+yz)$. 

In order to calculate the real emission part where up to two particles 
can be unresolved using the method 
of sector decomposition, there are several ways to go, ranging from 
the use of this method only as a check for integrals over subtraction 
terms calculated analytically to an almost completely numerical 
approach relying largely on this method. 
The former has been employed in \cite{Gehrmann-DeRidder:2003bm}, 
where it has been shown that any term appearing in 
a phase space integral of a $1\to 4$ matrix element in massless QCD
can be expressed as a linear combination of only four 
master integrals, where one of them is the 
"trivial" integral over the phase space alone.
In the notation introduced above, the remaining three master integrals are 
\begin{eqnarray}
R_6&=&(q^2)^{-2}\int d\, PS_4\,\frac{1}{(x_2+x_4+x_6)(x_3+x_5+x_6)}\\
R_{8,a}&=&(q^2)^{-4}\int d\, PS_4\,\frac{1}{x_2x_3x_4x_5}\\
R_{8,b}&=&(q^2)^{-4}\int d\, PS_4\,\frac{1}{x_2x_3(x_2+x_4+x_6)(x_3+x_5+x_6)}\;.
\end{eqnarray} 
Their analytical evaluation could be achieved by calculating $R_{8,a}$ 
explicitly and deriving the others from unitarity relations involving 
known results for three-loop two-point functions. 
The numerical results obtained by the sector decomposition algorithm, 
\begin{eqnarray*}
R_6&=&S_{\Gamma}(q^2)^{-2\varepsilon}\left[0.64498+7.0423\varepsilon+40.507\varepsilon^2
+{\cal O}(\varepsilon^3)\right]\\
R_{8,a}&=&S_{\Gamma}(q^2)^{-2-2\varepsilon}\left[\frac{5.0003}{\varepsilon^4}-
\frac{0.0013}{\varepsilon^3}-\frac{65.832}{\varepsilon^2}-
\frac{151.53}{\varepsilon}+ 37.552+{\cal O}(\varepsilon)\right]\\
R_{8,b}&=&S_{\Gamma}(q^2)^{-2-2\varepsilon}\left[\frac{0.74986}{\varepsilon^4}-\frac{0.00009}{\varepsilon^3}-
\frac{14.001}{\varepsilon^2}-\frac{52.911}{\varepsilon}-99.031 +{\cal
O}(\varepsilon)\right]\\
S_\Gamma &=& \frac{(4\pi)^{3\varepsilon}}{2^{11}\pi^5}
\, \frac{(q^2)^{-\varepsilon}}{\Gamma(1-\varepsilon)\Gamma(2-2\varepsilon)}
\;,
\end{eqnarray*}
agree with the analytical results within a numerical precision better than 
1\%.

The  algorithm can also be employed 
to avoid complicated analytical integrations over subtraction terms 
completely. 
The singularities can be extracted as outlined above, and the pole 
coefficients can be calculated numerically to a high precision to 
check their cancellation against the double virtual respectively 
one-loop virtual plus single-real-emission counterparts. 
The remaining functions are finite, and  the combination with an arbitrary 
(infrared safe) measurement function is straightforward as 
it does not hamper the numerical integration. 
In this way, fully differential Monte Carlo programs 
for $1\to n$ particle/jet processes can be constructed. 
A first step in this direction already has been undertaken in 
\cite{Anastasiou:2003gr}, where the contribution proportional to $N_f$ of 
$e^+e^-\to 2,3$ or 4 jets has been calculated using sector decomposition
techniques. 

Certainly, the automated sector decomposition algorithm applied to 
phase space integrals can also be useful 
in cases where massive particles are involved. 
As has been already proven by explicit examples in the case of loop 
integrals\,\cite{Binoth:2003ak}, masses do not present a principle 
problem for the method, 
but of course care has to be taken that thresholds do not 
destroy the numerical stability.

\subsection*{Acknowledgements}

We would like to thank the conference organisers for a stimulating and productive workshop.






\providecommand{\href}[2]{#2}\begingroup\raggedright\endgroup

\end{document}